%% file: main.tex
\documentclass[]{aa} 

\usepackage{graphicx}
\usepackage{txfonts}
\usepackage{multirow}
\usepackage[final]{hyperref}
\hypersetup{colorlinks=true, linkcolor=red,citecolor=blue,urlcolor=magenta} 

\usepackage{comment}
\usepackage{longtable}
\usepackage{rotating}
\usepackage{placeins}
\usepackage{xcolor}
\usepackage{booktabs}
\usepackage{pdflscape}
\usepackage{float}
\usepackage{tikz}

\newcommand{\mmc}{CH$_3$SH\xspace}
\newcommand{\met}{CH$_3$OH\xspace}
\newcommand{\et}{C$_2$H$_5$OH\xspace}
\newcommand{\etc}{C$_2$H$_5$CN\xspace}
\newcommand{\vc}{C$_2$H$_3$CN\xspace}
\newcommand{\dme}{CH$_3$OCH$_3$\xspace}
\newcommand{\mf}{CH$_3$OCHO\xspace}
\newcommand{\mic}{CH$_3$NCO\xspace}
\newcommand{\ad}{CH$_3$CHO\xspace}
\newcommand{\fmm}{NH$_2$CHO\xspace}
\newcommand{\2}{$_2$}
\newcommand{\3}{$_3$}
\newcommand{\5}{$_5$}

\newcommand{\scm}{cm$^{-2}$\xspace}
\newcommand{\kms}{\,km\,s$^{-1}$\xspace}

\begin{document}

   \title{Shocking Sgr\,B2\,(N1) with its own outflow:}
   \subtitle{A new perspective on segregation between O- and N-bearing molecules}

   \author{Laura~A.~Busch
          \inst{1}\thanks{Member of the \href{https://blog.mpifr-bonn.mpg.de/imprs/}{International Max\,Planck Research School\,(IMPRS) for Astronomy \& Astrophysics} at the Universities of Bonn and Cologne.}
          \and
          Arnaud~Belloche\inst{1}
          \and
          Robin~T.~Garrod\inst{2}
          \and
          Holger~S.~P.~M\"uller\inst{3}
          \and
          Karl~M.~Menten\inst{1}
          }

   \institute{Max-Planck-Institut f\"ur Radioastronomie, Auf dem H\"ugel 69, 53121 Bonn, Germany\\
              \email{labusch@mpifr-bonn.mpg.de}
              \and 
              Departments of Chemistry and Astronomy, University of Virginia, Charlottesville, VA 22904, USA
              \and
              I. Physikalisches Institut, Universit\"at zu K\"oln, Z\"ulpicher Str. 77, 50937 K\"oln, Germany
             }

   \date{Received ; accepted }

 
  \abstract  
  {}
   {Because studies on complex organic molecules (COMs) in high-mass protostellar outflows are sparse, we want to investigate how a powerful outflow, such as that driven by the exciting source of the prominent hot core Sagittarius\,B2\,(N1), influences the gas molecular inventory of the surrounding medium with which it interacts. Identifying chemical differences to the hot core unaffected by the outflow and what causes them may help to better understand molecular segregation in other star-forming regions.} 
   {We made use of the data taken as part of the 3\,mm imaging spectral-line survey ReMoCA (Re-exploring Molecular Complexity with ALMA). We studied the morphology of the emission regions of simple and complex molecules in Sgr\,B2\,(N1). For a selection of twelve COMs and four simpler species, spectra were modelled under the assumption of local thermodynamic equilibrium and population diagrams were derived at two positions, one in each lobe of the outflow. From this analysis, we obtained rotational temperatures and column densities. Abundances were subsequently compared to predictions of astrochemical models and to observations of L1157-B1, a position located in the well-studied outflow of the low-mass protostar L1157, and the source G$+$0.693$-$0.027 (G0.693), located in the Sgr\,B2 molecular cloud complex, which are 
   other regions whose chemistry has been impacted by shocks.}
   {Integrated intensity maps of SO and SiO emission reveal a bipolar structure with blue-shifted emission dominantly extending to the southeast from the centre of the hot core and red-shifted emission to the northwest. The morphology of both lobes is complex but can roughly be characterised by an emission component at a larger opening angle that contains the bulk of emission and narrower features. The wider-angle component is also prominently observed in emission of S-bearing molecules and species that only contain N as a heavy element, 
   including COMs, but also \met, \ad, HNCO, and \fmm. Rotational temperatures are found in the range of $\sim$100--200\,K. Abundances of N-bearing molecules with respect to CH\3OH are enhanced in the outflow component compared to N1S, a position that is not impacted by the outflow. A comparison of molecular abundances with G$+$0.693$-$0.027 and L1157-B1 does not show any correlations, suggesting that a shock produced by the outflow impacts Sgr\,B2\,(N1)'s material differently or that the initial conditions were different.}
   {The short distance of the analysed outflow positions to the centre of Sgr\,B2\,(N1) lead us to propose a scenario in which a phase of hot-core chemistry (i.e. thermal desorption of ice species and high-temperature gas-phase chemistry) preceded a shock wave. The subsequent compression and further heating of the material resulted in the accelerated destruction of (mainly O-bearing) molecules. Gas-phase formation of cyanides seems to be able to compete with their destruction in the post-shock gas. 
   The abundances of cyanopolyynes are enhanced in the outflow component pointing to (additional) gas-phase formation, possibly incorporating atomic N sourced from ammonia in the post-shock gas. To confirm such a scenario, chemical shock models need to be run that take into account the pre- and post-shock conditions of Sgr\,B2\,(N1). In any case, the results provide new perspectives on shock chemistry and the importance of the environment in which it occurs. }

    \authorrunning{L.A.~Busch et al.}
    \titlerunning{COM chemistry in the outflow of Sgr\,B2\,(N1)}
   \maketitle
 
\section{Introduction}

At the center of astrochemistry is the study of the formation and destruction pathways of molecules, whose knowledge is mandatory for an understanding how the chemistry of the interstellar medium evolves along with the star-formation process. Out of the variety of species that have been detected in the interstellar medium so far \citep[see, e.g.][]{McGuire22}, complex organic molecules \citep[COMs, carbon-bearing molecules of six or more atoms,][]{Herbst09} are of particular interest as they present the building blocks of more complex species from which life, as we know it from Earth, may have emerged. 
By now, COMs have been detected in the solid and gas phase towards a wide variety of sources that cover all the  stages of star formation (see \citet{Jorgensen20} for a recent review): in cold dark clouds \citep[e.g.][]{Taquet17,Agundez21,Zeng18}, prestellar cores \citep[e.g.][]{Bacmann12,Jimenez-Serra16}, protostellar environments \citep[e.g.][]{Belloche13,Jorgensen16,Pagani17}, protoplanetary disks \citep[e.g.][]{Walsh16,vdM21,Brunken22}, and small bodies in the Solar System \citep[][]{Altwegg19, Naraoka23}. 

COMs can be formed in the solid phase on the surface or in the ice mantle of dust grains, and in the gas phase. Gas-phase reactions are efficient in producing COMs mainly at high temperatures ($\gtrsim$100\,K) although some reactions may produce some COMs also at low temperatures \citep[][]{Balucani15}.
A substantial number of COMs is thought to form on dust grains and, subsequently, desorb thermally at a certain temperature or non-thermally.
Depending on the species, the production in the solid phase can start as early as during the prestellar phase, that is at extremely low temperatures during the collapse phase before the onset of protostellar heating, or later on when the protostar gradually heats its environment \citep[e.g.][]{Garrod22}. 
Detectable amounts of COMs in the gas phase in low-temperature environments such as prestellar cores are likely the consequence of mainly non-thermal desorption processes that release the molecules from the dust 
grain surfaces. These processes include interactions with cosmic rays or with the secondary ultraviolet photons that these cosmic rays produce upon interaction with the gas, or various chemical processes \citep[e.g.][]{Ruaud15,Shingledecker18,Jin20,Paulive21}. 

COMs may also desorb as a consequence of grain processing by the passage of a shock. For example, shock-chemistry  likely plays a substantial part in enriching the gas of clouds in the Galactic central molecular zone (CMZ) that are devoid of star formation with COMs and simpler species
\citep[e.g.][]{Requena-Torres06,Requena-Torres08}. In addition, enhanced cosmic-ray fluxes in the CMZ also impact the chemistry of these clouds \citep[e.g.][]{Indriolo2015}.
One of these CMZ clouds that has been the target for many recent follow-up studies on COMs is G$+$0.693$-$0.027 (G0.693 hereafter), which is located in the cloud complex Sagittarius\,B2 (Sgr\,B2 hereafter) and has no signs of active star formation. 
Over the past few years, the detection of many molecules, including COMs, several of them being even new interstellar detections, have made G0.693 one of the chemically richest sources in the Galaxy \citep[e.g.][]{Rivilla2021a,Rivilla2021,Rivilla22,Rivilla22PO,Rivilla2023,jimenez-serra2022,Zeng23}.  
Shocks may also be provoked by protostellar outflows that travel through the material of the parental cloud of their driving source. For example, this has been associated with the the detection of various COMs at positions that are exposed to the outflow of the low-mass protostar L1157 \citep[e.g.][]{Arce08}.

Thermal desorption is naturally expected to account for the high COM abundances observed in hot corinos and hot cores that surround low- and high-mass protostars, respectively. 
It was proposed that a COM can thermally desorb either alongside water at the desorption temperature of water, that is $\gtrsim$100\,K, or depending on the COM's individual binding energy, with which it sticks to the dust grain surface. 
Which of the processes occurs in three-phase astrochemical models (which distinguish the surface and bulk layers of the ice mantles) depends mainly on a molecule's ability to diffuse in the bulk ice and from the bulk ice to the surface of the grain \citep[e.g.][]{Garrod13} or its disability to do so \citep[][]{Garrod22}. Accordingly, if the COM reaches the outermost layer of the grain surface, it will desorb at its characteristic desorption temperature.

\begin{figure}
    \centering
    \includegraphics[width=0.5\textwidth]{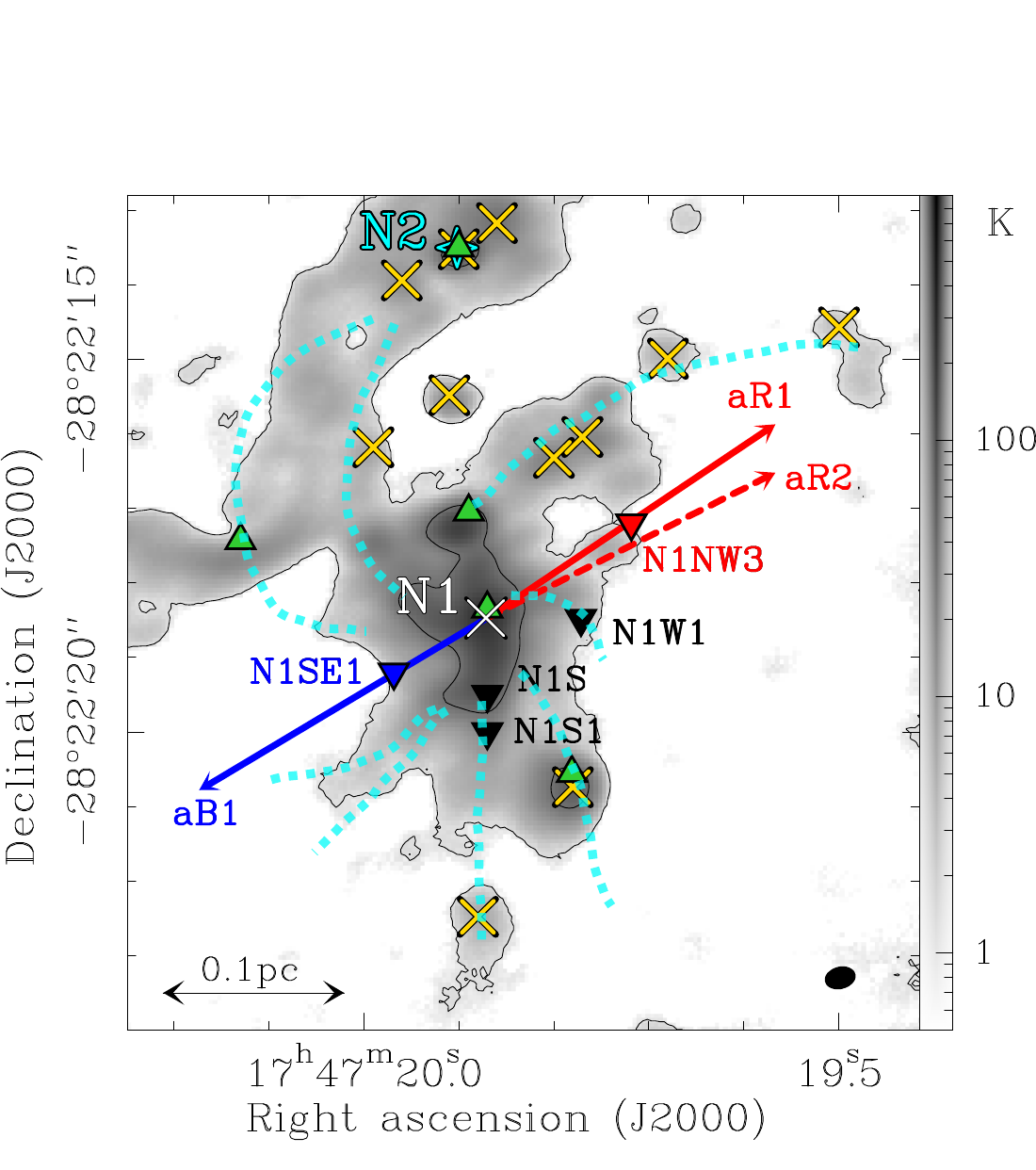}
    \caption{Continuum map at 99\,GHz extracted from the ReMoCA survey data of the central region of Sgr\,B2\,(N) in grey scale with contour steps 3$\sigma$ and $81\sigma$, where $\sigma=0.4$\,K. The centre of the main hot core N1 is marked in white and was determined based on the ReMoCA data in \citet{Busch22}. The position of the secondary hot core N2 is shown with a cyan tetragon reported by \citet{Bonfand17} on the basis of the EMoCA survey. Emission peaks of H{\small II} regions \citep{Gaume95,DePree15} are marked with green triangles, continuum sources identified by \citet[][]{Sanchez-Monge17} based on their 1\,mm ALMA data with yellow crosses. Filaments identified by \citet{Schwoerer19} based on the same 1\,mm (continuum and spectral-line) data are roughly indicated in cyan. The blue and red arrows labelled aB1, aR1, and aR2 indicate the outflow axes identified in this work based on SO emission (see Sect.\,\ref{ss:sio-so}). 
    Positions N1SE1 and N1NW3, which are located along these axes, were analysed in this work; positions N1S, N1S1, and N1W1, which are not associated with the outflow, were studied in \citet[][]{Busch22}, but have been revisited in this work. The HPBW is shown in the bottom-right corner. The map is not corrected for primary beam attenuation.}
    \label{fig:overview}
\end{figure}

In our previous study \citep[][Paper\,I hereafter]{Busch22}, we addressed exactly this question. We used the data of the ReMoCA survey \citep[Re-exploring Molecular Complexity with ALMA,][]{Belloche19} that were obtained with the Atacama Large Millimetre/submillimetre Array (ALMA) towards the massive star-forming region Sgr\,B2\,(N)orth, which is located in the Galactic centre region at a distance of 8.2\,kpc from the sun \citep{Reid19}. Figure\,\ref{fig:overview} provides an overview of observed sources and features in the region surrounding the two main hot cores N1 and N2.
Thanks to the subarcsecond angular resolution of ReMoCA, we resolved the COM emission in the main hot core  Sgr\,B2\,(N1) and derived abundance profiles of various COMs towards the south and west directions starting from $\sim$0.5\arcsec\,\,from the hot-core centre and going up to a distance of $d=5$\arcsec\,\,and 4\arcsec, respectively. This analysis included positions N1S ($d=1$\arcsec), N1S1 ($d=1.5$\arcsec), and N1W1 ($d=1.3$\arcsec), which are marked in Fig.\,\ref{fig:overview}. 
A steep increase in the observed abundance profiles of most COMs over one or more orders of magnitude at around $\sim$100\,K ($d\sim 3-4$\arcsec) suggests that the bulk of these molecules desorbs thermally at this temperature. A comparison with most recent chemical models performed by \citet[][]{Garrod22} revealed that the similarity in the COMs' desorption temperatures provides observational evidence for the thermal co-desorption process of COMs and water. 
Moreover, we discussed that the derived non-zero abundance values at temperatures $<$100\,K might result from non-thermal desorption processes or another thermal desorption process, for example as a consequence of reduced binding energies of COMs in water-poor outer ice layers that would be rich in CO at low temperatures. The observed abundance profiles of some COMs suggest that gas-phase formation purely or partly accounts for an increase in these COMs' abundances in the gas phase at temperatures above 100\,K. 

In this earlier work we focussed on the thermal desorption process. Therefore, to reduce the chance of contributions to COM abundances by non-thermal (shock-induced) desorption, we intentionally avoided to analyse positions that are associated with the outflow powered by Sgr\,B2\,(N1) \citep[see Fig.\,\ref{fig:overview} and][]{Higuchi15,Schwoerer20}. The goal of the present work is to study the chemistry in the outflow (or in regions impacted by the outflow) and to investigate its role in the formation, destruction, and desorption behaviour of mainly complex but also simpler molecules. 
In particular, we want to find out if the gas molecular composition in the outflow shows significant differences to that derived in Paper\,I. This includes a comparison of abundances between `outflow' positions N1SE1 and N1NW3 (see Fig.\,\ref{fig:overview} and Sect.\,\ref{ss:selection}) and `hot-core' positions N1S, N1S1, and N1W1 (Paper\,I).

An outflow driven by Sgr\,B2\,(N1) was long proposed to account for red- and blue-shifted line emission observed in spectra of simpler molecules such as SO, SO\2, SiO, HC\3N, but also, for example \etc or \vc \citep[][]{Lis93,Liu99,Belloche13}. A large number of spots showing intense maser emission in the 22\,GHz water line have been observed in the close vicinity of the hot core \citep[][]{McGrath04}. These are signposts of high-mass star formation, in particular, shock-related events such as protostellar outflows as these masers are the results of shock chemistry and are collisionally pumped \citep{Elitzur1989}. \citet[][]{Higuchi15} used the data of the EMoCA survey \citep[Exploring Molecular Complexity with ALMA,][]{Belloche16}, ReMoCA's predecessor, which has an angular resolution of $\sim$2\arcsec, to map SO\2 and SiO emission. These maps revealed a bipolar outflow, with a dynamical age of of $\sim$5\,kyr and a total mass of 2000\,$M_\odot$.
\citet[][]{Schwoerer20} studied the outflow of Sgr\,B2\,(N1) in emission of mainly simple molecules such as SiO, SO, SO\2, HNCO, and others, but also \met based on ALMA observations at 1.3\,mm with an angular resolution of $\sim$0.5\arcsec\, and $\sim$0.05\arcsec. 
The observed morphology of the outflow emission suggests that it may be interacting with material associated with the H{\small II} region located in the northeast and that it may be framed by some of the identified filaments \citep[see also][]{Schwoerer19}, provided that all sources lie at the same distance along the line of sight. \citet{Schwoerer20} also derived an outflow mass of 230\,$M_\odot$, which is a factor 10 lower than the value from \citet[][]{Higuchi15}, due to a different way of derivation, a total kinetic energy output of $\sim$10$^{48}$\,erg, a dynamical age of 3--7\,kyr, and a mass ejection rate of $\sim$0.05\,$M_\odot$\,yr$^{-1}$. Assuming a kinetic temperature of 250\,K and a core radius of 0.03\,pc, the author estimated a total luminosity of $\sim6\times10^6\,L_\odot$ for Sgr\,B2\,(N1), a total gas-and-dust mass of 2000\,$M_\odot$, and a stellar mass content of 800--3000\,$M_\odot$, based on which he proposed that multiple sources at the centre of Sgr\,B2\,(N1) drive outflows that appear to be one. In this case, the intriguing rather clear separation of blue-shifted emission in the southeastern lobe and red-shifted emission in the northwestern lobe would be largely fortuitous. Independently of the number of driving sources, the outflow of Sgr\,B2\,(N1) is one of the most massive and 
powerful protostellar outflows known to date.

In this work we want to study the impact of the outflow on the COM inventory in Sgr\,B2\,(N1) and how this compares to other sources in which the molecular content is influenced by shocks, which are L1157-B1, a region located in the blue-shifted lobe of the outflow driven by the low-mass protostar L1157, and G0.693. The comparison to the latter is of particular interest as it is located within the same cloud complex as Sgr\,B2\,(N1), exposed to likely similar physical processes as the positions in Sgr\,B2\,(N1) that are impacted by the outflow, however, G0.693 has a lower density. 
The article is structured as follows: Section\,\ref{s:obs} provides details on the observations and the data analysis including the LTE modelling of spectra and the derivation of population diagrams. In Sect.\,\ref{s:results} we present our results, which are discussed and compared with observational results of the other shock-dominated regions and with astrochemical models in Sect.\,\ref{s:discussion}. The conclusions are provided in Sect.\,\ref{s:conclusion}.

\section{Observations and method of analysis}\label{s:obs}
\subsection{The ReMoCA survey}
We made use of data that were obtained as part of the imaging spectral-line survey ReMoCA \citep[][]{Belloche19} towards Sgr\,B2\,(N) with ALMA. The phase centre at $(\alpha,\delta)_\mathrm{J2000}=(17^\mathrm{h}47^\mathrm{m}19\overset{s}{.}87,-28^\circ22^\prime16\overset{\arcsec}{.}00)$ is located north of Sgr\,B2\,(N1), halfway to the secondary hot core Sgr\,B2\,(N2).  Five observational setups (S1--S5) were observed, each delivering data in four spectral windows, covering the frequency range from 84 to 114\,GHz in total. Different antenna configurations yielded angular resolutions that vary from $\sim$0.75\arcsec\,in setup 1 to $\sim$0.3\arcsec\,in setup 5.
Further details on the observations as well as average rms noise levels for each spectral window can be found in Table\,2 of \citet[][]{Belloche19}. 
Details on the data reduction can also be found in the latter article. 
The size (HPBW) of the primary beam varies from 69\arcsec\,at 84\,GHz to 51\arcsec\,at 114\,GHz \citep[][]{ALMAc4} and the spectral resolution of the reduced spectra is 488\,kHz, which translates to 1.7--1.3\kms.

\subsection{LTE modelling with Weeds}\label{ss:weeds}

In order to identify molecules in the observed spectra and to determine the properties of their emission, we performed radiative transfer modelling with Weeds \citep[][]{Maret11}. Weeds is an extension of the GILDAS/CLASS software\footnote{\url{https://www.iram.fr/IRAMFR/GILDAS/}} and is used to produce synthetic spectra under the assumption of local thermodynamic equilibrium (LTE). Assuming LTE conditions in Sgr\,B2\,(N) is appropriate given the high volume densities of $\sim10^7$\,cm$^{-3}$ that have been derived towards the source \citep{Bonfand17}.

The modelling procedure is performed in the same way as in Paper\,I. Weeds requires five input parameters for each molecule: total column density, rotational temperature, size of the emission region, velocity offset with respect to the systemic velocity of the source, and linewidth (FWHM).
The last two parameters have been derived by applying one-dimensional Gaussian fits to optically thin and unblended transitions of a molecule.
Column density and rotational temperature were first selected by eye and adjusted subsequently, based on results that were obtained from a population diagram analysis (see Sect.\,\ref{ss:PD}). 
Following Paper\,I, the size of the emission region was fixed to a value of 2\arcsec, based on the assumption of resolved emission. 
The identification of a molecular species is validated when the synthetic spectrum correctly predicts each observed transition. By adding up the synthetic spectra of all individual species, we derive a combined model of molecules. More details on the modelling procedure can be found in \citet{Belloche16}. All Weeds parameters determined for each molecule at each selected position (see Sect.\,\ref{ss:selection}) are summarised in Tables\,\ref{tab:n1se1hc}--\ref{tab:n1nw3of}.
The modelling of synthetic spectra relies on spectroscopic information, which, for most parts, are taken from the CDMS \citep[Cologne Database for Molecular Spectroscopy,][]{CDMS} or the JPL \citep[Jet Propulsion Laboratory,][]{JPL} spectroscopy database. For some COMs, we provided an extended description on the laboratory background and on the vibrational spectroscopy in Paper\,I. 

\subsection{Population diagrams}\label{ss:PD}

To support the results of the LTE modelling, we derived population diagrams that yield rotational temperature, $T_\mathrm{rot}$, and column density, $N_\mathrm{col}$, of a molecule. This analysis is based on the following formalism \citep[][]{Mangum15}: 
\begin{align}
    \ln\left(\frac{N_u}{g_u}\right) = \ln\left(\frac{8\,\pi\,k_B\,\nu^2 \int J(T_B) \mathrm{d}\varv}{c^3\,h\,A_\mathrm{ul}\,g_u\,B}\right) = \ln\left(\frac{N_\mathrm{tot}}{Q(T_\mathrm{rot})}\right) - \frac{E_u}{k_B T_\mathrm{rot}},
\end{align}
where $N_u$ is the upper-level column density, $g_u$ the upper-level degeneracy, $E_u$ the upper-level energy, $k_B$ the Boltzmann constant, $c$ the speed of light, $h$ the Planck constant,  $B=\frac{\mathrm{source\,size}^2}{\mathrm{source\,size}^2\,+\,\mathrm{beam\,size}^2}$ the beam filling factor, $A_\mathrm{ul}$ the Einstein A coefficient, $N_\mathrm{tot}$ the total column density, and $Q$ the partition function. Integrated intensities in brightness temperature scale, $J(T_B)$, are obtained over a 
visually selected velocity range, d$\varv$, in the baseline-subtracted spectra.

Following the strategy of Paper\,I, we only use setups 4--5 because of their higher angular resolutions compared to setups 1--3. Only in the case of \fmm, setups 1--3 were used because there are not enough lines in the higher angular-resolution setups to construct a population diagram for this COM. Population diagrams for relevant species and positions (see Sect.\,\ref{ss:selection}) can either be found in Paper\,I or in Appendix\,\ref{app:popdiagrams}. Values for column density and rotational temperature can be found in Tables\,\ref{tab:n1se1hc}--\ref{tab:n1nw3of}. 
For each molecule and position, there exist two diagrams: the left panel that shows the original integrated intensities and the right panel, in which two correction factors (discussed below) have been applied to the data points. 
If there is contaminating emission from other species or from other velocity components of the same species, due to line blending,  
its contribution is subtracted from the value of integrated intensity, where we used the information of the complete Weeds model that was created based on the individual spectra of the molecules analysed here.

Moreover, some spectral lines may be affected by high optical depth. To account for this, the integrated intensities of both the observed and modelled transitions were multiplied with a correction factor, $\frac{\tau}{1-\mathrm{e}^{-\tau}}$ \citep[e.g.][]{Mangum15}. The opacity values were taken from our Weeds model for the respective transition. However, the Weeds model has only limited capabilities in treating high optical depths leading to an underestimation of the correction factor at very high values. Therefore, we did not consider a transition in the population diagram when the opacity exceeded a value of $\sim$2--3.
After applying these two corrections, some small scatter between the observed and modelled data points and amongst the observed points remains, which can have multiple reasons that we elaborated on in Sect.\,3.4 of Paper\,I.

In each population diagram, the data points follow a linear trend implying that the level distributions can be explained by a single temperature.
The error bars shown in the population diagrams only include the standard deviation coming from integrated intensities and a quadratically added additional contribution of 1$\sigma$, where $\sigma$ is the median noise level measured in channel maps of the continuum-removed data cubes taken from Table\,2 in \citet{Belloche19}, to account for the uncertainty in the continuum level (Paper\,I). 
We applied a linear fit to the observed data points to obtain the rotational temperature and column density. To avoid giving too much weight to the most intense or contaminated lines, the fit does not take into account the uncertainties of the data points. 

In some cases, a molecule may be detected but the number of available transitions is insufficient to derive a population diagram. By using a $3\sigma$ upper limit for the intensity of non-detected lines, 
we derive upper limits for the entries in the population diagrams to, in turn, obtain an upper limit on the temperature and an estimate of the column density. In two cases, the linear fit did not provide a reliable result, and so we fixed the temperature in the population diagram to obtain a column density value.

\begin{figure*}[htp]
    \includegraphics[width=.5\textwidth]{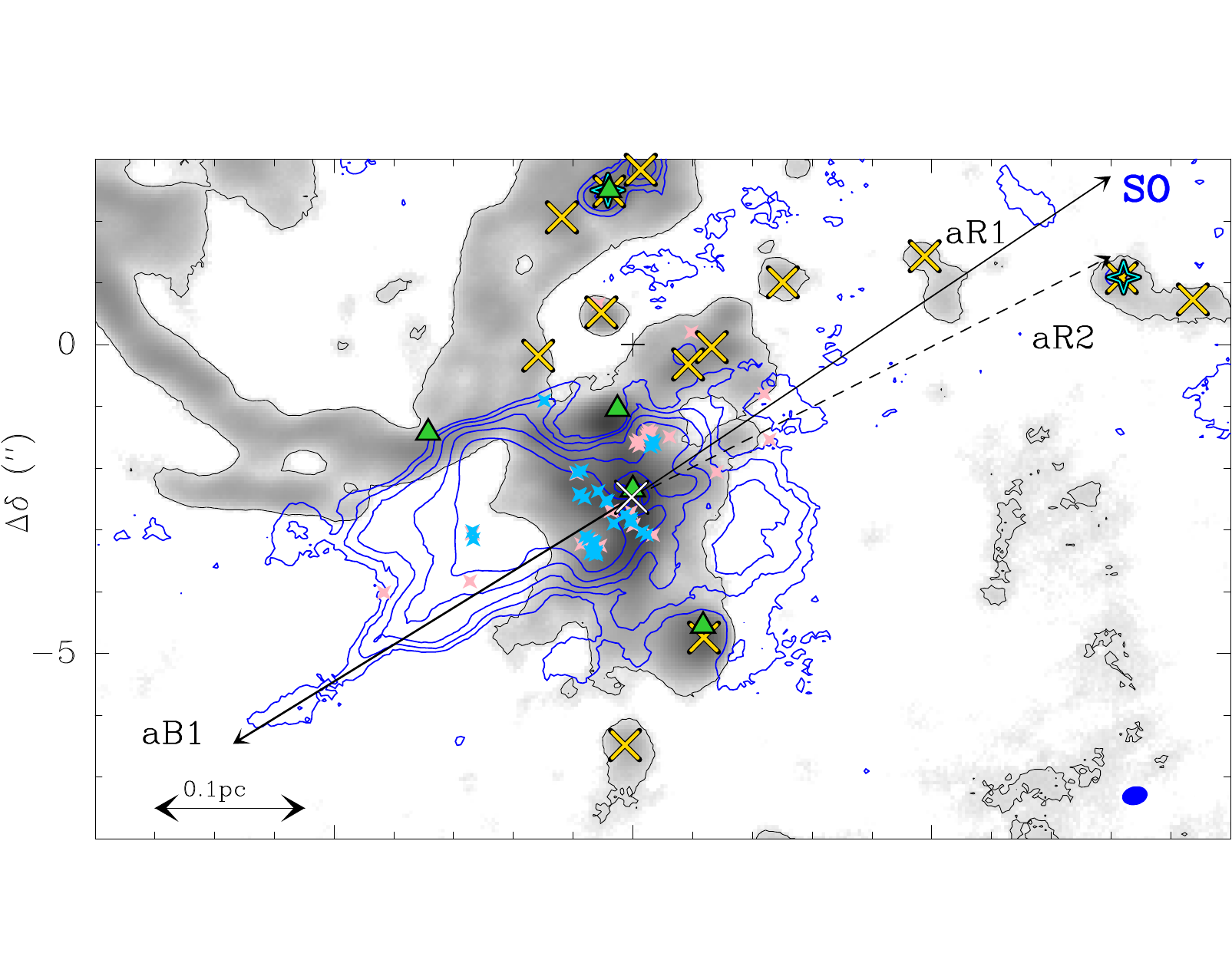}\hspace{0.2cm}
    \includegraphics[width=.5\textwidth]{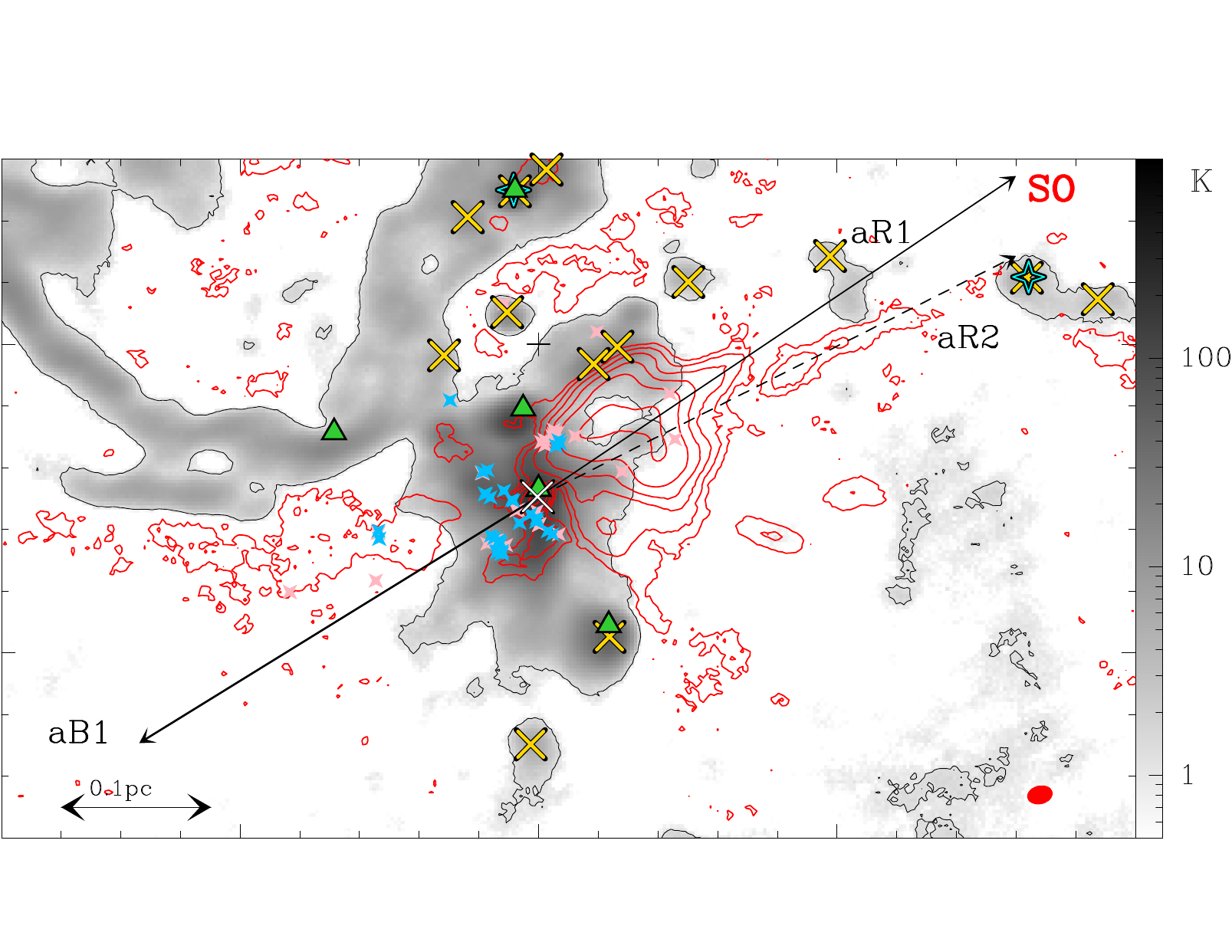}\\[-1.8cm]
    \includegraphics[width=.511\textwidth]{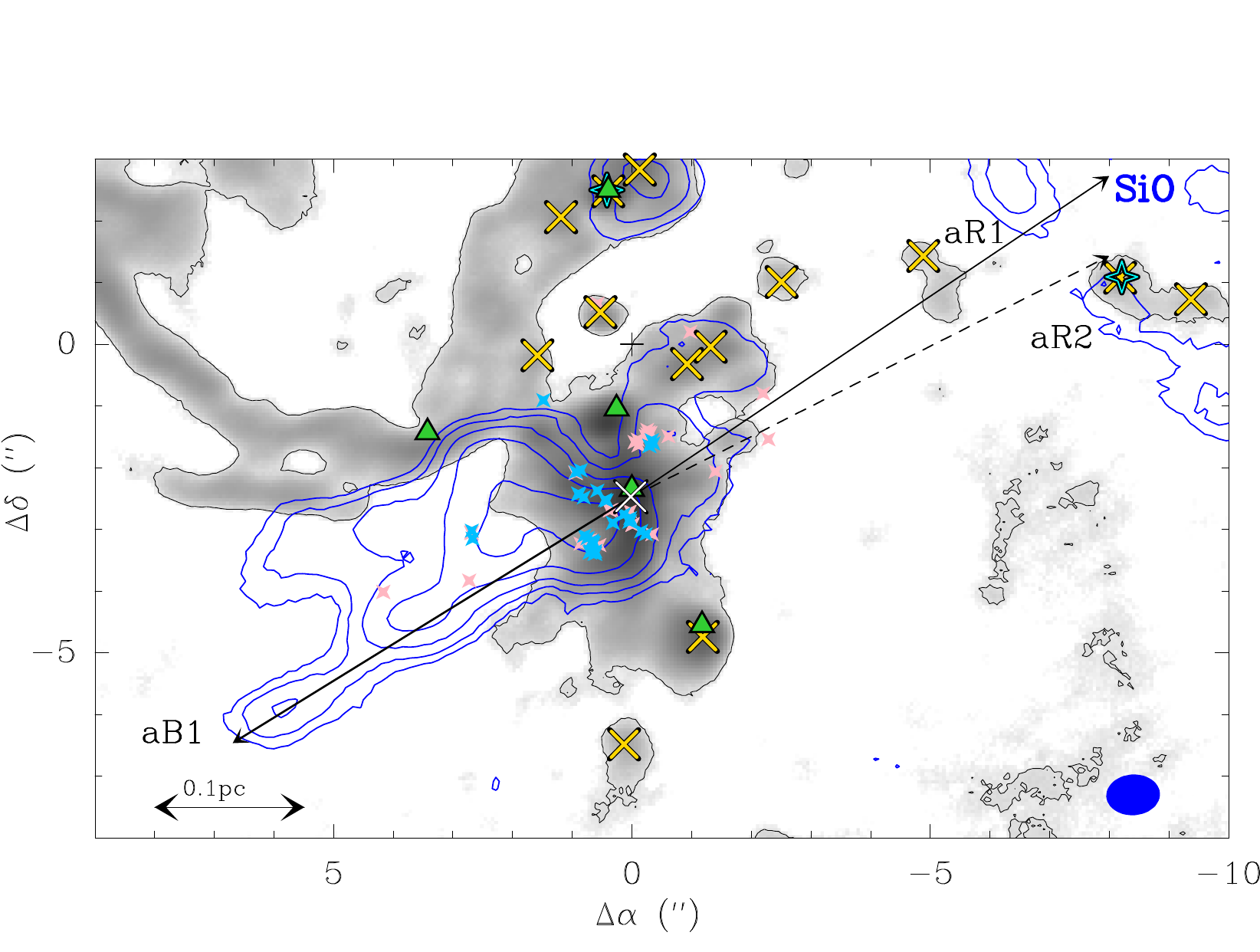}\hspace{0.01cm}
    \includegraphics[width=.5\textwidth]{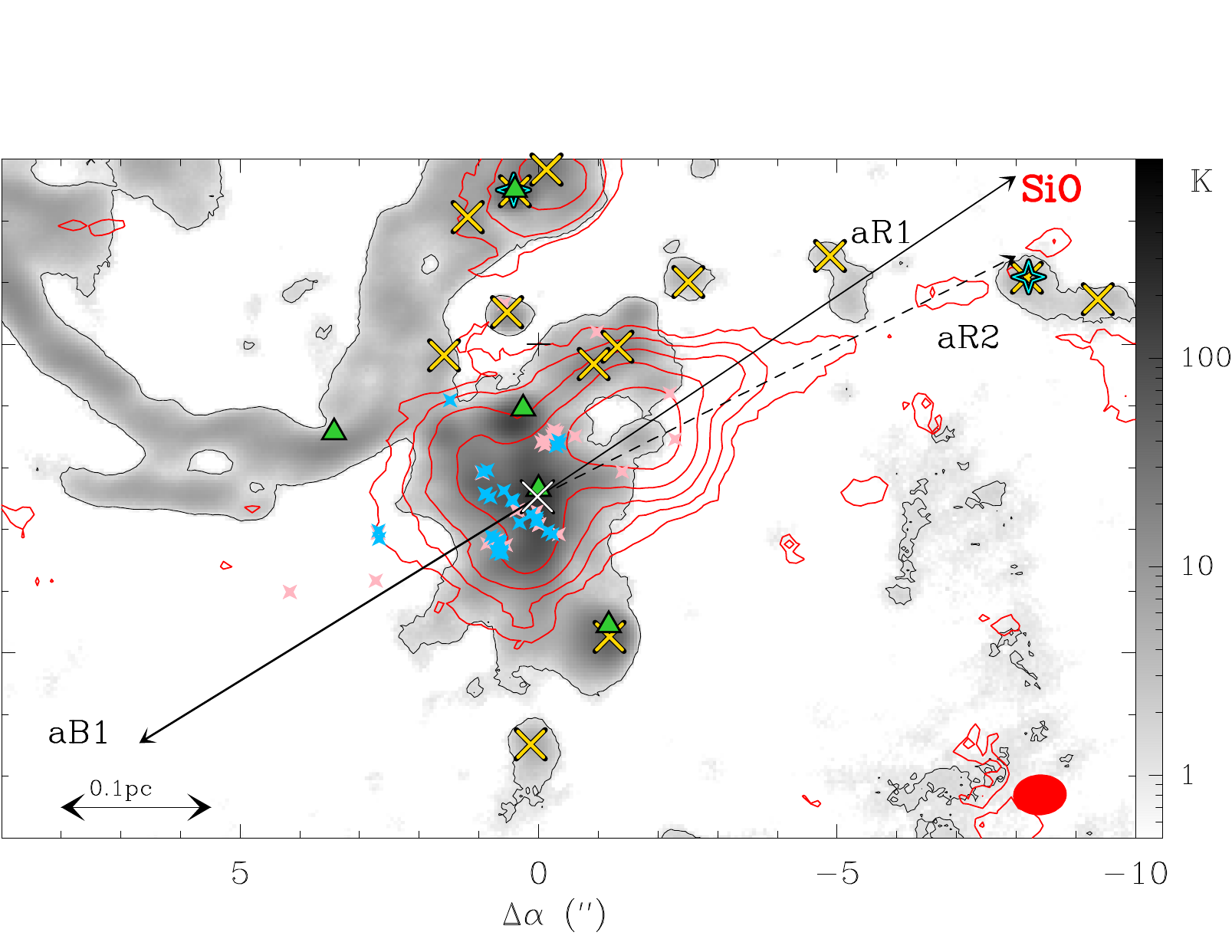}
    \caption{Continuum emission at 99\,GHz in grey scale overlaid by contours of SO $J=2_3-1_2$ at 99.3\,GHz (\textit{top}) and SiO $J=2-1$ at 86.85\,GHz (\textit{bottom}) integrated intensities. The inner integration limits vary such that channels that contain deep absorption are excluded. The outer integration limits are fixed at 15 and 115\kms for blue- and red-shifted SO emission, respectively, and 5 and 130\kms for blue- and red-shifted SiO emission, respectively (see also Fig.\,\ref{fig:spec_so+sio}).
    The blue and red contours start at 5$\sigma$ and then increase by a factor of 2, where $\sigma=10.5$ (SO, blue), 10.8 (SO, red), 4.5 (SiO, blue), and 5.8\,K\kms (SiO, red) and corresponds to an average noise level that was measured in the respective map. The black contour indicates the 3$\sigma$ level of the continuum emission (see Fig.\,\ref{fig:overview}).
    Based on the SO maps, we identify collimated features possibly tracing outflow axes that are shown as solid and dashed black arrows and labelled aB1, aR1, and aR2. The markers are the same as in Fig.\,\ref{fig:overview}. In addition to N2, the hot core N3 identified by \citet{Bonfand17} is marked with a cyan tetragon. 
    Blue and pink star markers indicate H\2O maser spots \citep[][]{McGrath04} with blue- and red-shifted velocities, respectively, with respect to $\varv_{\rm sys}\equiv62$\kms. The HPBW is shown in the bottom-right corner of each panel. The position offsets are given with respect to the ReMoCA phase centre (black cross). The maps are not corrected for primary beam attenuation.}
    \label{fig:so}
\end{figure*}
\begin{figure}[htp]
    \centering
    \includegraphics[width=0.49\textwidth]{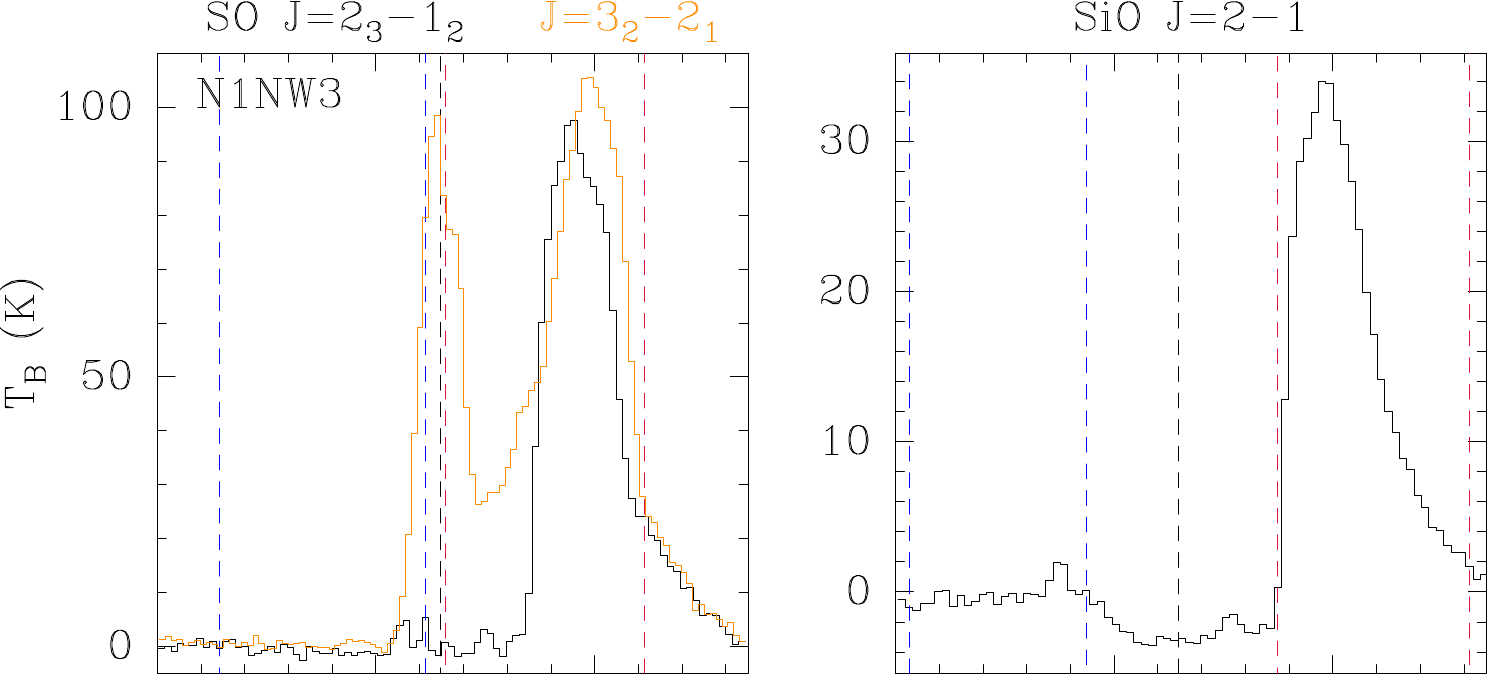}\\
    \includegraphics[width=0.49\textwidth]{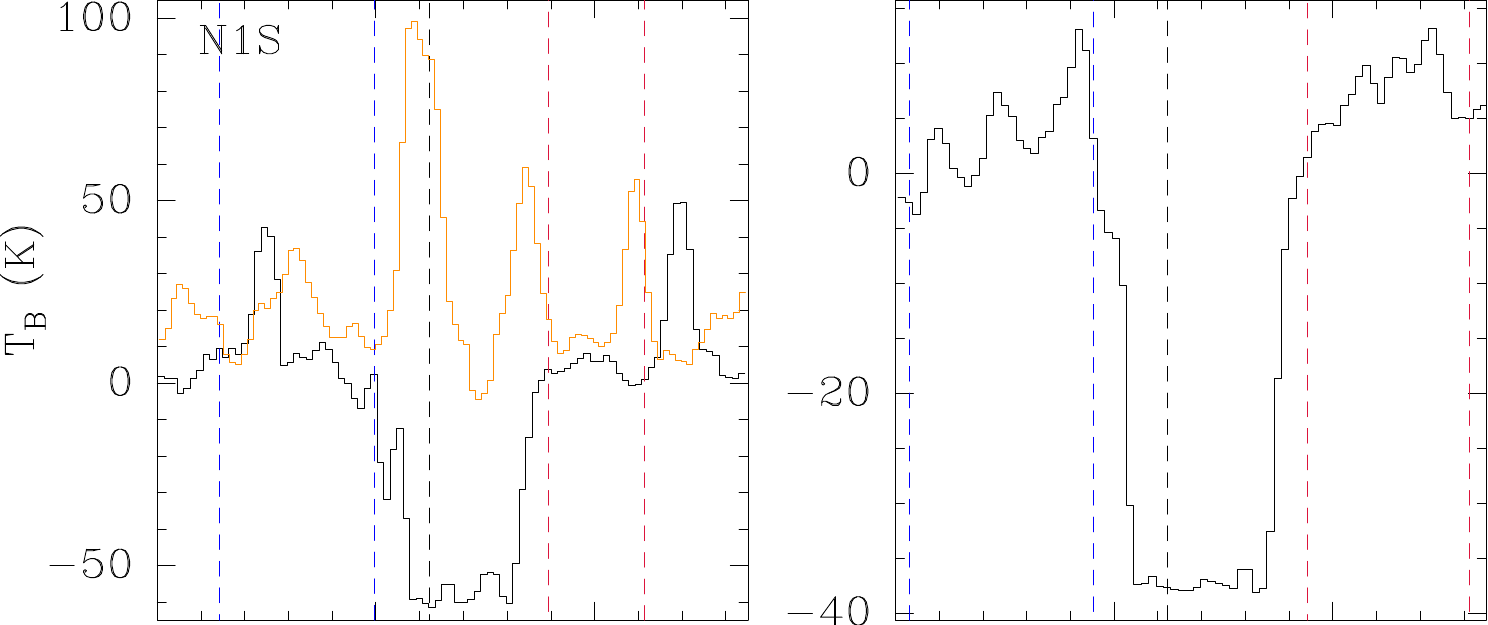}\\
    \includegraphics[width=0.49\textwidth]{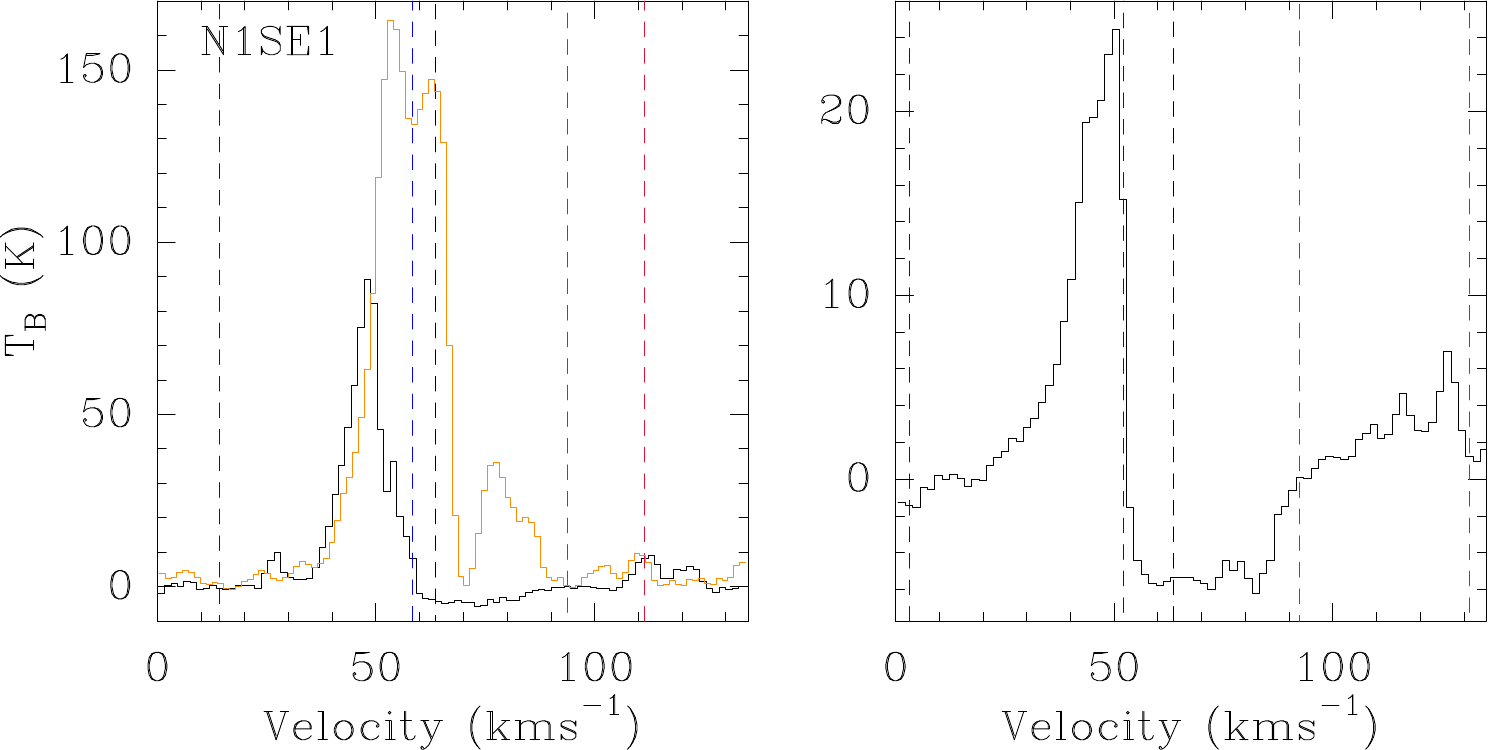}\\
    \caption{Spectra of SO $J=2_3-1_2$ at 99.3\,GHz, SO $J=3_2-2_1$ at 109.25\,GHz, and SiO $J=2-1$ at 86.85\,GHz towards a position in the red-shifted lobe (N1NW3, \textit{top row}), N1S (\textit{middle row}), and a position in the blue-shifted lobe (N1SE1, \textit{bottom row}). The black dashed lines mark the systemic velocities of 64.8\kms (N1NW3), 62.2\kms (N1S), and 63.6\kms (N1SE1), while the dashed blue and red lines indicate the inner and outer limits used to integrate the blue- and red-shifted emission shown in Fig.\,\ref{fig:so}. While the outer limits are fixed values, the inner limits are pixel-dependent and set such that channels that contain absorption features close to the systemic velocity are excluded.}
    \label{fig:spec_so+sio}
\end{figure}

In comparison to the radiative transfer models, a population-diagram analysis is affected by a few more uncertainties, which we discussed in detail in Paper\,I. For example, in the Weeds models the background continuum is taken into account in the 
equation of radiative transfer, while it is not (and cannot be properly) when fitting the population diagrams \citep[cf.][]{Goldsmith99}. Moreover, the correction factors that we apply to the observed and modelled values in the population diagrams depend on the Weeds model. On the one hand the opacity correction is an output of the Weeds model. On the other hand, although to some extent we can reduce the contribution of contaminating emission, it cannot be avoided entirely when deriving the population diagrams, while for the  Weeds models we ensure that the modelled intensity of a transition for any given molecule does not exceed the contribution of that transition to the observed spectrum. Therefore, our analysis is focused on the results derived from the radiative transfer modelling, while we use the population diagrams to support those results.

\section{Results}\label{s:results}
\subsection{Outflow morphology}
In the following, we investigate the morphology of the outflow as seen in emission of various molecules. In Sect.\,\ref{ss:sio-so}, we focus on typical shock-tracing molecules, namely SO and SiO. Then, to better take into account velocity and linewidth gradients across Sgr\,B2\,(N1), we computed linewidth- and velocity-corrected integrated emission (LVINE) maps, which are introduced in Sect.\,\ref{ss:lvineso}, for SO (Sect.\,\ref{ss:lvineso}) and other species (Sect.\,\ref{s:lvine}).

\subsubsection{SiO and SO emission}\label{ss:sio-so}

In the case of SiO, we analysed its $J=2-1$ transition at 86.85\,GHz, which is covered in observational setup 1, that is at lower angular resolution ($\sim$0.7\arcsec). 
For SO, we used its $J=2_3-1_2$ transition at 99.3\,GHz, which is covered in setup 5 with a two times higher angular resolution than for the SiO line ($\sim$0.3\arcsec). 
We present integrated-intensity maps of the two molecules in Fig.\,\ref{fig:so} on top of a continuum map at 99\,GHz which has contributions of both dust emission and free-free emission from ionised gas (see Paper\,I for a detailed description of the continuum). We integrated over the blue- and red-shifted emission and show the maps in separate panels. The pixel-dependent integration starts from the first channel that no longer shows absorption (starting from the systemic velocity, $\varv_{\rm sys}$)
and progresses up to the respective outer integration limits. For blue- and red-shifted SO emission, these outer limits are fixed at 15 and 115\kms, respectively, while for SiO emission, these are at 5 and 130\kms, respectively. 
To illustrate the choice of integration intervals, Fig.\,\ref{fig:spec_so+sio} shows the spectra of both transitions towards two positions in the outflow (N1NW3 and N1SE1, see Fig.\,\ref{fig:overview}) and one position that is primarily not associated with the outflow (N1S, see also Paper\,I). The positions N1NW3 and N1SE1 are further analysed in Sect.\,\ref{s:radTransfer}. The spectra reveal absorption at velocities close to $\varv_{\rm sys}$, but also prominent red-shifted (wing)emission towards N1NW3 and blue-shifted emission towards N1SE1, which is indicative of the outflow. In addition, a second SO transition at 109.25\,GHz ($J=3_2-2_1$, orange), which is only in emission, is shown for comparison of the line profiles. 
The integrated-intensity maps of Fig.\,\ref{fig:so} may contain some contaminating emission from other molecules as can be seen from additional spectral lines in the integration interval in Fig.\,\ref{fig:spec_so+sio}, especially for N1S. On the other hand, because the outer integration limits were set such that contaminating emission could be excluded at some positions, outflow emission at higher velocities may be missed at other positions (see, e.g. red-shifted SO emission at N1S and N1SE1).

Although the morphology has some complexity to it, blue-shifted emission is dominantly observed to the southeast, while red-shifted emission extends to the northwest. There is some overlap of both in the closest vicinity of the hot core's centre. The longest features labelled in the blue- (aB) and red-shifted (aR2) emission maps, which are at position angles of 120$^\circ$ and $-$64$^\circ$ east from north, respectively, starting from the continuum peak of Sgr\,B2\,(N1), stand out due of their strong collimation in the SO maps. They do not appear as narrow in the SiO maps due to the lower angular resolution. These features extend up to 0.3\,pc (blue) and 0.38\,pc (red). There was only one detached contour in the CO map shown by \citet{Schwoerer20} hinting at this spatially extended feature. The position angles correspond more or less to those stated by \citet[][]{Higuchi15}, who mapped the outflow in SiO emission using the EMoCA survey. Neither the high degree of collimation nor the maximum spatial extent seen here could be identified in the EMoCA data, due to the lower angular resolution and lower sensitivity of that survey.
Moreover, in contrast to the SiO maps shown by \citet{Higuchi15}, we are able to identify an additional feature (aR1) in the SO maps, which also has some degree of collimation but a smaller spatial extent than aR2. The position angles of features aR1 and aR2 differ only by a few degree, which possibly indicates precession of the outflow. There is also another blue-shifted feature extending to the east at a position angle of $\sim$95$^\circ$, whose possible origins are discussed in Sect.\,\ref{app:detailedmorph}, where the outflow morphology is described in more detailed also in comparison to other species.

As noted by  \citet[][]{Schwoerer20}, the blue- and red-shifted emissions seem to have multiple meeting points with continuum emission. The blue-shifted emission extending to the east follows the continuum (free-free) emission of the large H{\small II} region located in the northeast \citep[cf.][]{Gaume95}, while the emission along aB1 seems to be framed by one of the filaments identified by \citet[][see also Fig.\,\ref{fig:overview}]{Schwoerer19}. Similarly, the red-shifted emission seems to be embedded in the filamentary structure of the continuum emission. Additionally, in the northwest portion of mainly the SiO maps, the outflow of N3, another hot core \citep[][]{Bonfand17}, can be identified. The red-shifted collimated feature aR2 seems to extend up to the location of N3, with which it might be interacting if they were located at the same distance along the line of sight. 
To the north, there is also emission that is associated with the hot core Sgr\,B2\,(N2).

Despite the difference in angular resolution of their maps, both molecules trace generally the same morphology. Only, there is red-shifted emission in the SiO maps towards the south(east), which is not as prominently observed in the SO maps. Given that it still coincides with continuum emission, there might be some contamination coming from another species. We cannot be conclusive as this is the only SiO transition covered by ReMoCA.  In contrast, the SO emission reveals faint red-shifted emission to the east that follows the structure of the H{\small II} region, which may further be associated with the two H\2O masers that are observed at red-shifted velocities in this region (pink markers). In the map of blue-shifted SO emission, there is an extension towards the (south)west that is not observed in SiO.

\subsubsection{LVINE maps for SO}\label{ss:lvineso}

The molecular emission towards Sgr\,B2\,(N1) reveals gradients in both peak velocity (from $\lesssim$\,60\kms up to $\sim$70\kms, see \citet{Schwoerer19} and Paper\,I) and linewidth (from $\sim$3 to 12\kms).
In order to account for this when integrating intensities in Paper\,I, we first derived peak-velocity and linewidth maps for the region around Sgr\,B2\,(N1) (see Fig.\,B.2 in Paper\,I). For this purpose, we used a transition of ethanol, which remains sufficiently optically thin even at closest distances to the centre of the hot core, and a bright methanol line beyond the distances where ethanol is no longer securely detected. Subsequently, the peak velocities and linewidths were used to adjust the integration limits in each pixel of the LVINE maps. This LVINE method is an extension of the VINE method \citep{Calcutt18} in that we also consider the variations in linewidth. 
Here we used this method and the peak velocity and linewidth maps from Paper\,I to determine the inner integration limits for the integrated-intensity maps of the blue- and red-shifted emission. We first find the channels for which the velocity is $\varv_t\pm FWHM_t$ for the blue- and red-shifted emission, respectively, where $\varv_t$ and $FWHM_t$ are the peak velocity and linewidth of the template spectral lines (ethanol in the inner part, methanol in the outer part). To obtain the inner integration limits, we apply a shift of another two channels outwards to avoid as much emission from the hot-core component as possible. The outer integration limits are pixel-independent, fixed values that were defined based on a comparison of spectra as, for example, shown in Fig.\,\ref{fig:spec_so+sio}, but over a larger area.
Because emission of the SO and SiO transitions used for Fig.\,\ref{fig:so} suffer from strong self-absorption over large portions of the maps and over a large velocity interval, they cannot be used. In the case of SO, we instead
use its transition at 109.25\,GHz and show the LVINE maps for red-and blue-shifted emission in Fig.\,\ref{fig:lvine_coms}a. It reveals essentially the same morphology as in Fig.\,\ref{fig:so} with collimated features (labelled aB1, aR1, aR2 in Fig.\,\ref{fig:so}) but the bulk of emission showing a wide opening angle. Additional structures are discussed while comparing to other molecules in Sect.\,\ref{s:lvine} or in Appendix\,\ref{app:lvine}. In the case of SiO, there is no transition unaffected by self-absorption in the ReMoCA survey.

\subsubsection{Emission from other molecules}\label{s:lvine}

\begin{figure*}
    \begin{tikzpicture}[remember picture]
  
    \node[] at (-3.5,-2.9) {\includegraphics[width=0.312\textwidth]{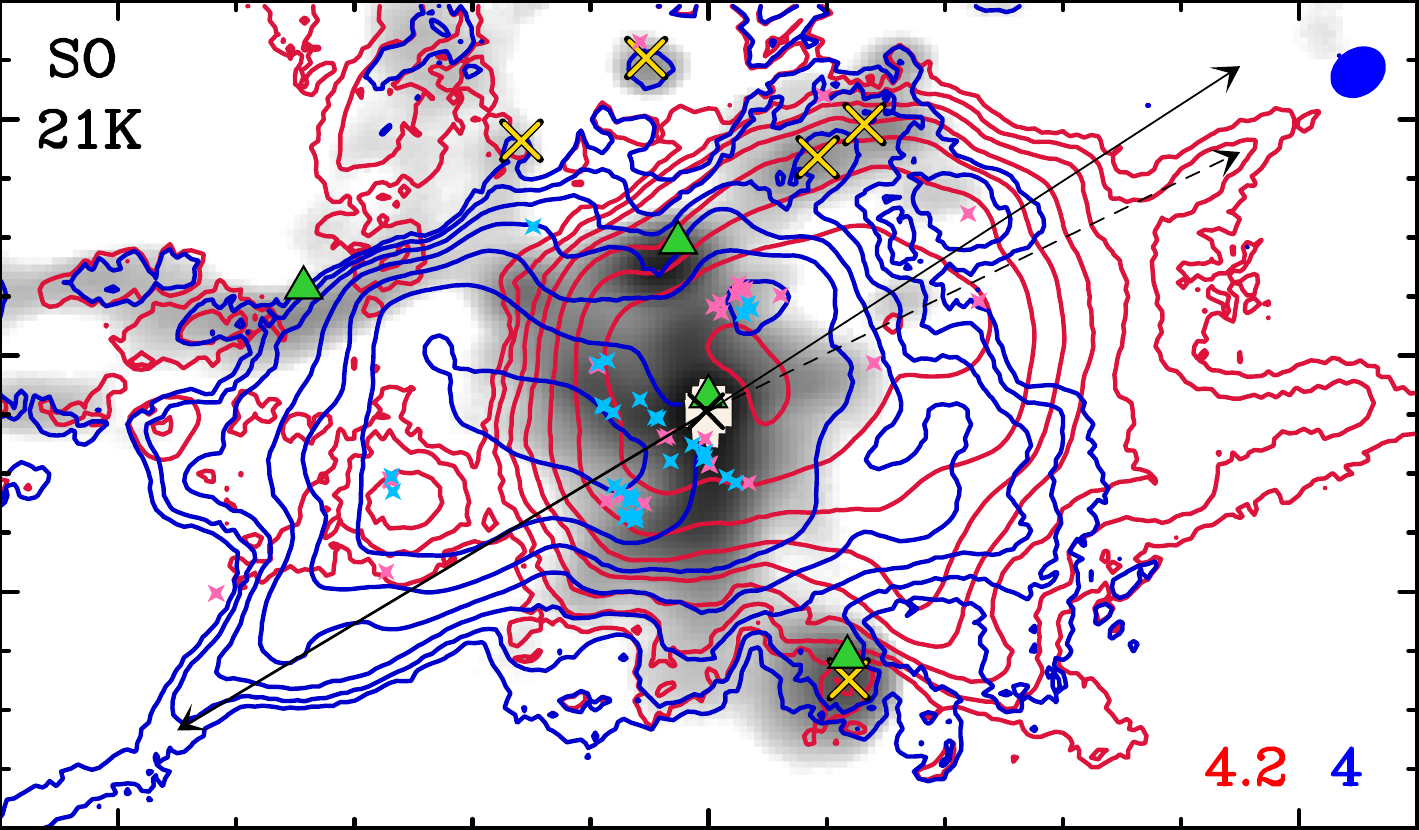}};
    \node[] at (2.35,-2.9) {\includegraphics[width=0.312\textwidth]{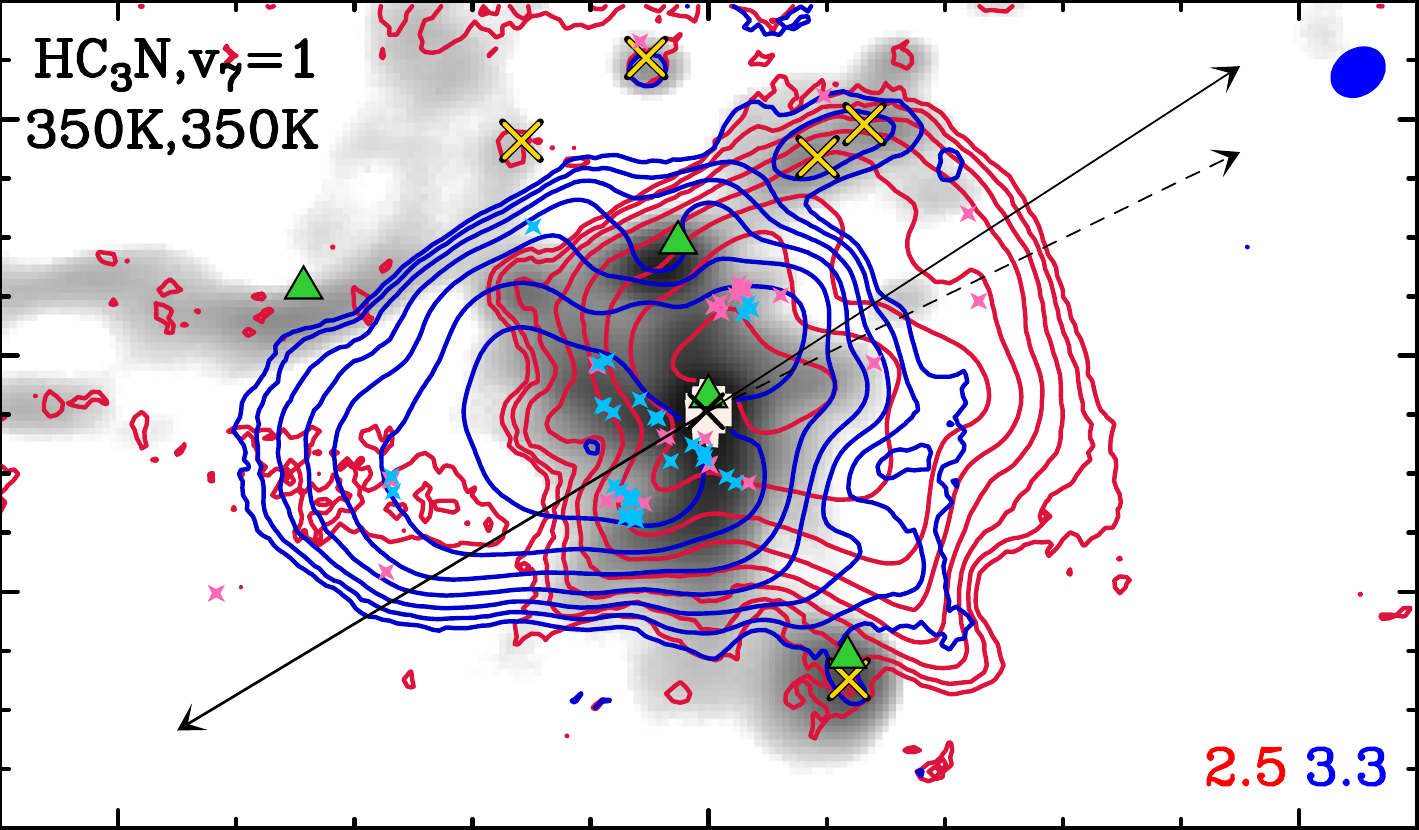}};
    \node[] at (8.2,-2.9) {\includegraphics[width=0.312\textwidth]{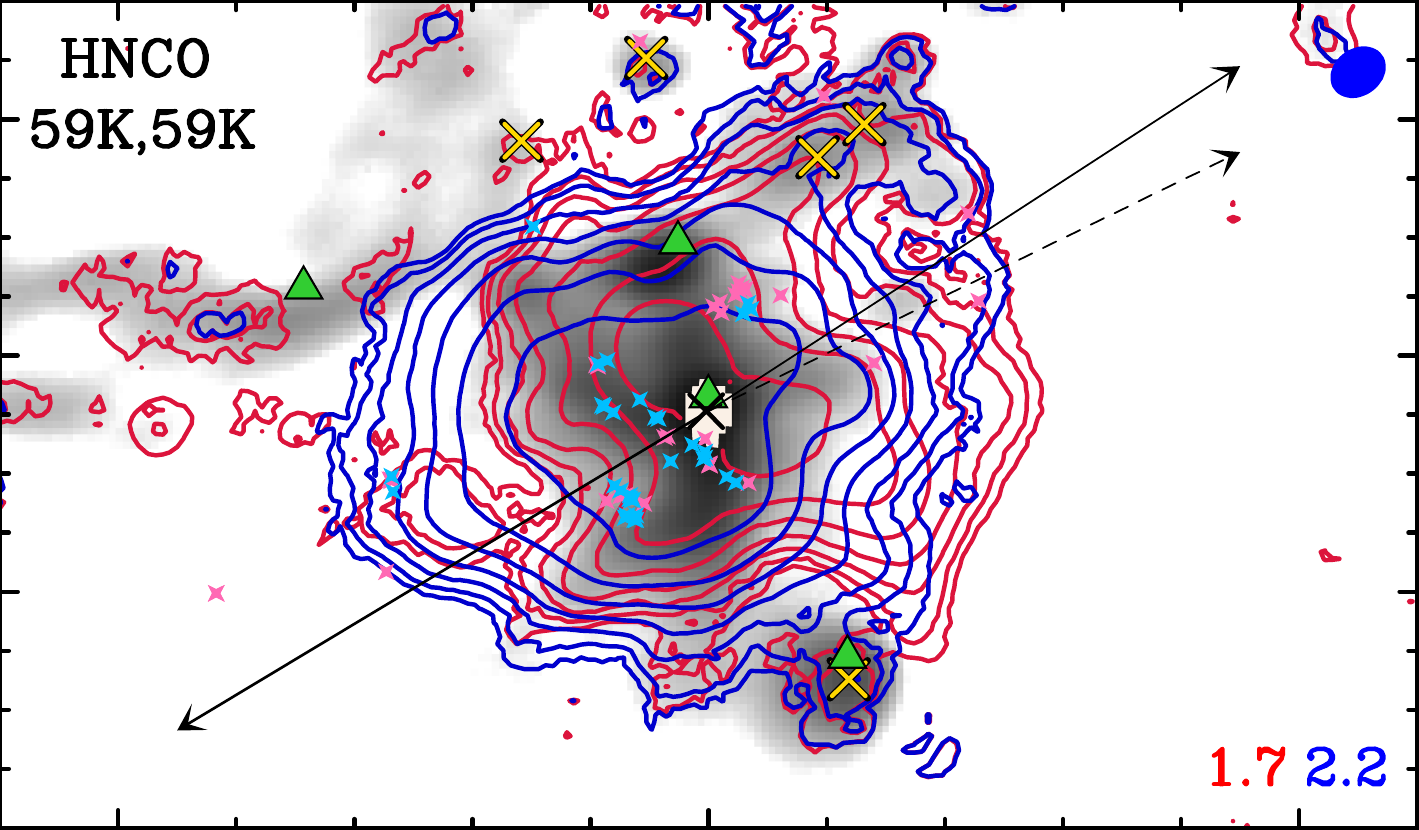}};
    \node[] at (-3.5,-6.3) {\includegraphics[width=0.312\textwidth]{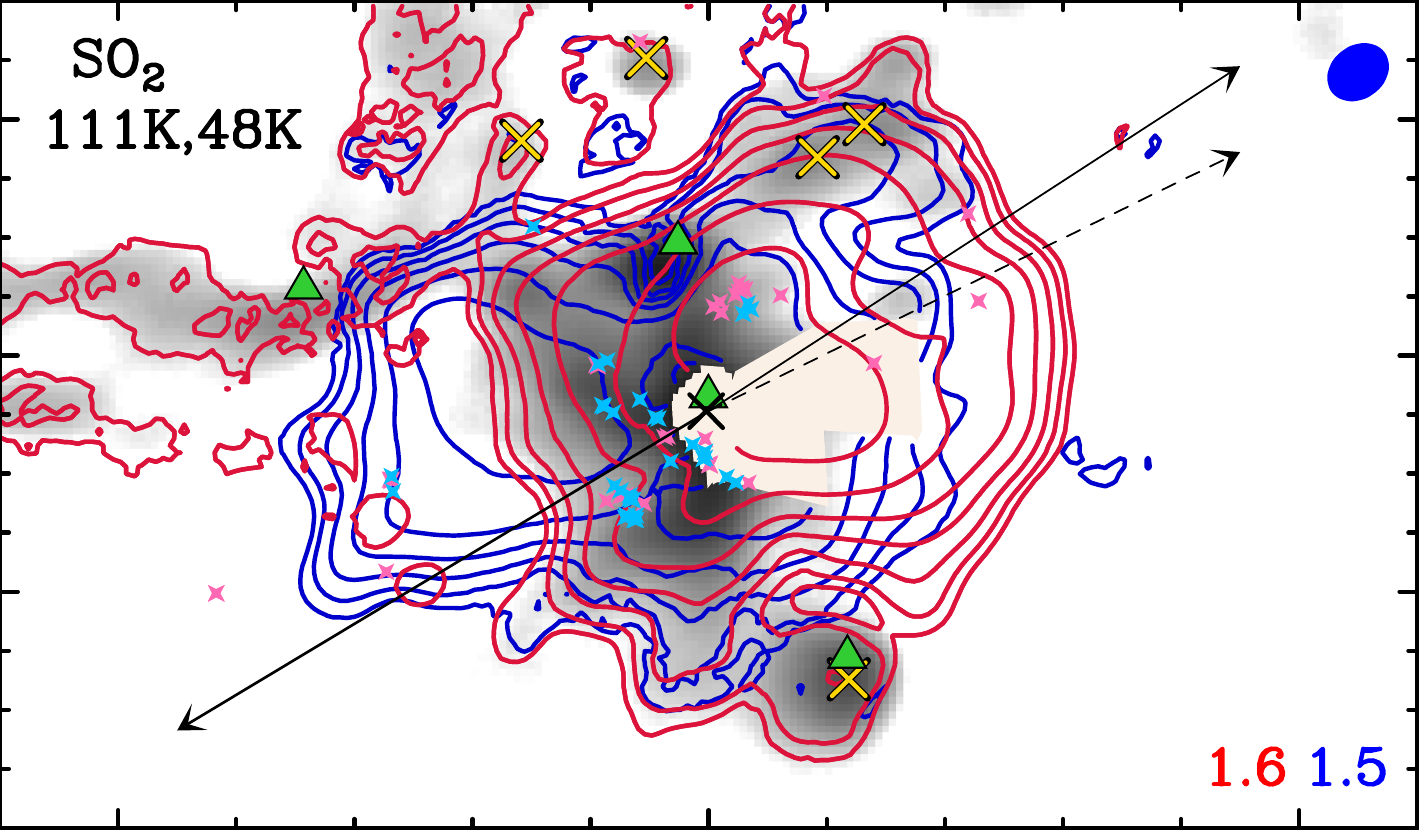}};
    \node[] at (2.35,-6.3) {\includegraphics[width=0.312\textwidth]{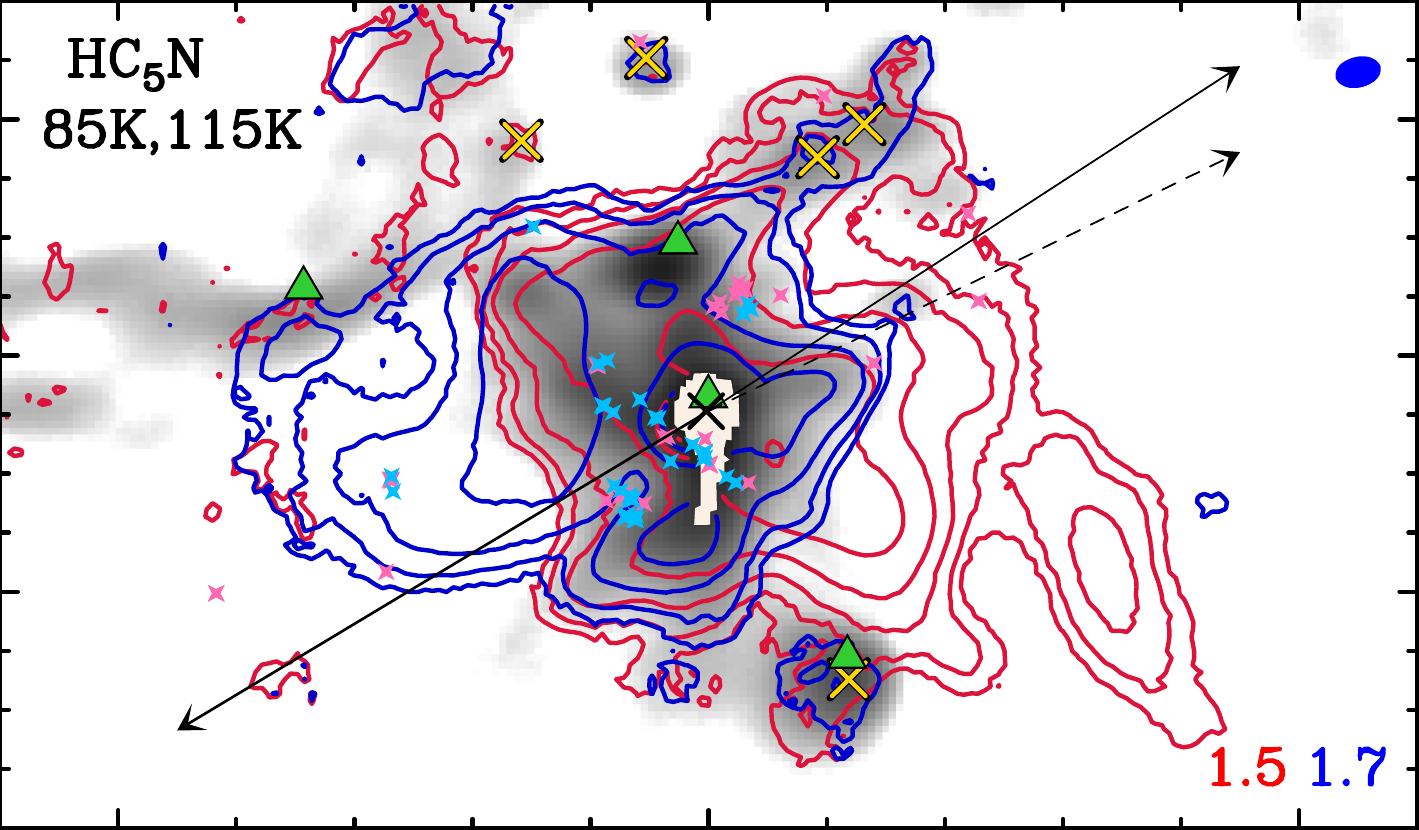}};
    \node[] at (8.2,-6.3) {\includegraphics[width=0.312\textwidth]{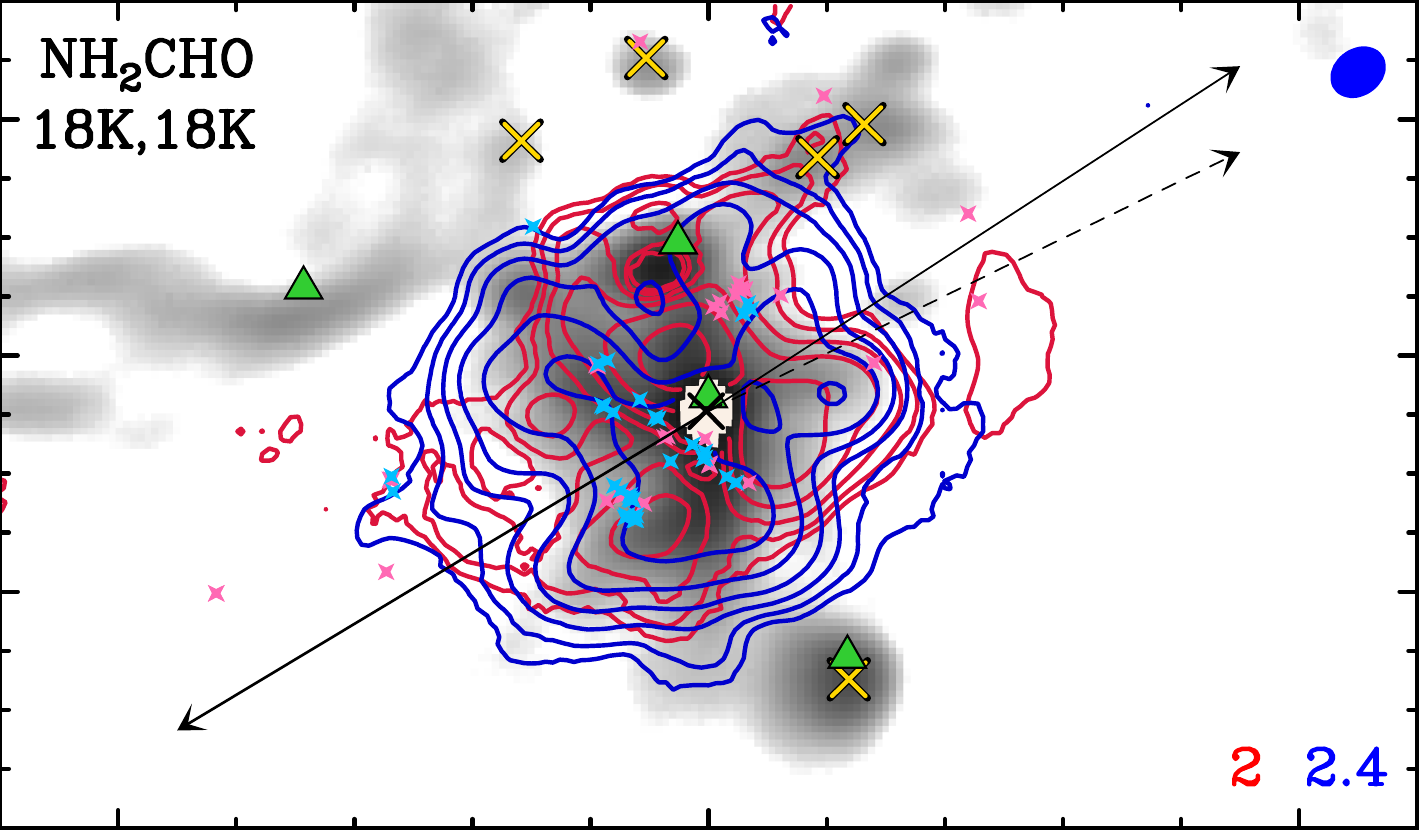}};
    \node[] at (-3.5,-9.7) {\includegraphics[width=0.312\textwidth]{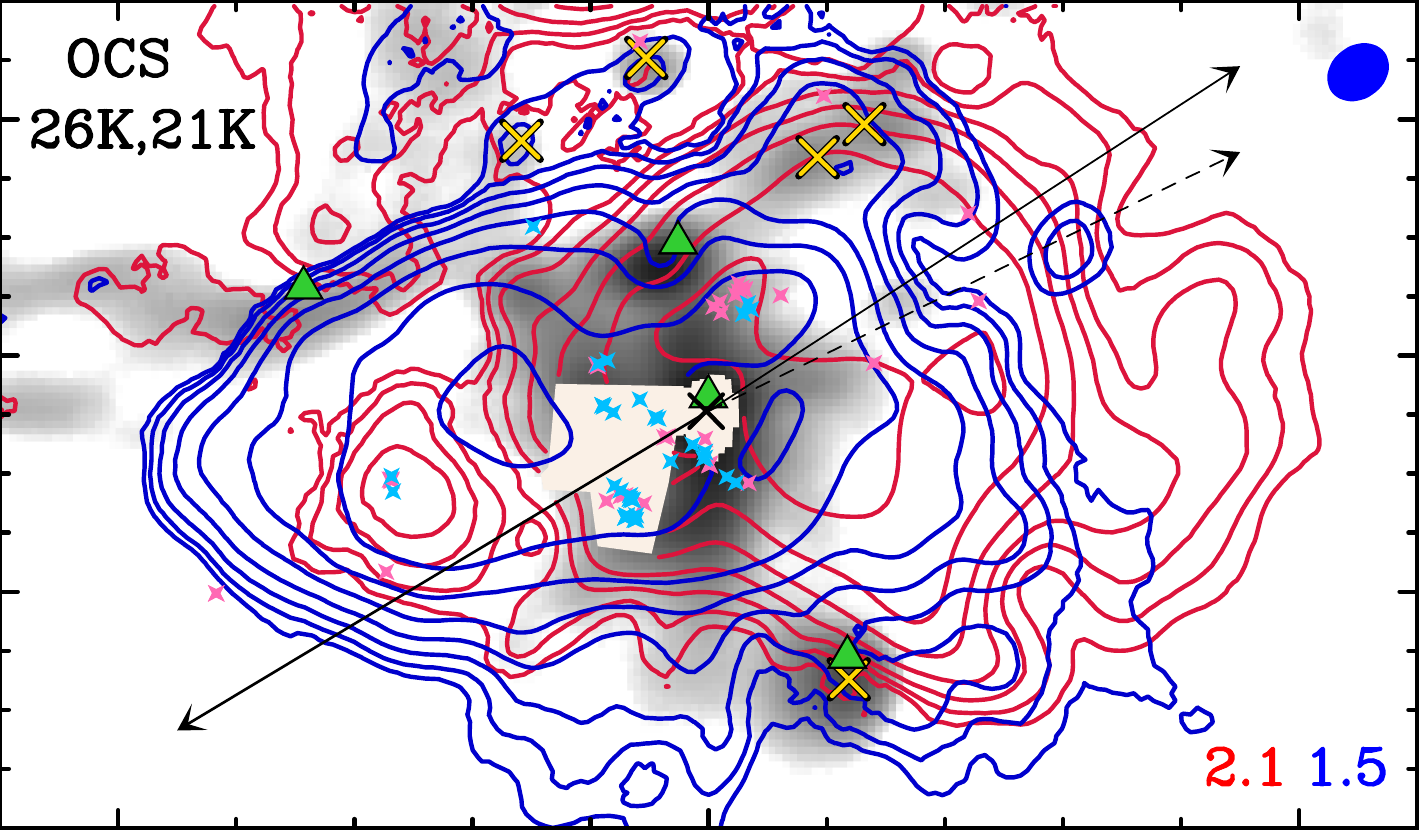}};
    \node[] at (2.35,-9.7) {\includegraphics[width=0.312\textwidth]{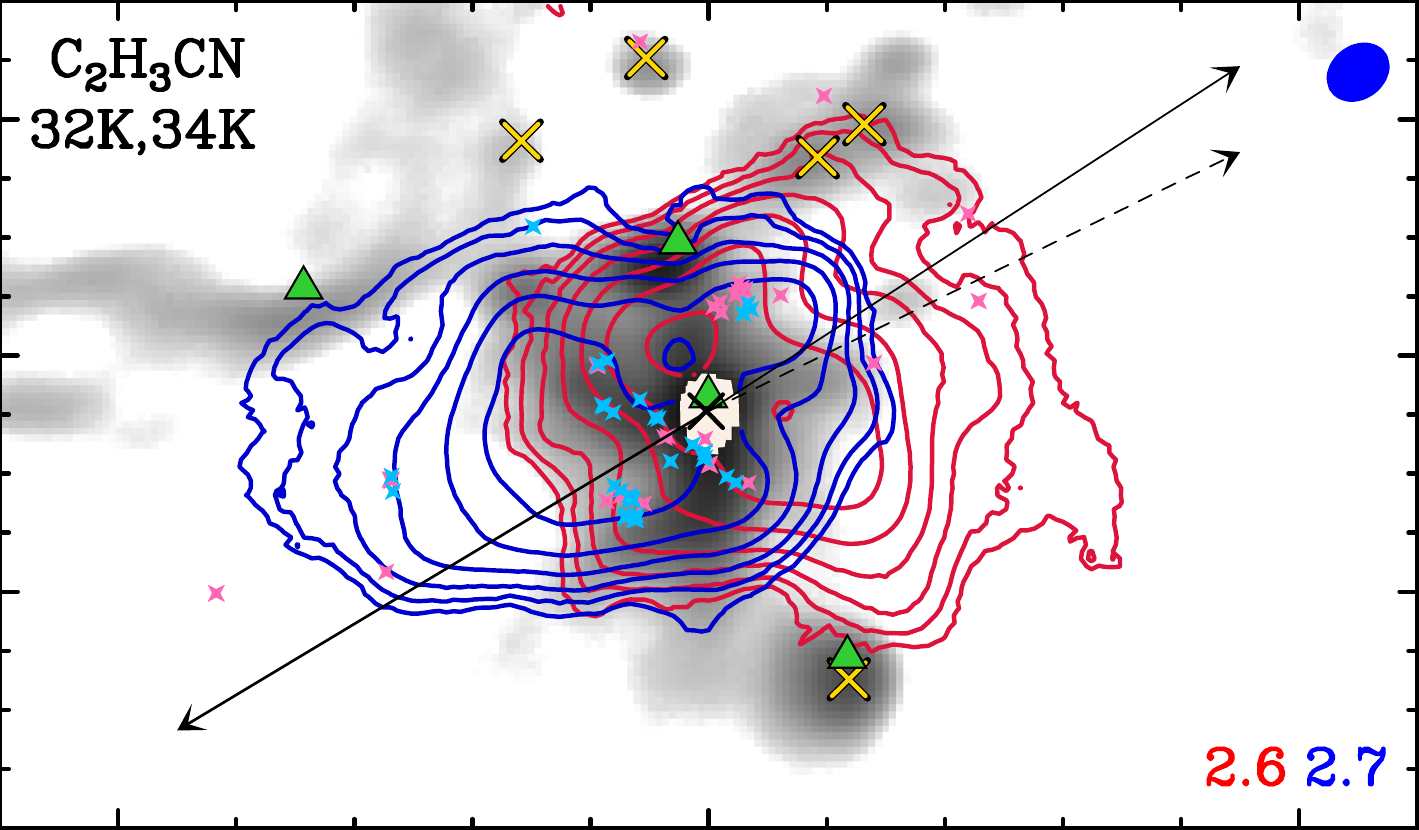}};
    \node[] at (8.2,-9.7) {\includegraphics[width=0.312\textwidth]{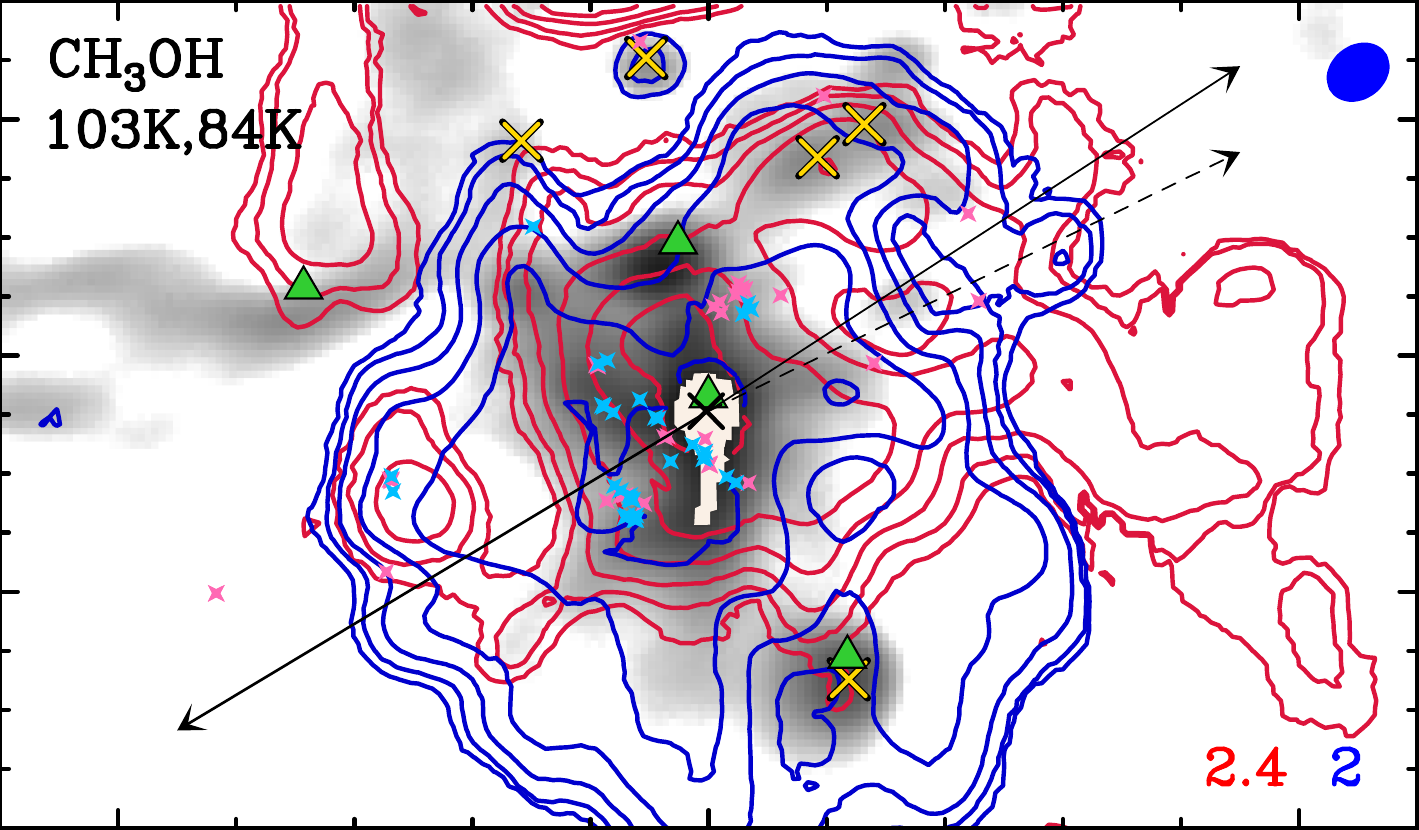}};
    \node[] at (-3.85,-13.39) {\includegraphics[width=0.35\textwidth]{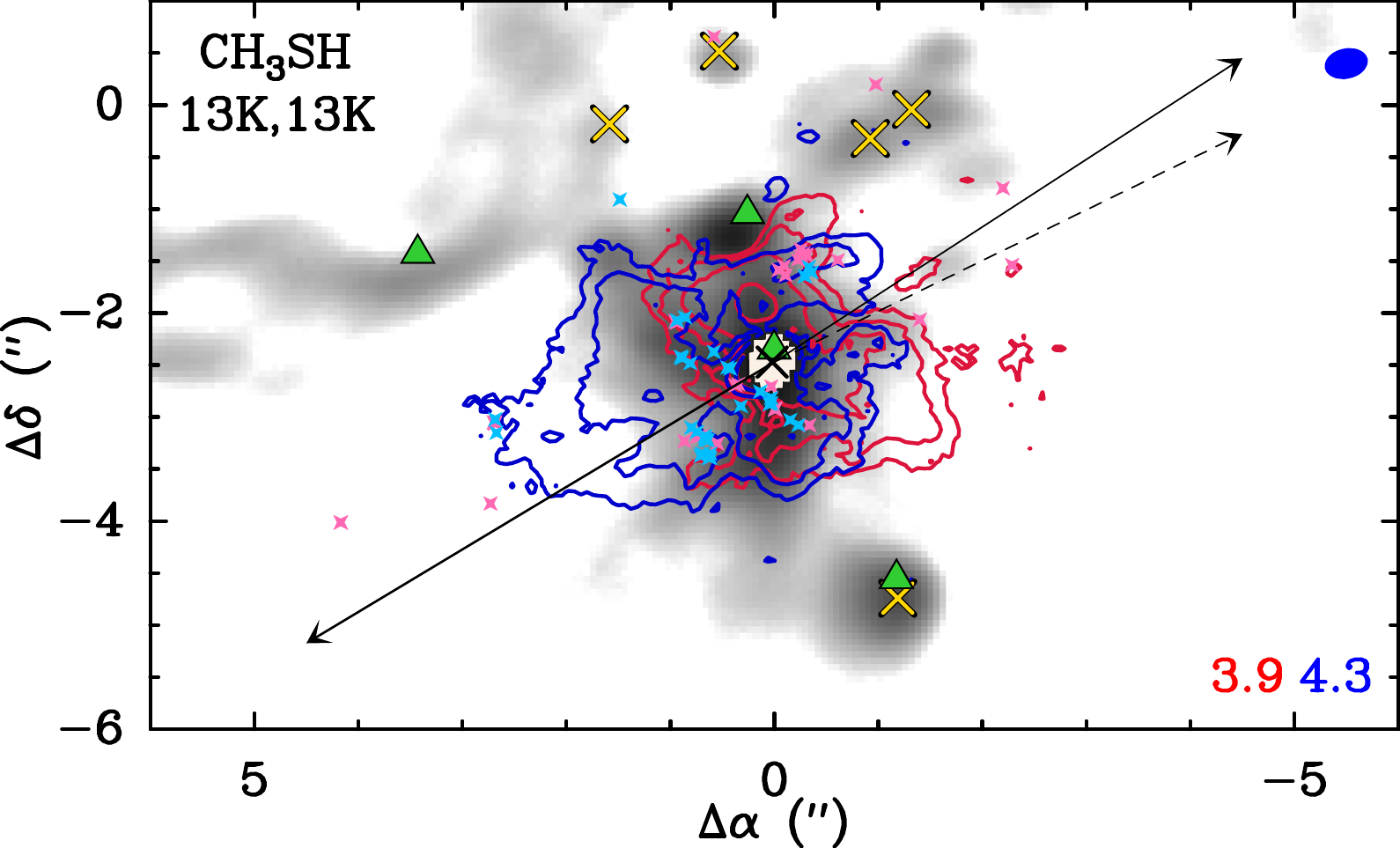}};
    \node[] at (2.35,-13.12) {\includegraphics[width=0.312\textwidth]{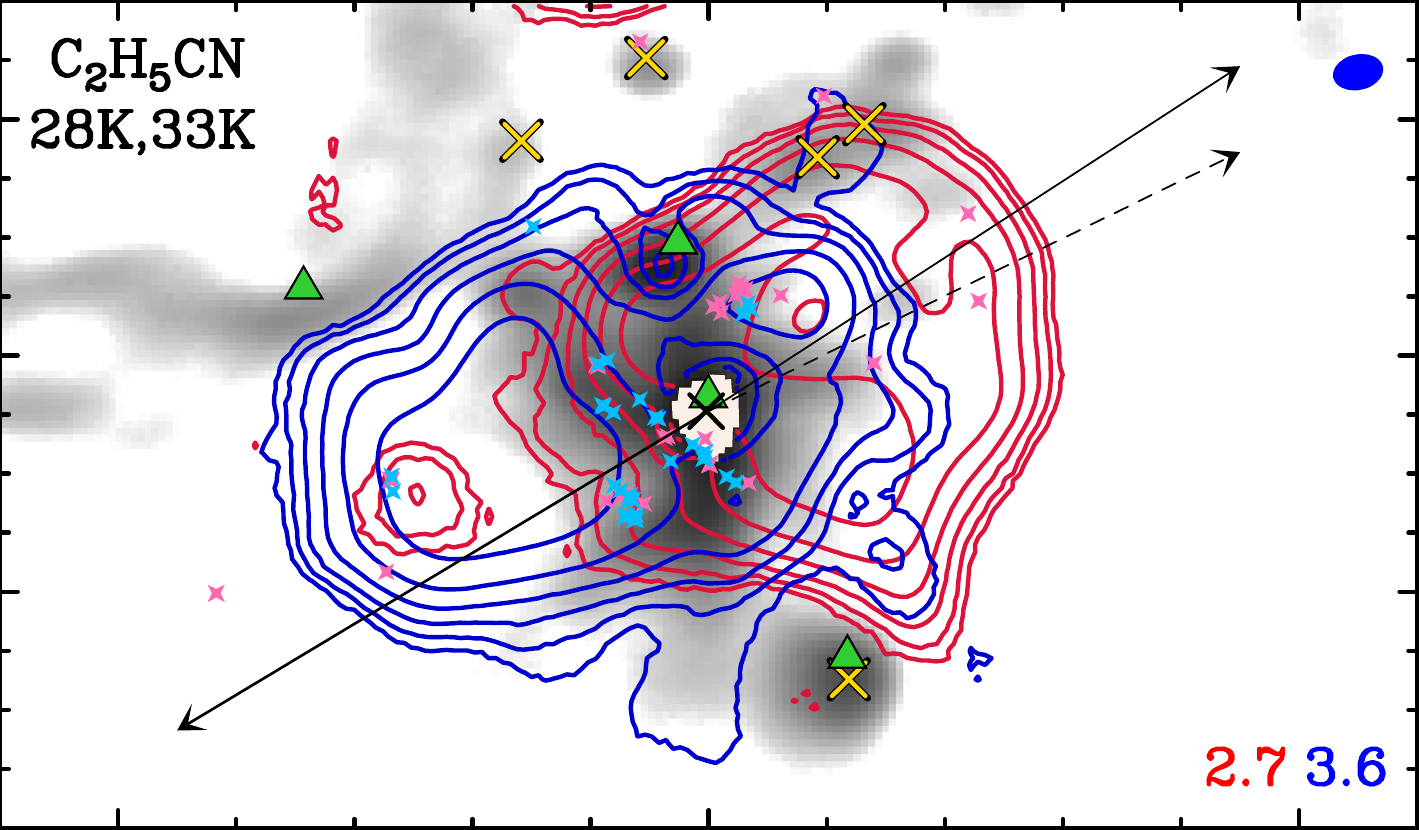}};
    \node[] at (8.2,-13.12) {\includegraphics[width=0.312\textwidth]{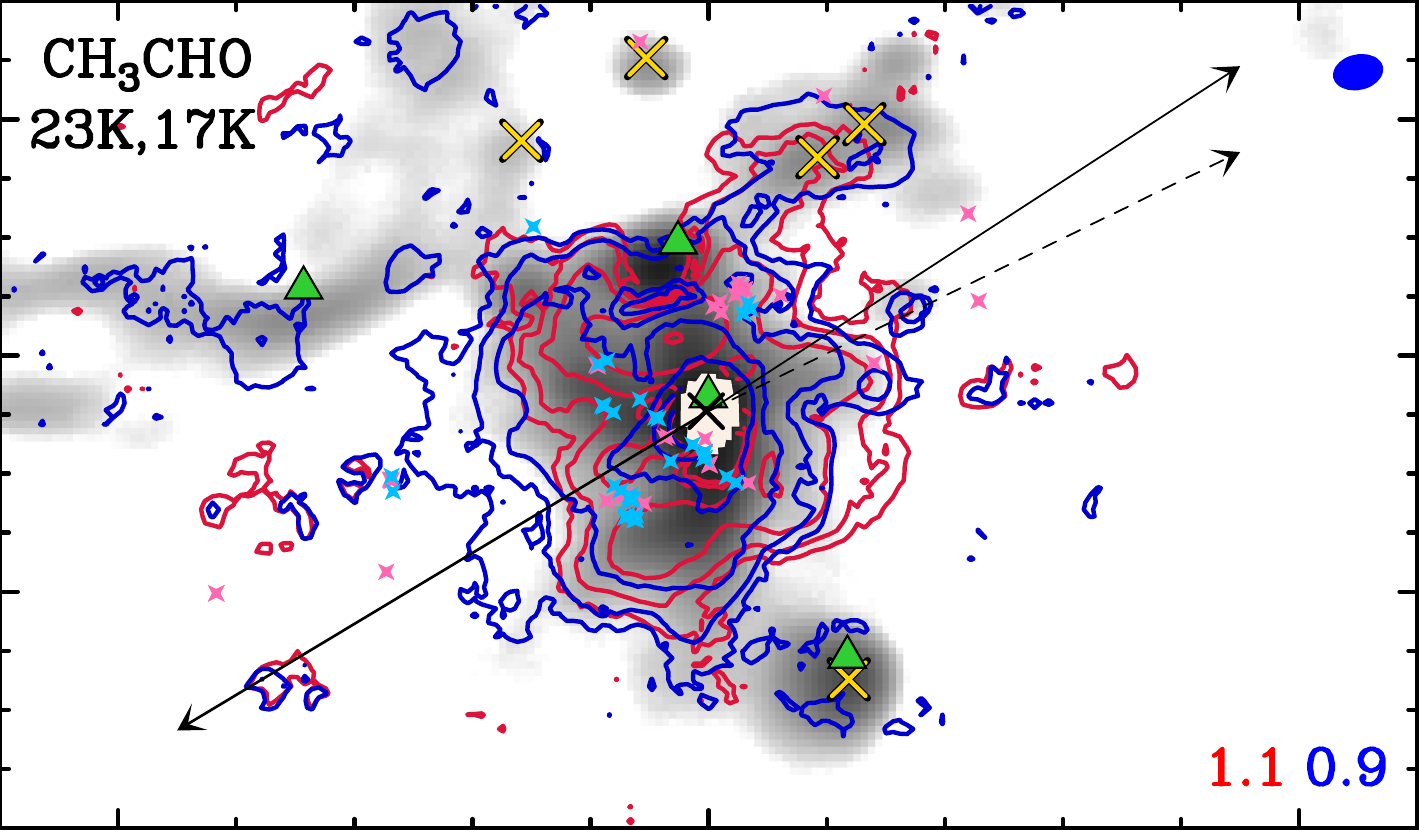}};
  
    \node[] at (-6.05,-4.1) {\textbf{(a)}};
    \node[] at (-6.05,-7.68) {\textbf{(b)}};
    \node[] at (-6.05,-11.08) {\textbf{(c)}};
    \node[] at (-6.05,-14.5) {\textbf{(d)}};
  
    \node[] at (-.2,-4.25) {\textbf{(e)}};
    \node[] at (-.2,-7.68) {\textbf{(f)}};
    \node[] at (-.2,-11.08) {\textbf{(g)}};
    \node[] at (-.2,-14.5) {\textbf{(h)}};
  
    \node[] at (5.65,-4.25) {\textbf{(i)}};
    \node[] at (5.65,-7.68) {\textbf{(j)}};
    \node[] at (5.65,-11.08) {\textbf{(k)}};
    \node[] at (5.65,-14.5) {\textbf{(l)}};
\end{tikzpicture}

    \caption{LVINE maps of blue- and red-shifted emission (blue and red contours, respectively) of S-bearing molecules \textit{(a--d)}, N-bearing molecules \textit{(e--h)}, and (N+O)- and O-bearing species \textit{(i--l)}. The contour steps start at 5$\sigma$ and then increase by a factor of 2, where $\sigma$ is an average noise level measured in an emission-free region in each map and is given in K\kms in the bottom-right corner in each panel. The grey scale in all panel shows the continuum emission at 99\,GHz (see Fig.\,\ref{fig:so}). The closest region around Sgr\,B2\,(N1) is masked out (beige areas) due to high frequency- and beam-size-dependent continuum optical depth (see Appendix\,C in Paper\,I). For OCS and SO\2, the masked region was extended due to contamination by emission from another species that was identified in their spectra.
    Markers and arrows are the same as in Fig.\,\ref{fig:so}. The upper-level energies of the transitions used to produce the maps are shown in the top-left corner in each panel. Other properties of the transitions and the outer integration limits are summarised in Table\,\ref{tab:trans}. The HPBW is shown in the top-right corner in each panel. The position offsets are given with respect to the ReMoCA phase centre.} 
    \label{fig:lvine_coms}
\end{figure*}
\begin{figure}
    \centering
    \includegraphics[width=0.49\textwidth]{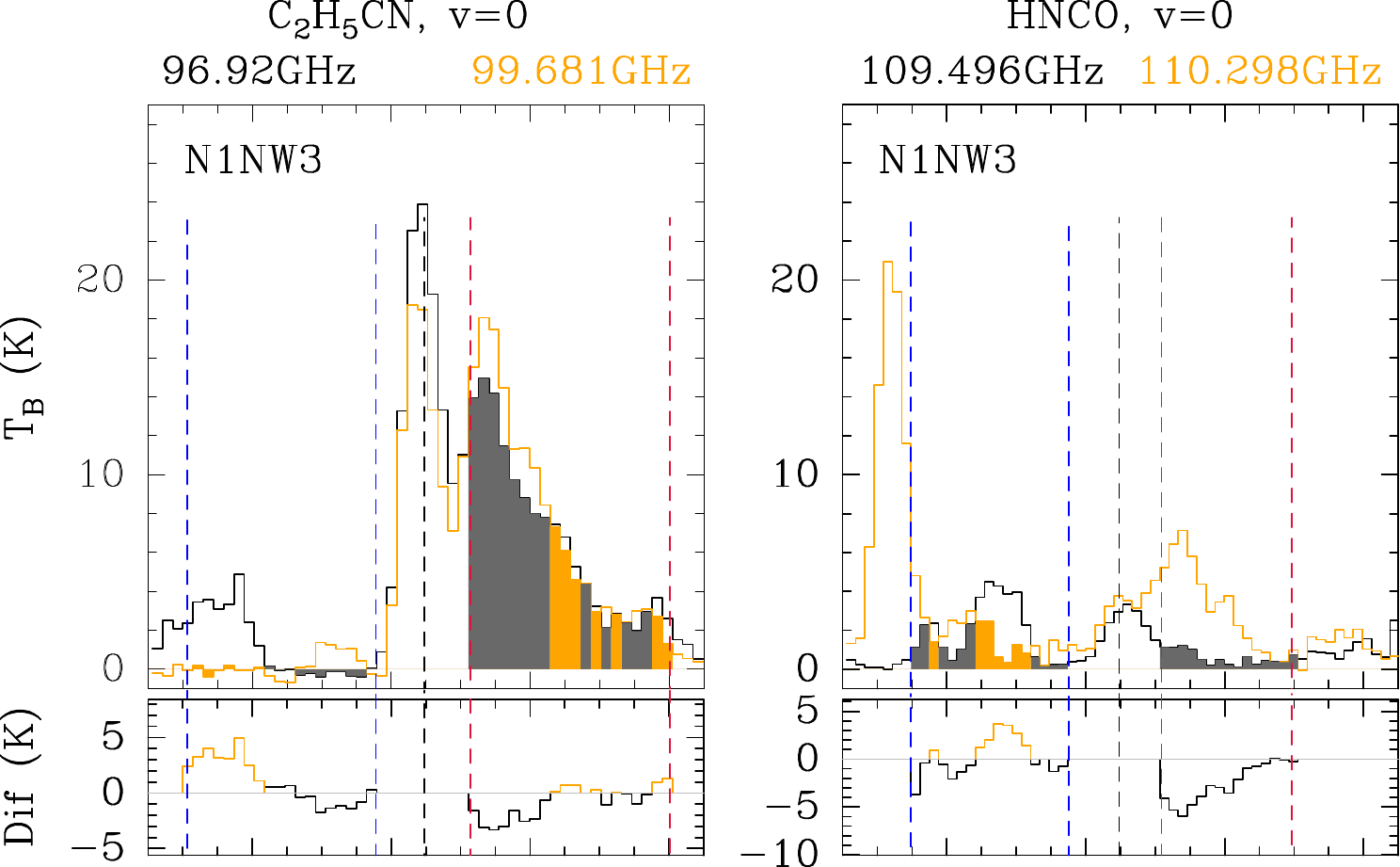}\\
    \includegraphics[width=0.49\textwidth]{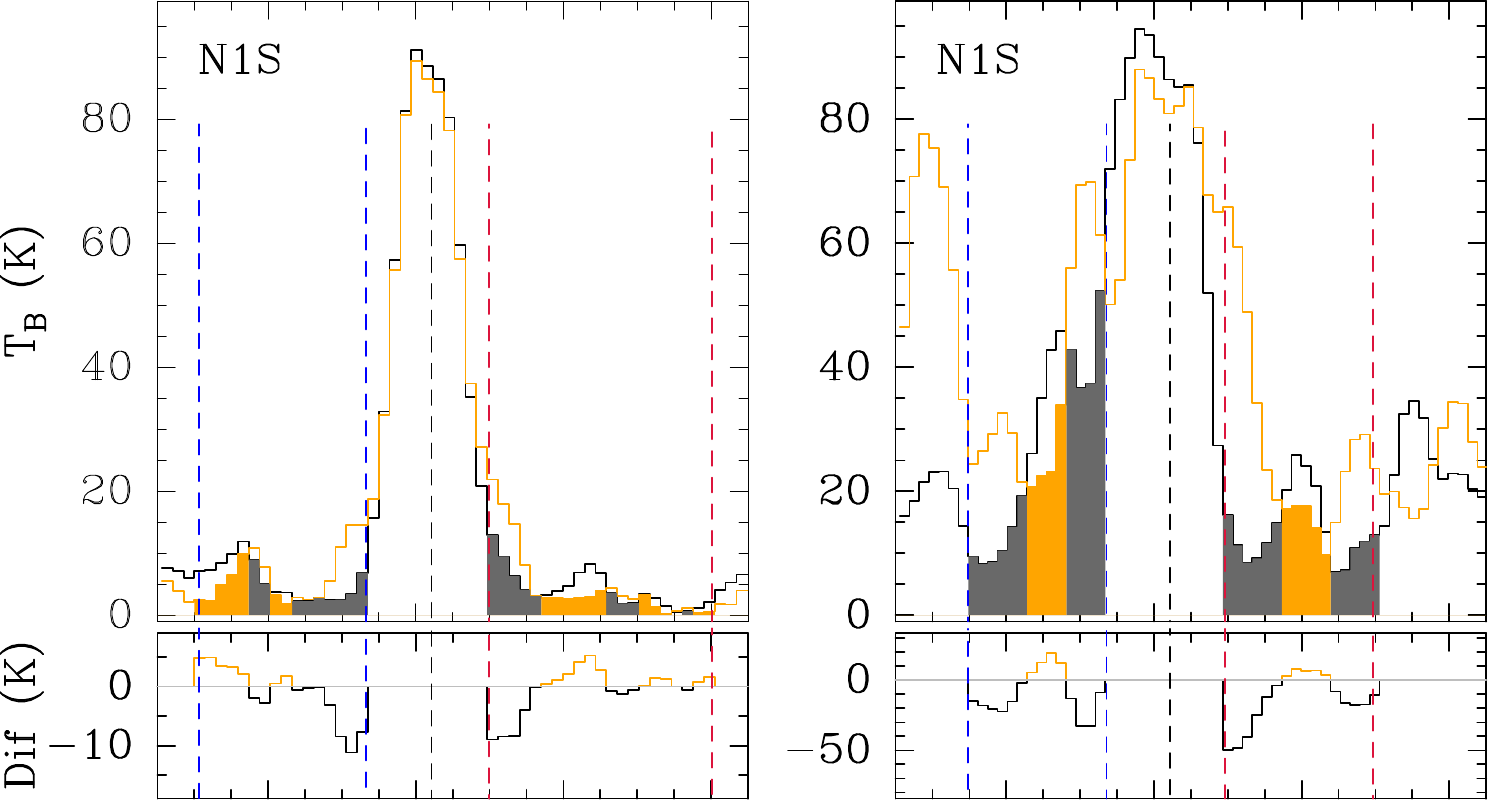}\\
    \includegraphics[width=0.49\textwidth]{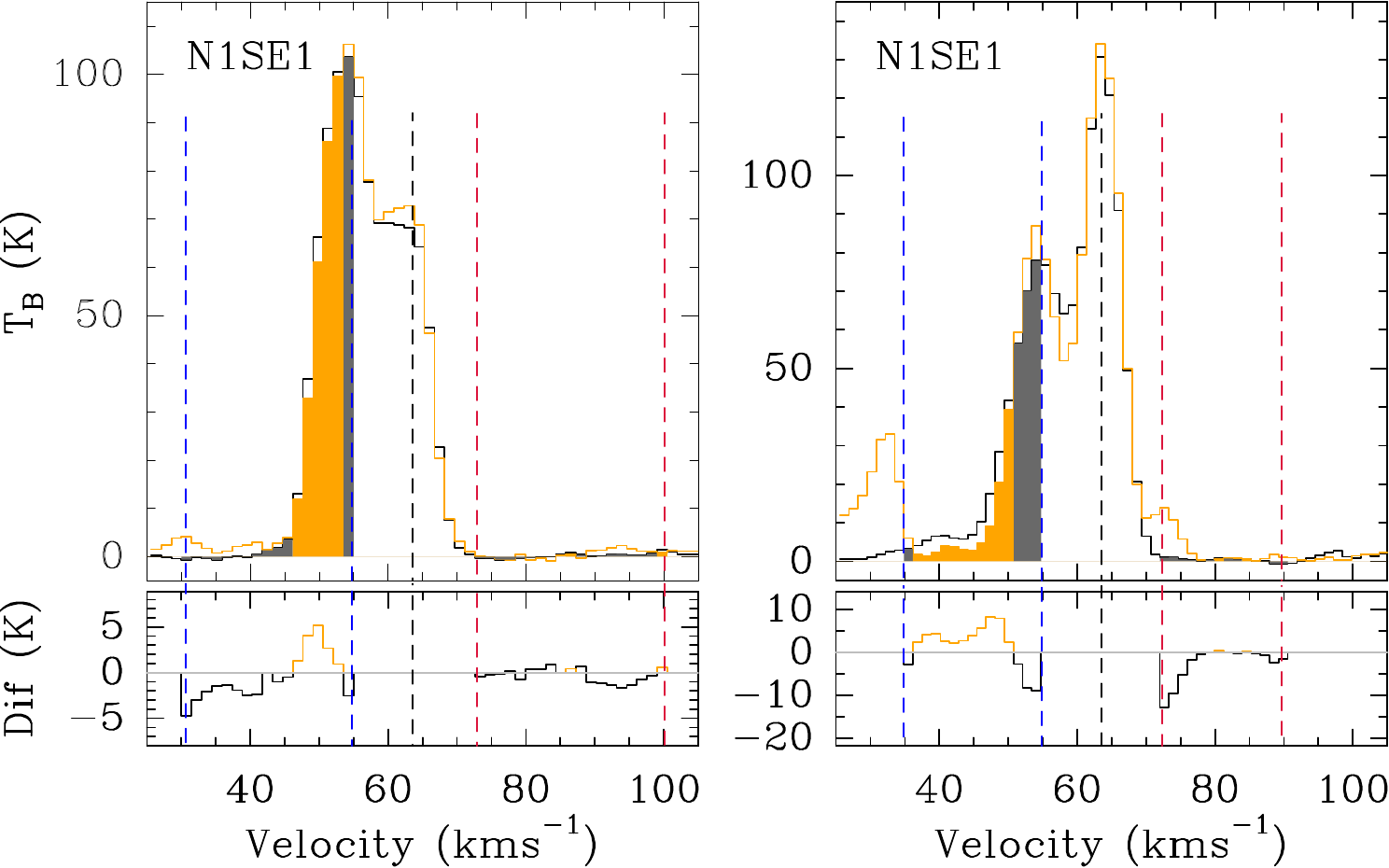}\\
    \caption{Spectra of two transitions of \etc and HNCO towards a position in the red-shifted lobe (N1NW3, \textit{top row}), N1S (\textit{middle row}), and a position in the blue-shifted lobe (N1SE1, \textit{bottom row}). The black and orange spectra show the transitions whose frequencies are given in the respective colour at the top. Additional line properties can be found in Table\,\ref{tab:trans}.  
    The black dashed lines are the same as in Fig.\,\ref{fig:spec_so+sio}, while the dashed blue and red lines indicate the fixed outer and pixel-dependent inner limits used to integrate the blue- and red-shifted emission shown in Fig.\,\ref{fig:lvine_coms}. The difference (Dif) of the two spectra ($black-orange$) within the integration intervals is shown below the respective panel. 
    The filled histograms and the colour of the difference spectra indicate which transition was used in each channel for the integration of the blue-shifted and red-shifted emission in order to minimise the contamination by other species.}
    \label{fig:lvine_spec}
\end{figure}

In addition to these typical outflow tracers, primarily blue-shifted components had been detected in lines of HC\3N and COMs such as \etc and \vc in the past \citep[e.g.][]{Belloche13}. While inspecting spectra extracted from positions along the blue-shifted lobe, we identified for multiple COMs one or more emission component(s) in addition to that associated with the hot core, that is at $\varv_\mathrm{sys}$. In general, these additional components are more prominent for N- and S-bearing species than for (N+O)- or O-bearing species, when detected at all for the latter.
As an example, Fig.\,\ref{fig:lvine_spec} shows spectra of \etc and HNCO towards the same positions as in Fig.\,\ref{fig:spec_so+sio}. In each panel, two transitions of the respective molecule are shown, which were selected based on their similar upper-level energies and Einstein A coefficients and thus have similar intensities. The difference of the two spectra ($black-orange$) inside the integration limits (dashed red and blue lines) is shown below the respective panel.
The profiles for \etc at N1S agree well for velocities close to the average systemic velocity of 62\kms. There is no clear hint of additional components or prominent wing emission within the velocity interval limited by the blue and red dashed lines, which would be indicative of the presence of an outflow. Most of the emission seen in these velocity intervals can likely be associated with other molecules, given that the black and orange spectra behave differently. This is also true for HNCO at this position, however, the contamination by emission from other molecules is more severe.
In contrast, at N1SE1 the two transitions agree well between 45 and 70\kms for both \etc and HNCO proving that the orange and black emissions that are observed in this velocity range come from the same molecule and are not contaminated by 
other species. Therefore, in addition to the component that is associated with the hot core at 63.6\kms at this position, there is a second one at blue-shifted velocities, which peaks at $\sim$54\kms. 
There is some contaminating emission in the orange spectra at $\sim$30\kms for \etc and at $\sim$32 and $\sim$72\kms for HNCO. At N1NW3, the two transitions of ethyl cyanide show similar line profiles between 60 and 100\kms, again with one `hot-core' component at 64.8\kms and one component at red-shifted velocities. The latter peaks at $\sim$74\kms and shows extended wing emission towards higher velocities. At blue-shifted velocities, there is some emission in the orange spectrum around $\sim$52\kms and in the black spectrum at 30--35\kms that is not seen in the other spectrum, respectively, again suggesting that this emission comes from another molecule. Emission of HNCO is much weaker at this position and the line profiles of the two transitions only agree within 60--67\kms. Emission at other velocities is most likely contamination from other molecules.

To determine the extent of  the blue- and red-shifted emission for different molecules, we compute LVINE maps (see Sect.\,\ref{ss:lvineso}). However, this is not an easy task because line emission is pervasive over large spatial scales, hence the risk of contamination by emission from another species is high, as was seen by the comparison of \etc and HNCO spectra in Fig\,\ref{fig:lvine_spec}. 
In these spectra we mark the peak velocities derived from the template line, $\varv_t$, with dashed black lines and the integration limits for the blue- and red-shifted emission with dashed blue and red lines, respectively. This shows that with the defined inner integration limits, we avoid (almost) all emission 
from the component at $\sim\varv_t$, however, in that way parts of the lower-velocity blue- and red-shifted emission are also partly excluded. Moreover, as described above, the velocity ranges defined for integration most likely contain contaminating emission from other species. In order to reduce the contamination in the LVINE maps for the blue- and red-shifted emission, we had to develop a different strategy, in which we made use of two transitions for each molecule to derive the LVINE maps. This procedure is explained in Appendix\,\ref{ass:lvine} and illustrated with the filled histograms (and coloured difference spectrum) in Fig.\,\ref{fig:lvine_spec}. 

In Fig.\,\ref{fig:lvine_coms} we show the final LVINE maps for a variety of COMs and simpler molecules, where the grey scale 
shows the continuum map at 99\,GHz as in Fig.\,\ref{fig:so}. 
%
Overall, the bipolar structure with blue-shifted emission extending to the (south)east and red-shifted emission to the (north)west can be identified for S- and N-bearing molecules. For O- and (N+O)-bearing species this  bipolarity is less striking, if evident at all.
None of the other molecules shows a secure detection of the narrow feature observed for SiO and both transitions of SO (labelled aB1, aR1, aR2 in Fig.\,\ref{fig:so}), which might be the result of the absence of the molecule at these far distances, but could also be an issue of insufficient sensitivity or excitation. 
At some positions that show bright continuum emission from the dense core, there is also a substantial overlay of blue- and red-shifted emission. Especially in this region, we cannot reliably determine whether the  outflow is the only origin of the emission, also given that some blue and red contours follow almost exactly the contour of the continuum emission, which makes the interpretation of the emission morphology more challenging in this region.  
Moreover, the filaments identified in the continuum emission (see Fig.\,\ref{fig:overview}) are also observable in blue- and red-shifted molecular emission \citep[][]{Schwoerer19}, which presents another difficulty in this decision. 
In addition to these general trends, a lot of structure is observed in the morphology of blue- and red-shifted emission that is described in further detail in Appendix\,\ref{app:detailedmorph}. To identify at which velocities the collimated features aB1, aR1, and aR2 appear and to denote additional low- or high-velocity features, each LVINE map is split into two, showing lower-velocity and higher-velocity emission (see Figs.\,\ref{fig:COMof}--\ref{fig:COMof3}), respectively. Moreover, velocity-channel maps are presented in Fig.\,\ref{fig:chan_maps}.

\begin{figure*}[thp!]
    \centering
    \includegraphics[width=.342\textwidth]{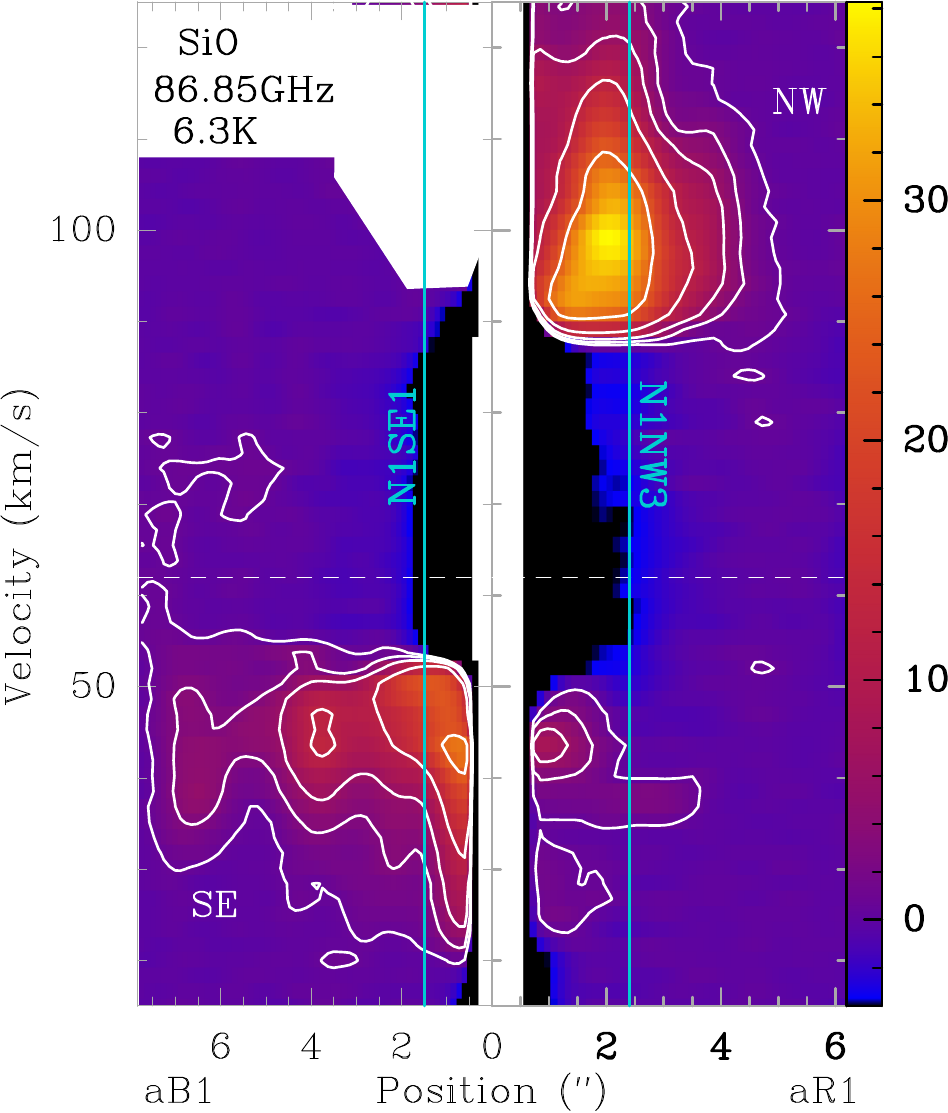}\hspace{0.1cm}
    \includegraphics[width=.302\textwidth]{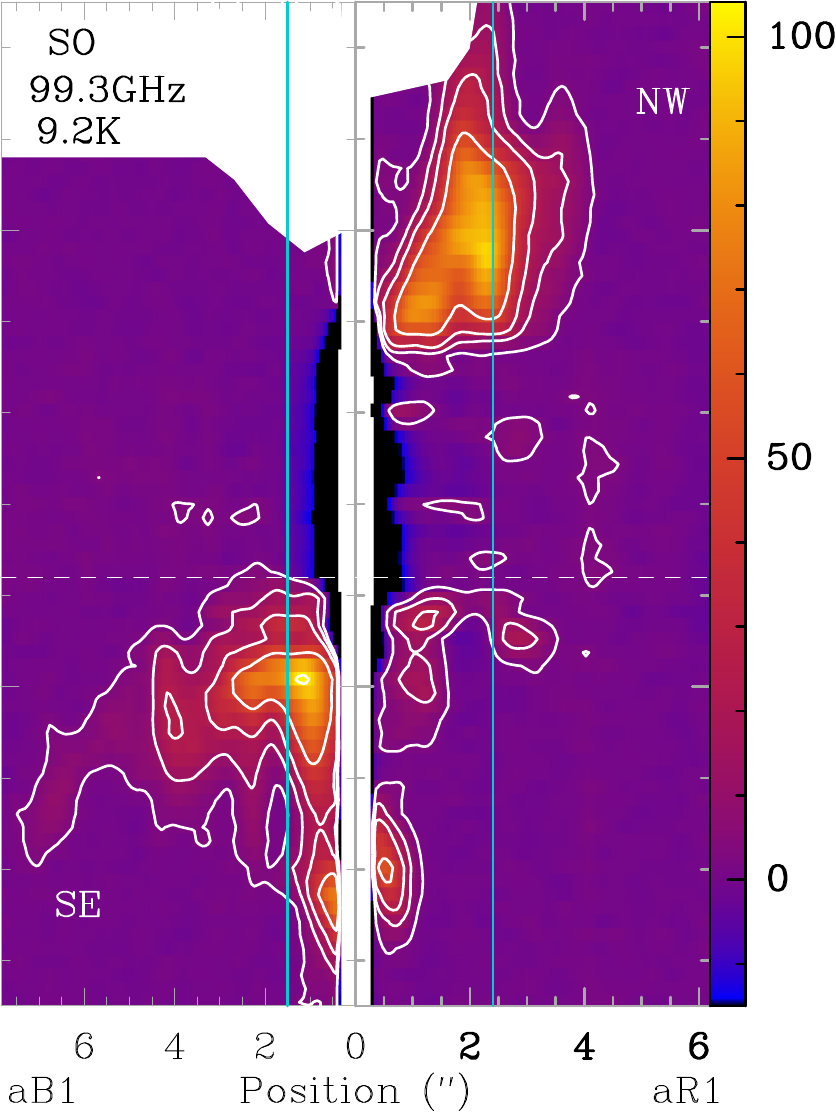}\hspace{0.1cm}
    \includegraphics[width=.318\textwidth]{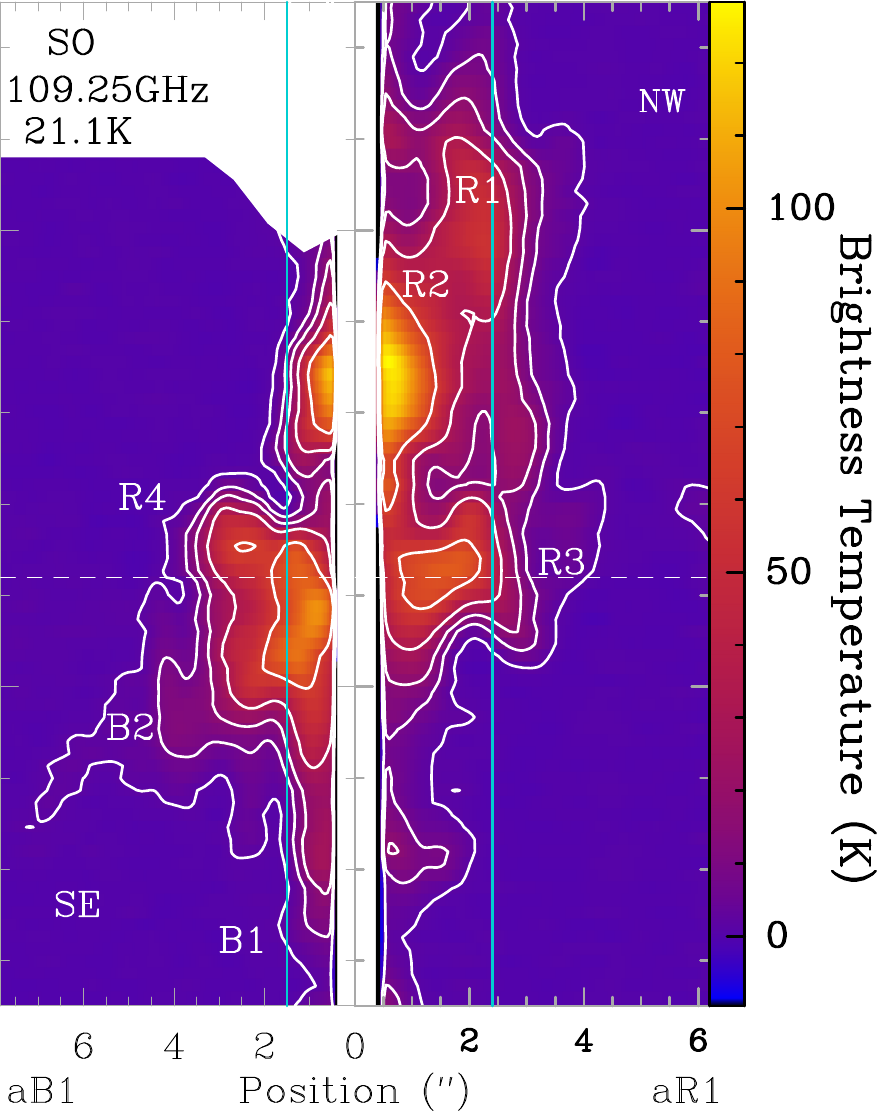}
    \caption{Position-velocity (PV) diagrams of SiO at 86.8\,GHz and SO at 99.3\,GHz and 109.252\,GHz taken along the solid black arrows labelled aB1 and aR1 in Fig.\,\ref{fig:so}. Here, the position labelled 0 corresponds to the centre of the hot core. Contours are at 5$\sigma$, 25$\sigma$, and then increase by a factor 2, where $\sigma=0.12$\,K for SiO and 0.5\,K and 0.32\,K for the two SO transitions, respectively, and was measured in an emission-free region in the respective data cubes. Pixels with intensities less than $-$30$\sigma$ are shown in black. The positions N1SE1 and N1NW3 are indicated with light-blue solid lines. Regions close to the centre and those containing contaminating emission from other species are masked in white. The white dashed line marks an average systemic velocity of 62\kms. The frequency and upper-level energy of the respective transition are written in the upper-left corner. Highlighted features in blue-shifted emission towards the southeast (SE, in the position-position maps) include B1 (elongated along velocity axis), B2 (elongated along both axes). Intensity peaks in red-shifted emission towards the northwest (NW) are labelled R1 and R2, and red-shifted emission close to the systemic velocity are labelled R3 (NW) and R4 (SE).}
    \label{fig:PV_SO}
\end{figure*}

\subsection{Position-velocity diagrams}

Figure\,\ref{fig:PV_SO} shows position-velocity (PV) diagrams of the two transitions of SiO and SO used in Fig.\,\ref{fig:so}, which have both an upper-level energy in temperature unit lower than $<$10\,K, and the second transition of SO with a higher upper-level energy of 21.1\,K. Positions close to the hot core's centre and those containing emission from other species are masked. The two transitions at lower upper-level energies present deep absorption close to the hot-core centre between velocities of $\sim$50 and 85\kms for SiO and in a slightly narrower velocity range for SO. The SO transition at 109\,GHz with slightly higher upper-level energy is not as heavily absorbed and presents emission over the whole velocity range from $\sim$25\kms to 120\kms.   
Towards the southeast along aB1, we highlight two features: B1 is spatially compact, but elongated along the velocity axis reaching highest blue-shifted velocities ($\lesssim$\,30\kms), while B2 is spatially the most extended feature that still reaches down to 30\kms. The latter represents the blue-shifted emission that extends farthest in the position-position maps (see, e.g. the velocity-channel map at 40\kms for SO in Fig.\,\ref{fig:chan_maps}). Given that B1 is more compact and faster than B2, the latter may present an older ejection event that decelerated while gaining a greater distance from the hot-core centre, which hints at episodic ejection. 
Due to its closeness to the centre, a feature in the position-position maps cannot unambiguously be assigned to B1.
For the red-shifted emission along aR1, there are at least two intensity peaks recognisable for the SO transition at 109\,GHz. There is a bright feature at $\sim$82\kms close to the hot-core centre (labelled R2), which is not prominent in the other two PV diagrams due to absorption at these velocities. The second peak, labelled R1, at $\sim$100\kms is observed for all transitions. These features likely correspond to intensity peaks P5 (for R2) and P6 (for R1) that were identified in the LVINE maps of SO and other species in Sect.\,\ref{app:detailedmorph}. Spatially extended emission at $\sim$55 to 65\kms labelled R3 (towards NW) and R4 (SE) can likely be associated with the hot core itself. The emission from the SO transition at 109\,GHz also reveals that within distances of $\sim$2\arcsec\, there is blue- and red-shifted in either direction even at velocities far from the systemic velocity. It is not entirely excluded that at least some of this emission comes from other species, especially at closest distances to the hot-core centre, where line emission is pervasive.

Figure\,\ref{fig:PV} shows PV diagrams of all the molecules for which we show LVINE maps in Fig.\,\ref{fig:lvine_coms} and, in addition, for \mic, \et, \mf, and \dme. The closest regions to the centre of the hot core and regions that are certainly contaminated with emission from other species are masked in white. We compare the distribution of emission of the various molecules to that of the SO transition at 109\,GHz, which is again shown in this figure in the upper left panel and its contours (black) in all other panels. 
Emission along B2 is clearly identified for all S- and N-bearing molecules, HNCO, and \met. Emission from all other O-bearing molecules, \fmm, and \mic remains compact in both the spatial and velocity domains. The highest-velocity feature B1 is, besides in SO emission, evident in emission of OCS, HC\3N, \vc, and \etc. It may also be seen for some other molecules such as SO\2 and HC\5N, however, we identified contaminating emission at these positions and velocities for these two molecules. Although the emission for the four molecules mentioned above can likely be associated with B1, we cannot definitely exclude that there may be contamination. 
At red-shifted velocities, R2 can be identified in emission of N- and simple S-bearing molecules, HNCO, and \met, for some more prominently than for others. However, the emission of all these molecules does generally not peak at velocities as high as for SO. Emission at R1 is evident for SO, SO\2, OCS, and HC\3N, faintly for \etc. Additionally, in the PV diagrams of SO, OCS, \etc, and some of the O-bearing COMs there may be some kind of cavity evident at around 70\kms and $\sim$1.5\arcsec\,\,in the northwestern direction.

\subsection{Radiative transfer analysis}\label{s:radTransfer}
In order to understand the impact of the outflow on the gas molecular inventory, we derived molecular abundances for two positions, one in each lobe of the outflow, and we compare them to results obtained previously in Paper\,I for the
hot core (Sect.\,\ref{ss:columns}) and later to results of other sources (Sect.\,\ref{dss:sources}). In this section, we first describe the process of selecting the positions and molecules for the analysis, before presenting the results derived from the radiative transfer modelling and population diagram methods. 

\subsubsection{Line and position selection}\label{ss:selection}
We adopted the selection of COMs from Paper\,I, which includes \met, \et, \dme, \mf, \ad, \vc, \etc, \fmm, and \mic. We added a few more species, which are CH\3CN, HC\3N, HC\5N, NH\2CN, and HNCO. Some of these may not have been analysed in Paper\,I because there were not enough transitions when considering setups 1--3 and 4--5 separately or due to high line optical depth out to large distances, or because the molecules' lines were too weak to be considered in the previous study. Moreover, we included the S-bearing species OCS and \mmc, both of which are expected to be indicators of shock chemistry. We do not analyse SO\2 here because there are not enough transitions available for a reliable radiative transfer modelling. We have not shown LVINE maps for CH\3CN, because the molecule's various $K$-ladder transitions for a given $J$ are blended. 

\begin{figure*}[htp]
    \centering
    \includegraphics[width=\textwidth]{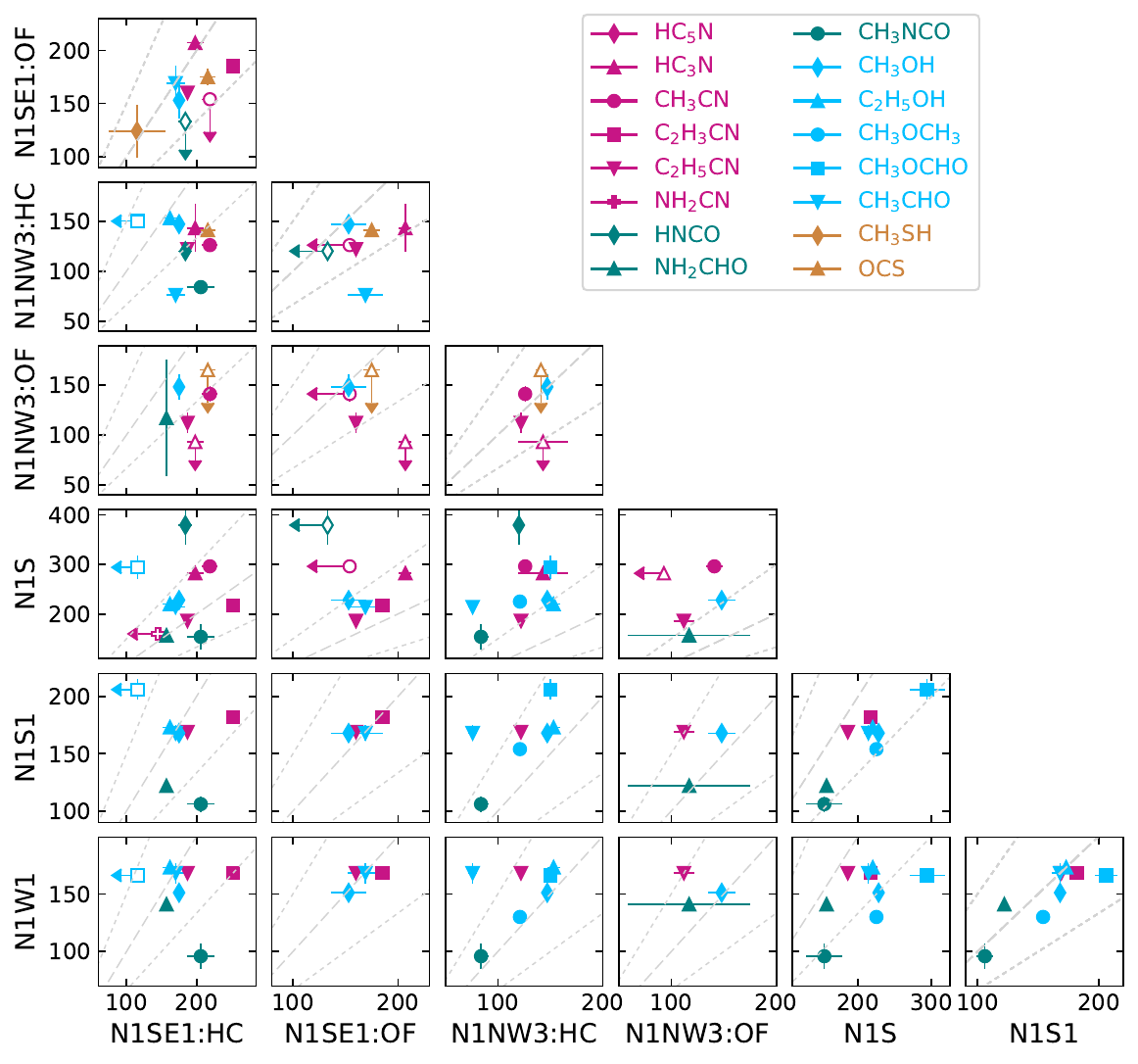}
    \caption{Rotational temperatures (in K) for various positions towards Sgr\,B2\,(N1) derived with the population diagram analysis. Pink markers indicate N-bearing species, teal markers (N+O)-bearers, blue O-bearers, and orange S-bearers. Arrows indicate upper limits. The grey dashed line shows where temperatures are equal. The two grey dotted lines indicate a factor 1.5 difference from unity.}
    \label{fig:corner_T}
\end{figure*}
\begin{figure}[htp]
    \includegraphics[width=.49\textwidth]{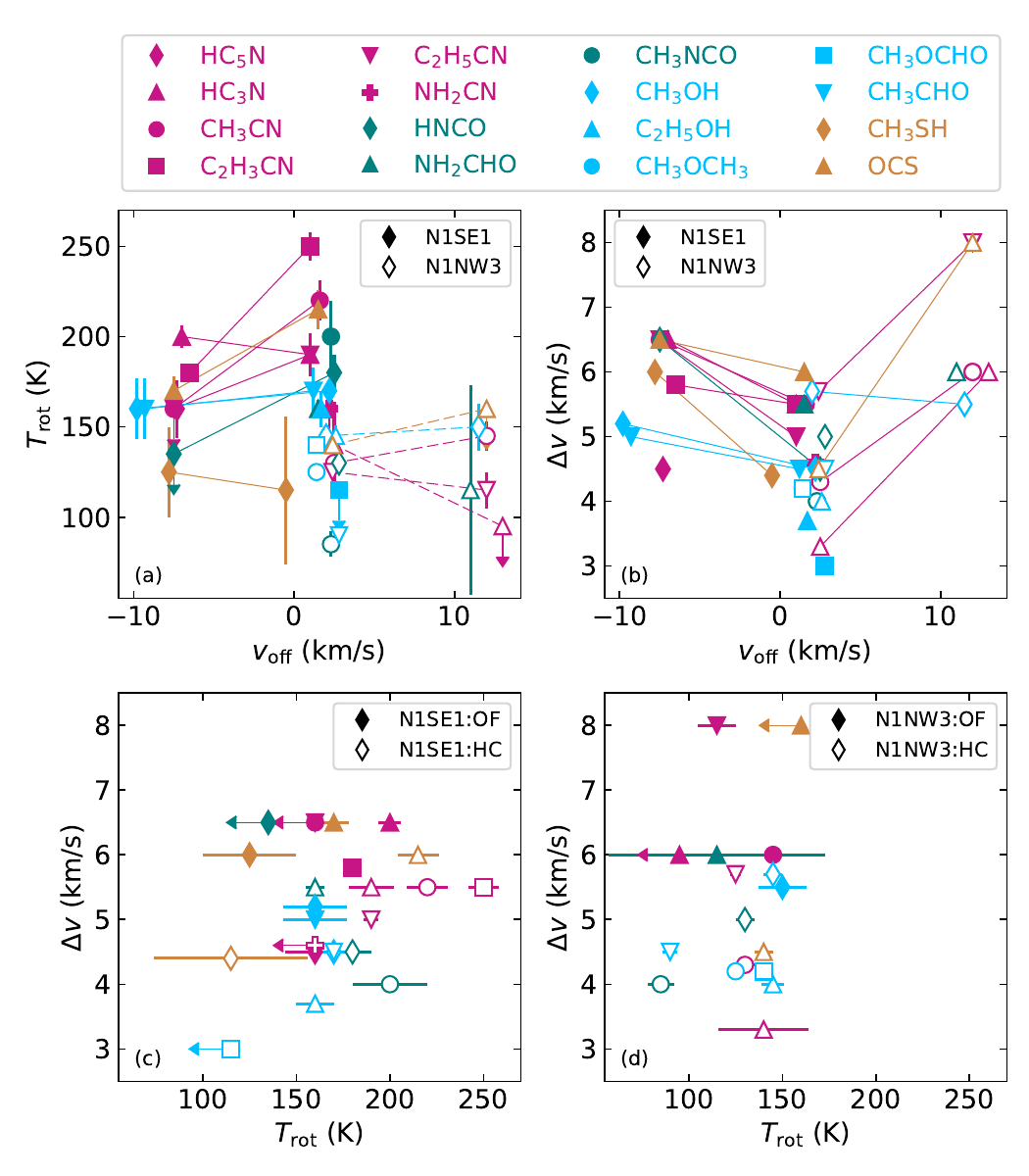}
    \caption{Correlation plots for Weeds model parameters: \textit{ Panel a:} Velocity offset $\varv_{\rm off}$ from the systemic velocity ($v_{\rm sys}=62$\kms) versus rotational temperature $T_{\rm rot}$. \textit{Panel b:} Offset $\varv_{\rm off}$ versus linewidth $\Delta\varv$ ($FWHM$). \textit{Panels c--d:} Temperature $T_{\rm rot}$ versus linewidth. Pink markers indicate N-bearing species, teal markers (N+O)-bearers, blue O-bearers, and orange S-bearers. Markers showing $|\varv_{\rm off}|$ values larger than 7\kms in (a) and (b) correspond to the blue- and red-shifted components of positions N1SE1 (filled markers) and N1NW3 (empty markers), respectively. In panels (c) and (d) empty markers present the hot-core component at N1SE1 and N1NW3, respectively. Uncertainties on the temperature values are taken from the results of the population diagrams. Arrows indicate upper limits. }
    \label{fig:kinematics}
\end{figure}

We looked for a position to the SE along the collimated high-velocity feature seen in SO emission labelled aB1 in Fig.\,\ref{fig:so} and chose N1SE1, which is at a distance of 1.5\arcsec\,from the continuum peak of Sgr\,B2\,(N1). The position naming is based on that started in Paper\,I and depends on the distance to the continuum peak. This position is sufficiently distant from the centre of the hot core to not be too contaminated by the pervasive line emission arising in the hot core itself, while showing intense blue-shifted emission, where detected (see Fig.\,\ref{fig:lvine_coms}). 
The position selection in the red-shifted lobe was more difficult to do, because the red-shifted emission is not as intense as the blue-shifted emission in general. We looked for a position along one of the outflow axes identified in the red-shifted SO emission maps, that is either along aR1 or aR2. 
Moreover, we searched for a position that clearly showed a red-shifted component in \met as we want to compute abundances with respect to this COM (among others) in the following in order to compare between the positions and with other sources. At a distance of $\sim$2\arcsec\,\,along aR1, there is a peak in the continuum map (see Fig.\,1 in Paper\,I) and in emission of some COMs that may be associated with another source. Closer to the centre of Sgr\,B2\,(N1), the component close to the systemic velocity and the red-shifted one become hardly distinguishable, which makes it difficult to model them. Therefore, we selected position N1NW3 at a distance of $\sim$2.5\arcsec along aR1. 
In the following, we distinguish the pair of velocity components at a given position by adding HC for the hot-core component close to the systemic velocity and OF for the supposedly outflow component at red- or blue-shifted velocities. 
In Figs.\,\ref{fig:specs_o}--\ref{fig:specs_nos} we show a selection of transitions for each selected molecule towards the two positions in order to validate the presence of at least one component in addition to the hot-core component.

The observed spectra of the selected species were modelled with Weeds (see Sect.\,\ref{ss:weeds}) and population diagrams were derived (see Sect.\,\ref{ss:PD}) for both components at N1SE1 and N1NW3 in order to obtain rotational temperatures and column densities. 
To compare with results that were previously obtained in Paper\,I, we additionally use column density values derived at positions N1S (at a distance of 1\arcsec\,from the hot core's centre to the south), N1S1 (1.5\arcsec\,to the south), and N1W1 (1.5\arcsec\,to the west). All positions are marked in Fig.\,\ref{fig:overview}.

\subsubsection{Temperatures and velocities}

First, we investigate whether the outflow positions N1SE1 and N1NW3 show significant differences in the derived rotational temperatures compared to the positions that are not exposed to the outflow and potential shocks that are associated with it.
In Fig.\,\ref{fig:Tweeds} we compare the rotational temperatures used to obtain the Weeds models with the results from the population diagrams at each position. The temperature values derived from the latter method deviate only marginally if at all from those used in the models, thereby validating the models.
In Fig.\,\ref{fig:corner_T} we compare the rotational temperatures of the various velocity components and positions in the hot core and in the outflow.
Assuming that positions N1SE1 and N1NW3 experience shocks due to their location in the outflow lobes, one might expect elevated temperatures as a consequence of these shocks.  However, in this regard the outflow positions do not stand out compared to the previously analysed positions (N1S, N1S1, and N1W1).
Interestingly, at N1SE1, temperatures derived for the blue-shifted component (N1SE1:OF) are lower than or similar to the values in the hot-core component (N1SE1:HC). In addition, N-bearing molecules tend to have slightly higher rotational temperatures with a mean of $\sim$210\,K (excluding NH\2CN) than the O-bearing molecules with $\sim$170\,K at N1SE1:HC. With only two O-bearing COMs detected at N1SE1:OF such a trend cannot be clearly identified.

In addition, we explore possible correlations between velocity, linewidth, and temperature in Fig.\,\ref{fig:kinematics}. 
Figure\,\ref{fig:kinematics}a shows temperatures in comparison with the velocity offset from the source velocity $\varv_{\rm sys}=62$\kms, where markers at $\varv_{\rm off}<-7$ and $>$11\kms correspond to the blue- and red-shifted emission, respectively. Each set of marker and colour represents one molecule. Filled markers represent N1SE1, unfilled markers N1NW3. When the molecule is detected in both the hot-core and outflow components at a given position, these two identical markers are connected by a line. This shows again that, in general, temperatures are lower in the outflow (OF) than in the hot-core component (HC) at N1SE1, except for \mmc and HC\3N, whose temperature values, however, have a higher uncertainty. 
A clear trend is not visible for N1NW3, only the much lower temperature for HC\3N in the outflow component stands out. 
Figure\,\ref{fig:kinematics}b shows the distribution of linewidths as a function of velocity offset, using the same marker and color scheme as in Fig.\,\ref{fig:kinematics}a. The outflow components generally show larger linewidths, which may arise from a higher degree of turbulence or blending of multiple narrower, unresolved velocity components. 
At N1NW3, the linewidth derived for methanol in the hot core seems to be larger than in the outflow, however, the outflow component is fairly weak in emission and may come with greater uncertainty, also in linewidth. 
Figures\,\ref{fig:kinematics}c and d show the distribution of line widths as a function of rotational temperature
for N1SE1 and N1NW3, respectively, however, no trend is visible.


\begin{figure}[tb]
    \centering
    \includegraphics[width=.243\textwidth]{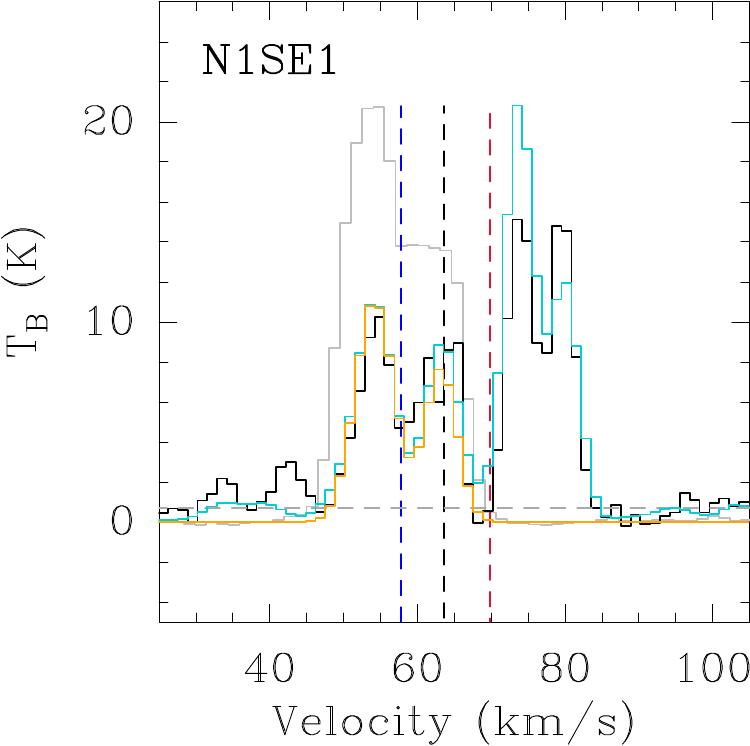}\hspace{0.3cm}
    \includegraphics[width=.215\textwidth]{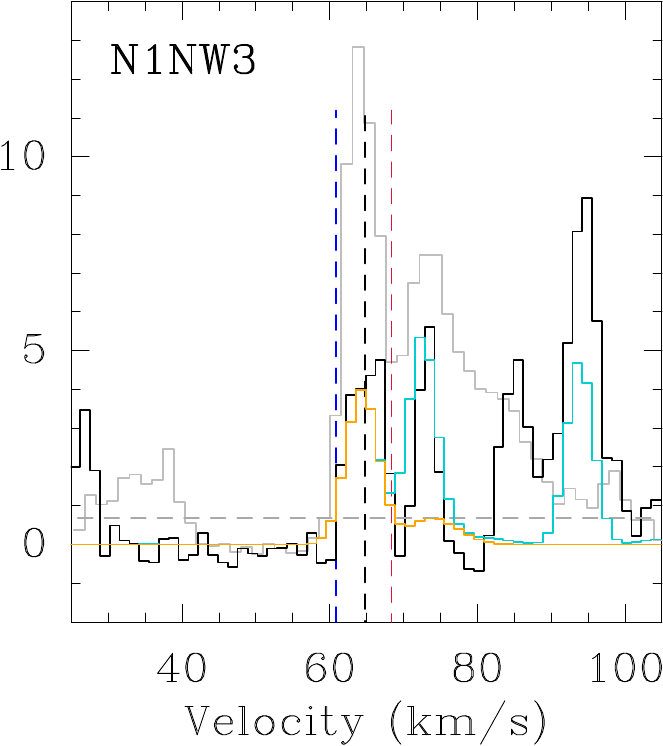}
    \caption{Observed spectrum of C$^{18}$O $J=1-0$ (black) overlaid by the respective Weeds models for both the outflow and hot-core components (orange) and the total Weeds model of all yet identified species at these positions (blue). For comparison, the grey spectrum shows the ethyl cyanide transition at 96.92\,GHz scaled down by a factor 5 for N1SE1 and a factor 2 for N1NW3 (see also Fig.\,\ref{fig:lvine_spec}). The vertical black dashed lines mark the systemic velocities, while vertical red and blue dashed lines show the pixel-dependent inner limits used to integrate the blue- and red-shifted emissions (see also Fig.\,\ref{fig:lvine_spec}). The grey horizontal line indicates the 3$\sigma$ level, where $\sigma=0.23$\,K.}
    \label{fig:specCO}
\end{figure}
\begin{figure*}[htp]
    \centering
    \includegraphics[width=\textwidth]{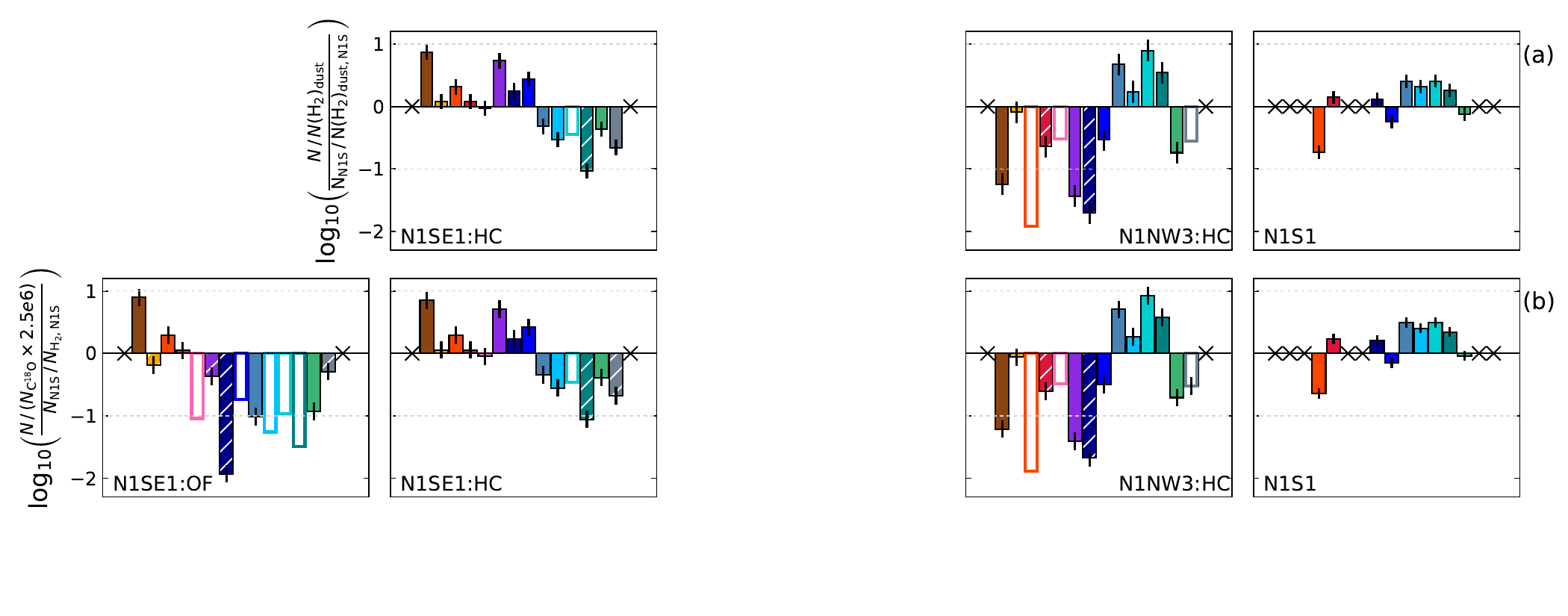}\\[-1cm]
    \includegraphics[width=\textwidth]{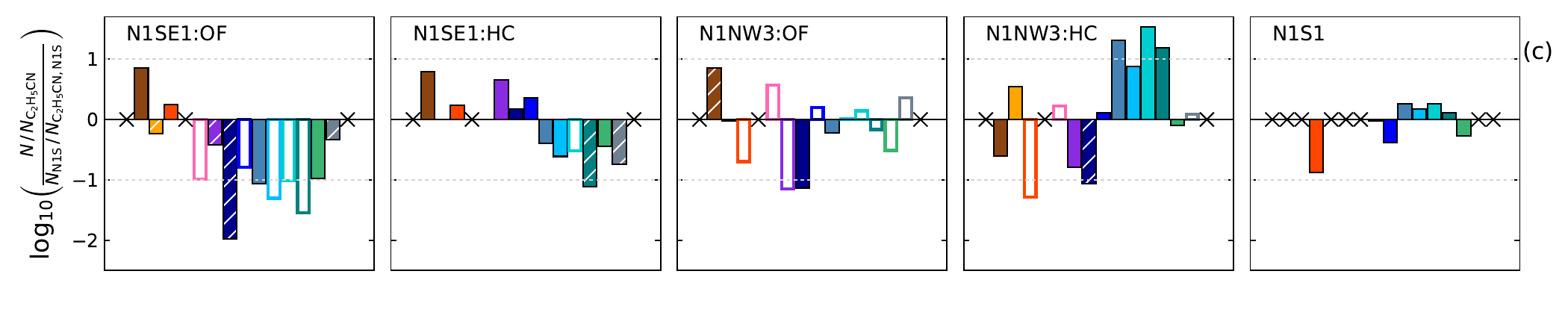}\\[-0.5cm]
    \includegraphics[width=\textwidth]{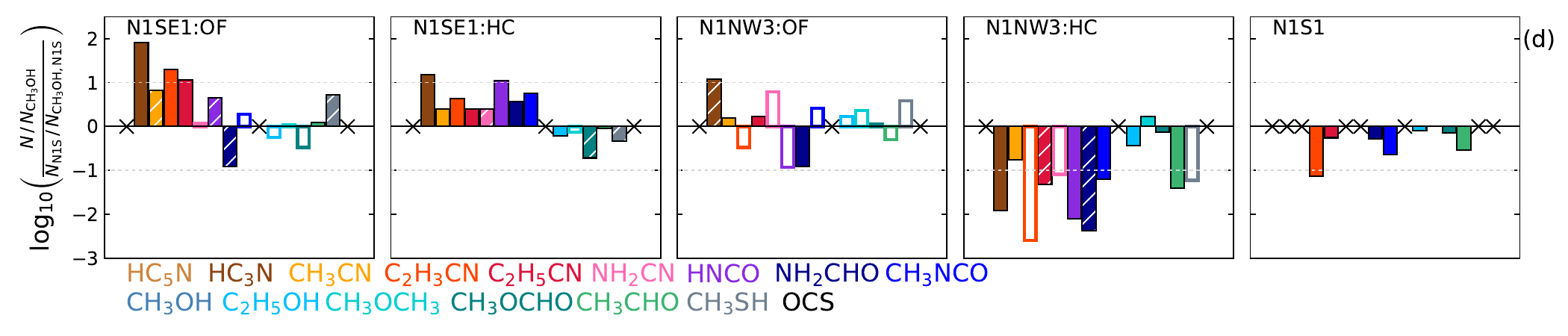}
    \caption{Comparison of the gas molecular content in the hot-core and outflow components at N1SE1 and N1NW3 amongst each other and with positions N1S and N1S1, which were analysed in Paper\,I. \textit{Rows a--b:}  Abundances with respect to H\2 normalised to the value derived for N1S, where H\2 column densities were derived from a) dust emission at 242\,GHz and b) C$^{18}$O for all components but N1S, for which the value from dust emission was used in both rows. Component N1NW3:OF is not shown, because C$^{18}$O is  difficult to identify due to contamination by another molecule. \textit{Rows c--d:} Same as (b), but abundances with respect to \etc and \met are shown, respectively. In all panels, hatched bars indicate when the rotational temperature in a population diagram (PD) was fixed (NH\2CN at N1SE1:HC and \mmc at N1S), the molecule was detected but a PD could not be derived (\fmm at N1SE1:OF), or only an upper limit for the rotational temperature in the PD was derived. Empty bars show upper limits. HC\5N and OCS are not shown as we do not have the column densities of the two molecules at N1S.}
    \label{fig:Xbars-norm}
\end{figure*}

\subsubsection{Column densities and abundances}\label{ss:columns}
We show the column densities that were used for the Weeds models in Fig.\,\ref{fig:Nbars}. The values are similar to those derived from the fit in the population diagrams underlining their reliability (cf. Tables\,\ref{tab:n1se1hc}--\ref{tab:n1nw3of}). Following Paper\,I, we multiplied the column densities by a temperature-dependent vibrational correction factor, when necessary.
Unfilled bars indicate upper limits on the column density, which were obtained by fixing all other parameters in the models to median values that were derived from other molecules. 
Figure\,\ref{fig:Nbars} shows in addition the column densities derived in Paper\,I for the positions N1S, N1S1, and N1W1 (see also Fig.\,\ref{fig:overview} to locate these positions). For molecules that were not analysed in Paper\,I, we computed a Weeds model and derived a population diagram (HC\3N, CH\3SH, and HNCO, see Fig.\,\ref{fig:PD_ch3sh}) or took values from other studies (2.8$\times$10$^{18}$\,\scm for CH\3CN and 2.6$\times$10$^{16}$\,\scm for NH\2CN from \citet{Mueller21} and \citet{Kisiel22} towards N1S, respectively). For positions N1S1 and N1W1, these molecules were not considered. 

To better compare the chemical compositions between the outflow and hot-core components and, later, between the results obtained for Sgr\,B2\,(N1) and those of other sources, we compute abundances with respect to H\2, \met, and \etc.
Deriving abundances with respect to H\2 is difficult because there may be dust emission for the hot-core components, however, this cannot simply be applied to the outflow components. Therefore, we estimated the H\2 column densities from C$^{18}$O $J=1-0$ emission by fixing all parameters in the Weeds model, except for the column density, to average values derived for other molecules for a given component. The observed spectrum and the corresponding Weeds model are shown in Fig.\,\ref{fig:specCO} for N1SE1 and N1NW3. There is a minor contribution from the blue-shifted component of HNCO,\,$v=0$ to the hot-core component of C$^{18}$O at N1SE1, otherwise, the C$^{18}$O spectral lines seem clean, although we cannot exclude further minor contamination by other species.
If the assumed rotational temperature in the Weeds models were higher or lower by 80\,K, the C$^{18}$O column densities would differ by at most a factor 2. 
Assuming that CO remains a good tracer of H\2 column densities at these positions impacted by the outflow, we multiply the C$^{18}$O column densities with a C$^{16}$O-to-C$^{18}$O ratio of 250\,$\pm$\,30 \citep{Henkel94} and a H\2-to-CO conversion factor of 10$^{4}$ to obtain H\2 column densities. This yields $N_{\rm H_2}$(N1SE1:HC)$\,= (1.0\pm 0.3)\times 10^{24}$\,\scm, $N_{\rm H_2}($N1SE1:OF)$\,= (1.6\pm0.5)\times 10^{24}$\,\scm, and $N_{\rm H_2}$(N1NW3:HC)$\,= (3.8\pm 1.2)\times 10^{23}$\,\scm. The errors include an uncertainty on the C$^{18}$O column density, where we assume $\Delta N_{\rm C^{18}O}=0.3N_{\rm C^{18}O}$, and the uncertainty on the CO isotopologue ratio.  
The values at N1SE1 are a factor 2--3 lower than the one at N1S1 shown in Fig.\,8 in Paper\,I, which is at the same distance from the centre of Sgr\,B2\,(N1) to the south, and a factor 10 lower than at N1S, which is 0.5\arcsec\,closer to the centre. The difference in H\2 column density between N1NW3:HC and N1W3, which are at the same distance, is less than a factor 2. The C$^{18}$O transition is not detected for N1NW3:OF because of a blend with stronger emission from HNCO,\,$v=0$. The H\2 column densities at N1S and N1S1 are taken from Table\,E.24 in Paper\,I, where for the latter we use the value derived from C$^{18}$O, while for N1S, the value derived from dust emission had to be used, because the C$^{18}$O transition was also seen in absorption whose contribution could not be determined. 

In Figs.\,\ref{fig:Xbars-norm}a and b, we show abundances with respect to H\2 normalised to the values at N1S. The H2 column densities were derived either from a) dust continuum emission at 242\,GHz \citep[][Paper\,I]{Sanchez-Monge17} or b) C$^{18}$O emission. For the former we assume that the dust temperature is equal to the rotational temperature of ethanol as done in Paper\,I, which yields $N_{\rm H_2}$(N1SE1:HC)$\,= (9.5\pm 2.7)\times 10^{23}$\,\scm and $N_{\rm H_2}$(N1NW3:HC)$\,= (4.0\pm 1.6)\times 10^{23}$\,\scm. The error bars include an uncertainty on the dust temperature of 20\% and on the continuum level, that is the baseline, of 1$\sigma = 1.67$\,K.  
The abundances with respect to H\2 revealed that \etc abundances are comparable at N1SE1, N1S, and N1S1. Therefore, we additionally show abundances with respect to this COM in Fig.\,\ref{fig:Xbars-norm}c. Figure\,\ref{fig:Xbars-norm}d shows abundances with respect to methanol as this COM is commonly used to compare the chemical inventories between different sources.\\
\\
\textit{Outflow (OF) components:} 
Abundances with respect to H\2 and \etc (Fig.\,\ref{fig:Xbars-norm}a--c) reveal that the O- and (N+O)-bearing molecules are less abundant in N1SE1:OF than at N1S, also N1S1, by factors of a few up to an order of magnitude, or even almost two orders of magnitude in the case of \fmm. The O-bearing molecules \et, \dme, and \mf are not even detected in the outflow components, only \met and \ad are. On the other hand, HC\5N is only securely detected in the outflow component at N1SE1. The molecule is not shown in Fig.\,\ref{fig:Xbars-norm} because it is not detected at N1S.
Abundances of CH\3CN and \vc at N1SE1:OF differ only slightly from the values at N1S, while HC\3N is more abundant.
At N1NW3:OF, only HC\3N, CH\3CN, \fmm, and \met are detected and follow similar trends as at N1SE1:OF, except that the \met abundance is more similar to N1S. 
The comparison with abundances with respect to \met presents a quite opposite trend, at least for N1SE1:OF, where abundances for N-bearing molecules, except for NH\2CN, are higher by roughly 1--2 orders of magnitude in this component than for N1S and N1S1. Also, \mmc abundances are enhanced at N1SE1:OF. However, \fmm abundances are lower in both outflow components compared to N1S, as was seen in abundances with respect to \etc. Abundances of HNCO are higher in N1SE1:OF and lower in N1NW3:OF than in N1S. \\
\\
\textit{Hot-core (HC) components:}
At N1SE1:HC, O-bearing molecules and \mmc are less abundant than at N1S and N1S1, similarly to the outflow component at this position but not as severely. Also similar to N1SE1:OF, abundances of N-bearing molecules, except for HC\3N, are comparable at N1SE1:HC and N1S. In contrast, (N+O)-bearing species are either similarly or more abundant in N1SE1:HC than in N1S, N1S1, and N1SE1:OF. The hot-core component at N1NW3 behaves differently compared to N1SE1:HC: in general, N- and (N+O)-bearing molecules are less abundant in N1NW3:HC, while O-bearing species are enhanced, except for \ad.
Abundances with respect to \met reveal an enhancement of N- molecules in N1SE1:HC compared to N1S and N1S1, similar to the outflow component, but not as prominent. Moreover, all (N+O)-bearing species have enhanced abundances in N1SE1:HC. Abundances of O-bearing species are comparable to the values found for N1S and N1S1. At N1NW3:HC, abundances with respect to methanol for N- and (N+O)-bearing molecules as well as \ad are lower than at N1S and N1SE1:HC by 1--3 orders of magnitude. Most O-bearing species, except for \ad, show similar abundances to N1S and N1S1. \\
\\
In summary, we do not only identify differences in the gas molecular inventory between the blue- and red-shifted components and all hot-core components analysed here, but also between the hot-core components at positions N1SE1 and N1NW3 and at N1S and N1S1. 
The most remarkable result are the much lower abundances of O-bearing molecules in the outflow component N1SE1:OF with respect to H\2 and \etc compared to N1S, while abundances of N-bearing molecules are comparable to N1S (except for HC\3N).
A similar trend is seen for the comparison between the hot-core component at this position N1SE1:HC and N1S. In contrast, (N+O)-bearing molecules are more abundant in N1SE1:HC and less abundant in N1SE1:OF than at N1S.
The comparison between abundances with respect to H\2 (or \etc) and to \met shows that, in this case, the latter may not be the best molecule to normalise to, as its abundance with respect to H\2 changes a lot between the different components. 

\section{Discussion}\label{s:discussion}

\subsection{Morphology of blue- and red-shifted emission}\label{dss1}

The blue- and red-shifted SiO and SO emission shown in Fig.\,\ref{fig:so} revealed a bipolar nature, 
however, especially at close distances to the centre of Sgr\,B2\,(N1), more structure is seen that is not simply bipolar. 
Because of the complex morphology, not only seen in the maps of SO and SiO, but also other molecules (see Figs.\,\ref{fig:lvine_coms}, \ref{fig:COMof}--\ref{fig:COMof3}, and \ref{fig:chan_maps}), it is challenging to disentangle the contribution of the outflow and other sources, such as the dense core itself or filaments, and to connect this to the gas molecular inventory that we have derived for blue- and red-shifted velocity components. 
Because of this complex structure, one might speculate whether Sgr\,B2\,(N1) is a site of an explosive outflow event. For example, this was observed in Orion \citep[e.g.][]{Zapata17}, where such an event is not only characterised by a large number of observed finger-like structures originating from a common location, but also emission from COMs has been detected in the surrounding regions embedded in Orion\,KL that are supposedly impacted by the explosion \citep[e.g.][]{Zapata11,Favre17,Pagani19}. \citet{Zapata17} reported on the observational differences of regular protostellar and explosive outflows. Although there are multiple finger-like structures  identified in the emission morphology of SO and SiO in Sgr\,B2\,(N1) (labelled aB1 and aR1--2 in Fig.\,\ref{fig:so} and, potentially, aB2 and aR3 in Fig.\,\ref{fig:COMof}), the rather clear separation of blue- and red-shifted emission with their spatial extension to the (south)east and (north)west, respectively, with at least some degree of collimation, suggests that we see a protostellar outflow. However, as also noted by \citet{Schwoerer20}, there might exist several protostars hidden from us by dust at the centre of Sgr\,B2\,(N1) that could each drive an outflow. Still, the clear bipolar structure may speak against such a scenario of multiple driving sources unless there is a mechanism that roughly aligns these outflows in the same direction. 

We highlighted three collimated features in emission of SO and SiO that we labelled aB1, aR1, and aR2 in Fig.\,\ref{fig:so}. Their high degree of collimation, their detection at extremely high blue- and red-shifted velocities (see also velocity-channel maps in Fig.\,\ref{fig:chan_maps}), and their extension to large distances from the hot-core centre let us conclude that at least these three structures can be associated with the outflow.
Determining any physical properties for these outflow features is difficult because of the highly uncertain inclination to the observer. The opening angle of the lobes is another factor of uncertainty due to the complex morphology of the outflow emission. Considering that blue- and red-shifted emission are fairly well separated, the inclination for the most collimated feature observed in SO have a value between 10$^\circ$ and 80$^\circ$. However, if we consider all blue-shifted emission to the southeast, which is observed with a much wider opening angle (up to 60$^\circ$), not only for SO, but also for all other molecules, the inclination may rather have a value between 30$^\circ$ and 60$^\circ$.
We estimate a maximum projected length along aB1 (feature B2 in the PV diagrams in Fig.\,\ref{fig:PV_SO}) of $\sim7\arcsec$, which corresponds to 0.3\,pc, and a projected velocity of $\varv_{\rm of}= \varv_{\rm sys}-\varv_{\max} \sim 62-35 = 27$\kms (cf. channel maps of SO in Fig.\,\ref{fig:chan_maps}). If this collimated feature was however inclined by 80$^\circ$ to the line of sight (i.e. almost edge on), the true velocity would be as high as 150\kms. On the other hand, if it were seen with an inclination of 10$^\circ$, its spatial extent could reach up to 1\,pc. These most extreme values of inclination result in a lower limit on the dynamical age of 2\,kyr and an upper limit of 57\,kyr. Similar values can be obtained for the red-shifted structure along aR1. Assuming an intermediate inclination of 45$^\circ$ yields an age of $\sim$10\,kyr. 
The second blue-shifted feature B1 identified in the PV diagrams, is spatially more compact (maybe 1\arcsec) but reaches higher velocities ($\varv_{\rm of}=37$\kms) and, hence, is younger with an age of 
at most 6\,kyr for an inclination of 10$^\circ$, but it could also be younger than 1\,kyr for higher inclinations. 

The blue-shifted component at $\sim$55\kms that we have analysed at N1SE1 can likely be associated with feature B2 in the PV diagrams shown in Fig.\,\ref{fig:PV}, that is the older one. On the other hand, the red-shifted component at N1NW3, which is only detected for a handful of molecules, cannot easily be assigned to one of the features identified in the PV diagram of SO, but rather coincides with more extended, less intense SO emission.
Understanding the impact of the outflow on the emission morphology of the various molecules studied here and on their abundances derived for the outflow and hot-core components at N1SE1 and N1NW3 is the subject of the next sections. 

\subsection{Comparison to chemical composition of other sources impacted by shocks}\label{dss:sources}
In this section we compare molecular abundances that we derived with those of other sources that are known for their organic chemistry substantially driven by shocks. We look at G0.693, which is a position located not far from Sgr\,B2\,(N) in the Sgr\,B2 molecular cloud complex. As many other CMZ clouds, G0.693 was found to be rich in COMs that are similarly or even more abundant than in known hot cores and corinos despite the absence of any sign of star formation \citep[][]{Requena-Torres06,Requena-Torres08,Armijos-Abendano15}. Instead, the extraordinary conditions in the Galactic centre (GC) region were made responsible, such as large-scale shocks and an enhanced cosmic-ray flux \citep[e.g.][]{Henshaw2022}. 
The richness in COMs across CMZ clouds and the only small variations in their abundances raise the question whether these represent the initial chemical conditions of the gas that eventually forms stars in the GC, such as in Sgr\,B2\,(N).
The 40 positions in the CMZ that had been observed in the original study by \citet[][]{Requena-Torres06} corresponded to peaks in SiO emission \citep[][]{Martin-Pintado1997}. Therefore, although shocks traced by SiO emission in the CMZ are ubiquitous, these 40 positions seem to experience either frequent or exceptionally strong shocks that may cause enrichments of COMs in the gas. For example, a cloud-cloud collision has been proposed as the driver of shocks in G0.693 \citep[][]{Zeng20}. In particular, this source was the focus of many follow-up studies that reported on several new molecular detections \citep[e.g.][and references therein]{jimenez-serra2022}. This does not mean, however, that these molecules are not present in other CMZ clouds with similar physical conditions. 

In addition to the fact that the positions studied by \citet{Requena-Torres06} may only represent the material that is most strongly impacted by shocks in the CMZ, it is also likely that the chemical composition of the gas phase presently measured at these positions has been modified by the shocks, which in turn means that the current gas-phase composition of these positions may not directly probe (after desorption induced by the shocks) the chemical composition of the dust grains during the prestellar phase in the CMZ. In this sense, we do not consider the chemical composition of G0.693 as representing the chemical composition of the dust-grain mantles in Sgr\,B2 during the prestellar phase. We rather see G0.693 as revealing the gas-phase composition after the impact of shocks on prestellar gas and dust, which motivates our comparison of the outflow components in Sgr\,B2\,(N1) to this source.

Based on observations with the IRAM\,30\,m telescope and their follow-up study with the Green\,Bank\,100\,m\,Telescope (GBT) towards positions in the Galactic centre region, including G0.693, \citet{Requena-Torres06,Requena-Torres08} derived column densities for O-bearing molecules. 
After that, other studies aimed to investigate further the chemical inventory of G0.693, for example, \citet{Armijos-Abendano15} who performed observations with the Mopra\,22\,m telescope and derived column densities for many molecules, \citet{Zeng18} who studied N-bearing molecules based on observations with the IRAM\,30\,m telescope and the GBT, and \citet{Rodriguez-Almeida21a} who used the IRAM\,30\,m and Yebes\,40\,m telescopes to study primarily S-bearing molecules. 
Molecular column densities that were derived towards G0.693 in the various studies are summarised in Table\,\ref{tab:G0693}. 

The low-mass protostar L1157-mm drives an outflow that produces shocks upon encounter with the ambient interstellar medium \citep{Bachiller1997}. Towards this outflow, multiple spots of shock-excited emission have been observed that reveal a rich and complex chemistry \citep[e.g.][cf. Table\,\ref{tab:G0693}]{Benedettini07, Arce08, Lefloch17}. Amongst these, the most studied one is L1157-B1, which is located in the blue-shifted outflow lobe. It has been the target of numerous observations at lower angular resolution (see column LR in Table\,\ref{tab:G0693}), from which abundances of many molecules were derived. Interferometric studies towards L1157-B1 revealed evidence for spatial and chemical segregation of different molecules. For the clump L1157-B1b, some abundances have been derived (see column HR (B1b) in Table\,\ref{tab:G0693}).

\begin{figure}[thp]
    \centering
    \includegraphics[width=.5\textwidth]{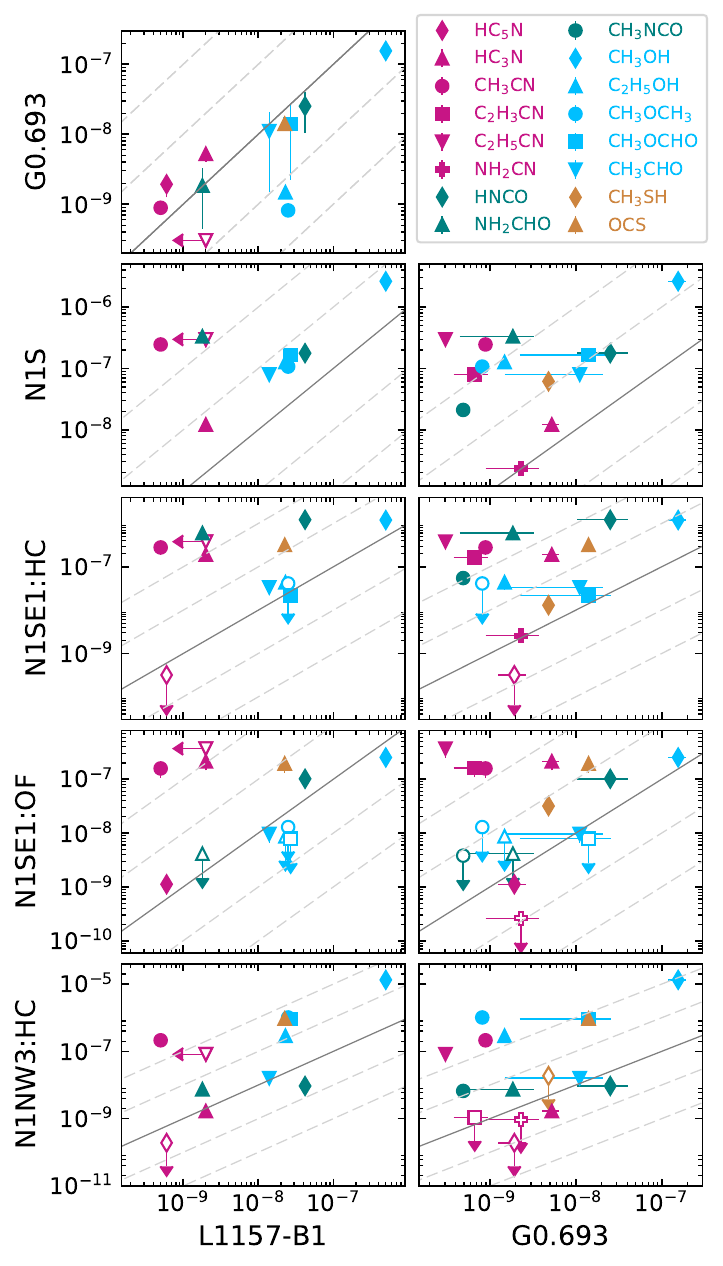}
    \caption{Molecular abundances with respect to H\2 for the outflow (OF) and hot-core (HC) components at N1SE1, for N1NW3:HC, and for N1S, G0.693, and L1157-B1. Pink markers indicate N-bearing species, teal markers (N+O)-bearers, blue O-bearers, and orange S-bearers. Unfilled markers with arrows indicate upper limits. The grey solid line shows where abundances are equal. The grey dotted lines indicate factors 10 and 100 from unity. }
    \label{fig:corner}
\end{figure}

In Fig.\,\ref{fig:corner} we compare molecular abundances with respect to H\2 between G0.693, L1157-B1, and the outflow positions N1SE1 and N1NW3 in Sgr\,B2\,(N1). The outflow component at N1NW3 is not shown as we could not derive a value for the H\2 column density. We use H\2 column densities of $1.35\times 10^{23}$\,\scm for G0.693 \citep{Martin08} and $2.0\times 10^{21}$\,\scm for L1157-B1 \citep{Lefloch17}. 
In Appendix\,\ref{ass:sources} we provide more details on the observations and the method of analysis used in the studies on these two sources. 
In the right column of Fig.\,\ref{fig:corner}, we show abundances with respect to H\2 towards G0.693, where we used the results of \cite{Zeng18} for N- and (N+O)-bearing species, except for \fmm, for which we used an updated value from \citet[][]{Zeng23}, those of \cite{Rodriguez-Almeida21a} for CH\3SH and \et, those of \cite{Requena-Torres08} for \met and \mf, and those of \citet{Requena-Torres06} for \ad. The study by \cite{Armijos-Abendano15} was performed at slightly lower angular resolution, hence, we only use their OCS column density.
In the cases where multiple studies reported on the column density of a molecule, we show this range of values as error bars in Fig.\,\ref{fig:corner}.
The larger difference in column density for HC\5N may result from different excitation temperatures that were used to model the molecule: \cite{Armijos-Abendano15} derived $\sim$32\,K from a rotation diagram analysis that is twice that of \cite{Zeng18}, who used the software \texttt{madcuba}. This may similarly apply to the results of NH\2CHO, however, it cannot explain the difference for \met as both studies assumed the same excitation temperature, however, \cite{Armijos-Abendano15} used only one transition to compute the column density. 
For L1157-B1, we used the values obtained from single-dish observations because more such data are available and because the spatial scales of a few 1000\,au probed by observations with the IRAM\,30\,m telescope towards L1157-B1 \citep[at a distance of 352\,pc,][]{Zucker19} are more similar to those probed with the ReMoCA survey towards Sgr\,B2\,(N1). The values are shown in the left column of Fig.\,\ref{fig:corner}. 

In the top panel of Fig.\,\ref{fig:corner}, we compare abundances with respect to H\2 between G0.693 and L1157-B1. Abundances of O- and (N+O)-bearing molecules are similar in both sources, except for \et and \dme, which are an order of magnitude less abundant in G0.693. Abundances of N-bearing molecules are generally higher in G0.693 by factors of a few. However, there is a tendency towards higher abundances for O-bearing molecules in comparison to N-bearing species in both sources, except for \et and \dme in G0.693. 
In Fig.\,\ref{fig:corner_met}, we additionally compare abundances with respect to \met, which reveals a similar trend as seen in Fig.\,\ref{fig:corner}.
The similarities in the molecular inventory, also noted by \citet[][]{Zeng18}, suggest that similar processes may drive the chemistry in the two sources. 

The detailed comparison of G0.693 and L1157-B1 with the positions in Sgr\,B2\,(N1) based on Fig.\,\ref{fig:corner} can be found in Appendix\,\ref{ass:sources}. The key points are summarised in the following.
No general trend between the two other sources and any position in Sgr\,B2\,(N1) can be observed. Abundances with respect to H\2 that were derived for N1S are all higher than in G0.693 and L1157-B1 by $\sim$1--2 orders of magnitude.  
At positions N1SE1 and N1S in Sgr\,B2\,(N1), the cyanides  (and HC\3N at N1SE1:OF) are more abundant by two orders of magnitude or more than in G0.693 and L1157-B1. 
The abundances of most O-bearing molecules at N1SE1:OF are likely lower than in the other two sources.
The lack of any correlation between the molecular abundances between the outflow position in Sgr\,B2\,(N1) and the other two sources suggests that a shock may impact the material differently or that the boundary physical conditions are different. A discussion on possible explanations for the chemical differences is provided in Sect.\,\ref{sss:scenario}. 

\subsection{Comparison to hot-core models}\label{dss:modelsRTG}

As in Paper\,I, we compare our results to recent astrochemical models performed by \citet[][G22 hereafter]{Garrod22}. As a 3-phase model it considers the chemistry within the bulk ice, the grain/ice surface, and the gas phase as well as
physical processes connecting one phase to the other. The physical setup compares to earlier models \citep[][]{Garrod08,Garrod13,Garrod17}, in which two stages are distinguished: a cold collapse phase (stage 1), that takes place before the onset of the protostellar heating, and a subsequent warm-up phase (stage 2). 
Stage 1 happens on a timescale of $\sim$10$^6$\,yr. During this time, the density increases from $3\times10^3$ to $2\times10^8$\,cm$^{-3}$, the dust temperature decreases from $\sim$15\,K to $\sim$8\,K, and the gas temperature is kept constant at a value of 10\,K. 
Subsequently, in stage 2 gas and dust temperatures increase in the same manner, as they are assumed to be well coupled during this stage. The density is fixed to the final value of stage 1. 
Three different timescales are regarded for stage 2, meaning that the warm-up to 200\,K either lasts $10^6$\,yr (slow), $2\times10^5$\,yr (medium), or $4\times10^4$\,yr (fast). The maximum temperature reached in the models is 400\,K. 
The main differences of the new models (G22) compared to the previous ones (amongst other changes) are the inclusion of non-diffusive chemical reactions and the absence of bulk diffusion for all species but H and H\2. The latter leads to the trapping of species (including COMs) in the water-dominated ice mantles of grains until water itself desorbs at a temperature $\gtrsim$100\,K.
This co-desorption process of molecules with water is associated with a steep increase in gas-phase abundances at this characteristic temperature. 
The resolved abundance profiles that we derived in Paper\,I revealed such a steep increase at $\sim$100\,K for multiple COMs (see Fig.\,15 in Paper\,I), based on which we could confirm that this process releases the bulk of these COMs to the gas phase in Sgr\,B2\,(N1). 
For COMs that did not show this steep increase we proposed a dominant formation route in the gas phase or a lack of sensitivity. 

In Appendix\,\ref{app:modelsRTG}, we provide a detailed comparison between the  modelled peak abundances in the slow and fast warm-up phases with the observed abundances towards both components at positions N1SE1, N1NW3:HC, N1S and N1S1 (derived in Paper\,I), and towards G0.693 and L1157-B1 that were discussed in Sect.\,\ref{dss:sources}. The comparison shows that none of the components is perfectly reproduced by any model (see Figs.\,\ref{fig:cfmodel} and \ref{fig:cfmodel_met}). In the following, we summarise some trends that can be deduced from this comparison regarding formation and destruction pathways for mainly O- and N-bearing molecules that play a role for the discussion on the impact of the outflow on the chemistry in Sgr\,B2\,(N1) in Sect.\,\ref{dss:implications}, also including some knowledge gained in Paper\,I. 

From Paper\,I, we learned that O-bearing molecules (\met, \et, \dme, \mf) are mainly produced in the solid phase and desorb thermally, in agreement with the models. \ad and \vc are efficiently produced in the hot gas phase and \etc showed evidence for both solid- and gas-phase formation, although the gas-phase formation of \etc is not (yet) included in the G22 models. Results for \fmm and \mic were less conclusive, but also likely suggest formation in both phases.
The detection of HC\5N in only the outflow component N1SE1:OF suggests that its formation may proceed more efficiently in the post-shock gas. According to the G22 model, there are two periods (at $\sim$200\,K and $\gtrsim$350\,K) during which the molecule is efficiently produced in the gas phase. This can be seen in Fig.\,\ref{fig:G22}, which shows qualitatively at what times and temperatures of the pre- and protostellar evolution the molecule is dominantly formed (green) or destroyed (blue) in the G22 models (of which this kind of figure is an output).
A dramatic temporal temperature increase caused by a shock (possibly no longer traced by the derived rotational temperatures in Sgr\,B2\,(N1)) may have accelerated the formation that is associated with the second period. At these high temperatures, atomic N is sourced from ammonia (NH\3) via H abstraction in the models and reacts with hydrocarbons (e.g. C\2H\5) to form HC\5N. In a similar way, HC\3N can be produced at high temperatures. We speculate that this may 
also be the case for the larger cyanides, but such a formation route is not included in the current models. Wing emission in NH\3 observed towards Sgr\,B2\,(N1) indicates that the molecule is present in the outflow \citep[][]{Mills2018}.

Ammonia is a widely used tracer of temperature in interstellar gas, especially also for CMZ clouds, where detections of metastable transitions with high upper-level energies ($E_u\gtrsim$\,500\,K) imply temperatures of $\gtrsim$\,300\,K \citep[e.g.][]{Huettemeister1995,Wilson2006,Riquelme2013,Mills2013,Candelaria2023}. These hot components are usually observed in addition to a colder component ($<$\,100\,K), which is traced by the lower-energy transitions, and maybe warm gas at temperatures in between. Although the hot NH\3 component is often associated with more tenuous gas than the core of the CMZ clouds, where most of the other molecules are located, \citet{Mills2013} found that the hot component rather follows the morphology and kinematics of HC\3N, which is associated with the denser gas. A systematic study on changes in the abundance of hot NH\3 and HC\3N or HC\5N as a function temperature may confirm the proposed chemical link between the species.

In Paper\,I, we found that the destruction of some O-bearing COMs seems to proceed more efficiently than predicted by the models. As a possible reason we stated that higher volume densities than assumed in the model may lead to an accelerated destruction, which remains to be confirmed by the models, however. If this were a consequence of higher densities, it may (partially) explain the reduced abundances of O-bearing molecules at N1SE1, assuming that a shock wave compresses the material through which it passes.
The possible impact of higher densities and (temporally) higher temperatures on the chemistry in the post-shock gas in Sgr\,B2\,(N1) are further discussed in Sect.\,\ref{dss:highnhight}.


\subsection{Comparison to shock models}\label{dss:shockmodel}

\citet{Burkhardt19} modelled the evolution of chemistry in a C-shock at a speed of 20\kms and distinguished between multiple regimes of chemistry. Within the first 50\,yr after the shock has passed, rapid formation of molecules on dust grains takes place due to enhanced mobility of species, shortly after which sputtering of grains releases the molecules to the gas phase. As a consequence, one would find the entire ice inventory in the gas phase at these times, which would mainly include methanol and other O-bearing species. 
The dust is heated faster than the gas ($T_\mathrm{dust,max}\sim 35$\,K) and the latter reaches its peak temperature around 1000\,yr ($T_\mathrm{gas,max}\sim 1000$\,K) after the shock has passed. This is when also enhanced gas-phase formation starts to proceed efficiently. Species that benefit from this include, for example, \ad and \fmm in the \citeauthor{Burkhardt19} models. Molecules that do not show a significant change in their gas-phase abundances at this stage are HNCO, \dme, and \met. After 10$^4$\,yr in the model, the environment has cooled back to its low initial temperature, which is when species start to return to the solid state. For some species such as \fmm, \ad, or \mf, gas-phase formation continues as the precursor molecules are still available. 

The physical conditions of the material that is impacted by the shock in this model are quite different from what we expect for Sgr\,B2\,(N1). Due to the close vicinity of, especially, N1SE1 but also N1NW3, densities and temperatures in the pre- and post-shock gas are likely higher than in the model, where the maximum post-shock density is $\sim3\times10^5$\,cm$^{-3}$. Therefore, we may also expect that thermal desorption, as a consequence of the protostellar heating, released ice species already in the pre-shock phase  
and we do not expect any redeposition phase to occur. Therefore, we would expect that a shock in Sgr\,B2\,(N1) may result in an increase of the gas temperature, compression of the material, and release of additional solid material from the dust grains that did not (yet) desorb thermally. 
When trying to sort the various higher-velocity features that we identified in the PV diagrams in Fig.\,\ref{fig:PV} in some sort of timeline, feature B2 could be an older ejection event, because it is spatially the most extended and its peak velocity with respect to the hot core is lower than that of the
faster and more compact feature B1. This additionally suggests that it may have decelerated. Therefore, post-shock gas-phase chemistry may be proceeding efficiently in B2 and enriching (or depleting) the gas phase with (of) species, including COMs. 
Accordingly, if B2 was an older ejection event, the peak sputtering phase may have gone by (if it happened at all), meaning that there was enough time to reduce abundances of molecules that are mainly the product of grain-surface chemistry, provided they do not have additional formation routes in the gas phase. From our observations in comparison with the G22 model, this would include the O-bearing COMs \met, \mf, \dme, but also NH\2CN and \mmc. 
Unfortunately, the behaviour of N-bearing species from the pre- to the post-shock phase was not studied in \citet{Burkhardt19}, which would be important to see whether enhanced gas-phase formation is expected, as it was indicated by the abundances with respect to \met in the outflow of Sgr\,B2\,(N1).
For the faster feature B1 we estimated a younger age (it could be younger than 1000\,yr if seen at high inclination). At this age, we might expect to find the source in the phase of rapid ice mantle evaporation. However, at these short distances to the centre of the hot core, all ice has likely been sublimated and further processed in the hot gas phase prior to any shock.  Therefore, it may not be surprising to observe this feature only securely in emission of SO, SiO, OCS, HC\3N, \vc, and \etc and not in emission of \met, \et, or \dme, which are products of grain-surface chemistry.
This could be supported by taking into account the results from Paper\,I, where abundance profiles of the latter three COMs decrease with shorter distances to the hot-core centre, while that of \vc increases, which we explained with destruction and formation of these COMs in the gas phase, respectively. 

\subsection{Implications for the chemistry in Sgr\,B2\,(N1)}\label{dss:implications}

The following discussion is focussed on position N1SE1, since not many molecules are detected in N1NW3:OF and the hot-core component seems to still be influenced by the nearby continuum source.
The behaviour of abundances with respect to H\2 (also \etc) suggests that O-bearing molecules might be less abundant (i.e. more efficiently destroyed) in N1SE1:HC, and even less in N1SE1:OF, than at N1S, while abundances of N-bearing species are similar in all components.
In both G0.693 and L1157-B1, the rotational temperatures derived for various molecules (10--30\,K) are generally much lower than the kinetic temperature \citep[50--140\,K,][]{Zeng18,Mendoza18,Codella20} determined for these sources. This is a consequence of low volume densities \citep[10$^{4-5}$\,cm$^{-3}$,][]{Armijos-Abendano15} and results in sub-thermal excitation. 
Rotational temperatures that we derived in both components at N1SE1 and N1NW3 are generally above 100\,K and reach up to 200\,K depending on the molecule. Hence, the volume density at these positions must be high. 
It may be that densities are even higher than towards N1S. This may be supported by the presence of numerous bright H\2O masers \citep{McGrath04} that need high densities to be excited \citep[$\sim$10$^9$\,cm$^{-3}$,][]{Hollenbach13}. 
At this point we are not certain about the exact relation between the two components at N1SE1 (and N1NW3) and the outflow. This discussion is taken up on in Sect.\,\ref{dss:morph}. 
In general, the passage of a shock (e.g. provoked by an outflow) is capable of compressing material and although temperatures are high in the outflow component at N1SE1, the values are lower than in the hot-core component at this position. This may suggest that the former could be more shielded, due to the higher densities, from external heating sources, provided that this temperature difference is not an effect of projection.

\subsubsection{Effects of a high volume density and high temperature}\label{dss:highnhight}

An enhanced volume density may accelerate the gas-phase production of molecules, provided the reactants are sufficiently abundant. For example, we mentioned in Appendix\,\ref{dss:model_n1se1} that, according to the G22 models, HC\3N and HC\5N can be produced from atomic N at temperatures of $>$350\,K. Atomic nitrogen at these temperatures is the product of the breakdown of ammonia through successive H abstraction, which may not only be involved in the formation of the cyanopolyynes but potentially also in the gas-phase formation of the complex cyanides. As the equivalent for O-bearing molecules, one might think of sourcing atomic O from H\2O or CO that may then be involved in the formation of larger molecules. However, the destruction reactions for these two molecules have much higher barriers than for NH\3, which makes them unlikely to proceed even in shocked gas. Even if atomic O was available, there are currently (almost) no reactions known that include atomic O to produce O-bearing COMs.
Hence, if an enhanced gas-phase formation is restricted to N-bearing molecules, due to the availability of reactants such as atomic N, then this explains why abundances of other products of gas-phase chemistry, such as \ad, are not enhanced. This may also apply to \dme and \mf as the G22 models include formation routes in the gas phase for these two COMs, which involve \met and \dme as reactants, respectively. 

On the other hand, in order to explain a decrease in the observed abundances of some (O-bearing) COMs towards higher temperatures steeper than in the model, we discussed in Paper\,I that densities higher than assumed in the model may cause a more efficient destruction of these COMs in the gas phase. Such a scenario may explain the lower abundances of O-bearing molecules relative to H\2 that we derived towards N1SE1. 
However, whether there are destruction mechanisms proceeding in higher-density gas that only apply to a specific group of molecules, that is O- and not N-bearing species, is uncertain. This may rather imply that a lack of gas-phase formation routes, able to compete against the destruction of O-bearing molecules, results in the segregation. 
Still, O-bearing molecules are less abundant than expected from N1S. Instead of the continuous destruction in the higher-density post-shock gas, reactions with energy barriers favouring certain products may have been able to proceed as a result of the brief temperature peak during the passage of the shock. For example, we may speculate that reactions that involve O-bearing COMs and produce H\2O tend to have lower barriers than reactions that involve N-bearing COMs and produce HCN, because the former have a higher exothermicity as water is a more stable molecule than HCN. However, this mechanism itself needs to be confirmed in models. Furthermore, these reactions likely need thousands of K to proceed efficiently, hence, this would require a shock stronger than in the model of \citet{Burkhardt19}, where $T_\mathrm{gas,max}=1000$\,K.

An important factor impacting the molecular content in the post-shock gas is dust. As described in Sect.\,\ref{dss:shockmodel}, the passage of a shock in a cold environment leads to the desorption of grain-surface species either due to the temporal temperature increase or because the dust grains get sputtered. Because we do not see an enhancement of species that are mainly the product of grain-surface chemistry (e.g. \et, \dme, and \mf), the relation between the dust and the shock may be different for the outflow of Sgr\,B2\,(N1). Because of the short distance of N1SE1 (and N1NW3) to the centre of the hot core, temperatures were likely high enough for thermal desorption to proceed in the pre-shock gas. Therefore, the gas phase has probably already been enriched with grain-surface species, but also with products of high-temperature gas-phase chemistry, before the shock hit the material. In that case, a shock may not have a great impact on the sublimation of ice-mantle species any more, but still on the release of less volatile species, such as Si to form SiO in the gas phase. 
Moreover, we do not expect the initial dust composition (set before warm-up) to be much different for the various positions because of the short distances of N1SE1 and N1NW3 to the centre of the hot core. Therefore, this cannot be the reason for chemical differences.

\subsubsection{A shock-preceding hot-core phase}\label{sss:scenario}

\begin{figure*}
    \includegraphics[width=1.25\textwidth]{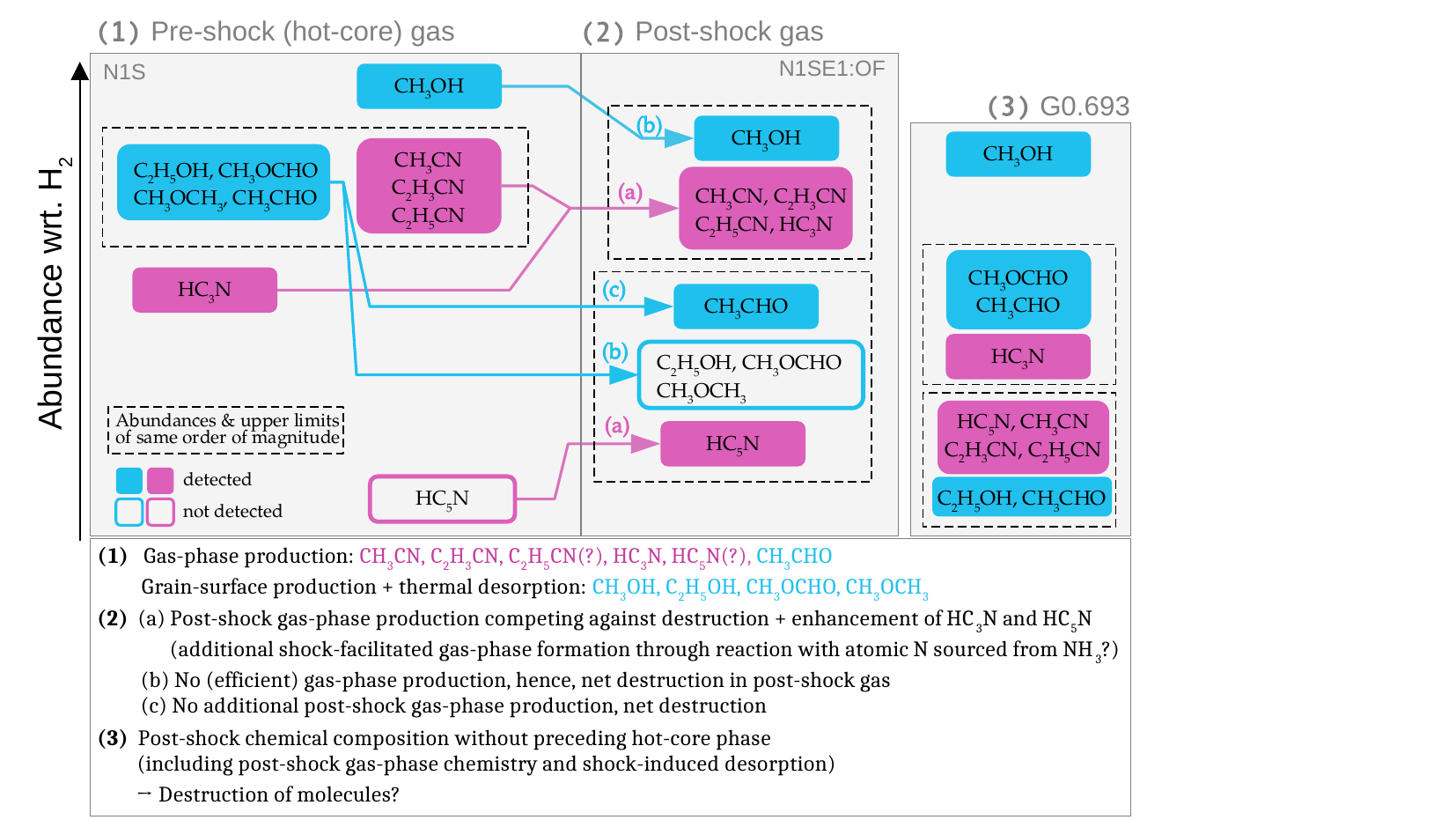}\\[-.3cm]
    \caption{Illustrative behaviour of abundances of O-bearing COMs (blue) and N-bearing molecules (pink, including cyanopolyynes and cyanides) in Sgr\,B2\,(N1) in (1) the pre-shock gas (here represented by N1S) and (2) the post-shock gas (N1SE1:OF) in comparison to (3) G$+$0.693$-$0.027. Involved physical processes or chemical reactions are discussed in detail in Sects.\,\ref{dss:modelsRTG}--\ref{dss:implications}; key points are provided below the figure. Boxes filled with colors indicate detections while empty ones represent upper limits. Dashed boxes enclose molecules with similar abundances (within an order of magnitude).}
    \label{fig:summary}
\end{figure*}
Based on the above discussion, the present-day gas chemical content of the outflow component at N1SE1 (in some way also the hot-core component) has probably been dominantly influenced by two processes:  a phase of hot-core chemistry, including thermal desorption of ice species and gas-phase chemistry, that preceded a shock impact, as well as compression and heating of the material by the shock. Accordingly, most of the ice species have already desorbed to the gas phase before the shock passed, meaning that the shock caused the release of primarily less volatile species. In the post-shock gas, some N-bearing molecules are efficiently formed, thereby, likely counteracting any destruction reactions, while there are no gas-phase formation routes for most O-bearing COMs. In contrast, these seem to be even more efficiently destroyed (when compared to N1S) either due to the increase in density or temperature, although the exact process is uncertain. Assuming that N1S is representative of pre-shock molecular composition in Sgr\,B2\,(N1) and N1SE1:OF of the post-shock composition, the impact of a shock on O- and N-bearing molecules in Sgr\,B2\,(N1) is schematically shown in Fig.\,\ref{fig:summary}.

This scenario may be able to explain several observed differences between Sgr\,B2\,(N1), G0.693, and L1157-B1 (see Fig.\,\ref{fig:corner}). For a direct comparison, G0.693 is also shown in Fig.\,\ref{fig:summary}.
Abundances of the cyanides are higher in both N1S and N1SE1 than in G0.693 and L1157-B1 suggesting that these species were initially produced in the hot pre-shock gas and, in the post-shock gas, formation and destruction may be competing, which holds up the high abundances. Given that G0.693 and L1157-B1 did not go through a phase of hot-core chemistry, this may explain the much lower abundances of cyanides compared to Sgr\,B2\,(N1). This may also apply to HC\3N and HC\5N in these sources. In Sgr\,B2\,(N1), HC\3N has a high abundance at N1S, meaning that it is also a product of hot-core chemistry (be it on grain surfaces or in the gas phase). However, it is about a factor 10 more abundant at N1SE1, and with that similarly enhanced as the cyanides compared to G0.693 and L1157-B1, maybe suggesting that there is an additional gas-phase formation route (possibly from atomic N). This also applies to N1NW3:OF where the HC\3N abundance is as high as for N1SE1. The non-detection of HC\5N in all hot-core components in Sgr\,B2\,(N1) and the similar abundance in N1SE1:OF, G0.693, and L1157-B1 may indicate that this species is not primarily a product of hot-core chemistry (at temperatures below 350\,K), but was rather efficiently formed later in the post-shock gas. 
Because O-bearing molecules generally tend to be similarly or more abundant than N-bearing molecules in G0.693 and L1157-B1, they may not be as efficiently destroyed as in N1SE1, which may be a consequence of the lower volume densities than at N1SE1 or of a less violent shock that did not facilitate certain gas-phase (destruction) reactions.
Especially for the comparison of G0.693 and Sgr\,B2\,(N1), taking into account the higher cosmic-ray flux and its influence on the chemistry in the different density environments may also be of importance.  

Ultimately, in order to confirm the proposed scenario in which gas and dust in Sgr\,B2\,(N1) are exposed to efficient protostellar heating before the passage of a shock, we would need to run shock models. 
The assumed physical conditions in the chemical models, with which we compared the observational results, do not exactly apply to those expected in the outflow of Sgr\,B2\,(N1). The G22 models do not include shocks and the models of \citet{Burkhardt19} were designed to predict the impact of the L1157 outflow on the ambient material, which is overall colder and less dense than in Sgr\,B2\,(N1). Therefore, we need shock models that take into account overall higher pre- and post-shock densities as well as higher initial temperatures. Moreover, the effect of hot-core chemistry in the pre-shock phase, including thermal desorption due to protostellar heating as well as gas-phase chemistry, should be investigated.
Finally, both the G22 models and the models of \citet[][]{Burkhardt19} assume a standard cosmic-ray ionisation rate. Exploring the impact of higher cosmic-ray ionisation rates, which are known to prevail in the CMZ \citep[e.g.][]{Indriolo2015}, on shock models would be interesting.


\subsection{Implications for the outflow morphology}\label{dss:morph}

Based on the integrated intensity maps shown in Figs.\,\ref{fig:so} and \ref{fig:COMof}--\ref{fig:COMof3}, it is difficult to disentangle structures that solely originate from the outflow. Although most of the hot-core emission can be avoided thanks to the LVINE method, some emission may still be included in the integration. Filamentary structures have also been identified \citep[see Fig.\,\ref{fig:overview} and][]{Schwoerer19} in dust continuum emission and in emission of some molecules at red- and blue-shifted velocities. 
The gas molecular content that we derived in N1SE1:OF along with rotational temperatures that are lower than in N1SE1:HC and the presence of H\2O masers suggest that this component may be characterised by a high volume density, possibly even higher than at N1S. Moreover, the similar behaviour of molecular abundances in N1SE1:HC and OF imply that the hot-core component at this position may also be influenced to some extent by what drives the chemistry in the outflow component. 
The molecular component of a protostellar outflow can often be subdivided into a fast molecular jet characterised by a high degree of collimation, a more loosely-collimated low-velocity outflow, and the walls of an outflow cavity \citep[e.g.][]{Tychoniec21}. Molecules that can be observed in a fast collimated jet are usually species such as H\2, SO, SiO, or CO, although the latter may also trace the outflow cavity; \citep[for an illustration, see the well-studied case of the low-mass protostellar outflow HH\,211 ][]{Gueth1999, Hirano2006}.  The collimated emission features  that we observe for SO and SiO (labelled aB1, aR1, and aR2 in Fig.\,\ref{fig:so}) and that remain visible up to very high blue- and red-shifted velocities may represent such a jet. The bulk of the emission from all analysed molecules in the outflow is observed at lower velocities and with a much wider opening angle suggesting that it may represent a kind of lower-velocity outflow, if all of the emission is indeed associated with an outflow. 
This wider-angle outflow may contain dominantly material that was entrained and accelerated by the jet as it propagates through the envelope or material that originates from some kind of disk and ended up in the outflow through disk winds. Because the behaviour of abundances in the hot-core component at N1SE1 is somehow similar to that of the outflow component, the former may be associated with some kind of cavity wall, whose density or temperature was increased by the shock, however, to a lesser extent than in the outflow component itself. This would not be a typical cavity wall, which is exposed to the UV radiation from the protostar and thus traced by photochemistry. 
This component in Sgr\,B2\,(N1) is still embedded in the hot core and shielded from UV radiation. However, it remains questionable how the shock impacted the density or temperature while the systemic velocity of this component does not show any signs of interaction with an outflow.
On the other hand, one might speculate whether N1SE1:HC represents the chemical composition prior to the shock impact, which would mean that this hot-core component was different from N1S to begin with. However, we cannot tell what kind of process would lead to such differences. 

\subsection{Implications for segregation of O- and N-bearing species in other sources}


Multiple observational studies revealed differences between species, in particular, O- and N-bearing molecules in their emission morphology or in values of rotational temperature and abundances. 
For example, early observations \citep[e.g.][]{Caselli93} revealed chemical differences between the hot core and the compact ridge in Orion. These two environments come with different physical properties \citep[e.g.][]{Tercero10}, where the hot core is hotter ($>$200\,K) and denser ($\gtrsim$10$^7$\,cm$^{-3}$) than the ridge ($\sim$100\,K and $\sim$10$^6$\,cm$^{-3}$) and shows high abundances of N-bearing molecules, while the compact ridge is dominated by emission of O-bearing molecules \citep[][]{Blake87,Crockett14,Pagani19}. Chemical models, that took into account the different observed physical conditions, predicted that hot gas-phase chemistry played a more important role in the hot core \citep[][]{Caselli93,Crockett15}. 
In addition, the outflow driven by the nearby young stellar object IRc2 was proposed to maybe have an impact on the observed molecular content \citep{Caselli93}.
More recently, the chemical segregation in the Orion\,KL region was explored with ALMA in the context of the explosive outflow event (already mentioned in Sect.\,\ref{dss1}) that happened in this region about 500--550\,yr ago \citep{Gomez2008} and is probably impacting the chemistry in the various sources contained in this region \citep[][]{Zapata11,Pagani17,Pagani19}. 
With this in mind, if the Orion hot core was indeed a typical hot core, that is, if it contained an internal heating source \citep{Wilkins22}, which is debated \citep[e.g.][]{Zapata11,KL17}, it might have gone through a phase of thermal desorption and high-temperature gas-phase chemistry before it started to interact with the shock originating from this explosion.
Similar to our observations, the hot-core region in Orion\,KL is bright in emission from, for example, the cyanides, HC\3N, and OCS, but also from \met and HNCO. As shown in earlier studies, most of these molecules reveal only weak emission, if any at all, in the compact ridge while emission of some O-bearing COMs such as \dme and \mf peak here. This could be more similar to what is observed in G0.693 and L1157-B1, also based on the lower temperatures associated with this source. However, other O-bearing molecules, for example \et, do not show emission as bright. The segregation between O-bearing COMs was studied by \citet[][]{Tercero18}, who proposed that the availability of different radicals (CH\3O and CH\2OH) as reactants might play a role. Besides the compact ridge and the hot core, there are more sources in this region that are bright in emission of some molecules and lack emission of others.
Ultimately, Orion\,KL reveals a high degree of chemical diversity and further investigation is needed to understand the impact of the explosive event on the chemistry in the various embedded sources and whether some of the involved processes resemble what we propose for the outflow position in Sgr\,B2\,(N1), where also the different timescales since the passage of a shock need to be considered.


\citet[][]{Allen17} derived the molecular composition towards four continuum sources (A, B1, B2, B3) in the high-mass star-forming region G35.20-0.74N and found that abundances with respect to H\2 of N-bearing species, especially \vc and \etc, are higher in sources A and B3 and lower, if they were detected at all, in sources B1 and B2. Amongst other ideas, the authors proposed a scenario in which  an enhanced gas-phase formation accounts for the enhancement in N-bearing molecules in A and B3, based on derived temperatures that are higher in these sources than in B1 and B2 (similar to the early Orion\,KL studies). 
Based on the interpretation of our observed results, where cyanides are produced in the hot gas phase, it may not be too surprising that their abundances are higher in the hotter sources B3 and A ($T>250$\,K) than in B1 and B2 ($T<180$\,K). 
However, in B3 also O-bearing molecules are generally more abundant than in the other sources, or at least similarly abundant. Again, based on our results, where O-bearing molecules are destroyed in the high-temperature gas phase, this may be the more puzzling result. 
However, the authors used average kinetic temperatures for their classification of the sources that they derived using CH\3CN emission. The actual range of temperatures shows that the sources could be more similar to each other. 
Recently, \citet{Zhang22} reported on a spiral-like structure around B1 in SO\2 emission, which is associated with a disk and which also coincides with the continuum peaks B3 and B2. In addition, sources A and B1 drive outflows \citep[e.g.][]{Zhang22} and there is also SiO emission associated with the filament in which all sources are embedded. Therefore, one or more sources could be impacted by shocks. For example, this may have heated up B3, which enabled the production of N-bearing species.

Observations towards the hot-core precursor G328.2551--0.5321 \citep[][]{Csengeri19} revealed an enhancement in abundances with respect to methanol of O-bearing molecules towards two position that the authors associated with a disk. The enhancement was proposed to result from accretion shocks, that sputtered the dust grains and, thereby, released  the molecules into the gas phase. On the other hand, N-bearing molecules showed a more spherical distribution that the authors associated with the thermal desorption of these species from dust grains. They proposed that this source may be in an early evolutionary stage. If we assume that this source is indeed of a younger age and is possibly on the verge of becoming a hot core similar to Sgr\,B2\,(N1), the explanation for the chemical segregation given by the authors together with the observed velocity structure and derived temperatures can hardly be connected to any scenario that we proposed in Paper\,I or here in this work. 

Another mechanism, that has been introduced as a potential driver of chemical segregation is carbon-grain sublimation. \citet{vantHoff20} proposed that a rise in temperature to $>$300\,K provoked by an accretion burst may lead to the thermal sublimation of less-volatile carbon-rich species from dust grains that will preferably react with N-bearing species in the gas phase. The possible additional release of ammonium salts 
could enhance the amount of nitrogen (and N-bearing species) that would be available for reactions. 
Observationally, this would manifest in intense and compact emission of cyanides, where these high temperatures are reached, and in enhanced abundances of these species \citep[see also][]{Nazari2023}. Furthermore, rotational temperatures of N-bearing species should be elevated compared to O-bearing molecules.
At N1SE1 in Sgr\,B2\,(N1) temperatures likely exceeded 300\,K temporarily as a consequence of a shock and the values that we derived tend to be slightly higher for N-bearing species (cf. Fig.\,\ref{fig:corner_T}). Abundances of the cyanides with respect to H\2 are similarly high in N1S and N1SE1 suggesting that either abundances remain more or less unchanged or that enhanced gas-phase formation competes against their destruction in the post-shock gas. We proposed that atomic N sourced from NH\3 may be involved in the gas-phase formation. Nonetheless, it is not excluded that carbon-grain sublimation contributes reactants for the formation of cyanides. 
However, because the high temperatures required for this process result from a violent event that is an accretion burst or the impact of a protostellar outflow, refractory carbonaceous material (and ammonium salts?) could also be released non-thermally, for example through grain sputtering, leading to the same observational features.

\subsection{Comparison to extragalactic sources}
\begin{figure}
    \centering
    \includegraphics[width=.45\textwidth]{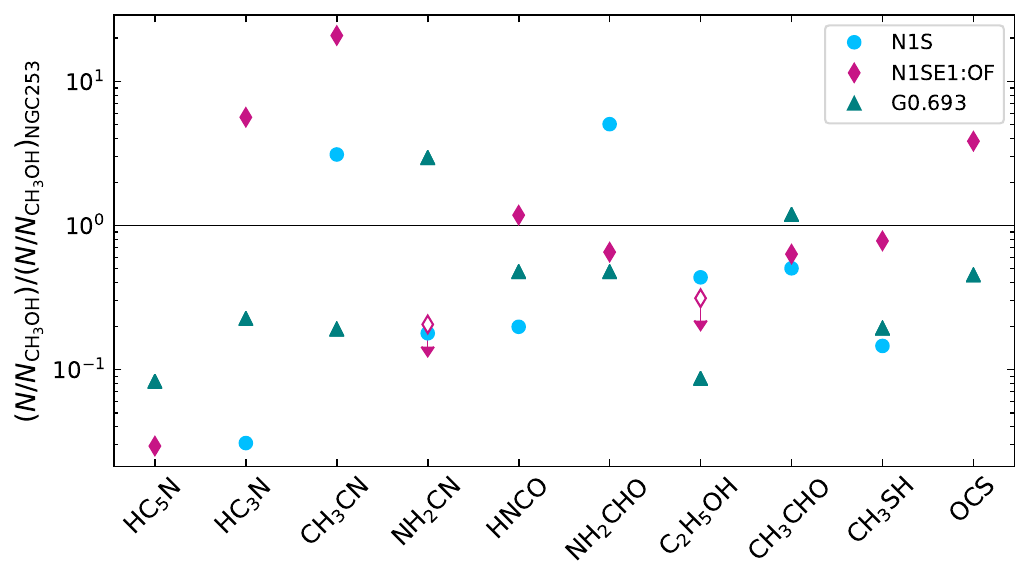}
    \caption{Abundances with respect to \met derived towards N1SE1:OF and N1S in Sgr\,B2\,(N1) and G0.693 normalised by the values observed towards the central molecular zone of the nearby starburst galaxy NGC\,253 \citep[][]{Martin2021}. Empty markers with arrows indicate upper limits.}
    \label{fig:ngc253}
\end{figure}

The results obtained towards CMZ sources, such as Sgr\,B2\,(N1) and its outflow or G0.693 can also be used to shed light onto the chemistry of clouds in the central regions of other galaxies, given that these may similarly be exposed to enhanced cosmic-ray fluxes and shocks with or without star formation. For example, unveiling the chemical composition towards the central  molecular zone of the starburst galaxy NGC\,253 is the goal of the ALCHEMI \citep[ALMA Comprehensive High-resolution Extragalactic Molecular Inventory,][]{Martin2021} survey. Figure\,\ref{fig:ngc253} compares abundances with respect to \met towards N1SE1:OF and N1S in Sgr\,B2\,(N1) and G0.693 normalised by the values observed towards NGC\,253. The values for NGC\,253 were taken from \citep{Martin2021}, who derived these from the data observed with the Atacama compact array (ACA) of ALMA at a resolution of 15\arcsec\,(260\,pc at a distance of 3.5\,Mpc). The scatter between NGC\,253 and Sgr\,B2\,(N1) as well as G0.693 is large and no clear trend can be identified. Nonetheless, the results of the higher angular-resolution data observed with the 12\,m ALMA array may resolve chemical segregations and reveal similarities to Sgr\,B2\,(N1) or G0.693.   

\section{Summary}\label{s:conclusion}
We aimed to study the impact of a high-mass protostellar outflow on the chemistry of COMs and a few simpler molecules in the ambient medium. 
Our target source, Sgr\,B2\,(N1), which is the most massive hot core embedded in the high-mass star-forming protocluster Sgr\,B2\,(N), has long been suspected to drive a powerful outflow, based on the detection of multiple velocity components and wing emission in molecular spectra \citep{Lis93,Liu99} and the presence of bright  water masers \citep[][]{McGrath04}. Later, the outflow was mapped and studied in emission of some, mainly simpler molecules \citep{Higuchi15,Schwoerer20}. 
Using the high-angular resolution and high sensitivity data obtained as part of the ReMoCA spectral line survey, we derived integrated intensity maps for SO and SiO as well as as well as four other simple species and twelve COMs. For one position in the blue-shifted lobe (N1SE1) and one in the red-shifted lobe (N1NW3), rotational temperatures and abundances with respect to H\2, \etc, and \met were derived for, in total, 16 species using LTE radiative transfer models and population diagrams. The resulting abundances for the outflow (OF) and hot-core components (HC) at both positions (with a greater focus on N1SE1) were compared to results obtained previously in Paper\,I (i.e. hot core positions not impacted by the outflow, mainly N1S), the gas molecular content of G0.693 and L1157-B1, and predictions of hot-core models and shock models. The main results of this work are summarised in the following:
\begin{itemize}
    \item The morphology of the SO and SiO maps shows that blue-shifted emission extends towards the southeast and red-shifted emission towards the northwest. Although this bipolarity can be identified, the morphology reveals complex structure, with various emission peaks that appear at different velocities. Most striking are highly collimated features in both blue- and red-shifted emission that are observed at high velocities, which may be associated with molecular jets.  However, the bulk of emission is found at lower velocities with a larger opening angle. 
    \item The blue- and red-shifted emission of other S- and N-bearing molecules reveal a similar bipolar structure as SiO and SO following the lower-velocity, wider-angle emission in large parts. This component may be associated with a low-velocity molecular outflow that consists of material entrained by the jet or pushed by some kind of disk wind. The (N+O)-bearing molecules HNCO and \fmm show more compact blue- and red-shifted emission similar to \ad. Methanol reveals extended emission, which does not simply follow the bipolar structure seen for other molecules. 
    \item Rotational temperatures derived from the 16 selected molecules are not too different from those obtained for N1S and go rarely below 100\,K in both the outflow and hot-core components at N1SE1 and N1NW3.
    \item Abundances with respect to H\2 and \etc show that purely O-bearing molecules are less abundant in N1SE1:HC and even less in N1SE1:OF than in N1S, while most values for N-bearing molecules are similar in these components. However, HC\3N is more abundant in both components at N1SE1 and in N1NW3:OF than in N1S and HC\5N is only detected in N1SE1:OF. 
    \item The high rotational temperatures in N1SE1:HC and OF together with the presence of H\2O masers in this region suggest that volume densities are high. 
    The compression of material can likely be explained by a shock passage.
    \item While the gas molecular inventories of G0.693 and L1157-B1 reveal similar trends, where O-bearing molecules tend to be more abundant than N-bearing species, there is no correlation with any component in Sgr\,B2\,(N1). This suggests that a shock impacted the chemistry in Sgr\,B2\,(N1) in a different way, likely because the initial conditions in the pre-shock gas were different and present-day volume densities are likely higher at N1SE1 in Sgr\,B2\,(N1). 
\end{itemize}
We propose a scenario for N1SE1 in which the pre-shock material went through a phase of hot-core chemistry, during which thermal desorption and high-temperature gas-phase formation enriched the gas, before a shock, provoked by any interaction with the outflow from Sgr\,B2\,(N1), hit the material. This means that the shock is not the main driver of ice sublimation, but still of the release of less volatile species. In addition, the shock led to compression of material and possibly a temporary rise in temperatures to hundreds of K if not 1000\,K or more, which has a great impact on the chemistry in the post-shock gas. 
Accordingly, the segregation of particularly O- and N-bearing molecules in Sgr\,B2\,(N1) and its outflow and the chemical differences to G0.693 and L1157-B1 may be explained as follows:
\begin{itemize}
    \item Cyanides were formed in the hot gas phase at N1SE1 prior to the shock. Gas-phase formation in the post-shock gas competes with the molecules' destruction keeping the abundances at pre-shock values. 
    \item Abundances of HC\3N and HC\5N are higher at N1SE1 than N1S, suggesting that these species benefit from post-shock gas-phase formation, which may involve atomic N that was sourced from the destruction of NH\3.
    \item Oxygen-bearing molecules do not possess gas-phase formation routes neither in the pre- nor post-shock gas that could compete against their destruction. Moreover, they seem to be destroyed more efficiently, however, the exact process is uncertain.
    Methanol is still observed because it is generally abundant. Acetaldehyde is produced in the gas phase, which does not seem to entirely uphold its pre-shock abundances. 
\end{itemize}
In order to confirm these trends, we would need to run shock chemical models that take into account higher initial densities and temperatures than typically assumed and a pre-shock gas that has already been enriched with molecules during a phase of hot-core chemistry before any shock impact. In any case, this study provides new perspectives on shock chemistry as our results differ from the more 'classical' picture, in which the post-shock gas is enriched by grain-surface species and some products of enhanced gas-phase chemistry. Accounting for the higher cosmic-ray ionisation rate that is prevalent in Sgr\,B2 would also be important.

\begin{acknowledgements}
We thank Elisabeth Mills, who acted as the referee of this paper, for her valuable comments and suggestions that improved the presentation and clarity of this article. 
This paper makes use of the following ALMA data: ADS/JAO. ALMA\#2016.1.00074.S. ALMA is a partnership of ESO (representing its member states), NSF (USA), and NINS (Japan), together with NRC (Canada), NSC and ASIAA (Taiwan), and KASI (Republic of Korea), in cooperation with the Republic of Chile. The Joint ALMA Observatory is operated by ESO, AUI/NRAO, and NAOJ. The interferometric data are available in the ALMA archive at https://almascience.eso.org/aq/. Part of this work has been
carried out within the Collaborative Research Centre 956, sub-project B3, funded by the Deutsche Forschungsgemeinschaft (DFG) – project ID 184018867. 
The authors would like to thank the developers of the many Python libraries, made available as open-source software, in particular this research has made use of NumPy \citep[][]{numpy}, matplotlib \citep[][]{matplotlib}, and SciPy \citep[][]{scipy}.
\end{acknowledgements}

\bibliographystyle{aa} 
\bibliography{refs.bib} 

\include{appendix}

\end{document}

%% file: appendix.tex
\begin{appendix}

\section{Population diagrams}\label{app:popdiagrams}
\begin{figure}[]
    \centering
    \includegraphics[width=0.47\textwidth]{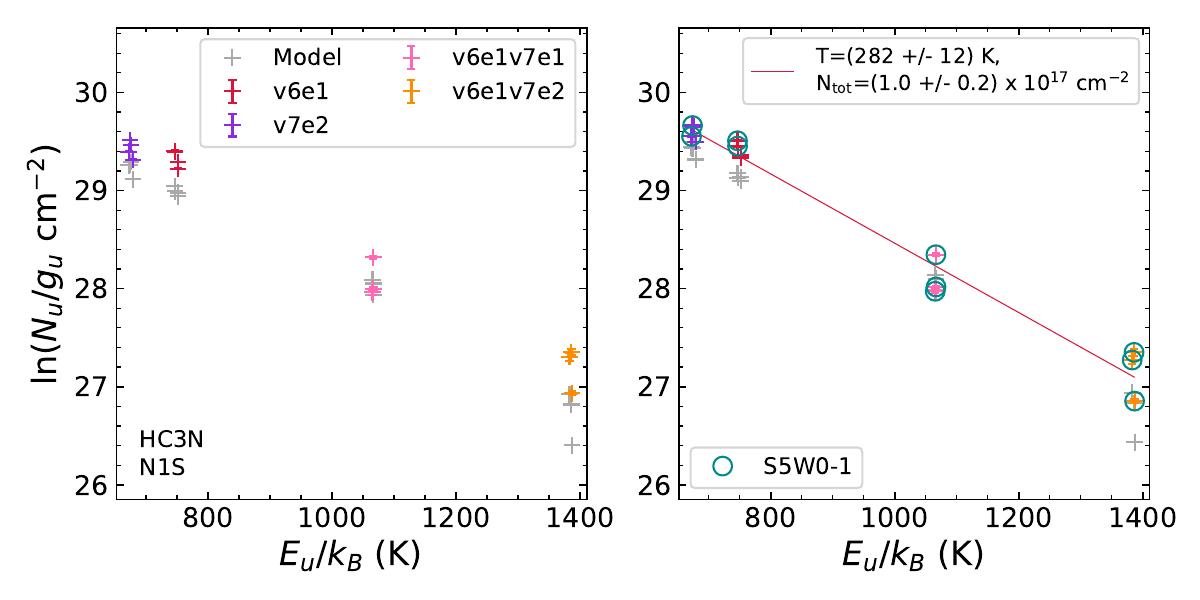}\\
    \includegraphics[width=0.47\textwidth]{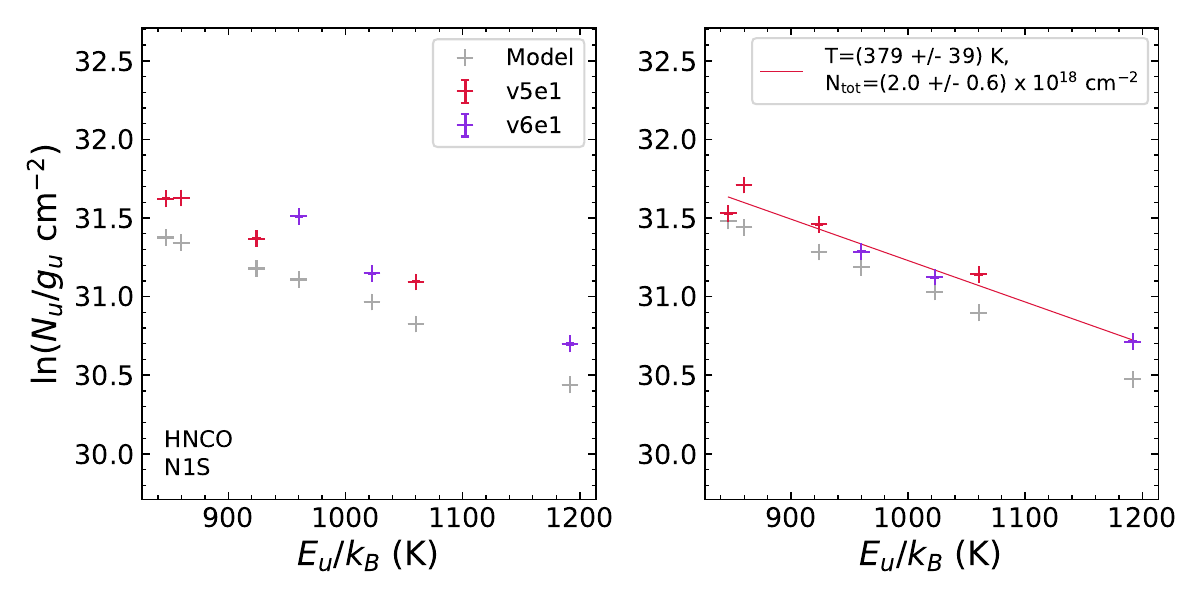}\\
    \includegraphics[width=0.47\textwidth]{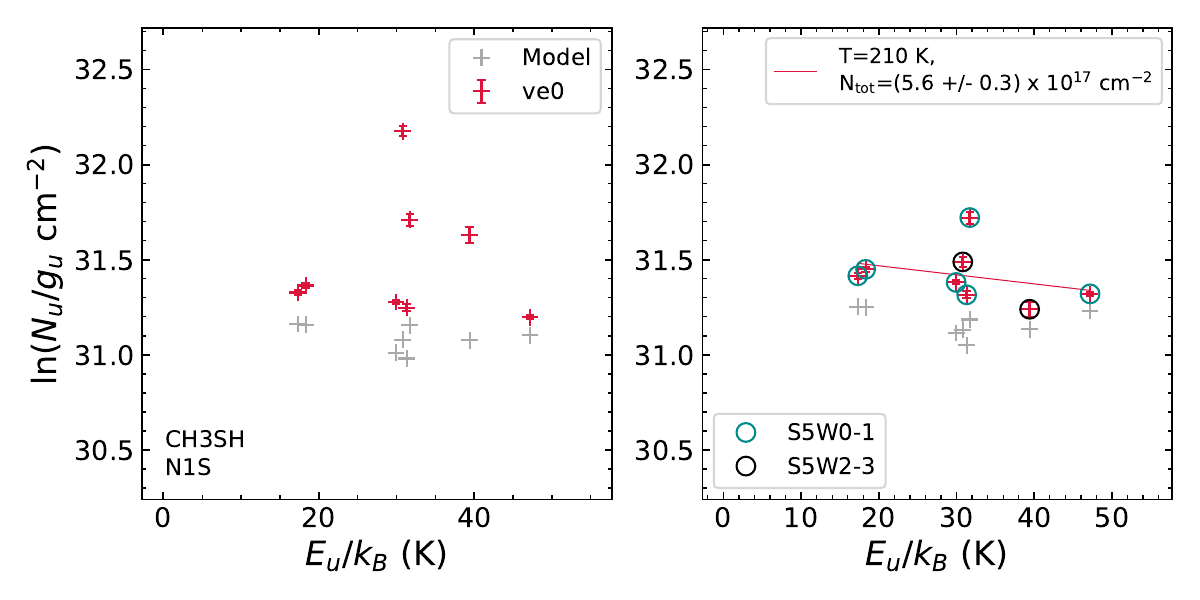}
    
    \caption{Population diagrams for HC\3N, HNCO, and \mmc at position N1S. Observed data points are shown in various colours as indicated in the top-right corner of the respective left plot, while synthetic data points are shown in grey. No corrections are applied in the respective left panels, while in the right plots corrections for opacity and contamination by other molecules have been considered for both the observed and synthetic populations. The red line is a linear fit to all observed data points (in linear-logarithmic space), where for \mmc the temperature was fixed. The results of the fits are shown in the respective right panels. Teal and black circles show observed data points from spectral windows 0--1 and 2--3 of observational setup 5, respectively, as indicated in the bottom-left corner in the right panels.}
    \label{fig:PD_ch3sh}
\end{figure}

In Paper\,I, we derived population diagrams of ten COMs towards multiple positions south- and westwards from the centre of the hot core. In order to compare the results derived here and those for position N1S, we had to derive a few additional population diagrams for HC\3N, HNCO, and \mmc towards this position, which were not analysed in Paper\,I, and show them in Fig.\,\ref{fig:PD_ch3sh}. Population diagrams of various molecules towards N1SE1 are presented in Fig.\,\ref{fig:PD1_n1se1} and towards N1NW3 in Fig.\,\ref{fig:PD1_n1nw3}.

\begin{figure*}[htp]
    \includegraphics[width=0.48\textwidth]{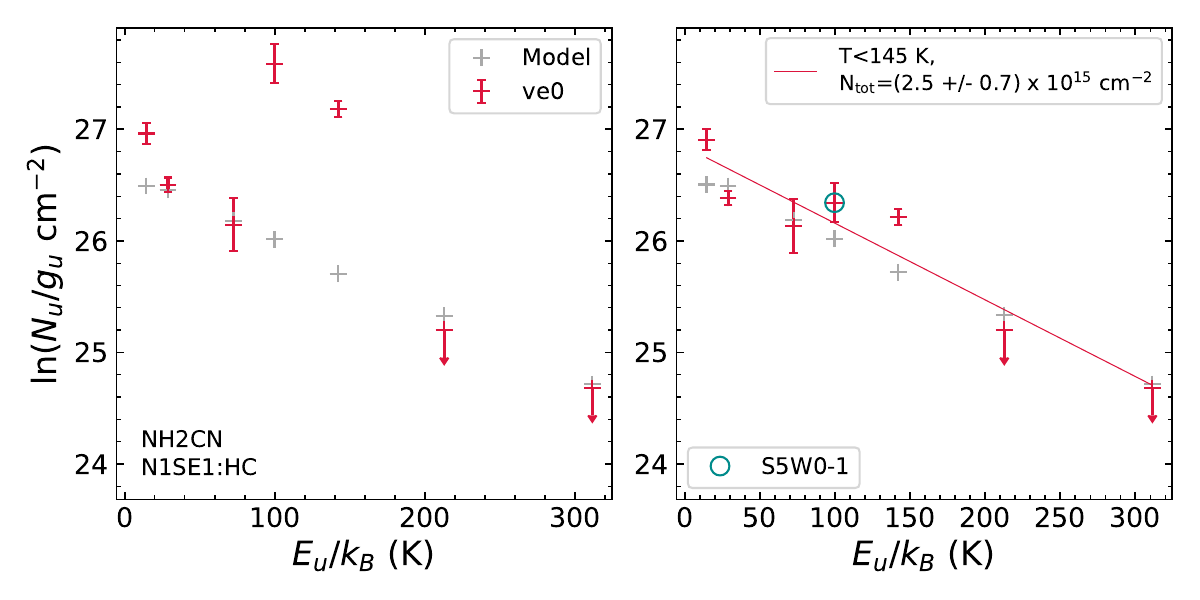}
    \includegraphics[width=0.48\textwidth]{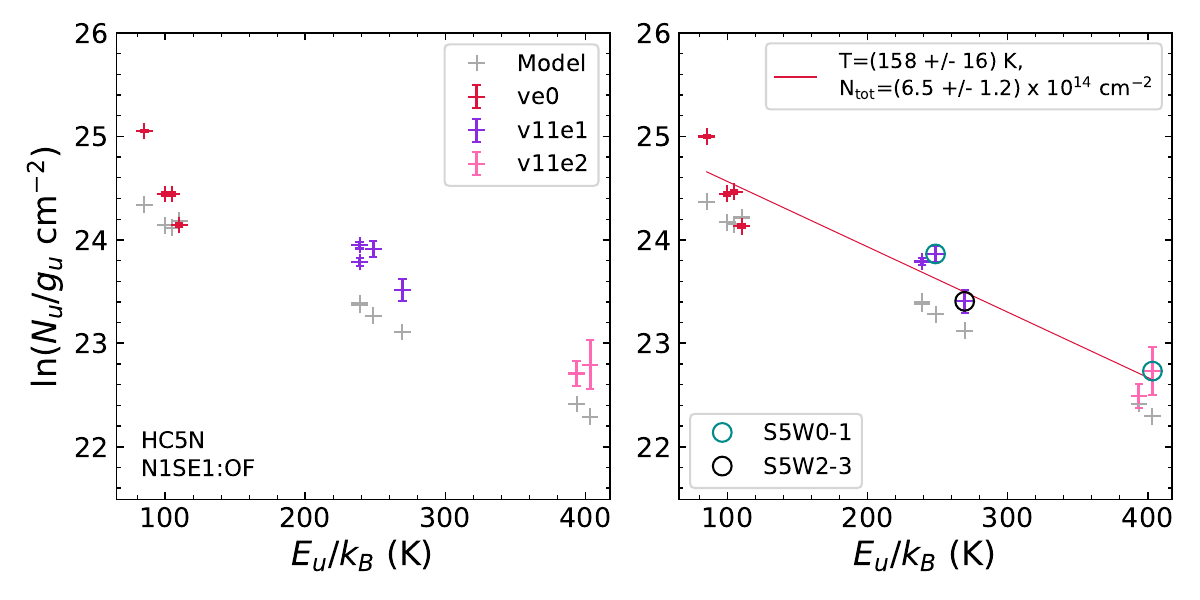}\\
    \includegraphics[width=0.48\textwidth]{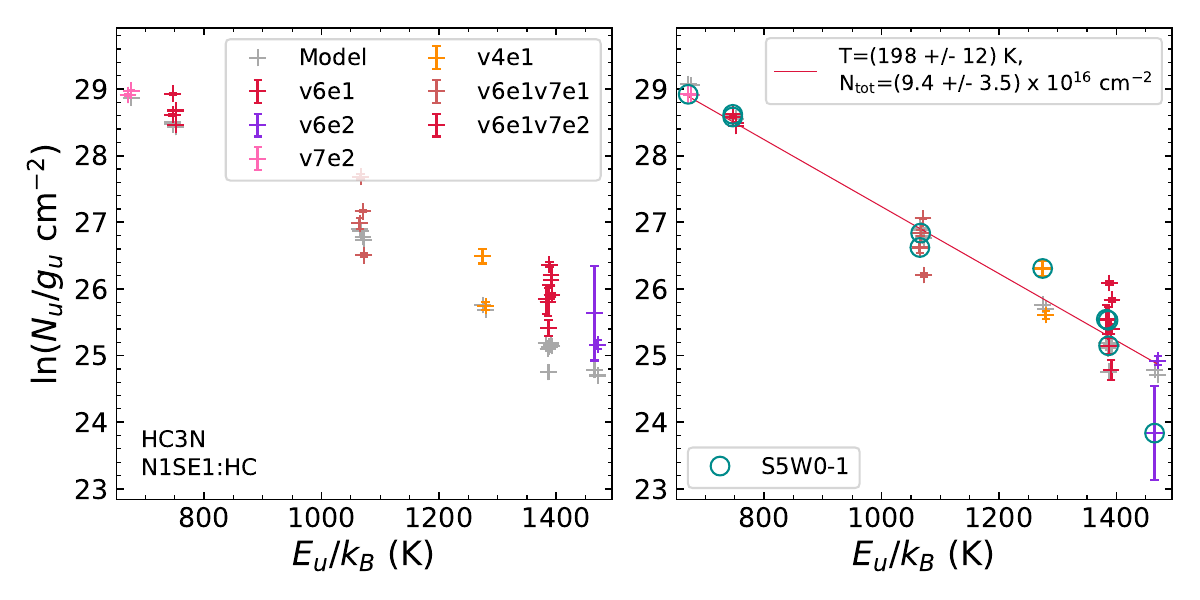}
    \includegraphics[width=0.48\textwidth]{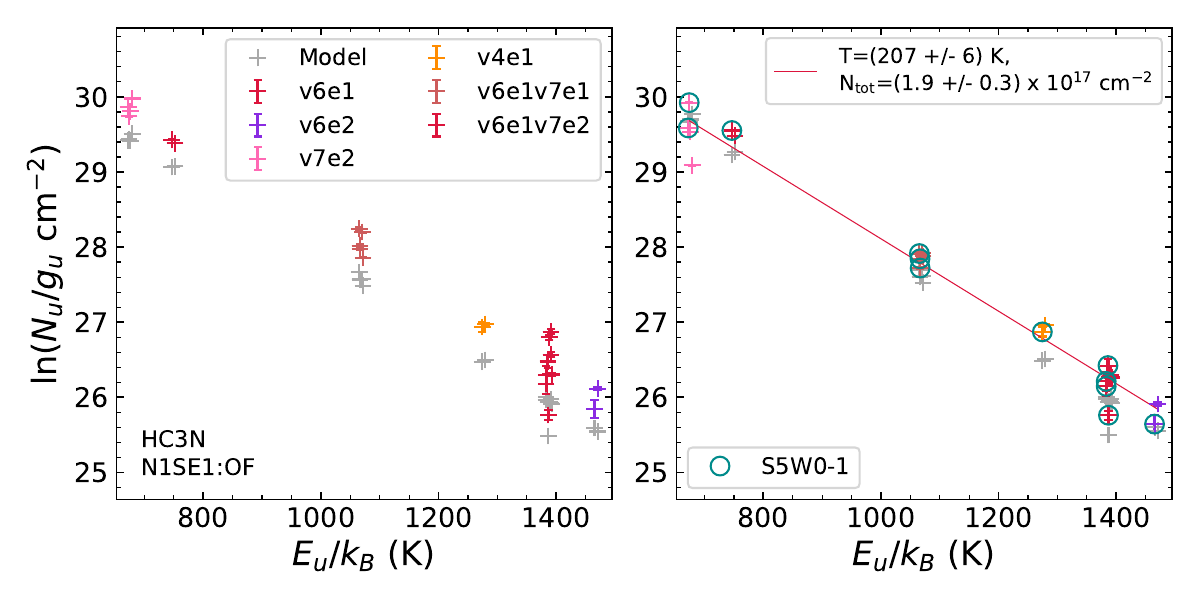}\\
    \includegraphics[width=0.48\textwidth]{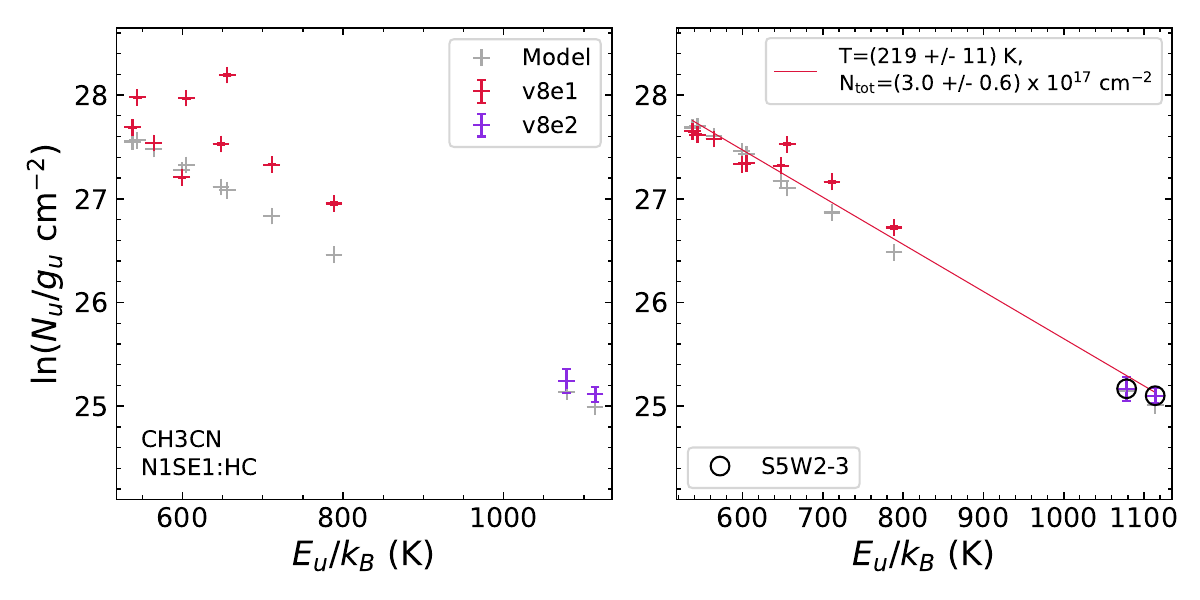}
    \includegraphics[width=0.48\textwidth]{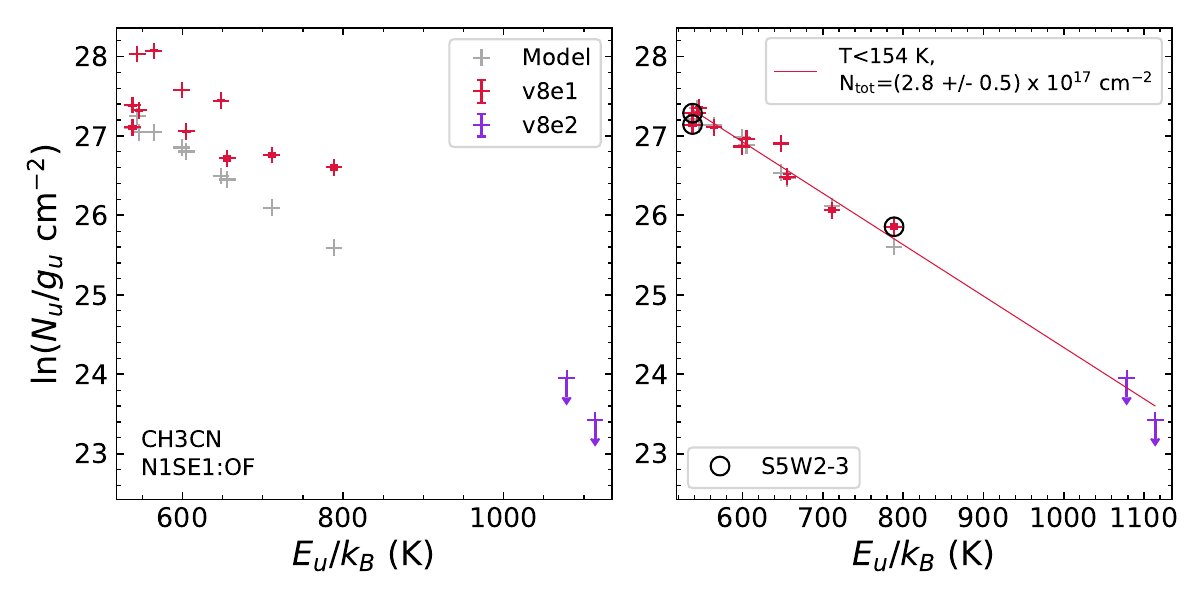}\\
    \includegraphics[width=0.48\textwidth]{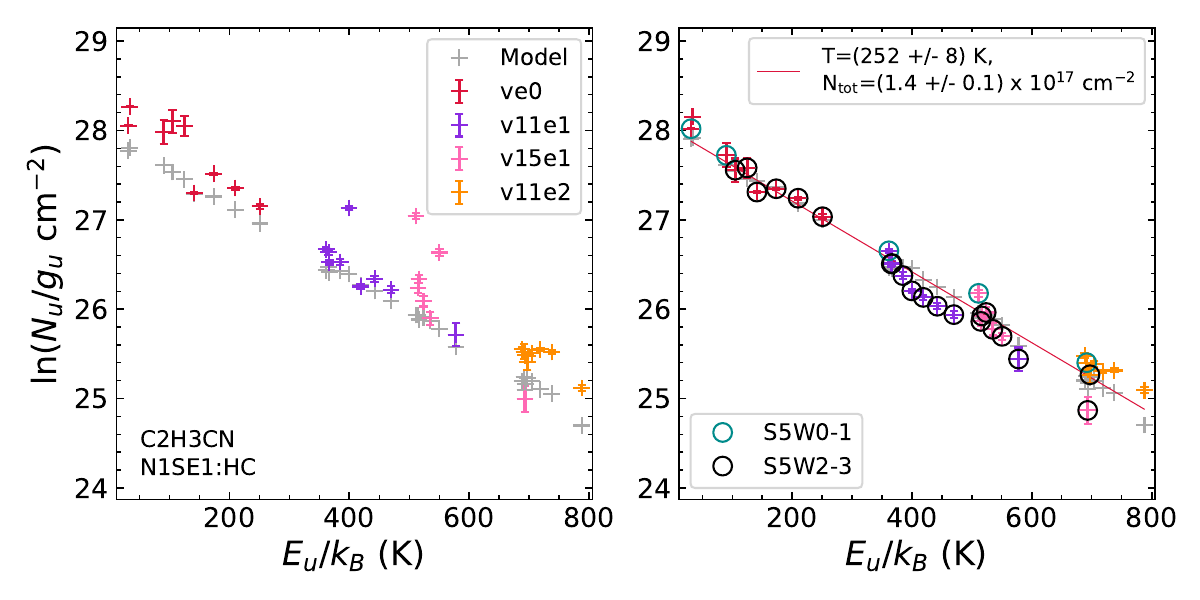}
    \includegraphics[width=0.48\textwidth]{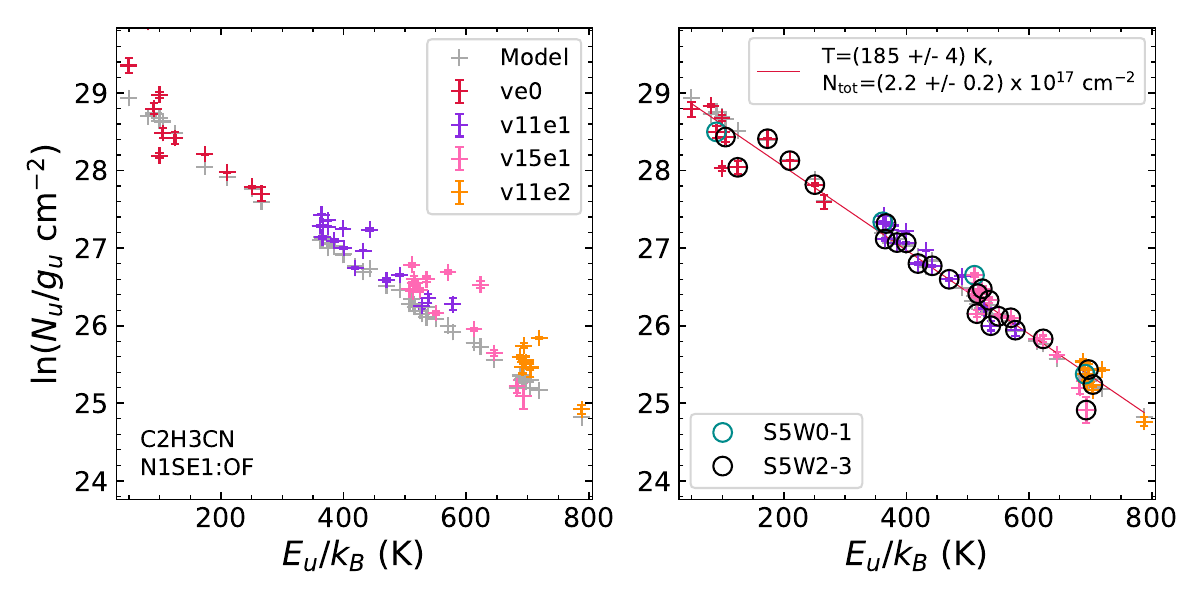}\\
    \includegraphics[width=0.48\textwidth]{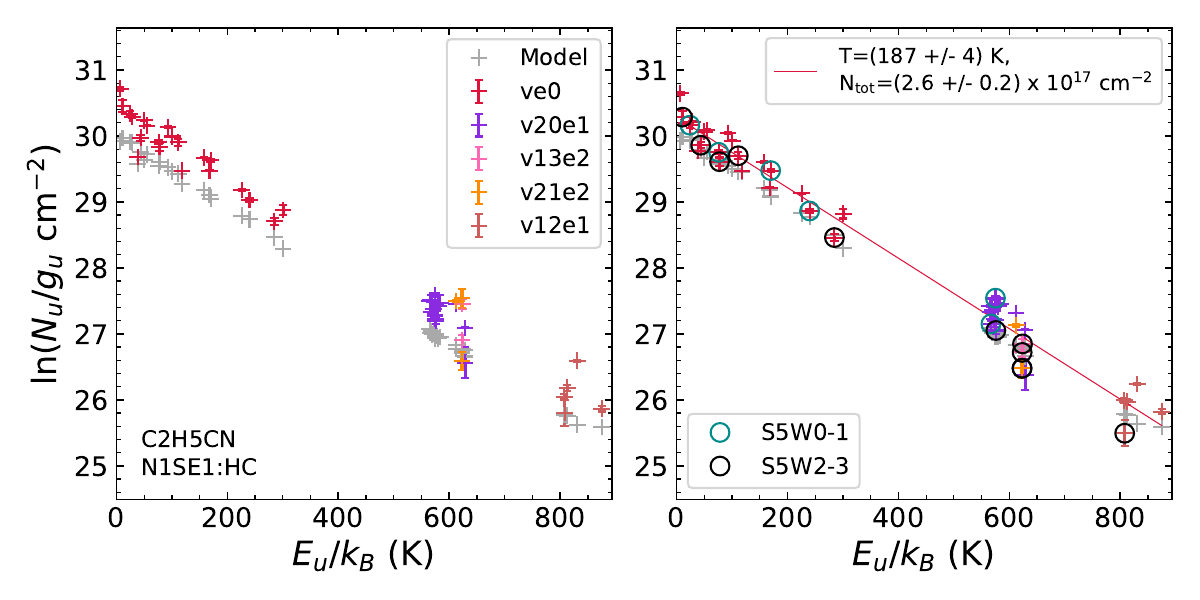}
    \includegraphics[width=0.48\textwidth]{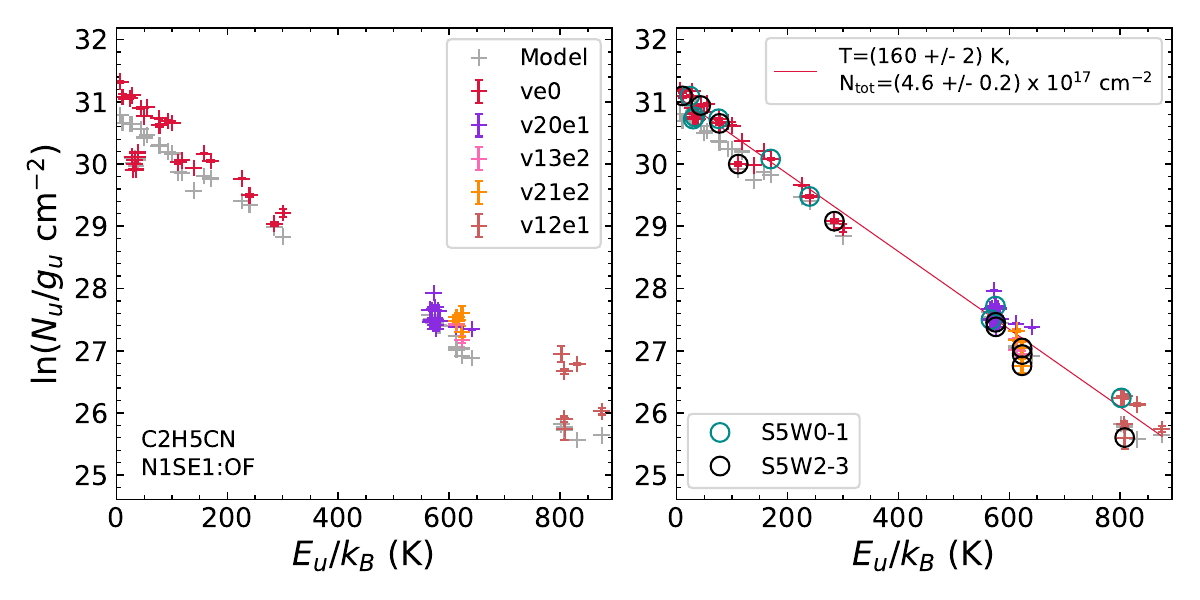}\\
    
    \caption{Same as Fig.\,\ref{fig:PD_ch3sh}, but for multiple molecules at position N1SE1 for the hot-core component (N1SE1:HC) and the outflow component (N1SE1:OF).}
    \label{fig:PD1_n1se1}
\end{figure*}
\begin{figure*}[htp!]
\addtocounter{figure}{-1}
    \includegraphics[width=0.48\textwidth]{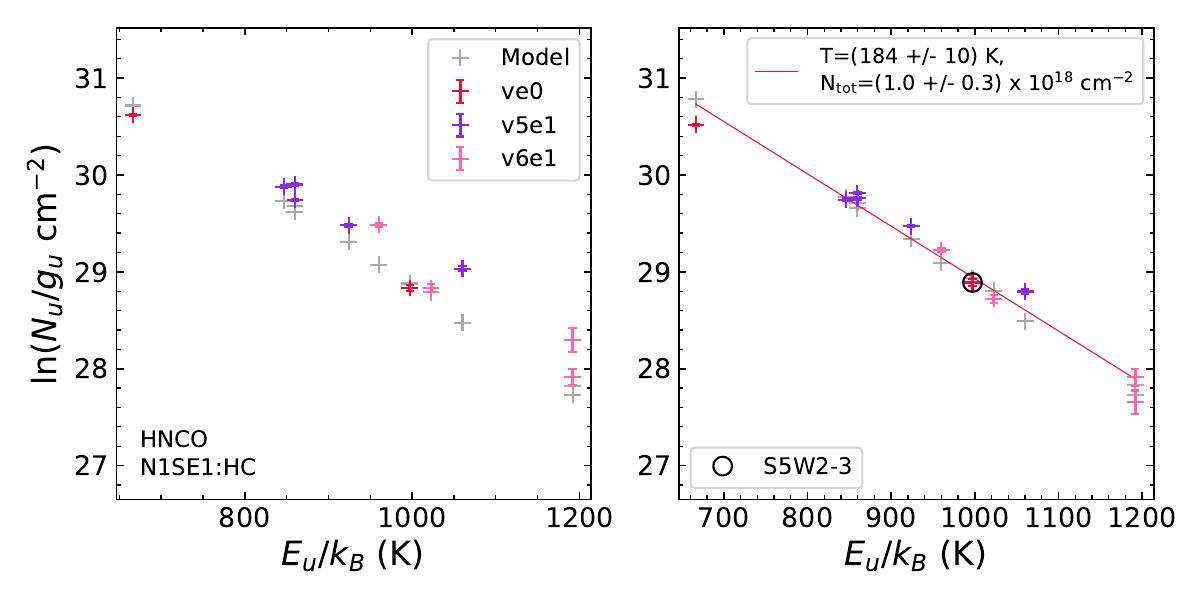}
    \includegraphics[width=0.48\textwidth]{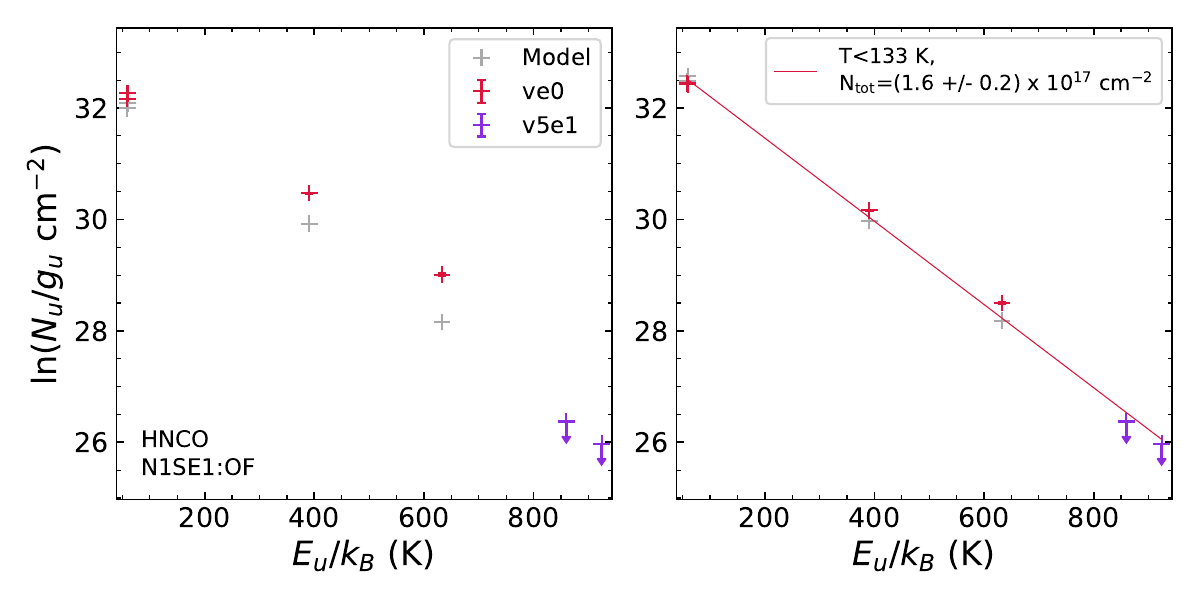}\\
    \includegraphics[width=0.48\textwidth]{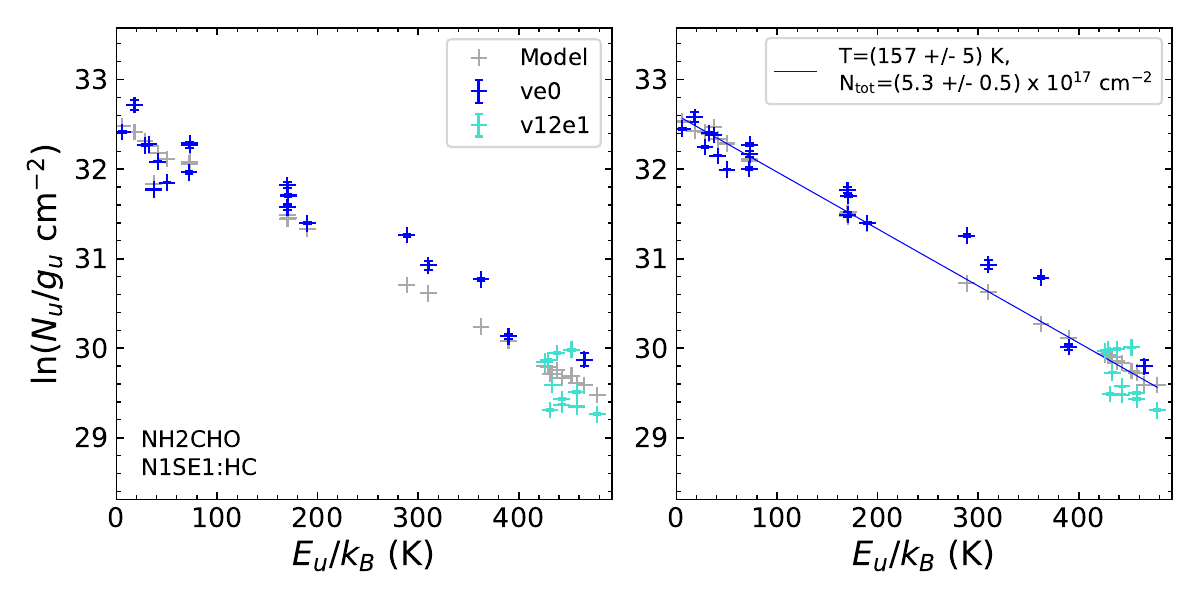}
    \includegraphics[width=0.48\textwidth]{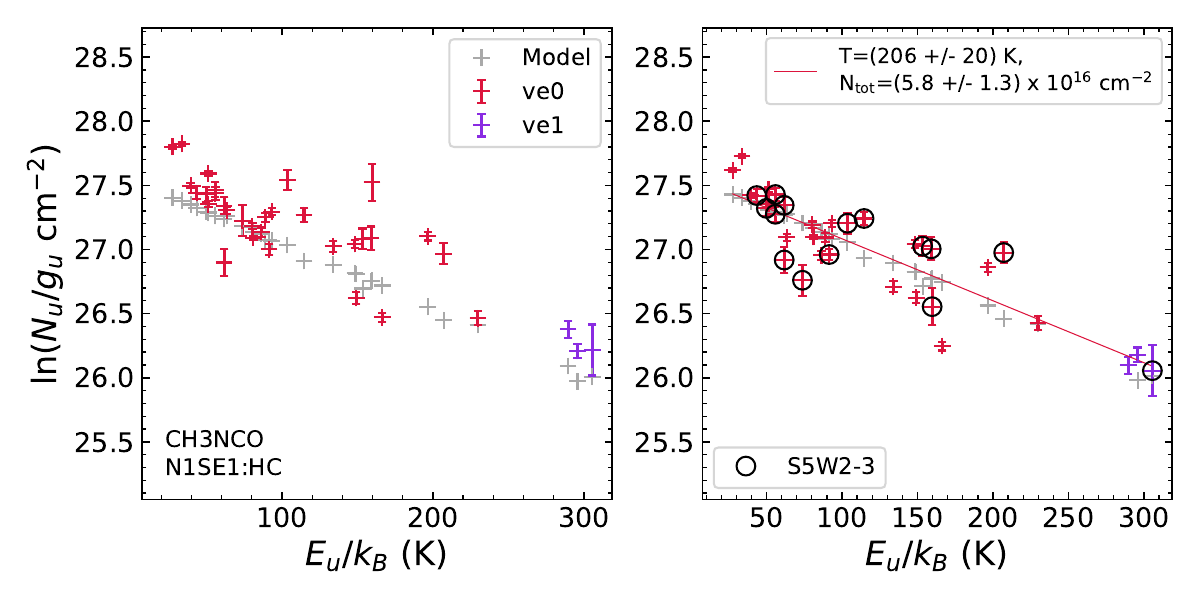}\\
    \includegraphics[width=0.48\textwidth]{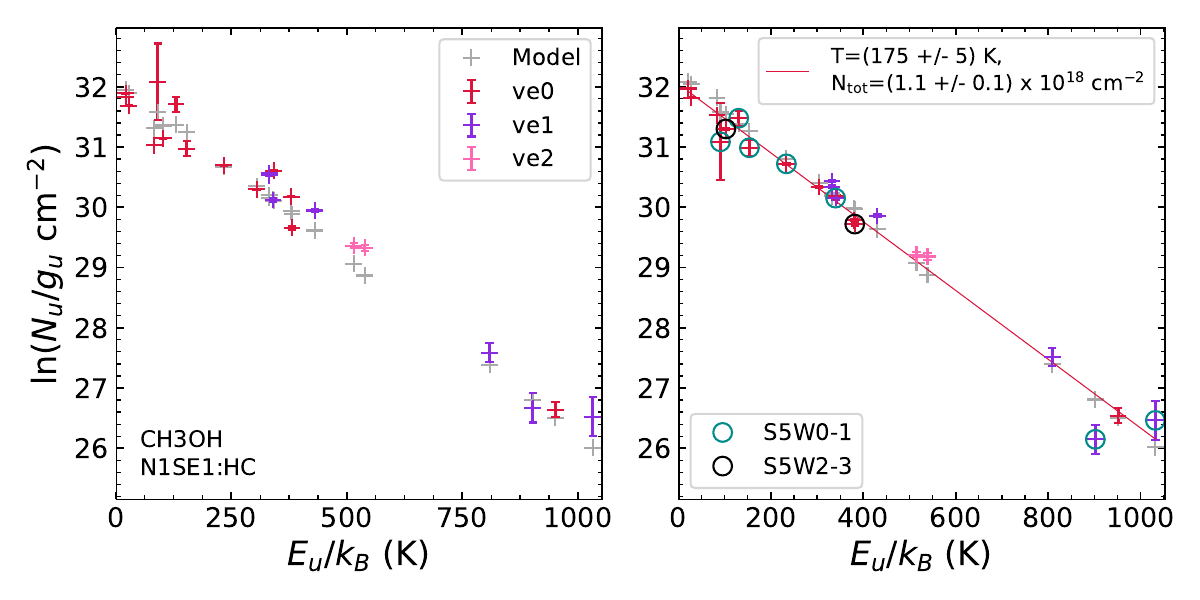}
    \includegraphics[width=0.48\textwidth]{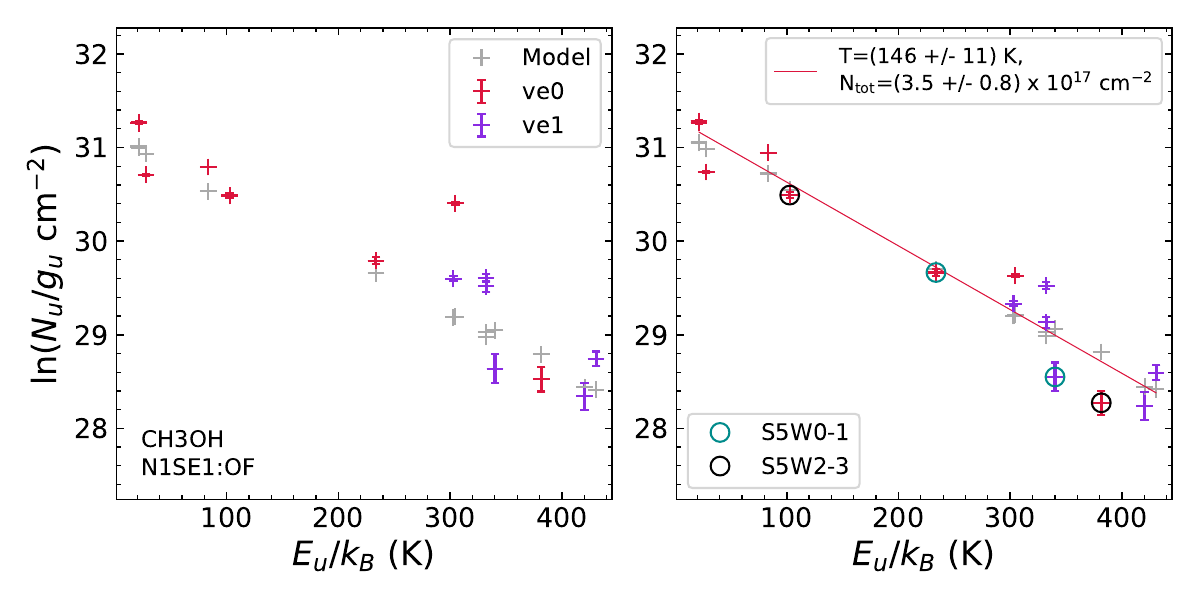}\\
    \includegraphics[width=0.48\textwidth]{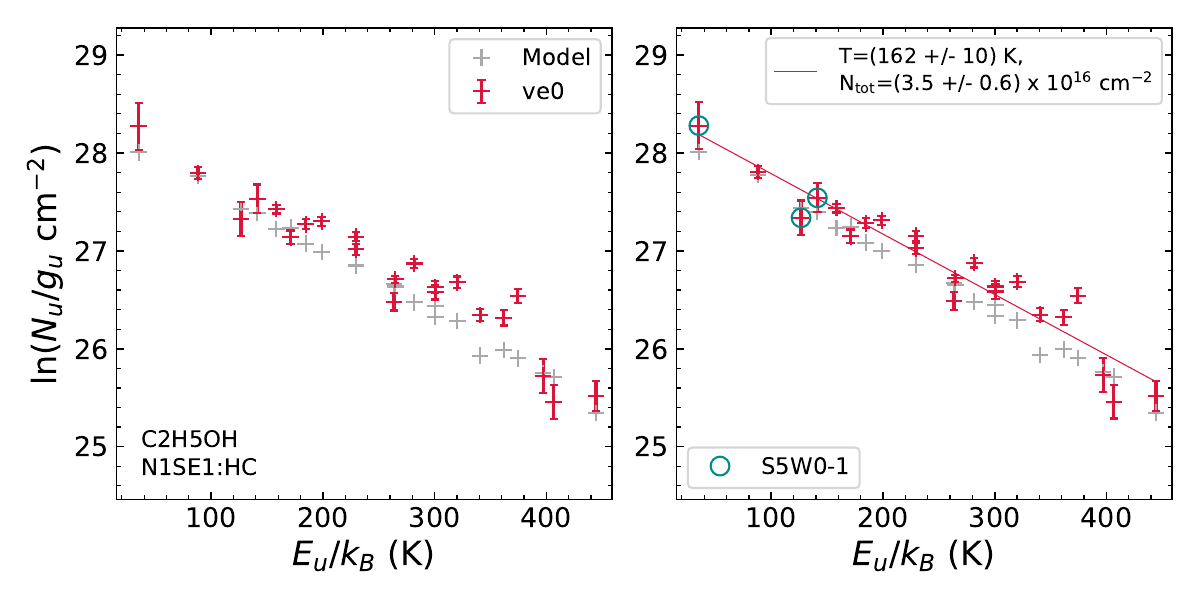}
    \includegraphics[width=0.48\textwidth]{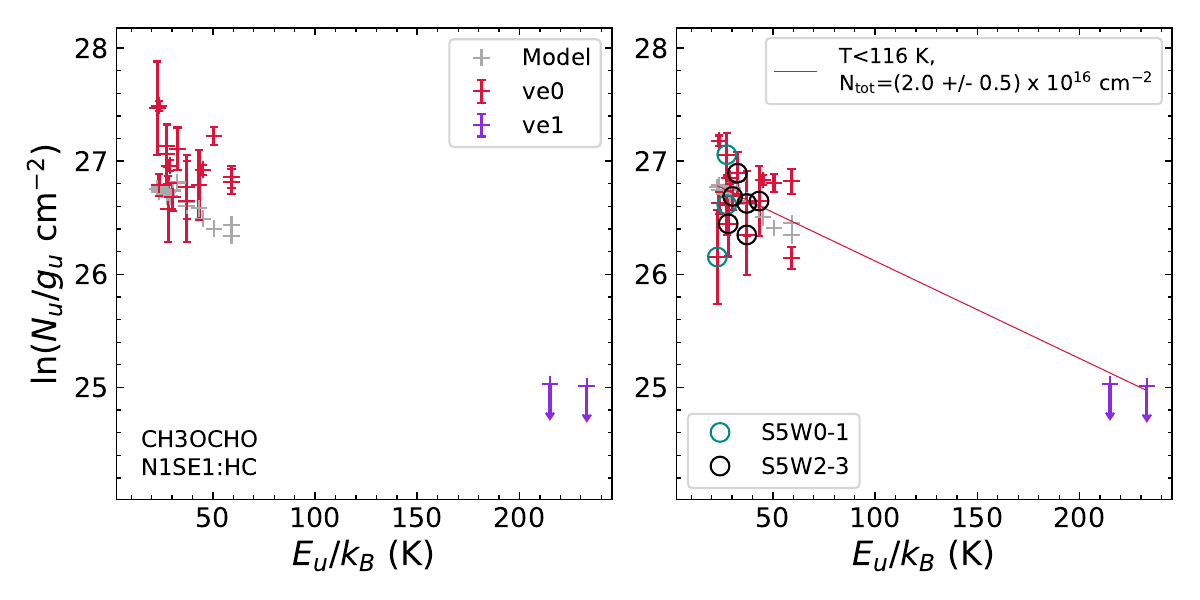}\\
    \includegraphics[width=0.48\textwidth]{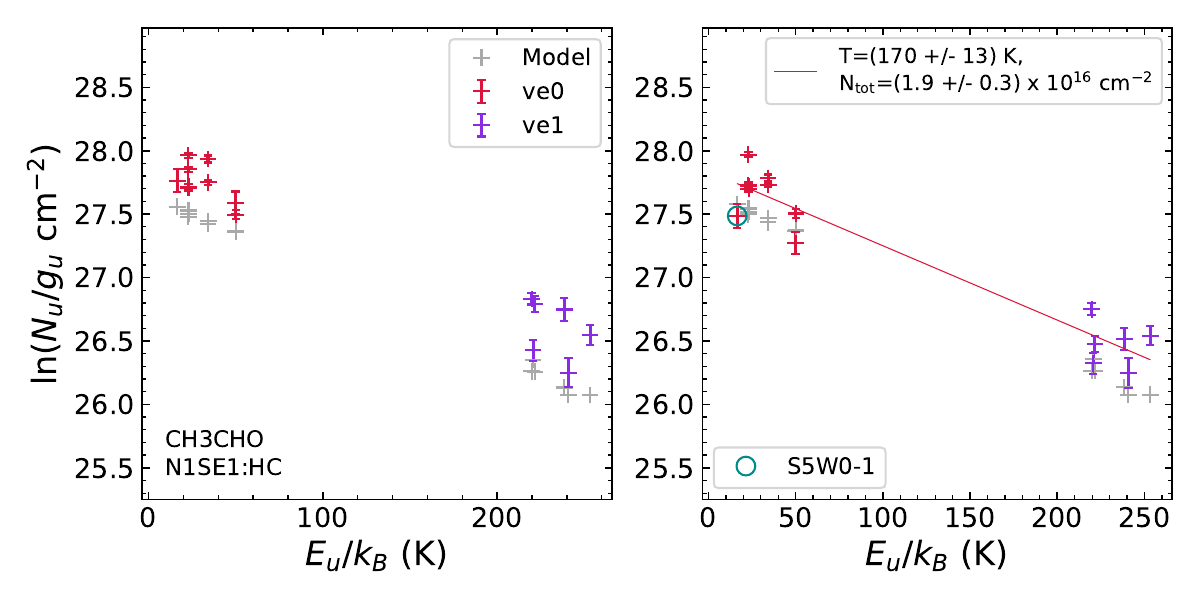}
    \includegraphics[width=0.48\textwidth]{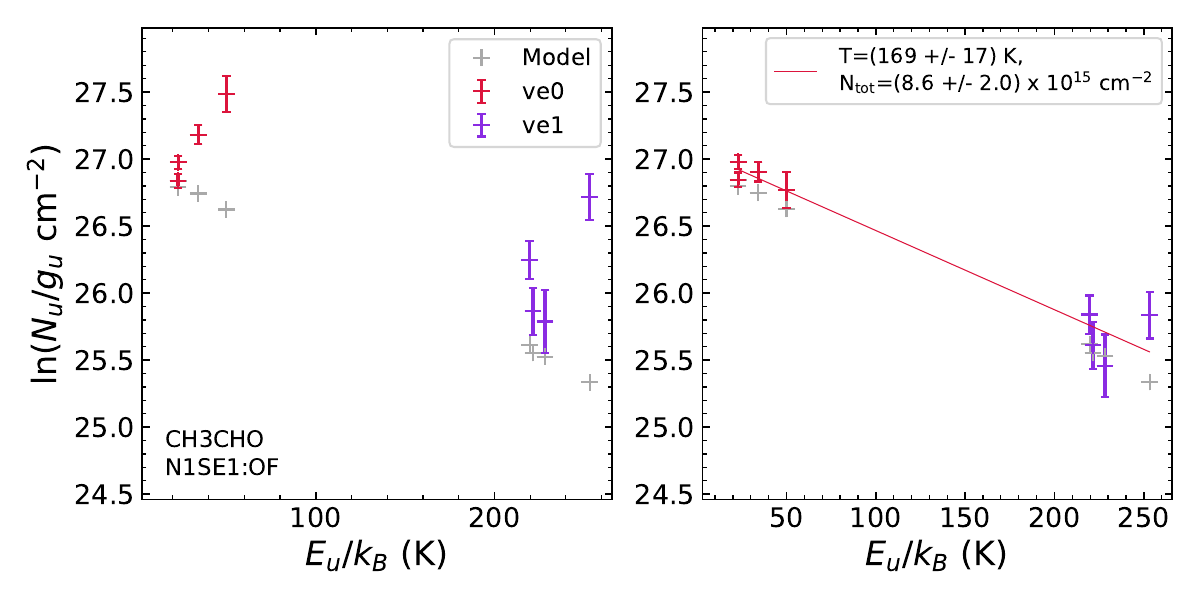}\\
    
    \caption{continued.}
    \label{fig:PD2_n1se1}
\end{figure*}
\begin{figure*}[htp!]
\addtocounter{figure}{-1}
    \includegraphics[width=0.48\textwidth]{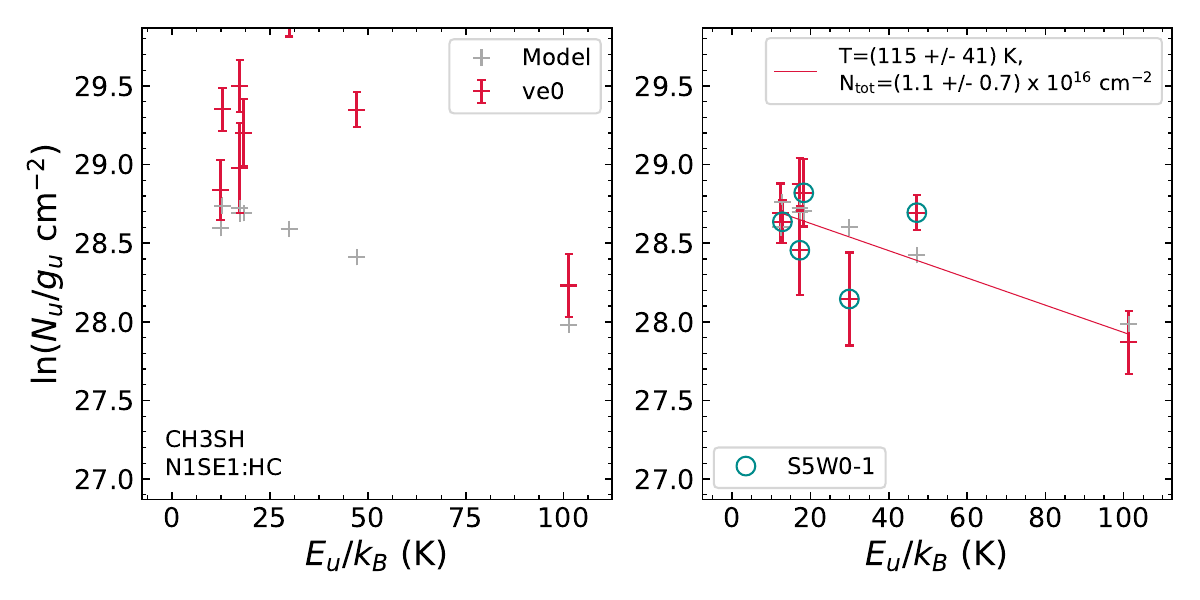}
    \includegraphics[width=0.48\textwidth]{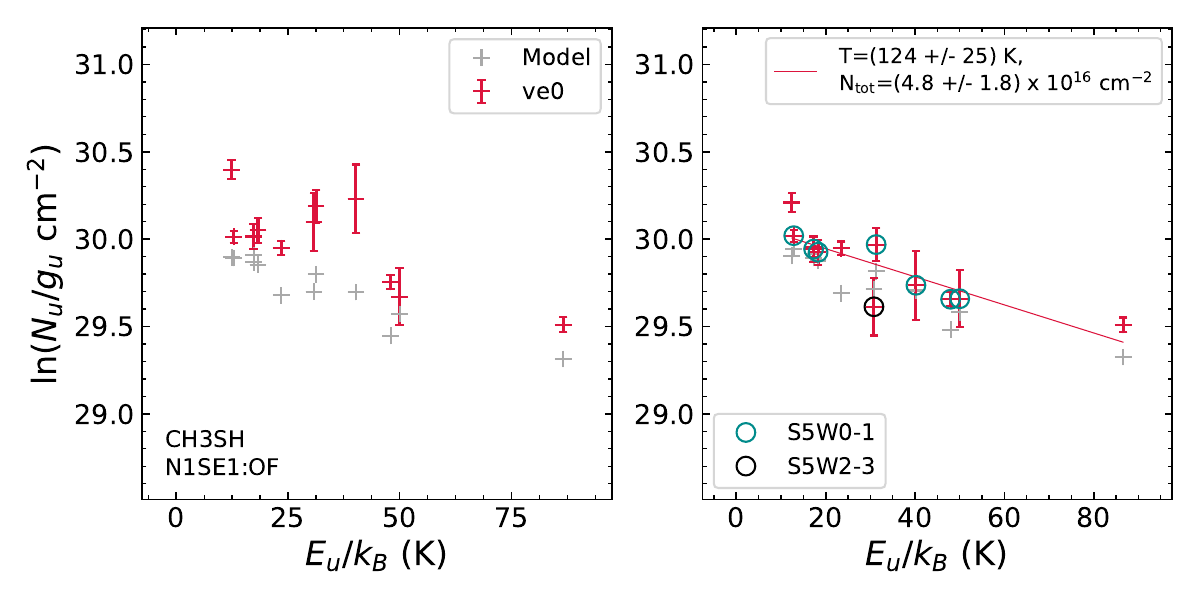}\\
    \includegraphics[width=0.48\textwidth]{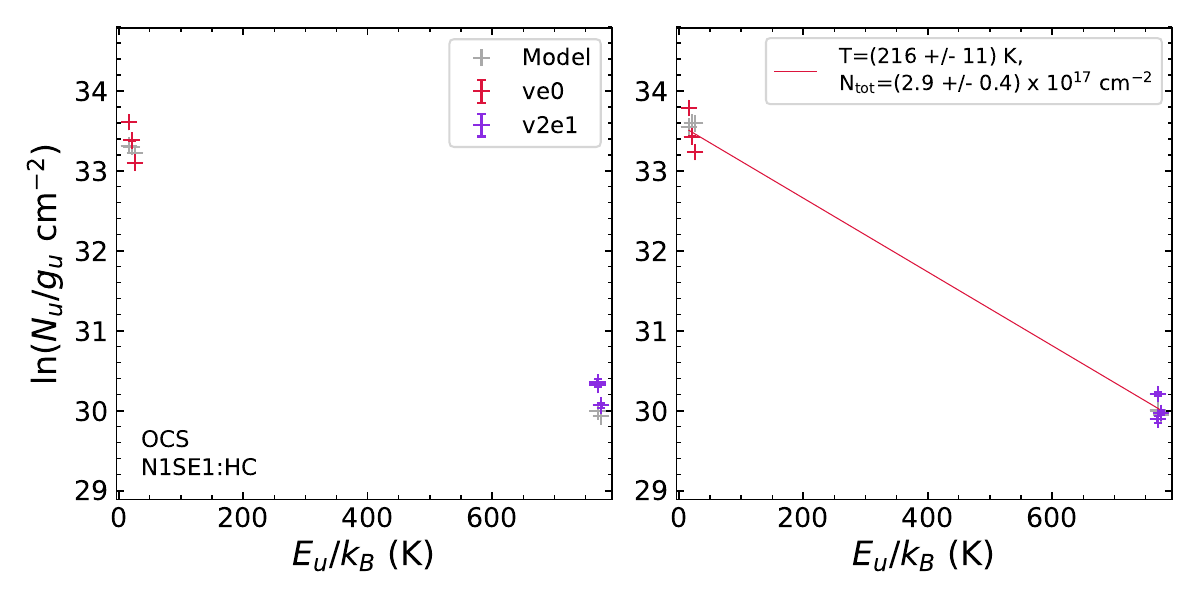}
    \includegraphics[width=0.48\textwidth]{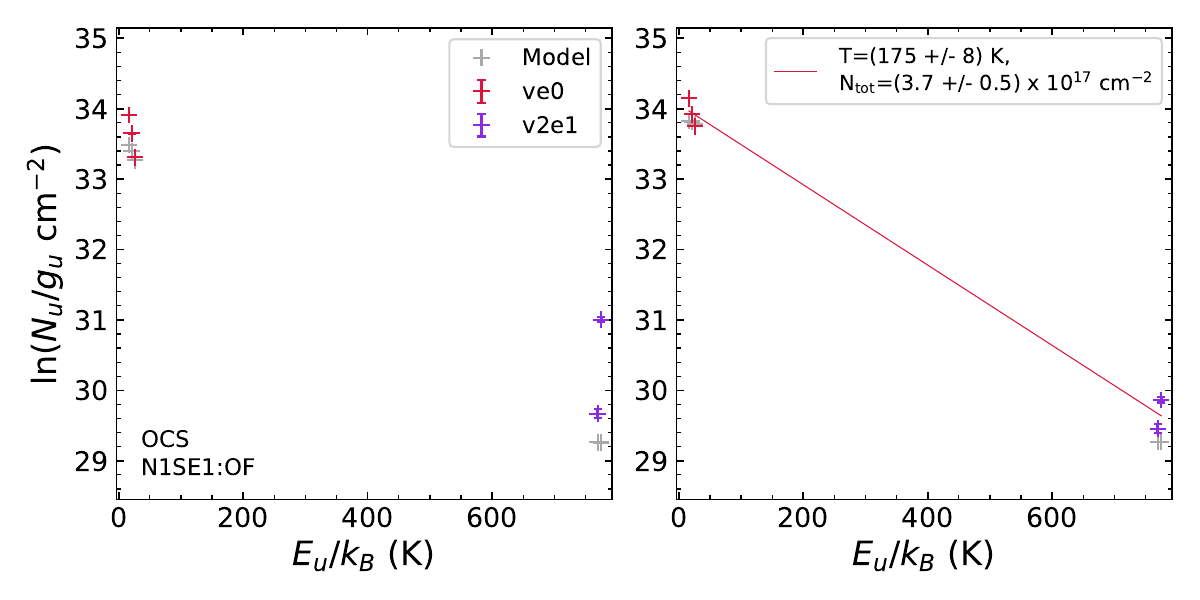}
    
    \caption{continued.}
    \label{fig:PD3_n1se1}
\end{figure*}
\begin{figure*}[htp!]
    \includegraphics[width=0.48\textwidth]{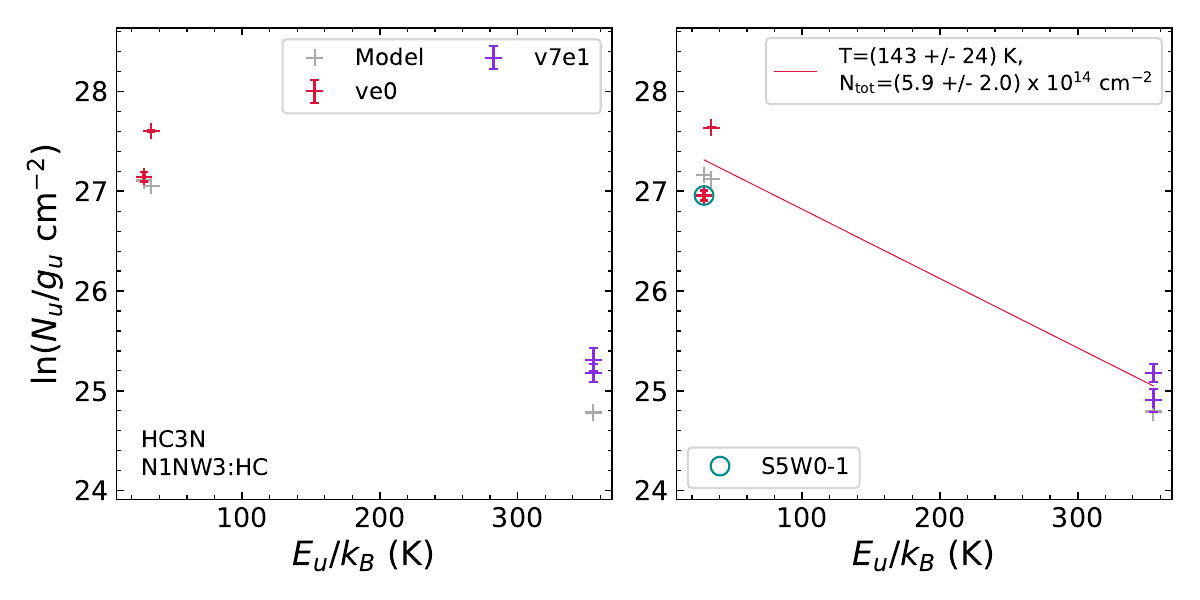}
    \includegraphics[width=0.48\textwidth]{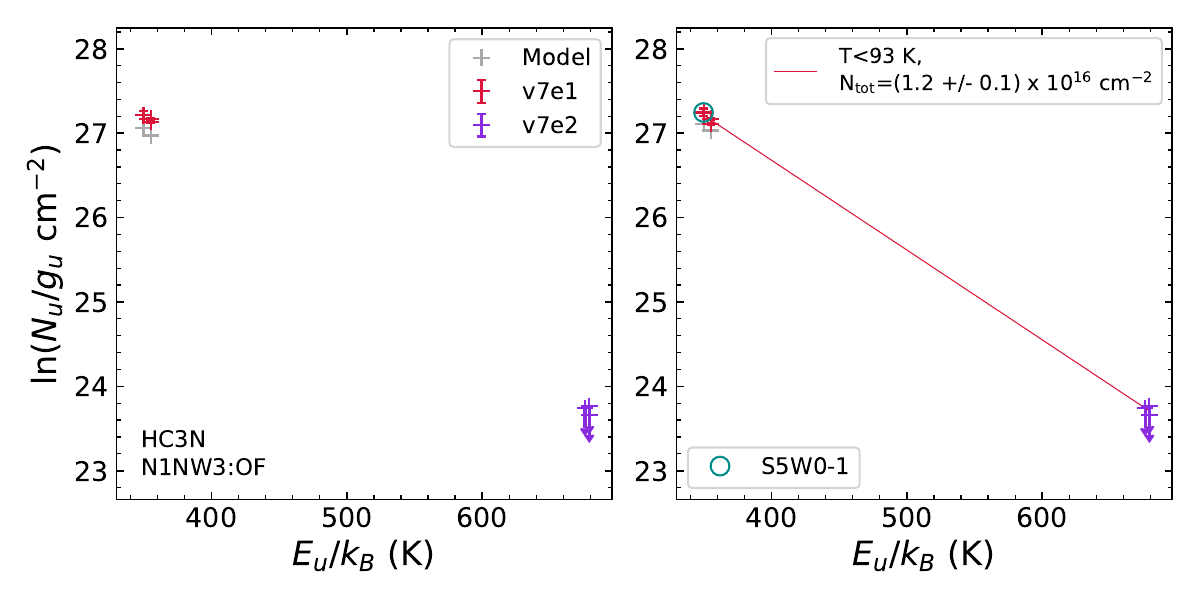}\\
    \includegraphics[width=0.48\textwidth]{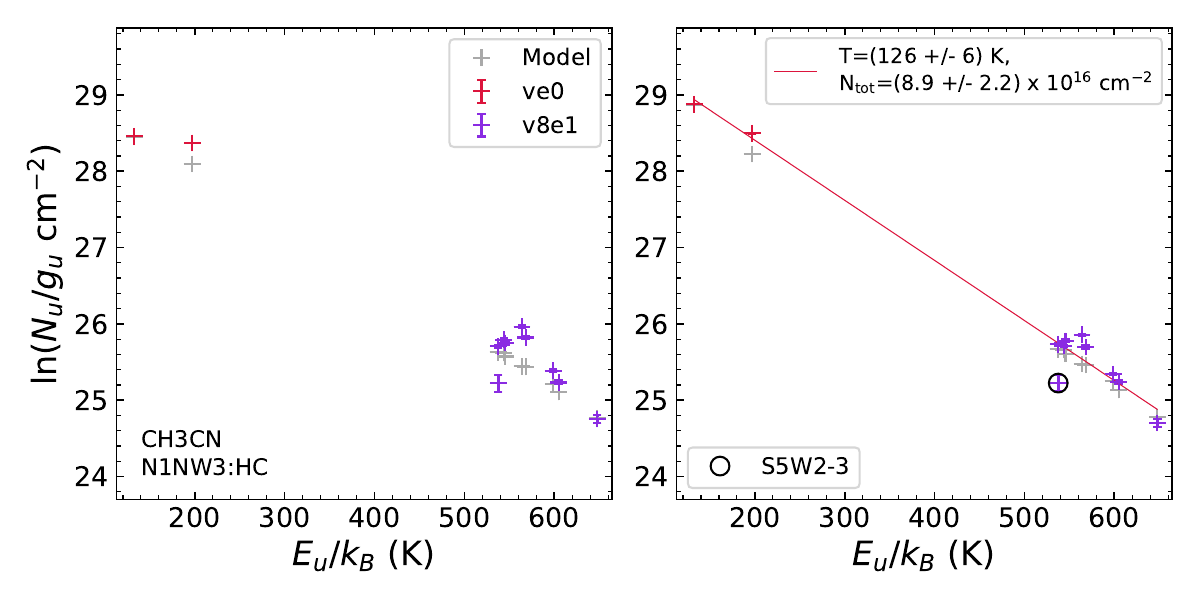}
    \includegraphics[width=0.48\textwidth]{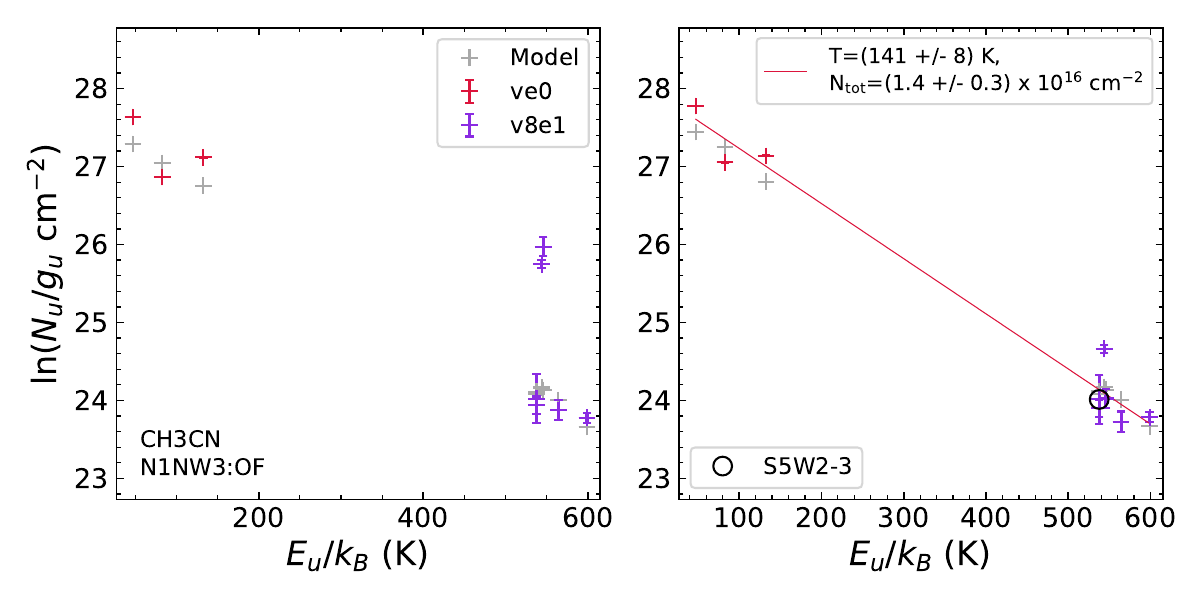}\\
    \includegraphics[width=0.48\textwidth]{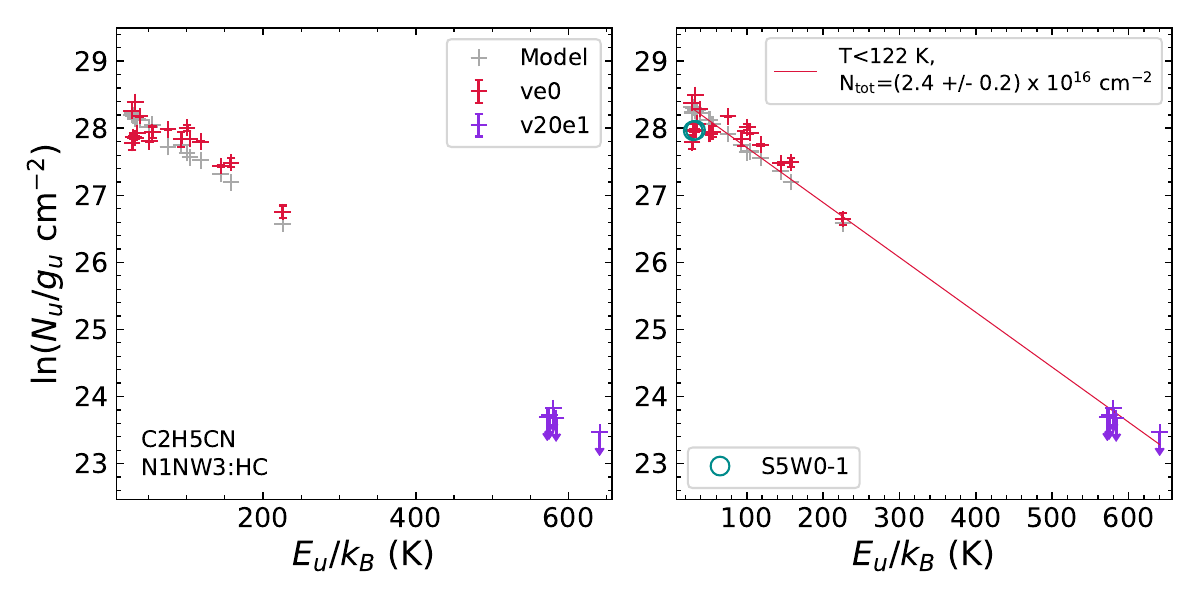}
    \includegraphics[width=0.48\textwidth]{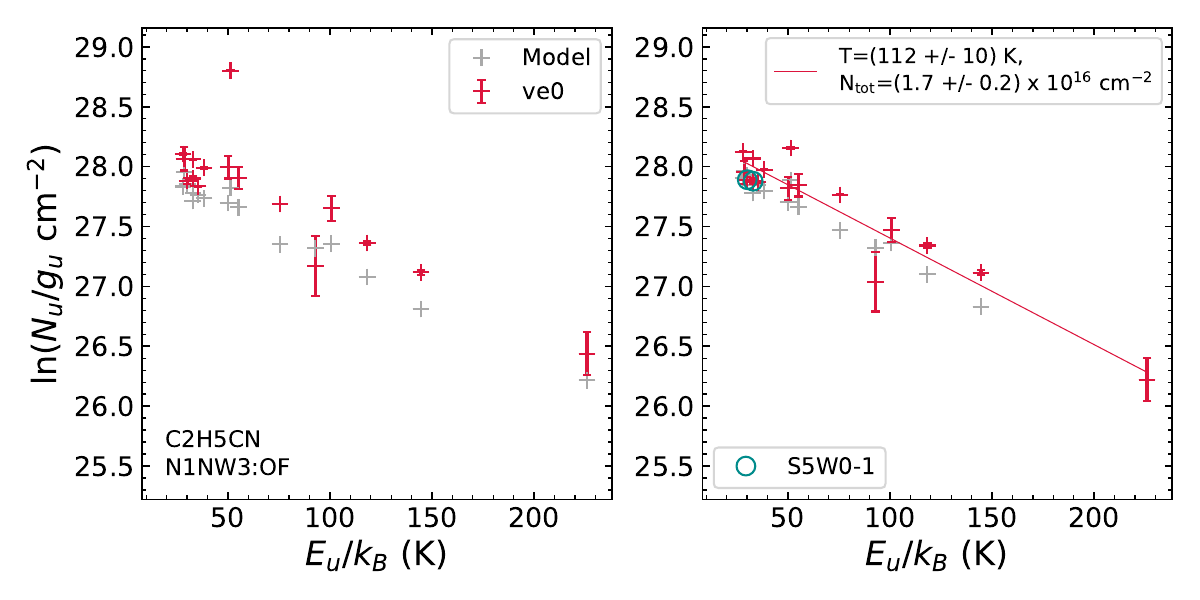}\\
    \includegraphics[width=0.48\textwidth]{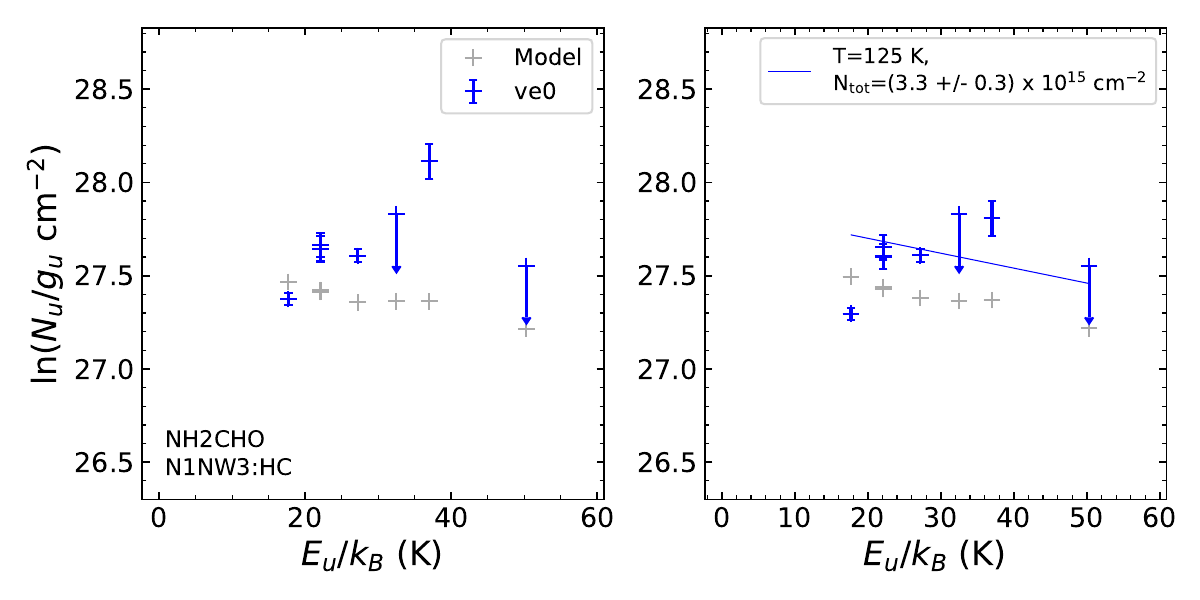}
    \includegraphics[width=0.48\textwidth]{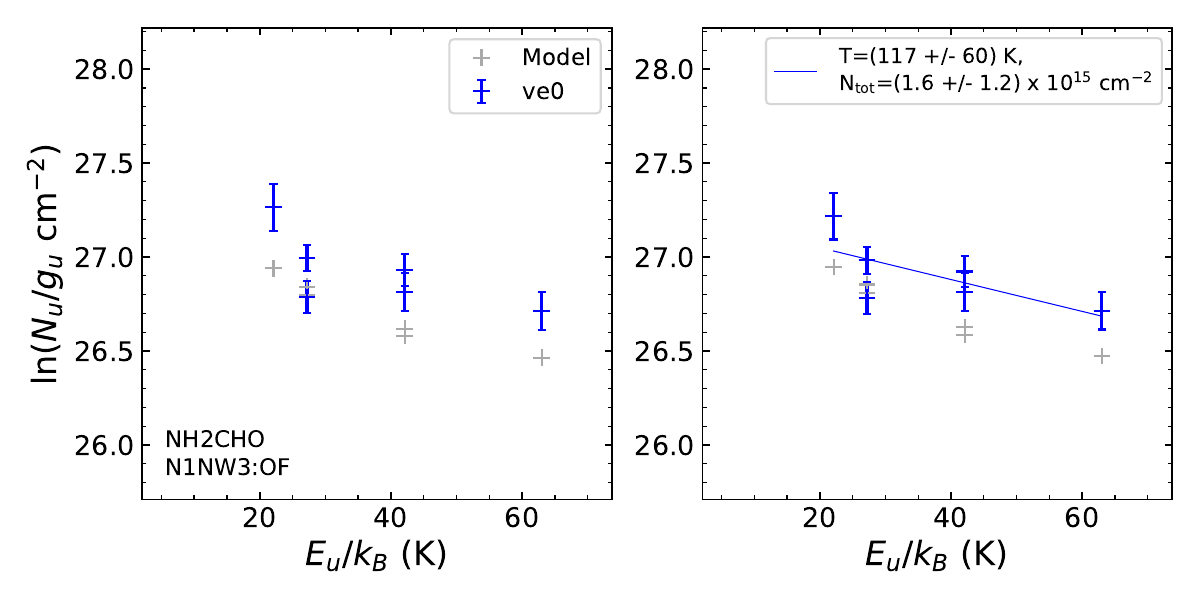}\\
    \includegraphics[width=0.48\textwidth]{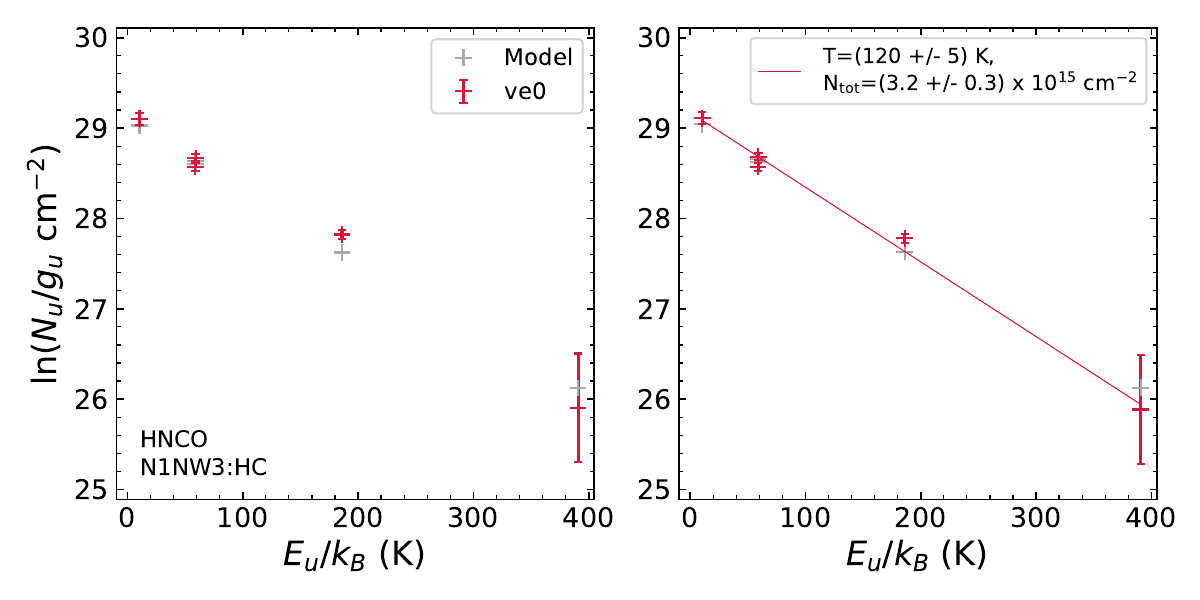}
    \includegraphics[width=0.48\textwidth]{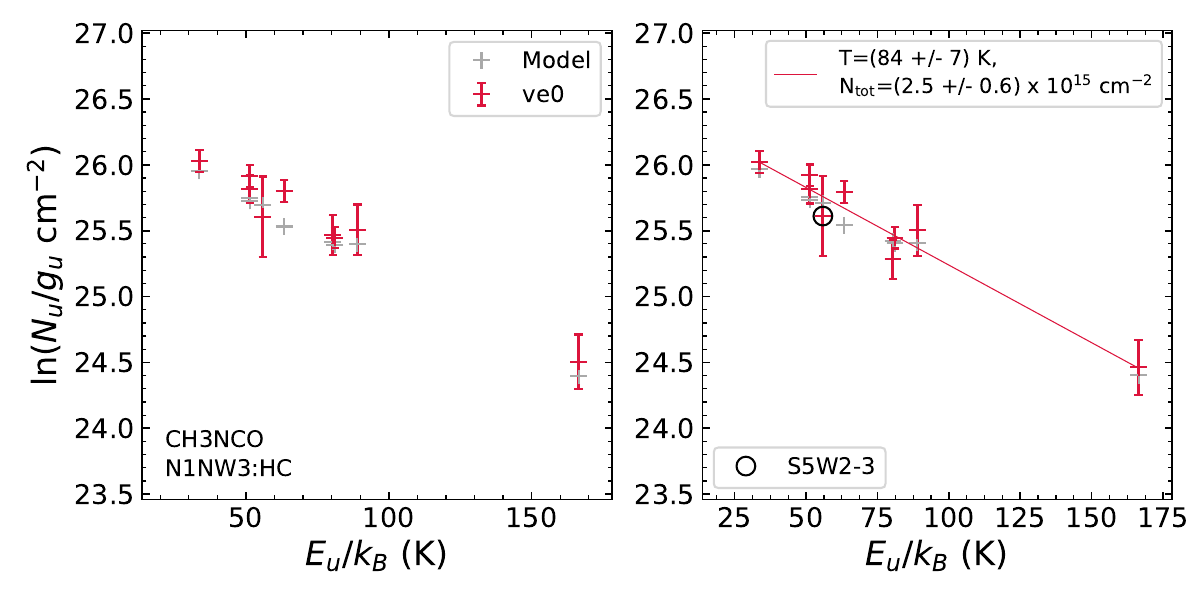}\\
    
    \caption{Same as Fig.\,\ref{fig:PD_ch3sh}, but for multiple molecules at position N1NW3 for the hot-core component (N1NW3:HC) and the outflow component (N1NW3:OF).}
    \label{fig:PD1_n1nw3}
\end{figure*}
\begin{figure*}[htp!]
\addtocounter{figure}{-1}
    \includegraphics[width=0.48\textwidth]{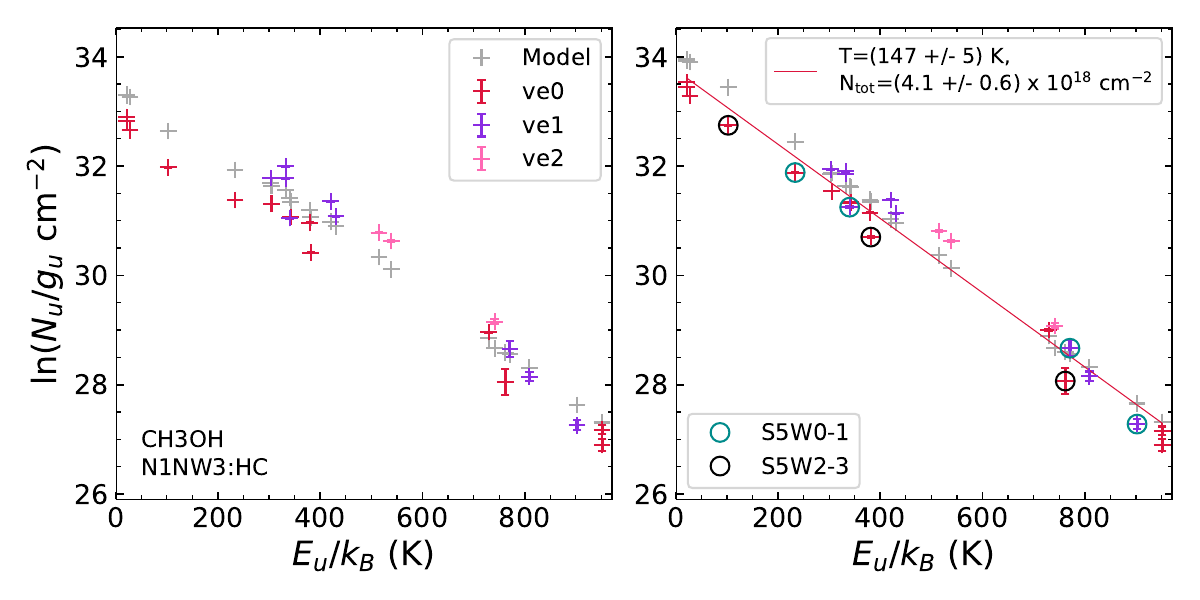}
    \includegraphics[width=0.48\textwidth]{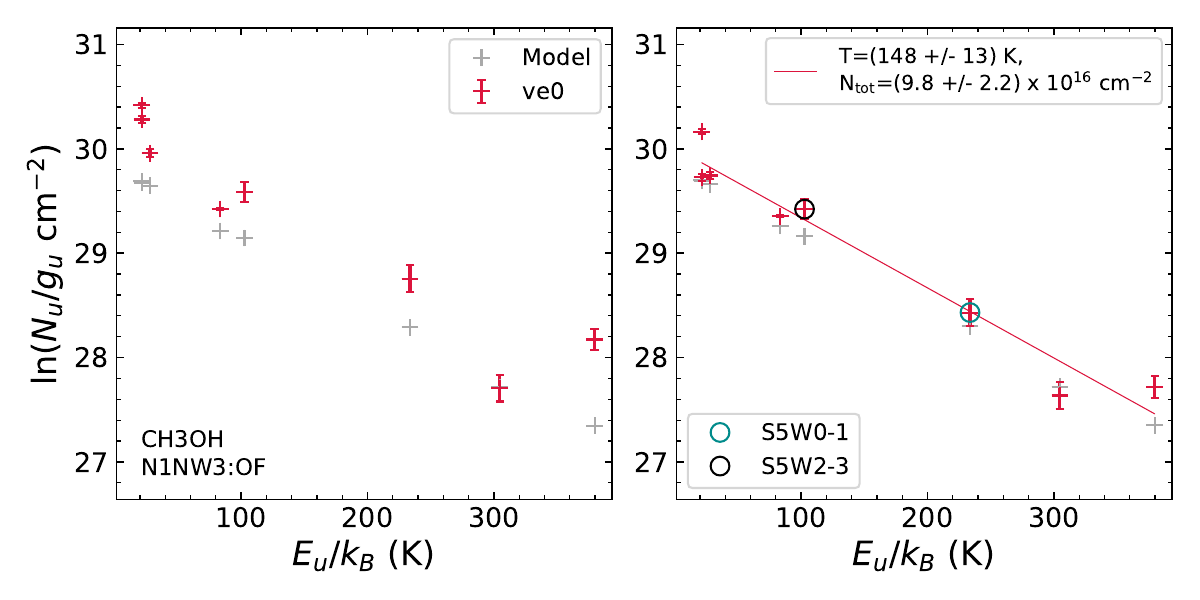}\\
    \includegraphics[width=0.48\textwidth]{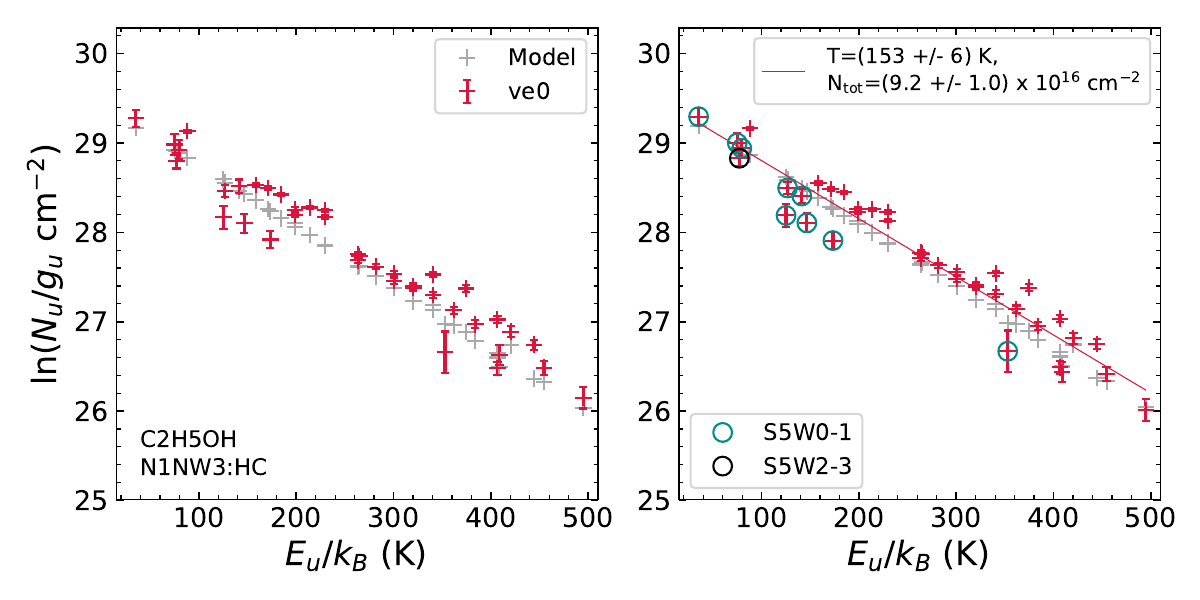}
    \includegraphics[width=0.48\textwidth]{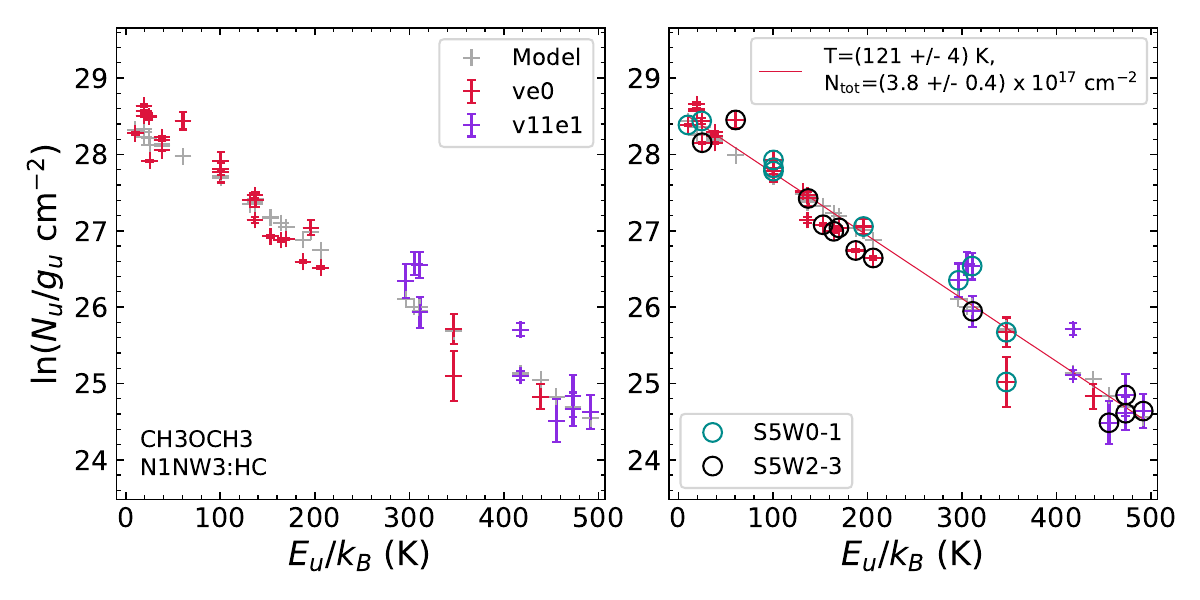}\\
    \includegraphics[width=0.48\textwidth]{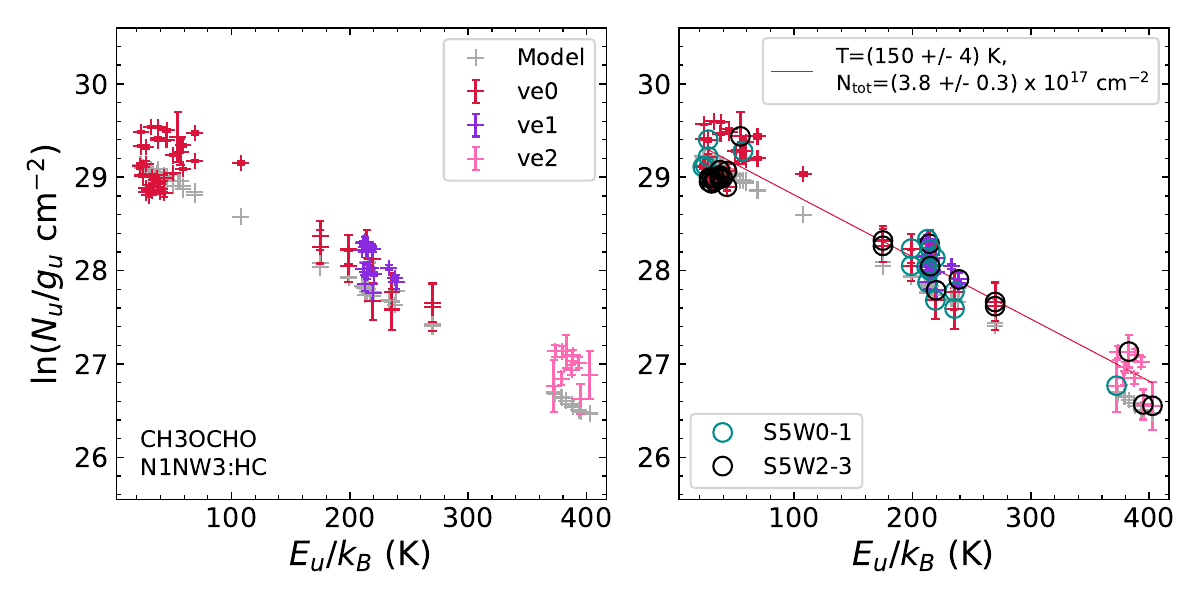}
    \includegraphics[width=0.48\textwidth]{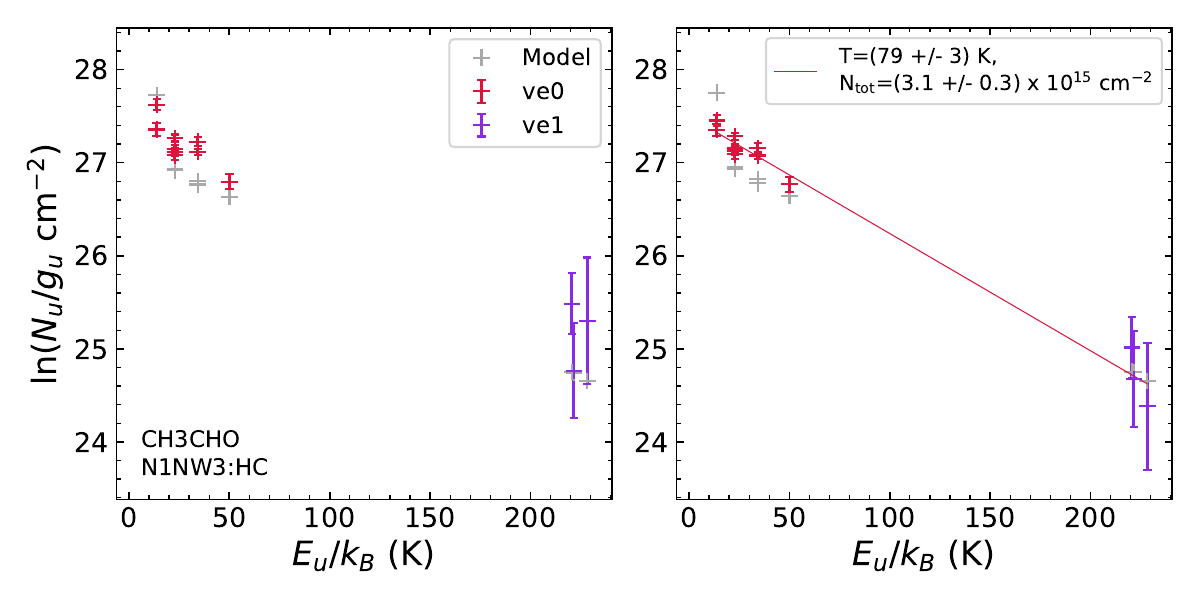}\\
    \includegraphics[width=0.48\textwidth]{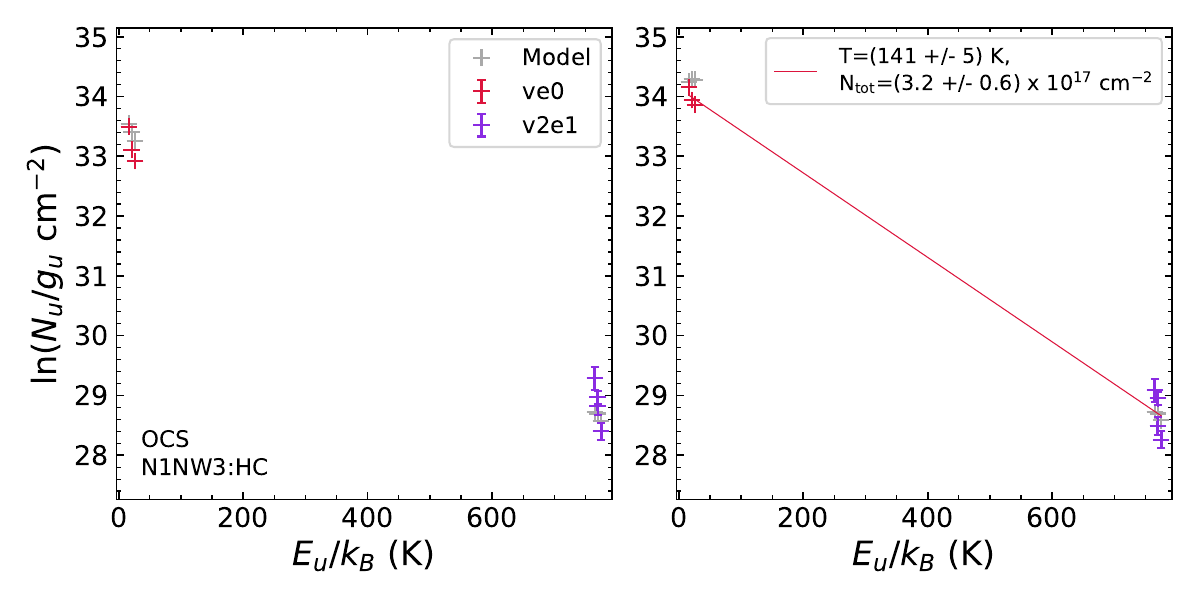}
    \includegraphics[width=0.48\textwidth]{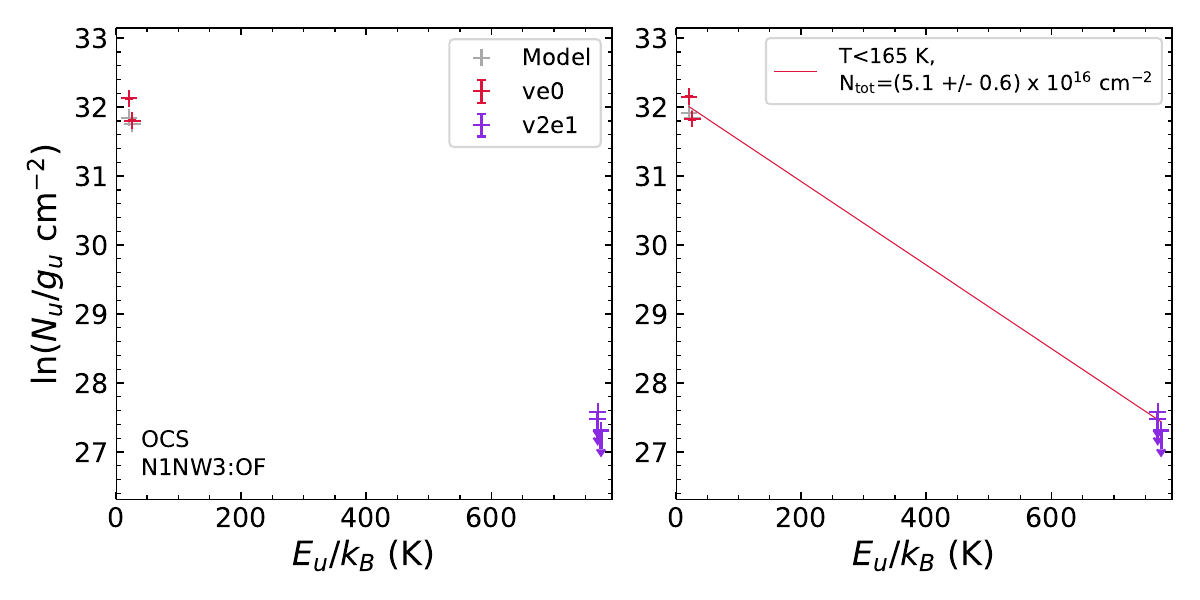}
    
    \caption{continued.}
    \label{fig:PD2_n1nw3}
\end{figure*}

\section{Additional tables}
Tables\,\ref{tab:n1se1hc}--\ref{tab:n1nw3of} summarise the parameters of the LTE radiative-transfer modelling performed with Weeds and of the population-diagram analysis for each COM and each of the four velocity components, respectively.

\input{tables/tab_n1se1hc.tex}
\input{tables/tab_n1se1of.tex}
\input{tables/tab_n1nw3hc.tex}
\input{tables/tab_n1nw3of.tex}

\section{LVINE maps for outflow emission}\label{app:lvine}

\subsection{Integration strategy}\label{ass:lvine}
\input{table_lvine}
Visualising the outflow emission is challenging due to the pervasive line emission close to the hot core's centre that can lead to contamination and the, in some cases, complex line profiles.
For each molecule, we made use of two transitions of similar upper-level energies and Einstein A coefficients, because these have similar intensities. The transitions with their properties are listed in Table\,\ref{tab:trans}. By comparing both spectra, we can thus identify emission that comes from other molecules, based on the assumption that such contaminating emission is unlikely to appear at exactly the same velocities in both spectra, as illustrated in Fig.\,\ref{fig:lvine_spec}. We compared the spectra of both transitions over an area of $8\arcsec\times 4$\arcsec around Sgr\,B2\,(N1) by extracting spectra every 0.5\arcsec. Although these transitions were taken from the higher angular-resolution setups 4 and 5, the mean beam sizes can still vary by up to a factor of 1.6 in size or 2.8 in area. Therefore, in the cases where the two selected transitions are covered in two different data cubes, the cube of higher angular resolution was smoothed with a two-dimensional Gaussian kernel to the resolution of the other cube. Overplotting both spectra within the same velocity range helped us identifying which part of the blue- and red-shifted emission belongs to the molecule of interest and what is contamination from other species. Based on these spectra, the integration strategy was developed. 

The procedure is based on the method of linewidth- and velocity-corrected integrated emission (LVINE) maps, according to which the integration limits in each pixel are adjusted based on variations of peak line velocities and linewidths measured in the spectra of a reference molecule or a set of reference 
molecules. We used this method in Paper\,I to determine the integration limits for the hot-core component for each molecule.
Therefore, maps of peak velocity, $\varv_t$, and linewidth, $FWHM_t$, were derived which are shown in Fig.\,B.2 in Paper\,I, using two template transitions. We selected one transition of ethanol at 108.44\,GHz with an upper level energy $E_u/k$ of 88\,K for positions close to the centre of Sgr\,B2\,(N1) as it remains sufficiently optically thin. At larger distances, that is when the ethanol line is too weak, we used a transition of methanol at 95.91\,GHz and an upper level energy of 21\,K. Where neither ethanol nor methanol are detected above an intensity threshold of 5$\sigma$, where $\sigma$ was taken from Table\,2 in \citet{Belloche19} for the respective observational setup, we assigned a fixed value of peak velocity and linewidth to that pixel. 
In this work, the LVINE method was utilised to define the starting points (from the hot-core component) for the integration over the blue- and red-shifted emission. Because ethanol does not show blue- or red-shifted emission and methanol emission remains dominated by the hot-core emission, we took the peak velocities and linewidths from the template maps derived in Paper\,I. 
However, during the inspection of the spectra of the two methanol transitions, we noticed that for some pixels at greater distances, we used the fixed values of peak velocity and linewidth although the methanol transition was still detected. This means that the methanol line at these positions was not properly fitted during the derivation of the peak-velocity and linewidth template maps. Therefore, we improved the procedure and used the updated peak-velocity and linewidth maps to produce Figs.\,\ref{fig:lvine_coms} and \ref{fig:COMof}--\ref{fig:COMof3}. Since these adjustments are made beyond distances of 3\arcsec\, from the hot core's centre, the maps shown in Fig.\,4 of Paper\,I are not severely affected. 

The channels where the velocity is $\varv_t\pm FWHM_t$ were first computed for each pixel. We pushed these limits by two additional channels away from the source velocity to define the inner integration limits of the blue- and red-shifted emission, in order to reduce the contamination from the hot-core component. In that way, however, it is not excluded that we lose some low-velocity outflow emission close to $\varv_{\rm sys}$. 
The outer integration limits are set to a fixed value based on the comparison of the spectra of the two transitions for each molecule such that no outflow emission is missed and no contaminating emission, if identified and if possible to avoid, is added to the integration. 

For each molecule, the integration strategy uses both selected transitions such that for a given channel the 
transition with the lowest intensity is used. This is demonstrated with the filled histograms in Fig.\,\ref{fig:lvine_spec}. In that way, we reduce the contribution of contamination by other species. The procedure is slightly more complicated when the intensity of one or both transitions is below 2$\sigma$, where $\sigma$ is the respective noise level in the spectrum, which can be different if the transitions are observed in two observational setups of different sensitivity. In theses cases, we use the spectrum obtained with better sensitivity for integration. The following scheme, which is applied channel-wise, presents the procedure in detail:
\begingroup
\allowdisplaybreaks
\begin{align}
    &\text{if } \sigma_1 \leq \sigma_2 \text{ then} \nonumber \\
    &\quad\text{if } T_{2,c} < 2\sigma_2 \text{ then} \nonumber \\    
    & \quad\quad \text{if } T_{1,c} < 2\sigma_1 \text{ then} \nonumber \\  
    & \quad\quad\quad \text{use } T_{1,c} \nonumber \\
    & \quad\quad \text{else if } 2\sigma_1 \leq T_{1,c} < 2\sigma_2 \text{ then} \nonumber \\ 
    & \quad\quad\quad \text{use } T_{1,c} \nonumber \\
    & \quad\quad \text{else if } T_{1,c} \geq 2\sigma_2 \text{ then} \nonumber \\  
    & \quad\quad\quad \text{use } T_{2,c} \nonumber \\
    &\quad\text{else if } T_{2,c} \geq 2\sigma_2 \text{ then} \nonumber \\
    & \quad\quad \text{if } T_{1,c} \leq T_{2,c} \text{ then} \nonumber \\ 
    & \quad\quad\quad \text{use } T_{1,c} \nonumber \\   
    & \quad\quad \text{if } T_{1,c} > T_{2,c} \text{ then} \nonumber \\   
    & \quad\quad\quad \text{use } T_{2,c}, \nonumber  
\end{align}
\endgroup
where $\sigma_{1,2}$ are the noise levels in spectra 1 and 2 either taken from Table\,2 in \citet{Belloche19} or measured in the smoothed data cubes and $c$ indicates that this runs over each channel within the integration limits. If $\sigma_1 > \sigma_2$, the conditions apply to $T_1$ and $T_2$ vice versa.
Although the method reduces contamination, it does not avoid everything if in a given channel both spectra contain some contamination.

\subsection{Details on the outflow morphology}\label{app:detailedmorph}

Although the outflow reveals an overall bipolar morphology, there is a lot of additional structure in the form of intensity peaks and more extended features that are described in more detail in the following. 
Figures\,\ref{fig:COMof}--\ref{fig:COMof3} show again the LVINE maps of the molecules in Fig.\,\ref{fig:lvine_coms}, but the presentation is differently.
The right-most panels in Figs.\,\ref{fig:COMof}--\ref{fig:COMof3} are the same as in Fig.\,\ref{fig:lvine_coms} showing integrated intensities over the full range of blue- and red-shifted velocities as indicated by the dashed coloured lines in Fig.\,\ref{fig:lvine_spec}. This also applies to the left panels (labelled `all v'), but red- and blue-shifted emissions are shown in separate panels. 
In addition, we show maps of only the lower- (low-v) and higher-velocity (high-v) emission. The low-velocity maps show  emission integrated from the inner integration limits up to half the number of channels used for the full integration interval, while the high-velocity maps include emission from the other half down or up to the outer integration limits 
Therefore, this division into the low- and high-velocity regimes is again pixel-dependent and each panel in Figs.\,\ref{fig:COMof}--\ref{fig:COMof3} shows maps of integrated intensities over different velocity ranges for each molecule, which has to be kept in mind when comparing the emission morphologies of the molecules in the following.
Because the only available transition for SiO at 86.85\,GHz severely suffers from absorption, the LVINE method could not be applied, however, we show SiO maps in Fig\,\ref{fig:COMof2}. For the all-v maps, intensities were integrated as in Fig.\,\ref{fig:so}, that is ignoring channels with absorption. The intermediate integration limit was taken from SO.  
In addition to the LVINE maps, Fig.\,\ref{fig:chan_maps} shows velocity-channel maps of eight molecules to study changes in the emission morphology within smaller velocity intervals.

\subsubsection{Blue-shifted emission}
\begin{figure*}[h!]
    \centering
    \includegraphics[width=0.865\textwidth]{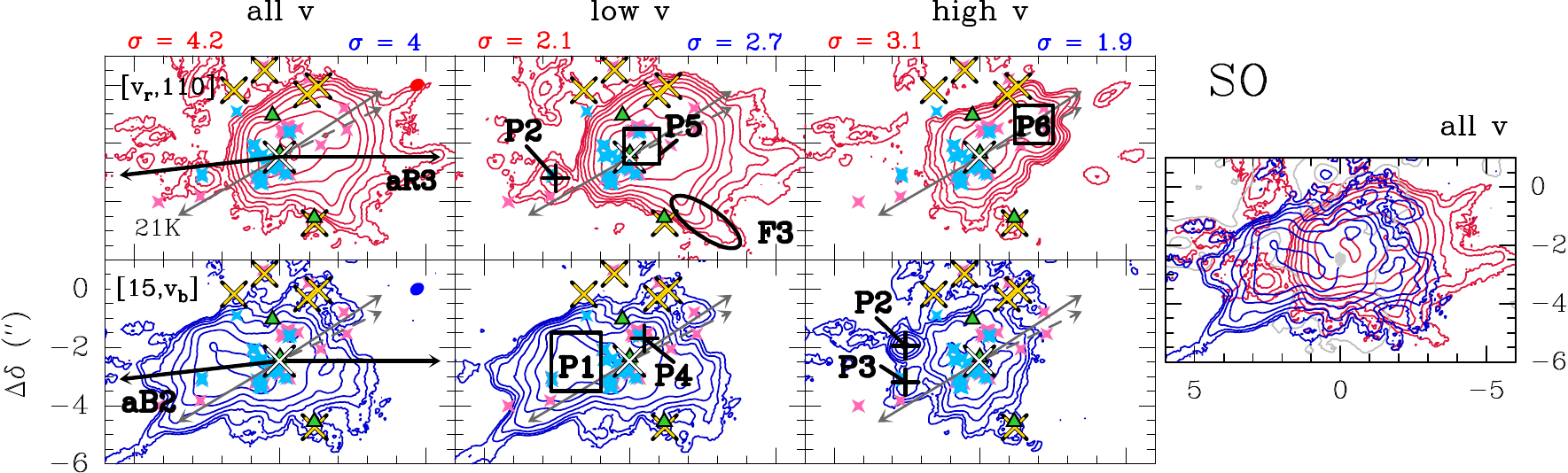} 
    \includegraphics[width=0.865\textwidth]{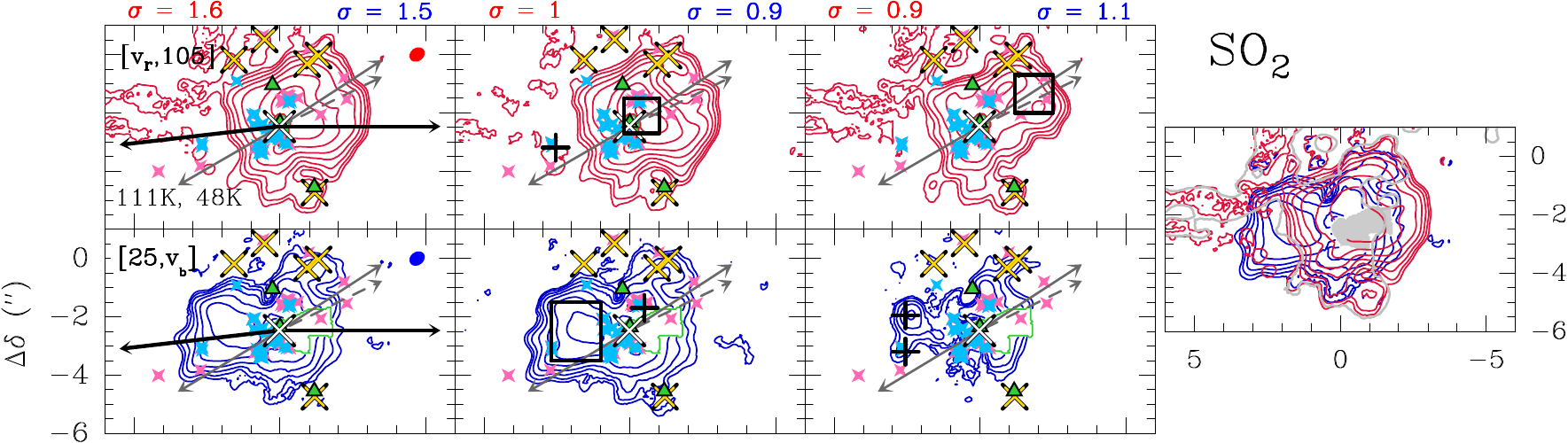} 
    \includegraphics[width=0.865\textwidth]{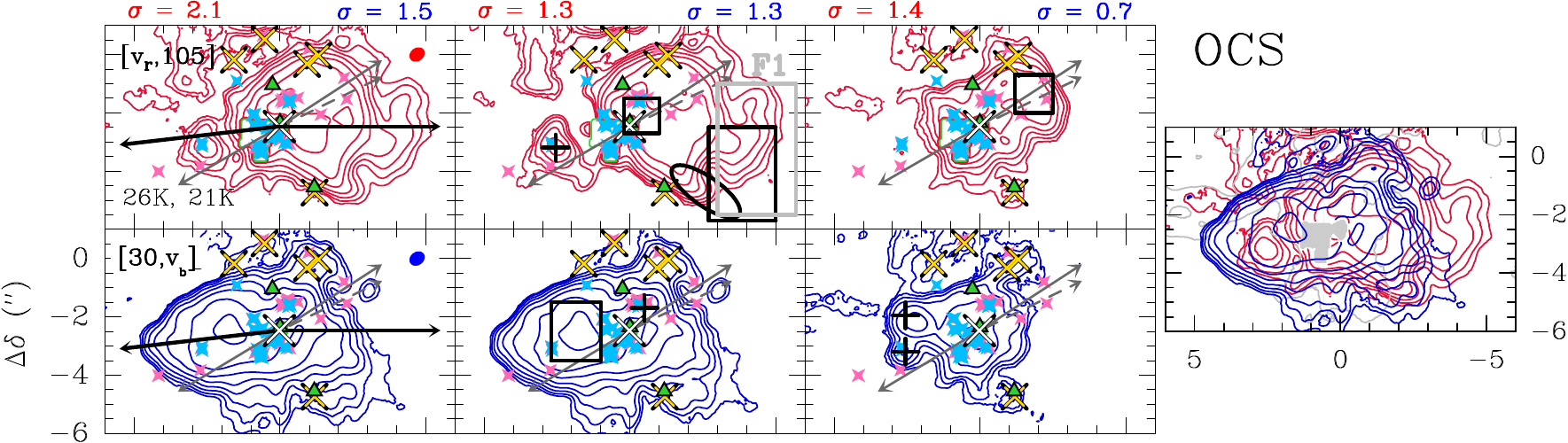} 
    \includegraphics[width=0.865\textwidth]{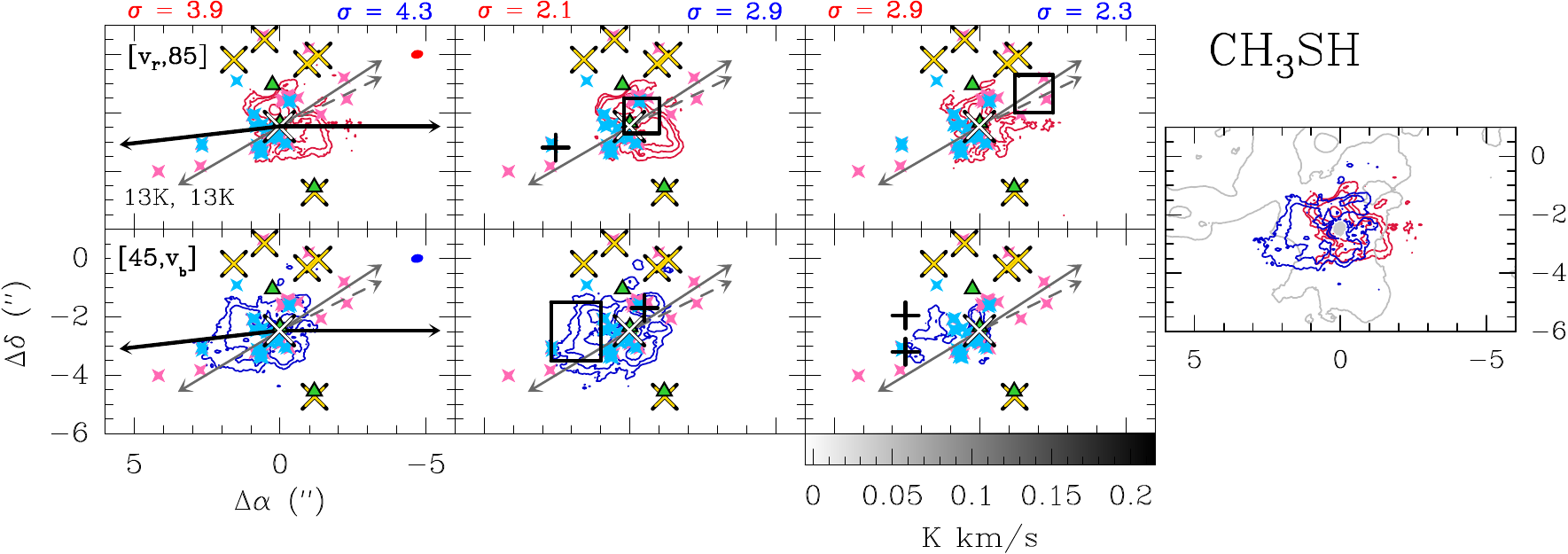} 
    \caption{LVINE maps of blue- and red-shifted emission of S-bearing molecules (blue and red contours, respectively). 
    The left-most panels show maps in which intensities were integrated over the complete velocity interval (all v), where the fixed, pixel-independent outer integration limits (in \kms) are shown in the upper left corner. The maps in the second column (low v) show intensities integrated over half the total velocity interval with velocities closer to $\varv_{\rm sys}$, while in the maps in the third column (high v), intensities were integrated over the other half up or down to the outer integration limits. In the right-most panels, the all-v maps of blue- and red-shifted emission are overlaid as in Fig.\,\ref{fig:lvine_coms}. The contour steps start at 5$\sigma$ and then increase by a factor of 2, where $\sigma$ is the average noise level measured in an emission-free region in each map and is given at the top of each panel in K\kms. The grey contour indicates the 3$\sigma$ level of the continuum emission at 99\,GHz. The closest region around Sgr\,B2\,(N1) is masked out due to high continuum optical depth that depends on frequency and beam size (see Appendix\,C in Paper\,I). For OCS and SO\2, the masked region was extended due to contamination by emission from another species that was identified in their spectra. Masks are either shown in white within a green contour or as grey-shaded regions.
    Grey arrows indicate the directions of the collimated features labelled aB1, aR1, and aR2 in Fig.\,\ref{fig:so}. 
    Coloured markers are the same as in Fig.\,\ref{fig:so}. Identified intensity peaks P1--P6 and elongated features F1--F3, as well as additional bipolar-like extensions labelled aR3 and aB2 are drawn in black. 
    The upper-level energies of the transitions used to produce the maps are shown in the lower-left corner of the top-left panels, respectively. Other properties of the transitions are summarised in Table\,\ref{tab:trans}. The HPBW is shown in the top-right corner of each each panel in the left-most panels. The position offsets are given with respect to the ReMoCA phase centre.}
    \label{fig:COMof}
\end{figure*}

\begin{figure*}[htpb]
    \centering
    \includegraphics[width=.855\textwidth]{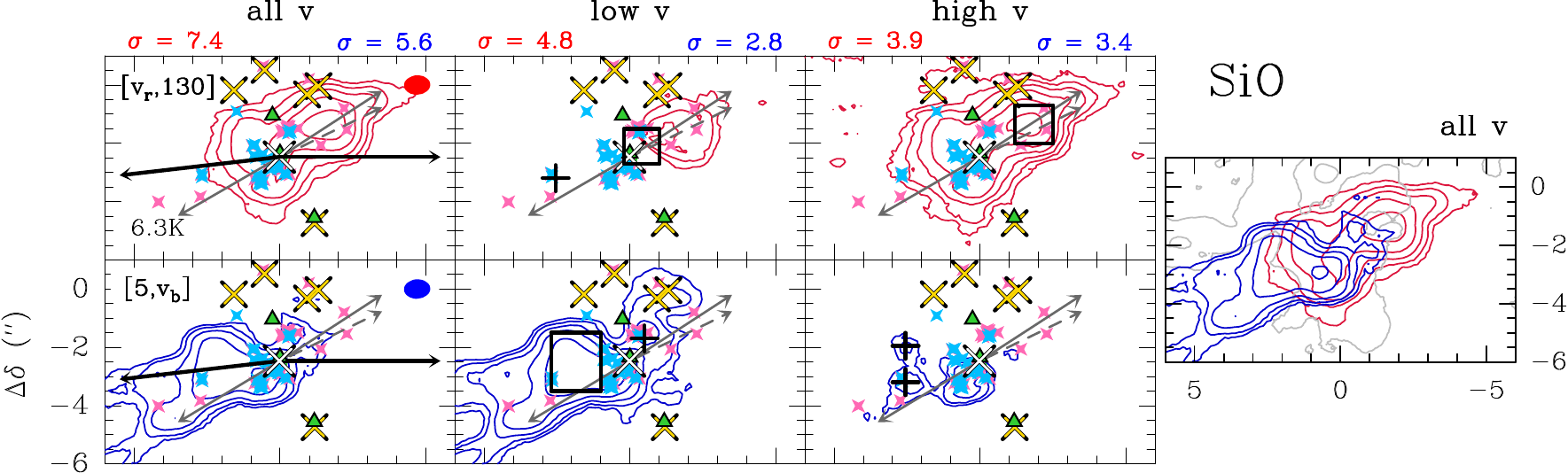}
    \includegraphics[width=0.855\textwidth]{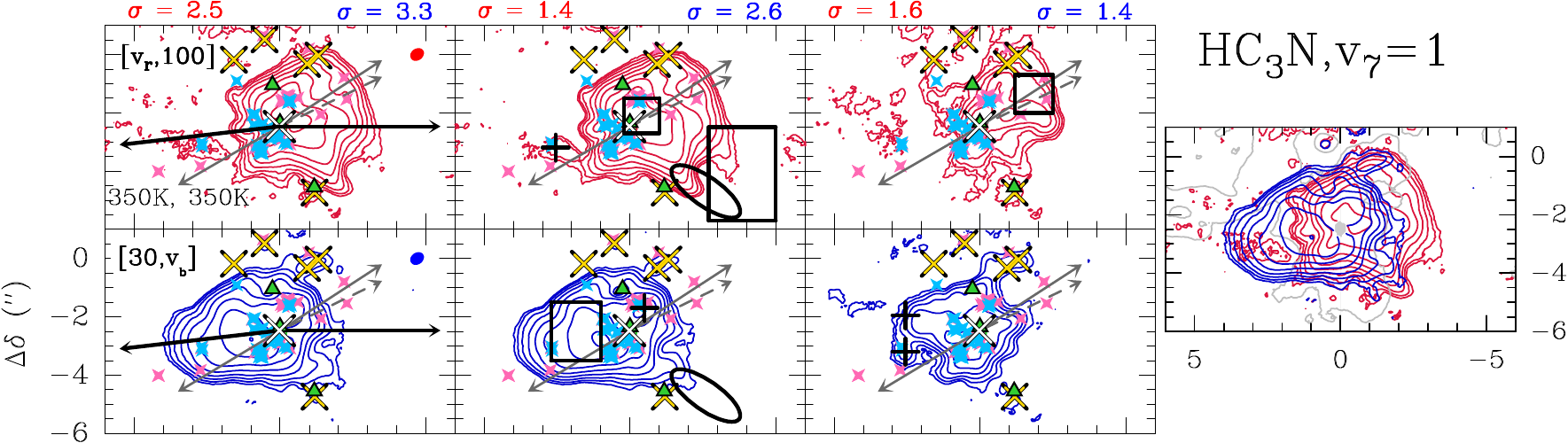} 
    \includegraphics[width=0.855\textwidth]{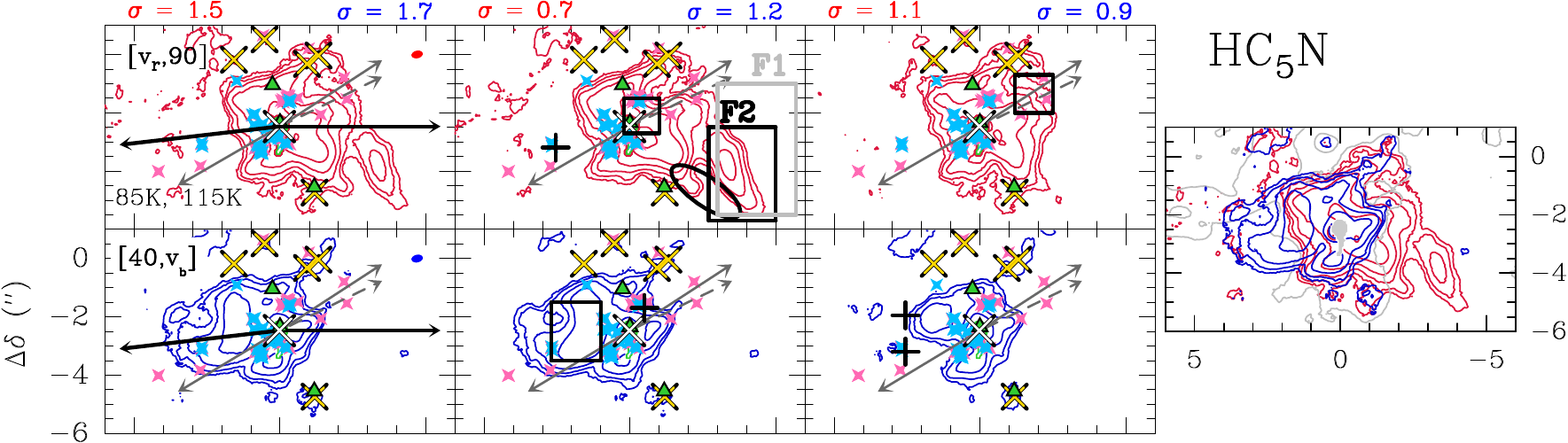} 
    \includegraphics[width=0.855\textwidth]{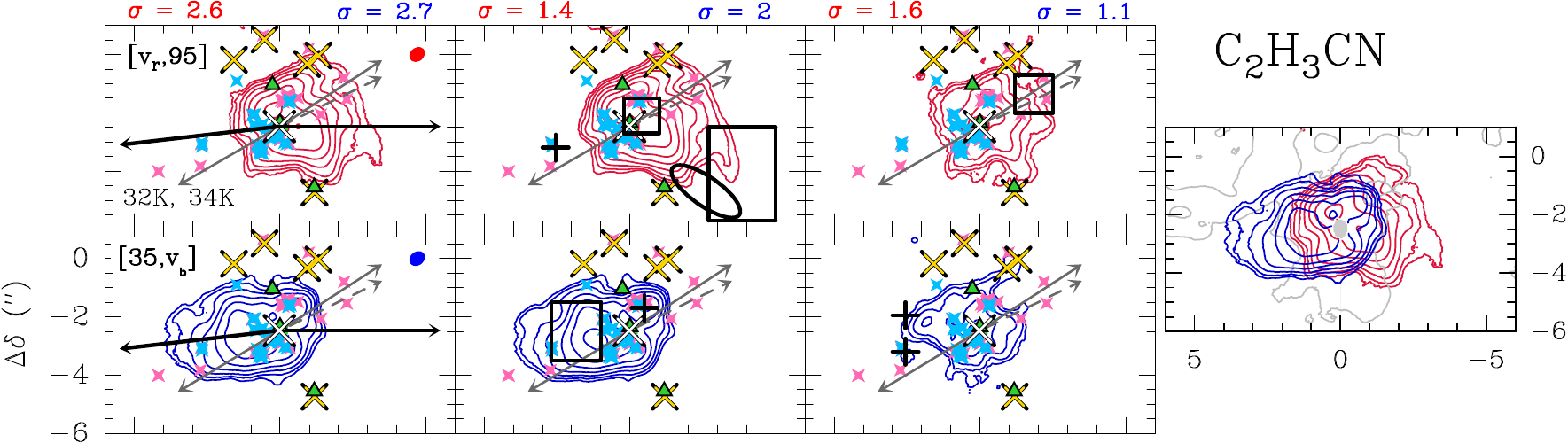} 
    \includegraphics[width=0.855\textwidth]{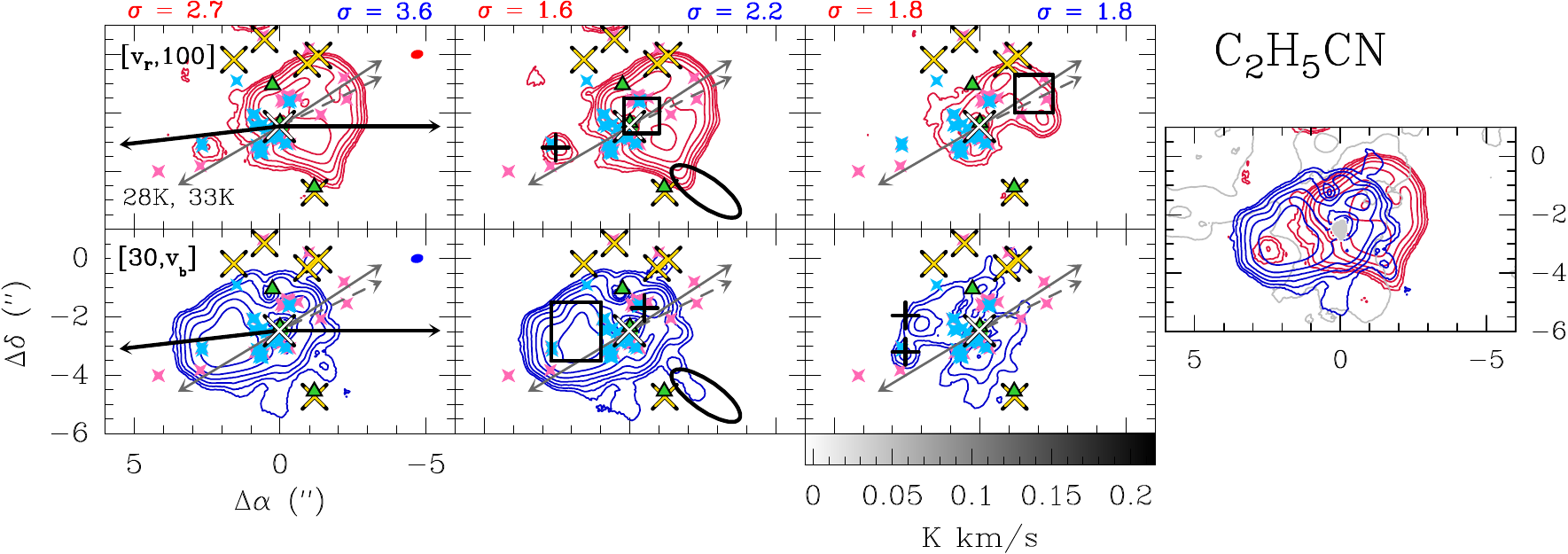} 
    \caption{Same as Fig.\,\ref{fig:COMof}, but for N-bearing molecules. The LVINE method could not be used for SiO, because of absorption features. The all-v maps were computed as in Fig\,\ref{fig:so}, while the intermediate integration limits that divide the spectrum into the low-v and high-v portions at each pixel was taken from SO.} 
    \label{fig:COMof2}
\end{figure*}

\begin{figure*}[h!]
    \centering
    \includegraphics[width=0.89\textwidth]{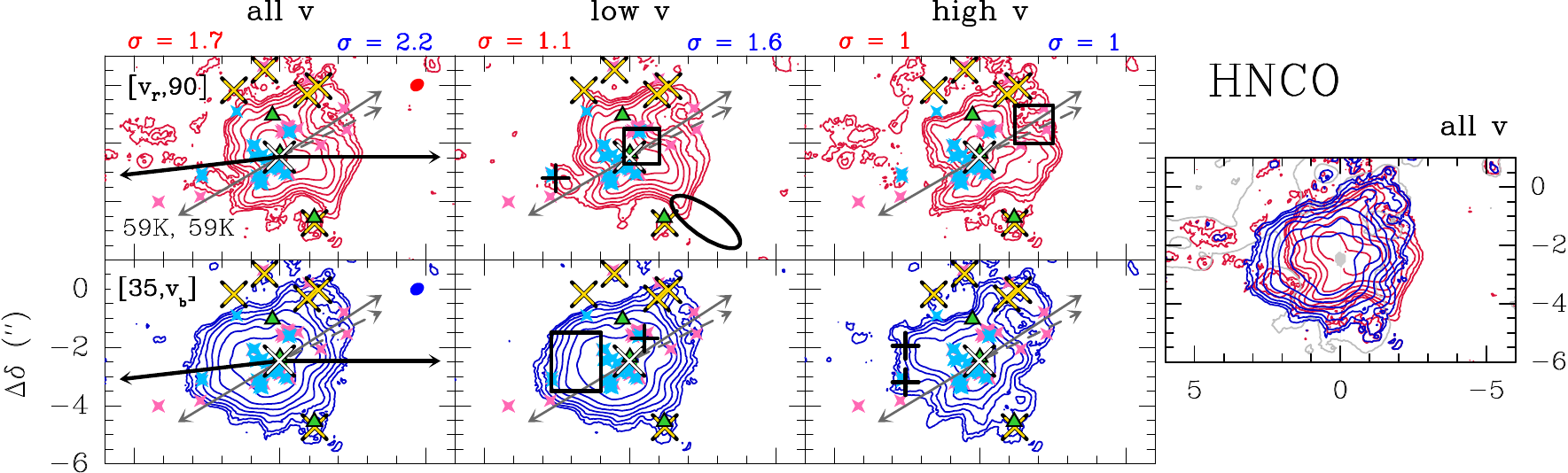} 
    \includegraphics[width=0.89\textwidth]{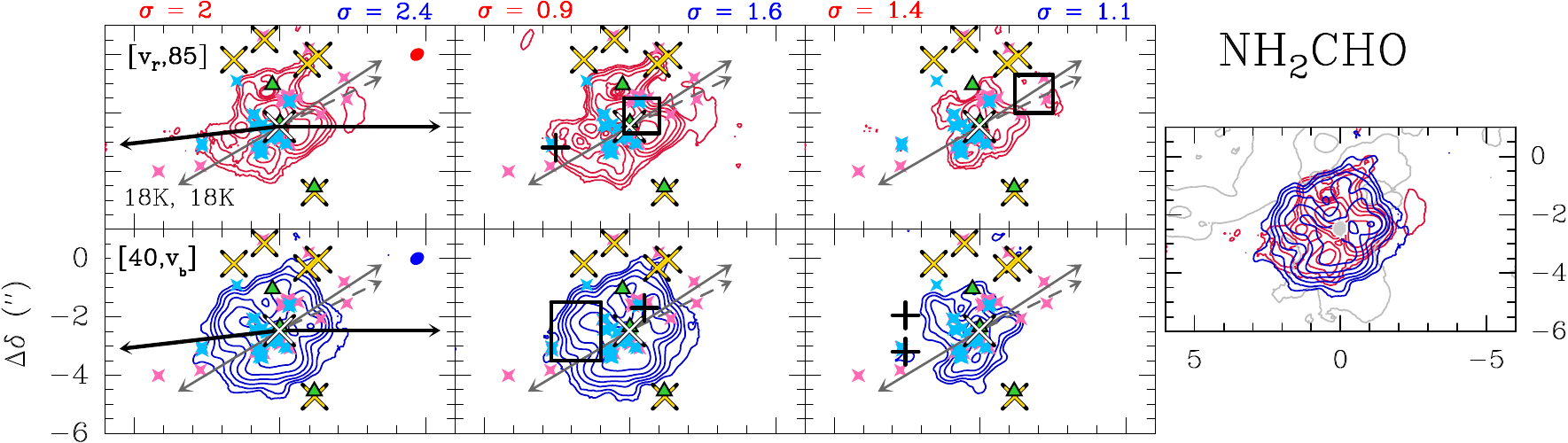} 
    \includegraphics[width=.89\textwidth]{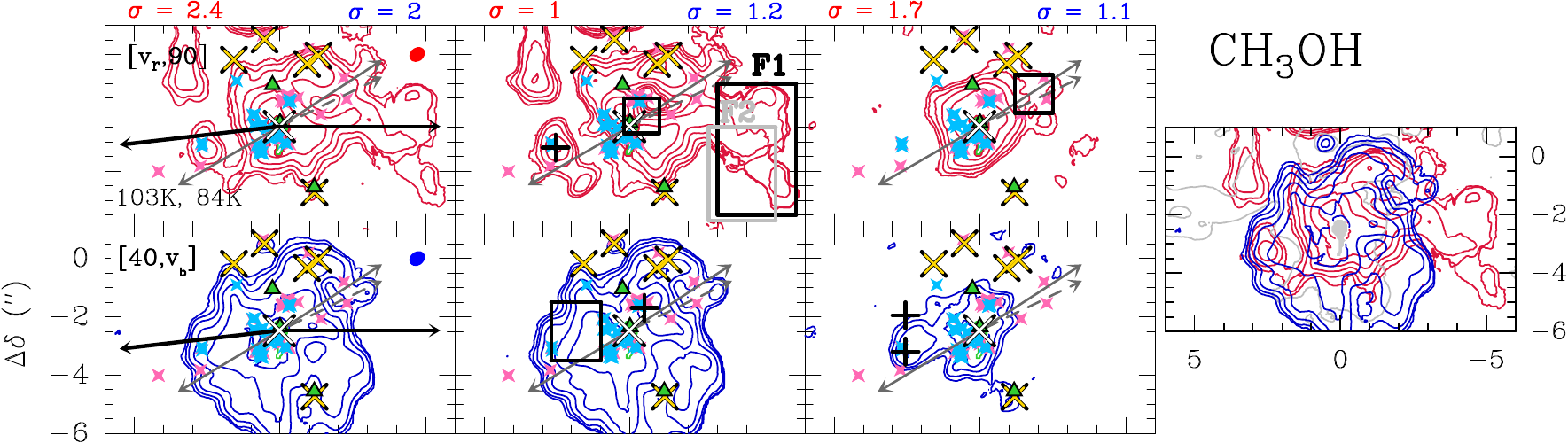}
    \includegraphics[width=.89\textwidth]{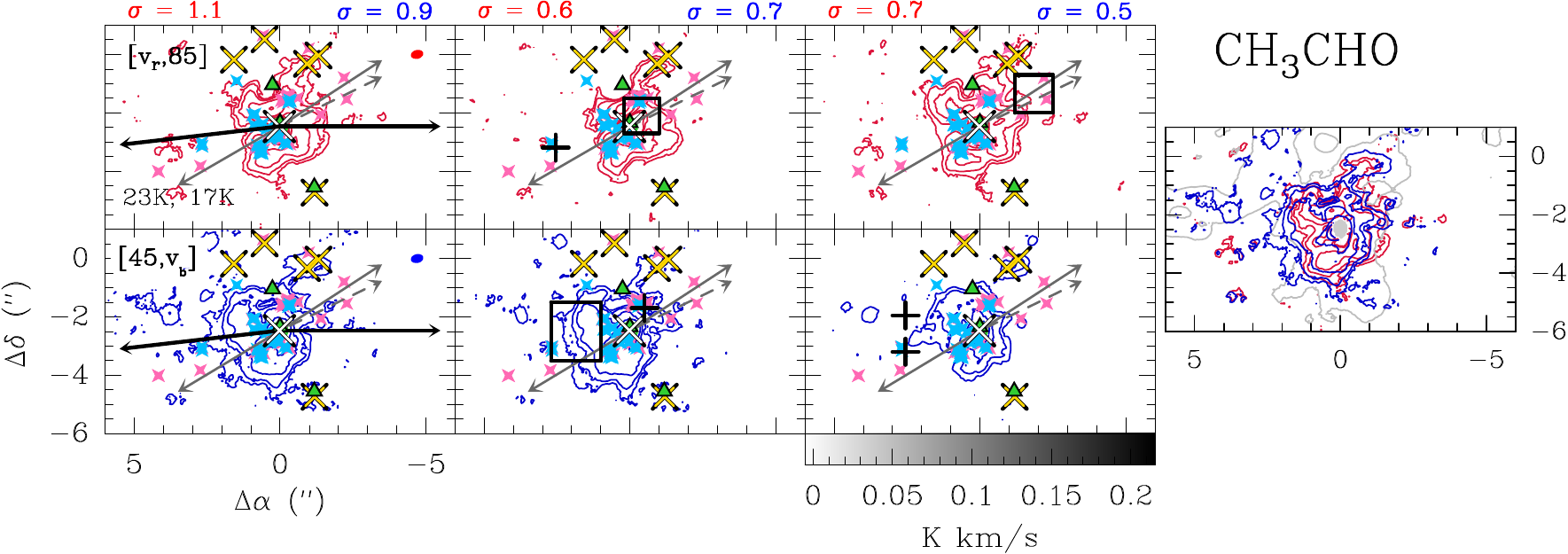}
    \caption{Same as Fig.\,\ref{fig:COMof}, but for (N+O)- and O-bearing molecules.} 
    \label{fig:COMof3}
\end{figure*}

In addition to SO and SiO, extended blue-shifted emission is observed for SO\2, OCS, \mmc, HC\3N, HC\5N, \vc, and \etc, all of which are either N- or S-bearing species, and follows the wider-angle emission of SO and SiO with farthest extension mainly identified in the low-velocity maps. 
The blue-shifted emission from HNCO and \fmm does spatially extend beyond that of the dense core continuum emission (as indicated with the 3$\sigma$ contour), however, is more compact (especially \fmm) than for the N- and simple S-bearing species. 
\met and \ad reveal extended blue-shifted emission, however, the morphology seems to rather resemble that of the hot-core emission (cf. Fig.\,4 in Paper\,I).
None of the molecules shows the collimated features that we identified in the SO (and SiO) maps in Fig.\,\ref{fig:so} and labelled aB1, aR1, and aR2. There is another nearly bipolar set of blue- and red-shifted collimated features extending from east to west that can be identified in the SO maps in Fig.\,\ref{fig:COMof} (labelled aB2 and aR3). Collimated, blue-shifted emission along aB2 is prominently seen in SiO as well, but its red-shifted emission does not
trace aR3. The bipolar structure aB2-aR3 could be another outflow axis, however, we cannot be certain about it as the red-shifted feature aR3 is not clearly observed in Fig.\,\ref{fig:so}. Moreover, the blue-shifted emission aB2 runs along the free-free continuum emission of the extended H{\small II} region in the northwest (see Fig.\,\ref{fig:so}) and may also be associated with it in some way. 

Looking into more detail, the peak of SO (and SiO) blue-shifted emission at low velocities (P1) 
more or less coincides with the peaks of OCS, SO\2, and the N-bearing molecules. Emission from O- and (N+O)-bearing molecules and \mmc seems to peak more towards the centre of the hot core. 
In addition, SO reveals a peak of high-velocity blue-shifted emission (P2), 
which is almost co-located with a peak in SiO, SO\2 and OCS emission, although the peak of the latter is slightly shifted to the southwest from this SO peak. The N-bearers HC\3N, \vc, \etc and HNCO also show a peak
of high-velocity blue-shifted emission but clearly shifted by $\sim$0.8\arcsec\,\,to the southwest. \fmm emission spatially extends in this direction, \met as well, but in a different direction than \fmm. For \mmc and \ad there is no clear connection with this peak, which could also be a sensitivity issue. This SO peak does not coincide with a continuum peak, it is rather framed by the continuum (free-free) emission in the north where the H{\small II} region seems to meet the emission associated with the hot core. According to \citet{Schwoerer19}, there are supposedly also filaments impinging on Sgr\,B2\,(N1) in this region that may be involved in the occurrence of this SO peak in some way (see also Fig.\,\ref{fig:overview}), when it meets the denser core emission around this position.

There is another peak of high-velocity blue-shifted emission (P3), 
which is clearly seen in emission of the simple S-bearing molecules, SiO, HC\3N, \etc, HNCO, \met, indicated by one contour for \mmc, \vc and \fmm, and not seen for HC\5N and \ad. For HC\5N, this peak may be hidden in the low-v maps due to the smaller overall integration interval. This peak is located at the tip of the high-intensity contours in the low-v panels. All molecules that show this peak also show red-shifted emission above 3$\sigma$ at this position, except SiO, which may be a consequence of absorption. In addition, there are three water masers associated with this peak, two at blue-shifted and one at red-shifted velocities.
On the other hand, there is another peak in blue-shifted emission for many molecules towards the northwest (P4), 
which also coincides with the location of a cluster of water masers. 
Generally, the majority of water-maser spots can be associated with blue- or red-shifted emission peaks of the molecules studied here.


\subsubsection{Red-shifted emission}
There are two peaks of SO red-shifted emission: almost right at the centre of the hot core (P5) and at higher velocities at $\sim$2\arcsec\,\,to the northwest (P6). The farther, high-velocity peak seems to be embedded in a cavity of continuum emission. 
The morphologies of SiO, OCS, SO\2, HC\3N, and \etc emission closely follow that of SO, except for the collimated features seen in SO further along aR1--aR3. Moreover, the emission of SO\2 and HC\3N peaks more towards the edges of the most intense SO emission at P6, particularly where it seems to be framed by the continuum emission, 
provided that we are not biased by optical depth. 
Red-shifted \vc emission shows the same trend in the low-v map, but it does not clearly show the high-velocity peak, and HNCO and HC\5N show spatially extended red-shifted emission towards the northwest, but also over the whole hot-core region. Red-shifted emission of \mmc, \fmm, and \ad is only observed towards the more central SO peak.
\met emission shows the extension to the northwest, but also additional structures that we refer to as F1 in Fig.\,\ref{fig:COMof3} and that are not seen for other molecules.
As mentioned above, there is a prominent peak \textbf{(P2)} coinciding with a peak in high-velocity blue-shifted emission and water masers, that is seen for the majority of molecules, for some more prominently than for others.

Another interesting feature is seen in the red-shifted HC\5N emission that we label F2 in Fig.\,\ref{fig:COMof2}, which is not observed as prominently in any of the other maps. It is evident in maps of HC\3N ($v_7=1$) and \vc, however, not with the spatial extent as for HC\5N. It does not coincide with the F1 feature seen in emission of \met, it is adjacent to it in the south, however.
This feature is observed in other transitions of HC\5N, hence, it is not caused by contamination from other molecules in the shown map. We cannot be conclusive on the nature of this feature and why it appears only this prominently in emission of this molecule. 
There is another feature (F3) that runs almost parallel to F2, which is observed in the maps of SO, OCS, the N-bearing molecules, and HNCO. This feature may be associated with one of the filaments identified by \citet[][ see also Fig.\,\ref{fig:overview}]{Schwoerer19}, but this is not certain. 

It is interesting to note that there is a clear lack of emission towards the H{\small II} region K3 (green triangle north of the hot-core centre) in the maps of blue-shifted SO, SO\2, OCS, HC\3N, \vc, and \etc emission. In contrast, for \ad and \fmm, this lack of emission is rather indicated in the red-shifted emission maps. \met, HNCO, HC\5N, and \mmc do not reveal a clear signature in any map. 

\subsection{Velocity-channel maps}
\begin{figure*}
    \centering
    \includegraphics[width=.97\textwidth]{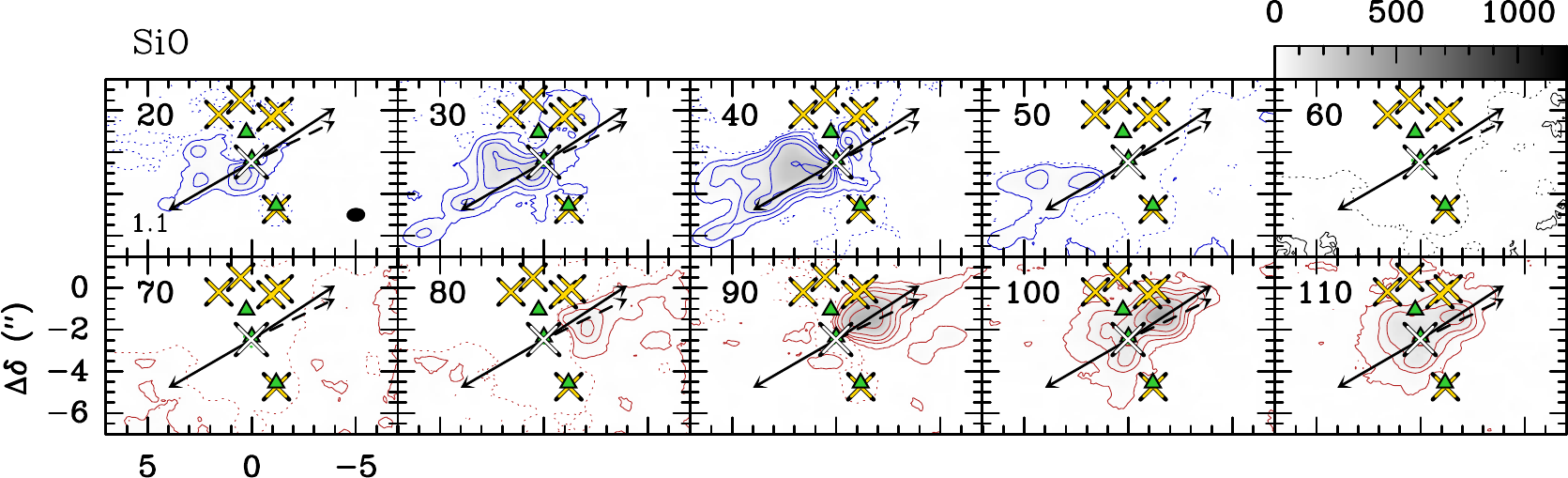}\\[0.1cm]
    \includegraphics[width=.97\textwidth]{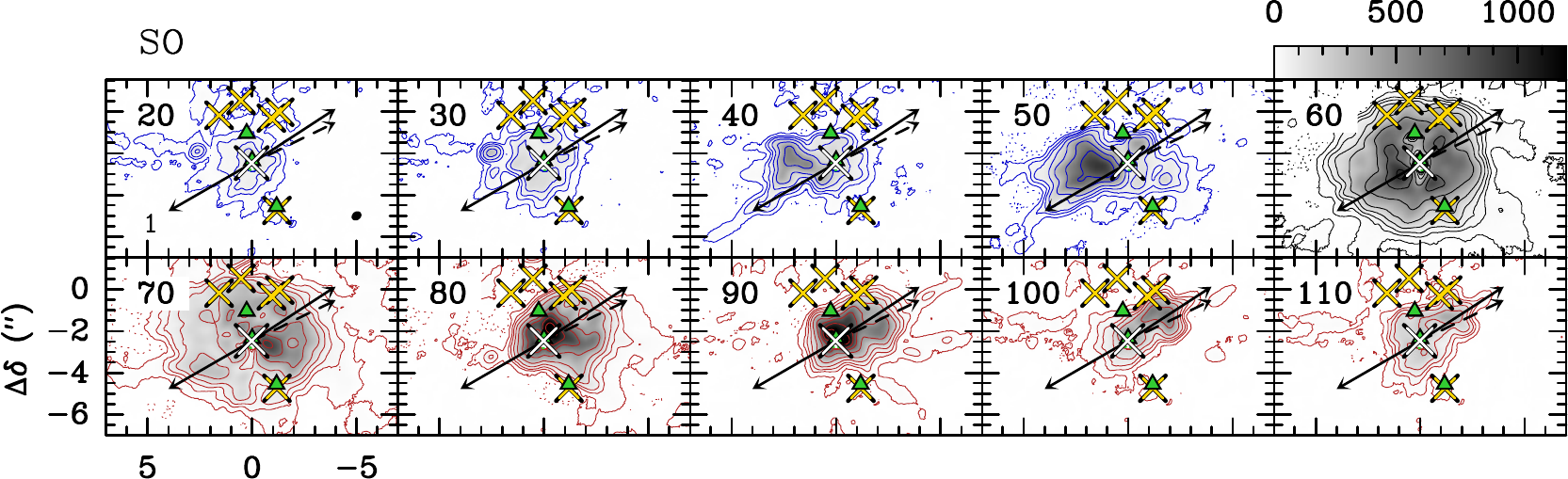}\\[0.1cm]
    \includegraphics[width=.97\textwidth]{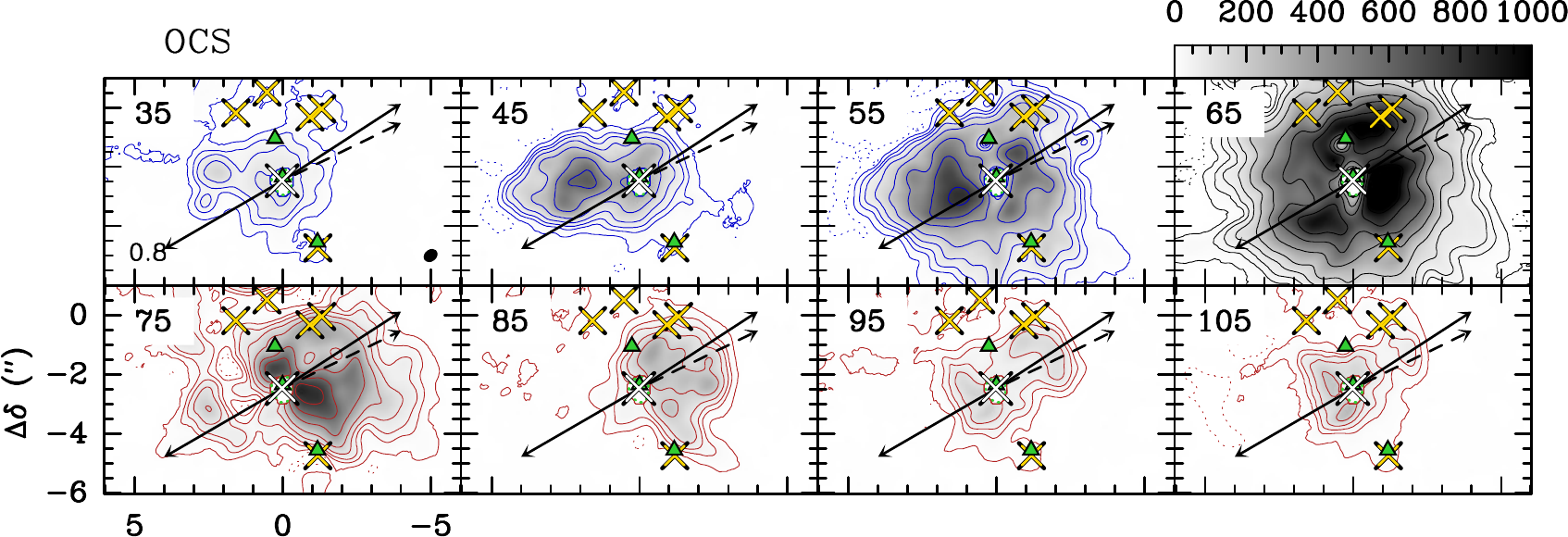}\\[0.1cm]
    \includegraphics[width=.97\textwidth]{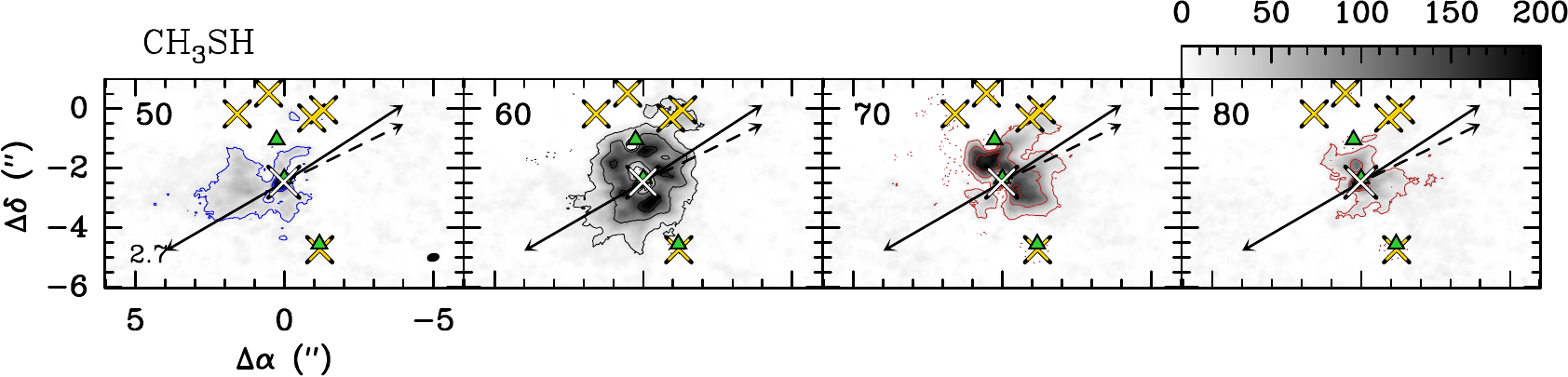}
    \caption{Velocity-channel maps of various molecules (in grey scale (K\kms) and contours). The integration strategy is the same as for the maps in Figs.\,\ref{fig:lvine_coms} and \ref{fig:COMof}-\ref{fig:COMof3} (i.e. using two transitions to minimise contamination by other molecules, except for SiO and SO (109.25\,GHz)), but in each panel the intensities were integrated over an interval of only 10\kms. The respective central velocity in \kms is shown in the upper left corner. The start and end velocities correspond to the outer integration limits used in Fig.\,\ref{fig:lvine_coms}. For all molecules but SO, the contours are $-$6$\sigma$, 6$\sigma$, 30$\sigma$, and then increase by a factor of 2, where $\sigma$ in K\kms is written in the lower left corner of the lowest-velocity map and was computed using $\sqrt{N}\times rms\times \Delta\varv$, where $N$ is the number of channels that was integrated over, $rms$ was taken from Table\,2 in \citet{Belloche19}, and $\Delta\varv$ is the channel separation in \kms. The SO contours start at 10$\sigma$ and then increase by factor of 3. Blue contours indicate blue-shifted emission, red contours red-shifted emission, and black contours emission close to the source systemic velocity. Masked regions, arrows, and markers are the same as in Fig.\,\ref{fig:so}. The HPBW is shown in the bottom-right corner of the lowest-velocity panel.}
    \label{fig:chan_maps}
\end{figure*}
\begin{figure*}
\addtocounter{figure}{-1}
    \centering
    \includegraphics[width=.95\textwidth]{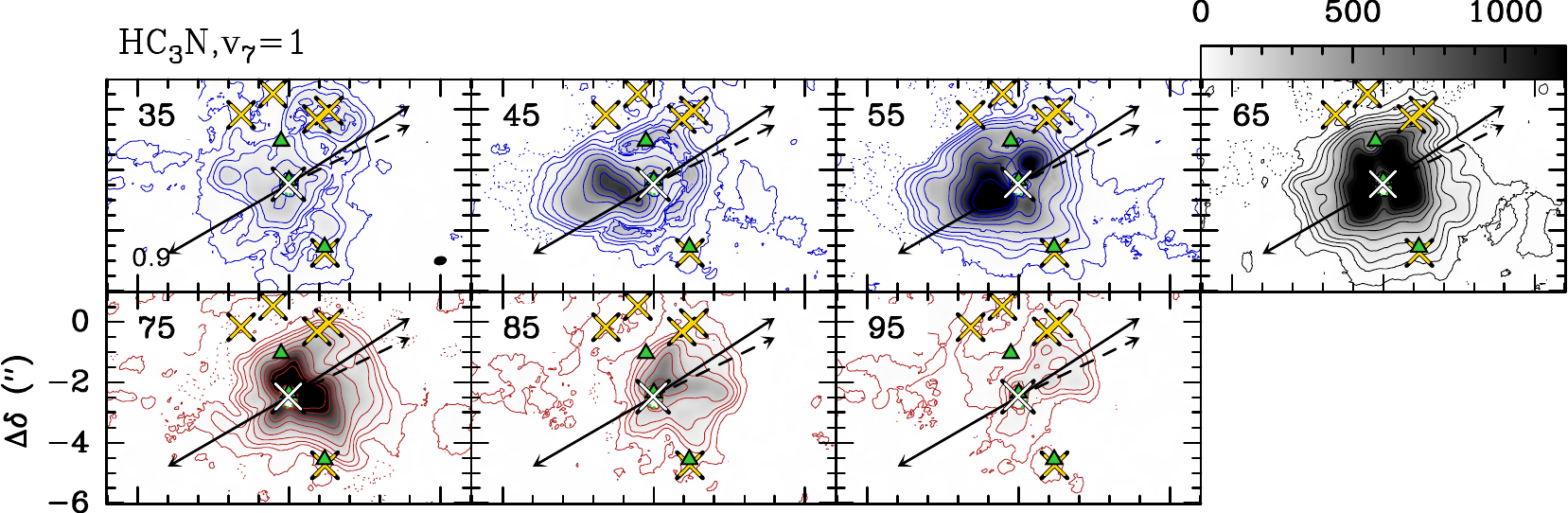}\\[0.2cm]
    \includegraphics[width=.95\textwidth]{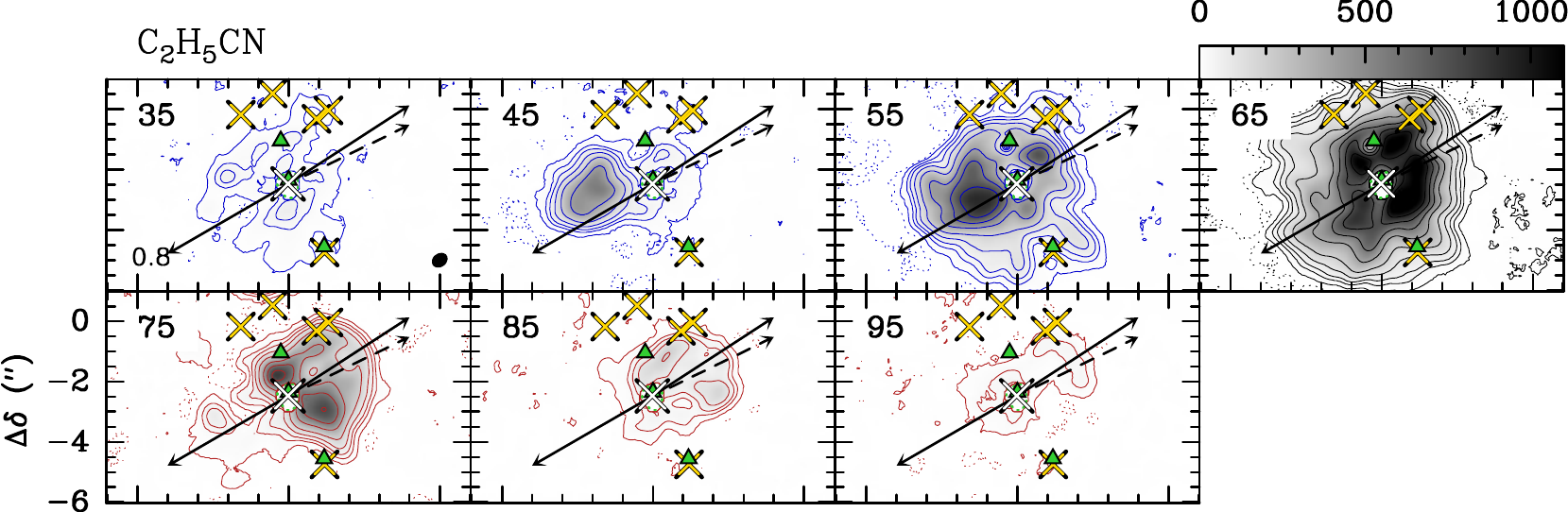}\\[0.2cm]
    \includegraphics[width=.95\textwidth]{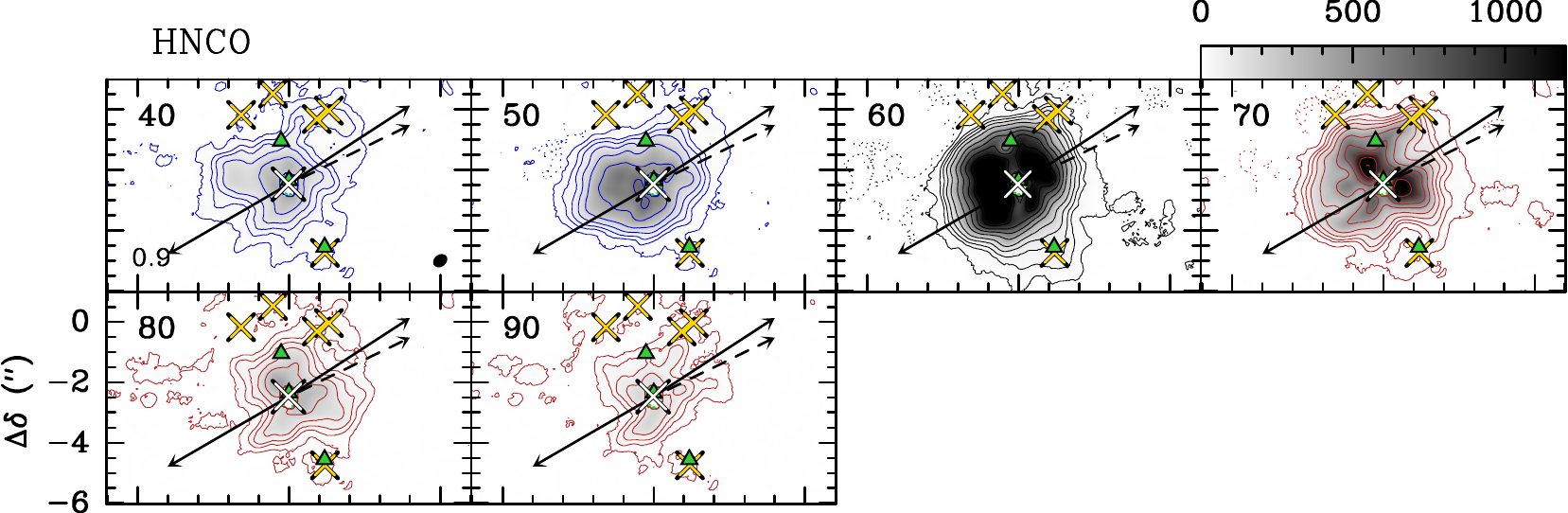}
    \includegraphics[width=.95\textwidth]{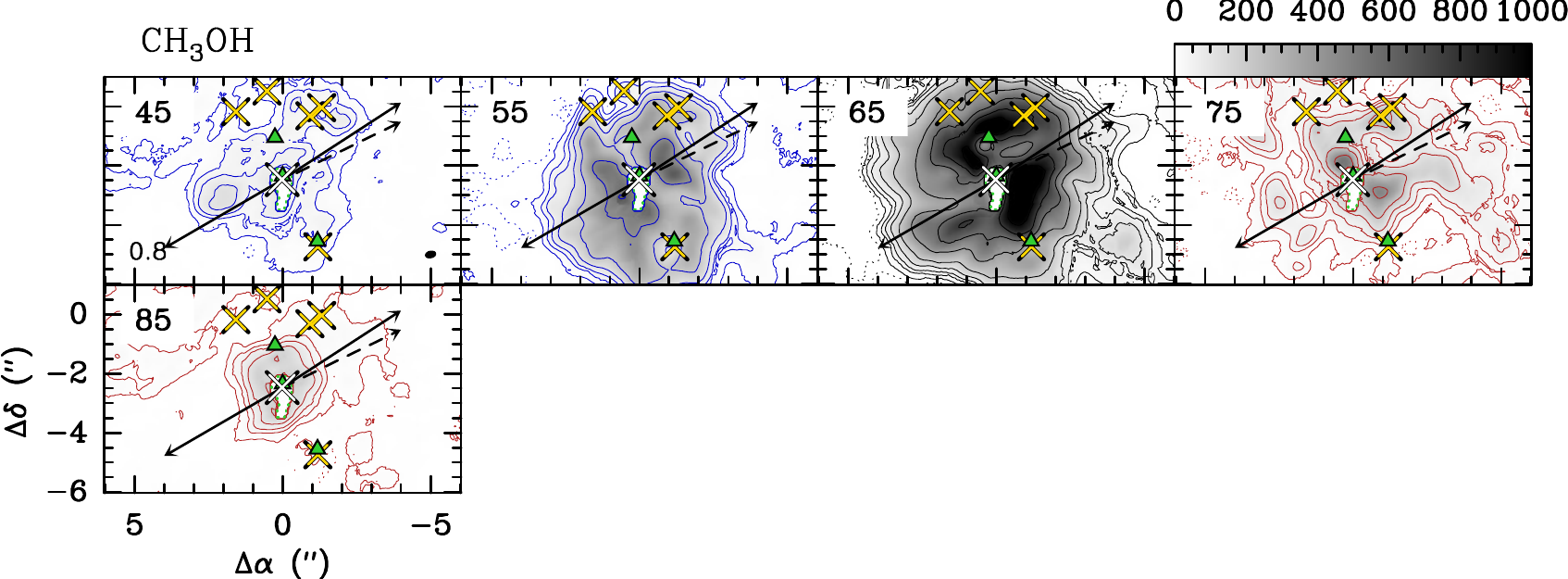}
    \caption{continued.}
\end{figure*}

In Fig.\,\ref{fig:chan_maps} we show velocity-channel maps for a selection of molecules, which includes SiO, SO (109.25\,GHz), OCS, \mmc, HC\3N, \etc, HNCO, and \met. 
Each channel map contains integrated intensities over a 10\kms velocity interval, where the central velocity is indicated in each map, and the complete velocity range that is shown corresponds to the one used for the LVINE maps. The same integration strategy was applied, where we use two transitions of a molecule to minimise the contamination by other molecules, except for SiO and SO.

We see the same trends as in the integrated intensity maps: strong blue-shifted emission for the simple S-bearing molecules, HC\3N, \etc, and HNCO. \mmc
shows weaker but spatially extended emission. 
The emission of \met is spatially extended, but does not follow the morphology of the other molecules. A similar classification of the molecules applies to red-shifted emission extending to the northwestern direction. 
The most collimated feature in the blue-shifted SiO and SO emission along arrow aB1 is prominent in the maps at 40--50\kms, for SiO even at 20--30\kms while the red-shifted feature along aR2 appears at 80--90\kms. At higher velocities, the finger-like feature along aR1 can be observed. The collimated features along aR3 and feature F3 become visible at 90\kms for SO. For OCS, HC\3N, and \etc, feature F3 is observed at 75--85\kms. 

\section{Additional figures}

\begin{figure*}
    \includegraphics[width=.264\textwidth]{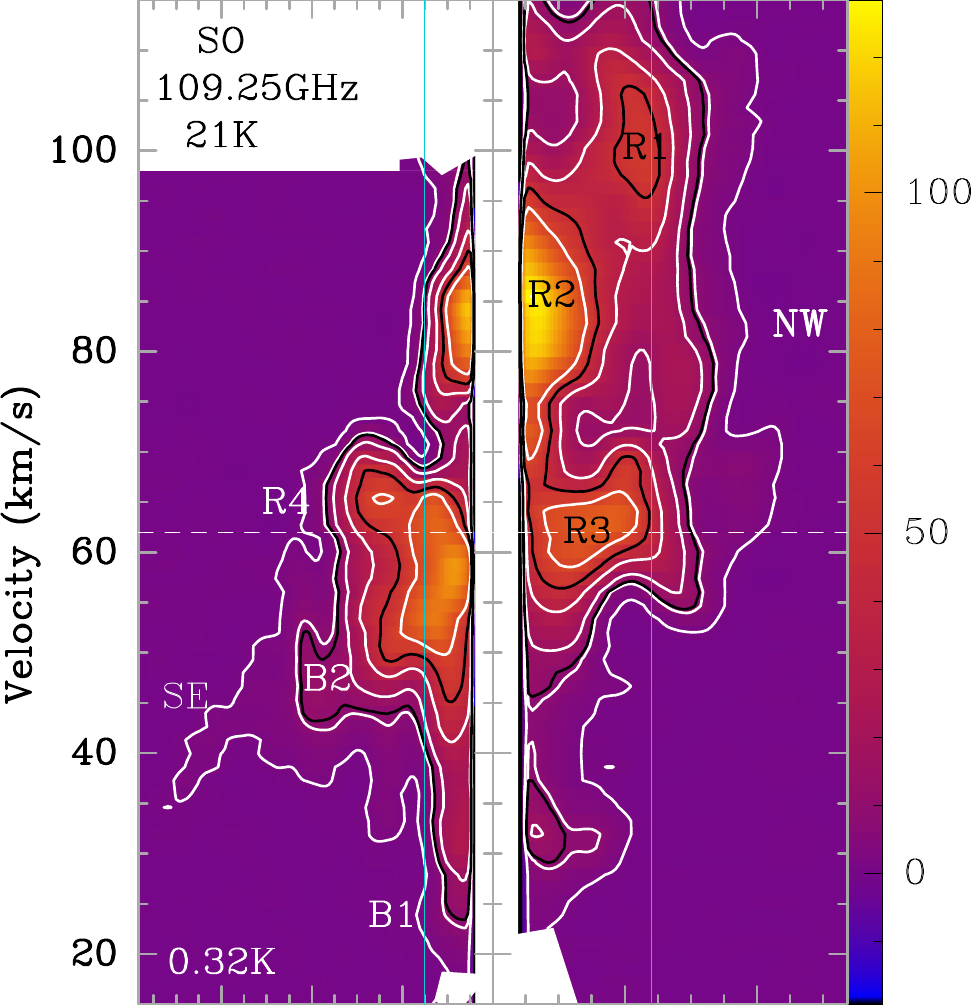}\hspace{0.1cm}
    \includegraphics[width=.219\textwidth]{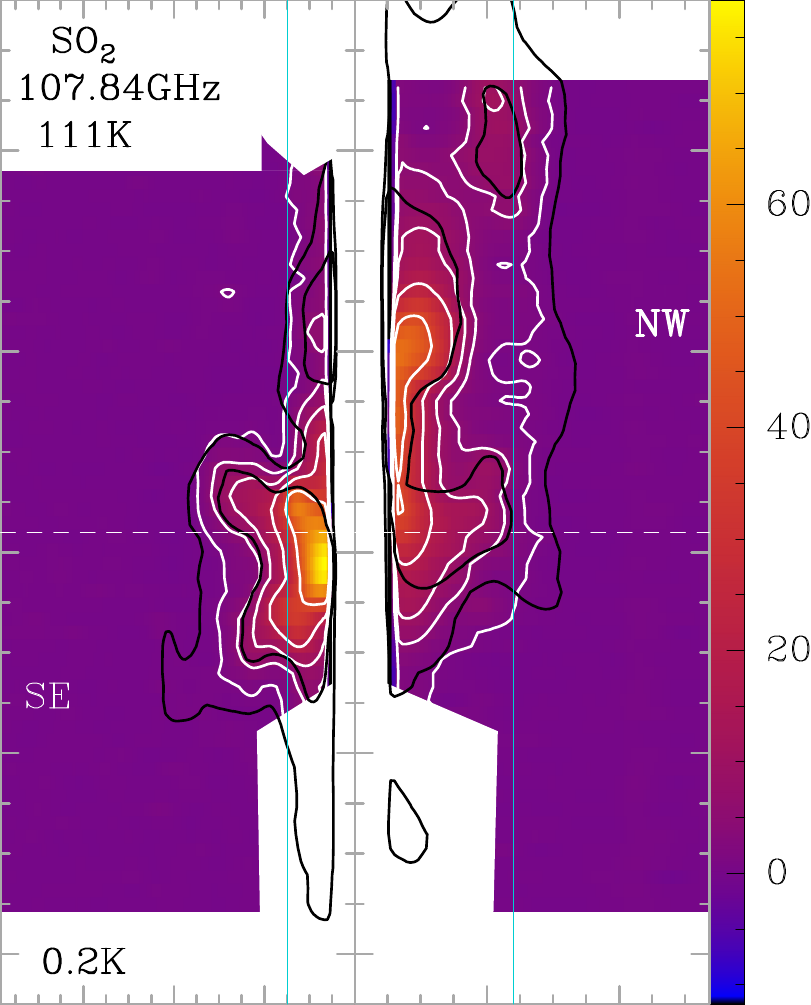}\hspace{0.23cm}
    \includegraphics[width=.227\textwidth]{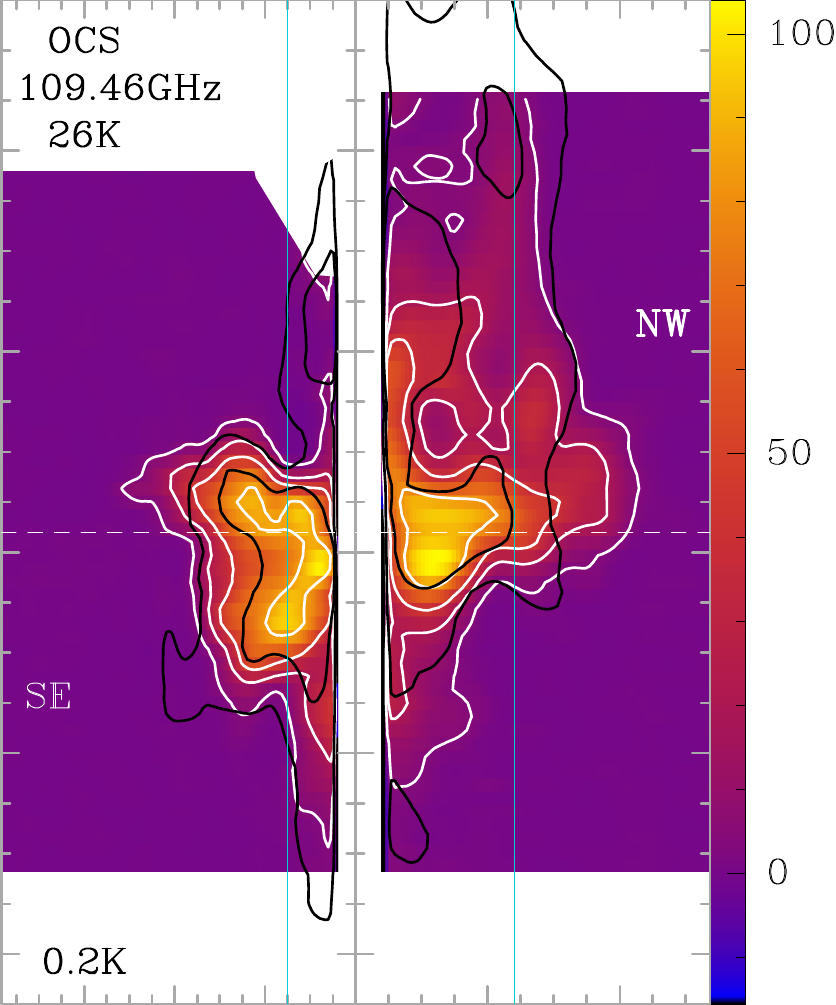}\hspace{0.1cm}
    \includegraphics[width=.219\textwidth]{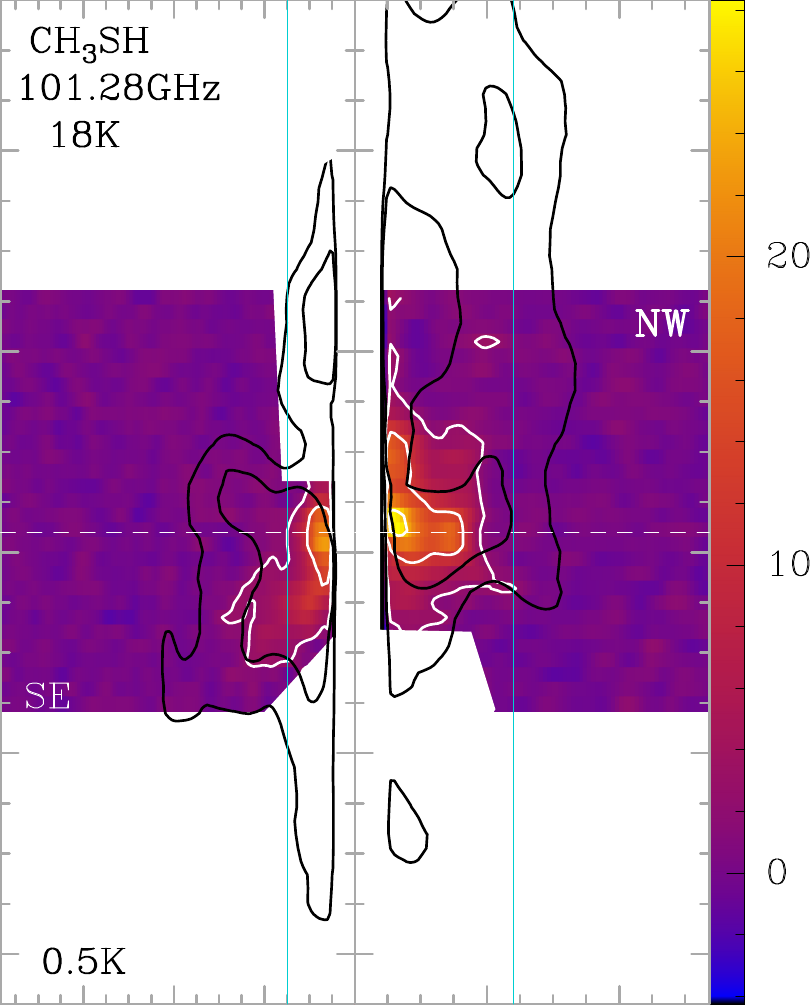}\\
    \includegraphics[width=.264\textwidth]{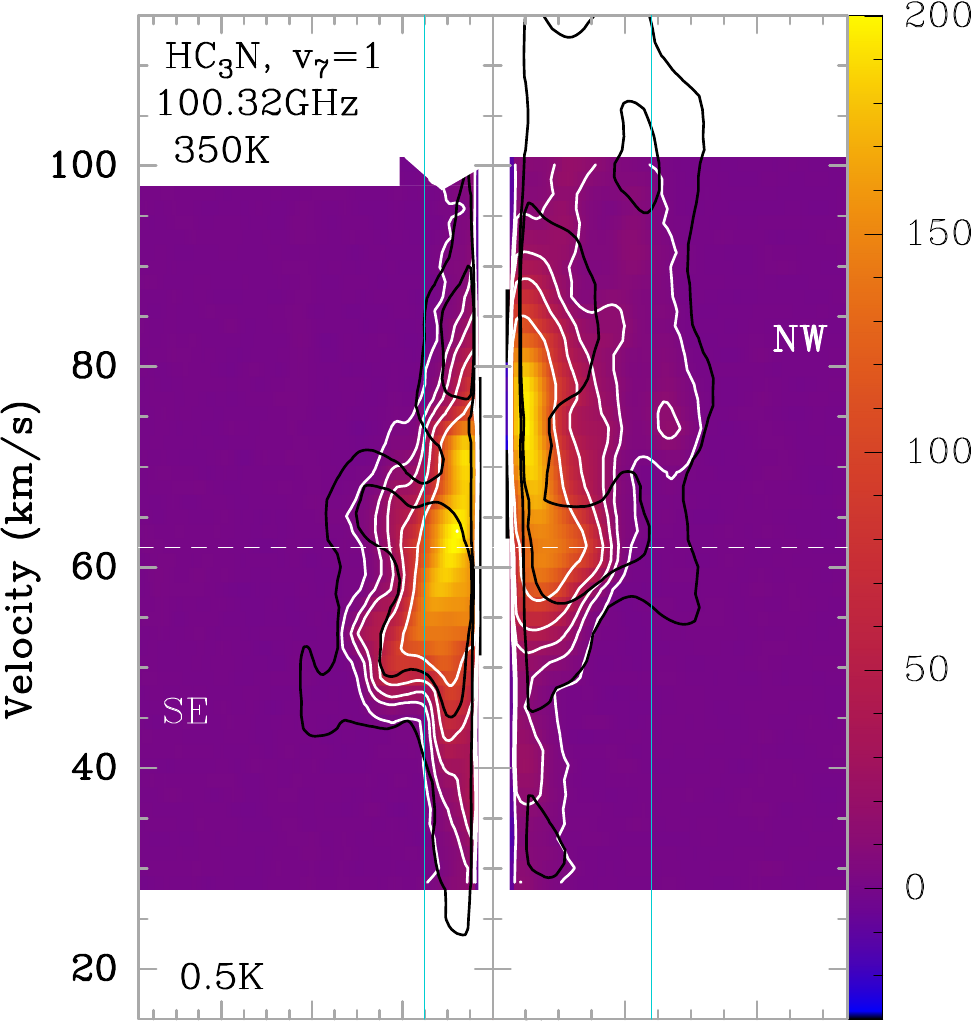}\hspace{0.1cm}
    \includegraphics[width=.219\textwidth]{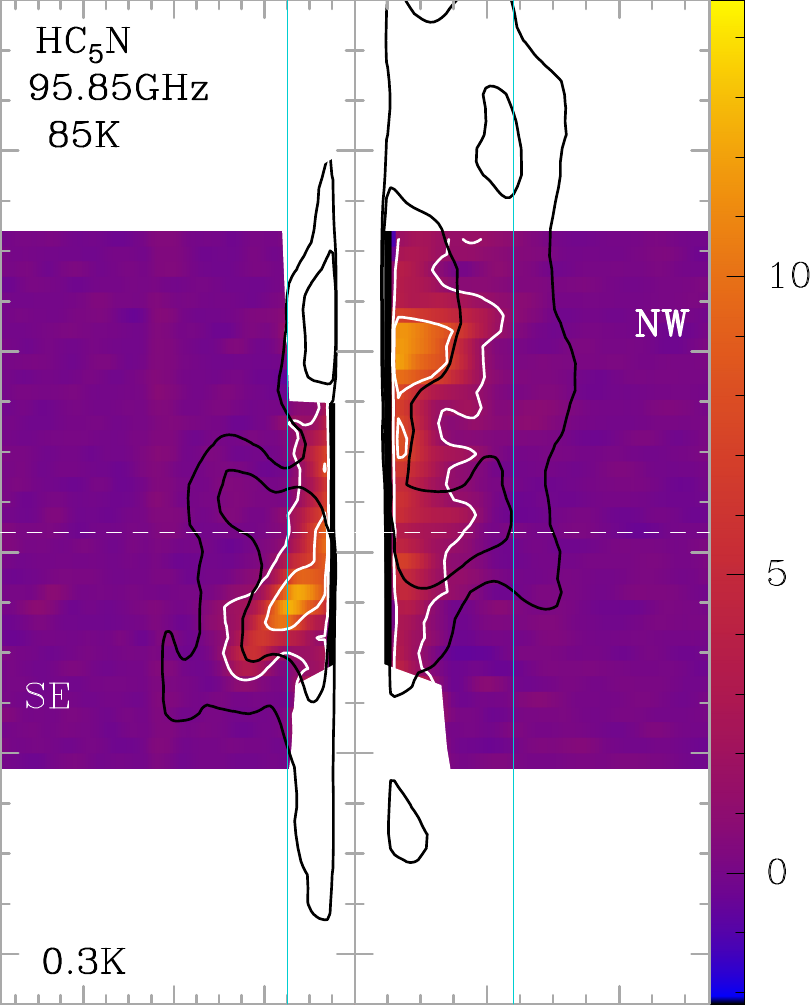}\hspace{0.23cm}
    \includegraphics[width=.227\textwidth]{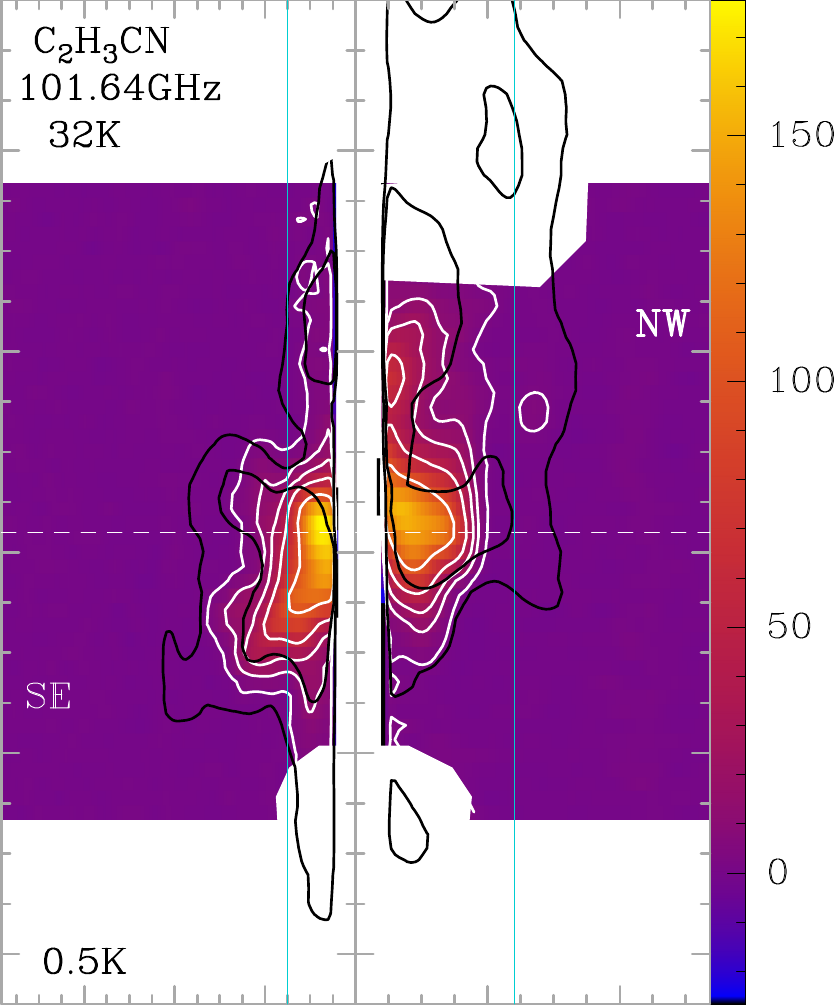}\hspace{0.1cm}
    \includegraphics[width=.227\textwidth]{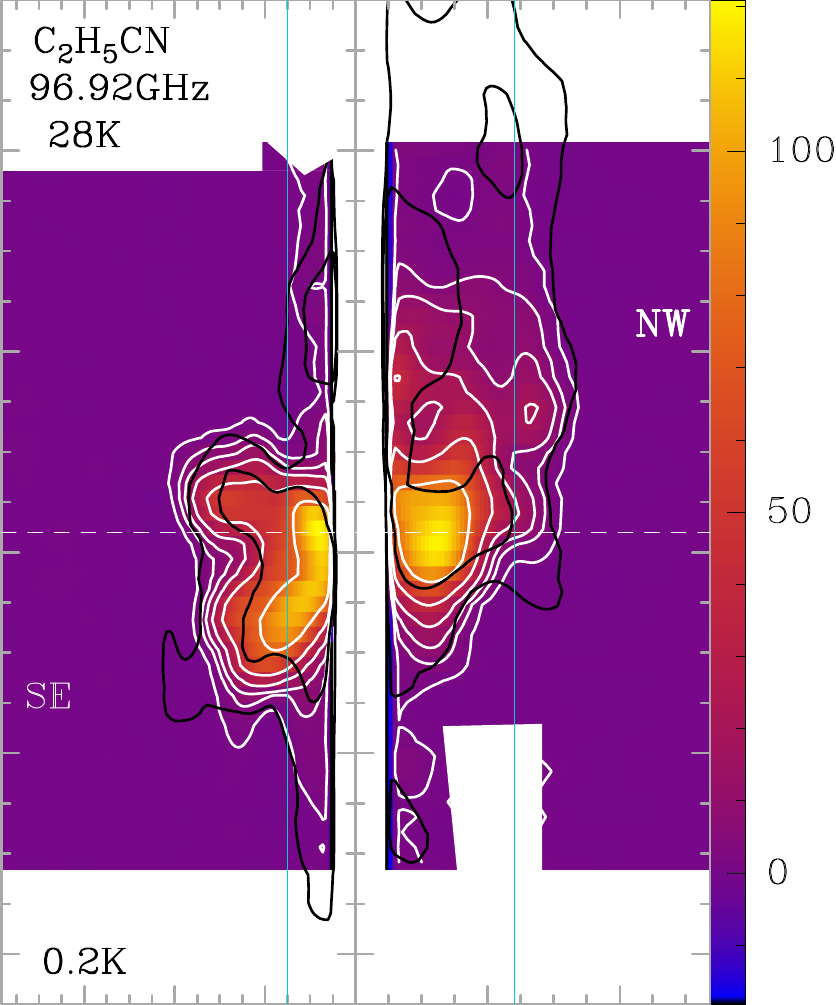}\\
    \includegraphics[width=.264\textwidth]{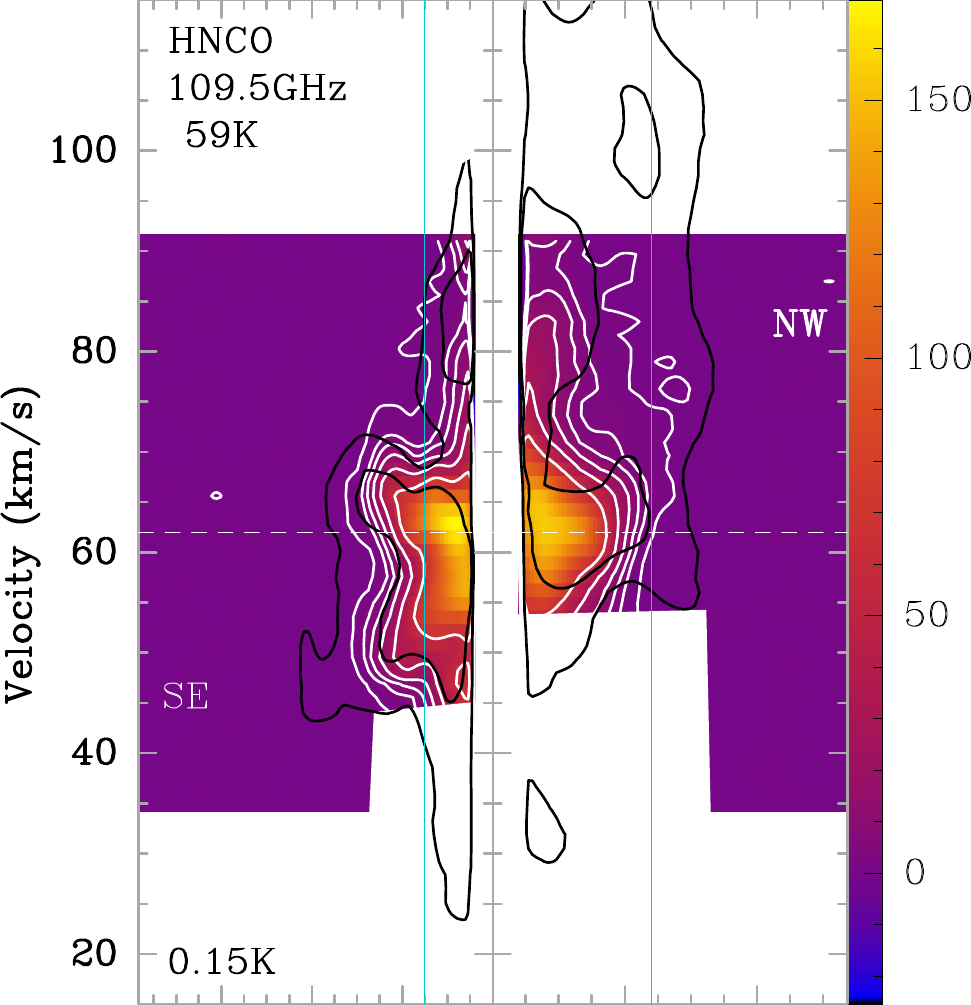}\hspace{0.1cm}
    \includegraphics[width=.227\textwidth]{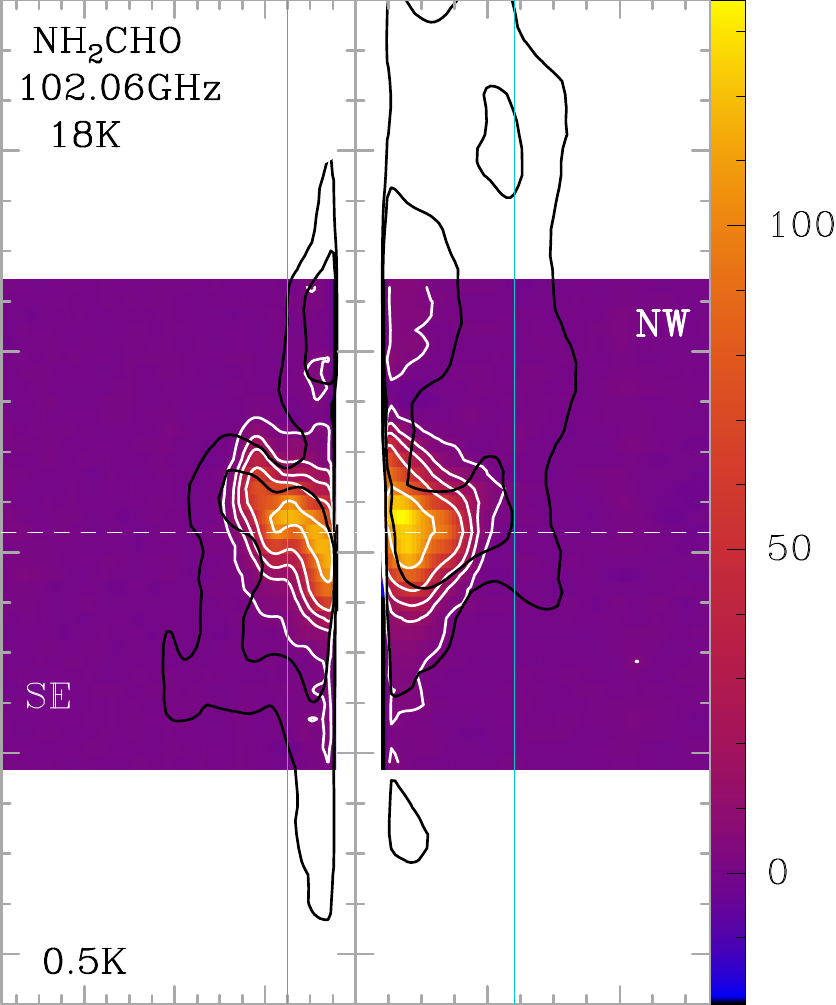}\hspace{0.1cm}
    \includegraphics[width=.219\textwidth]{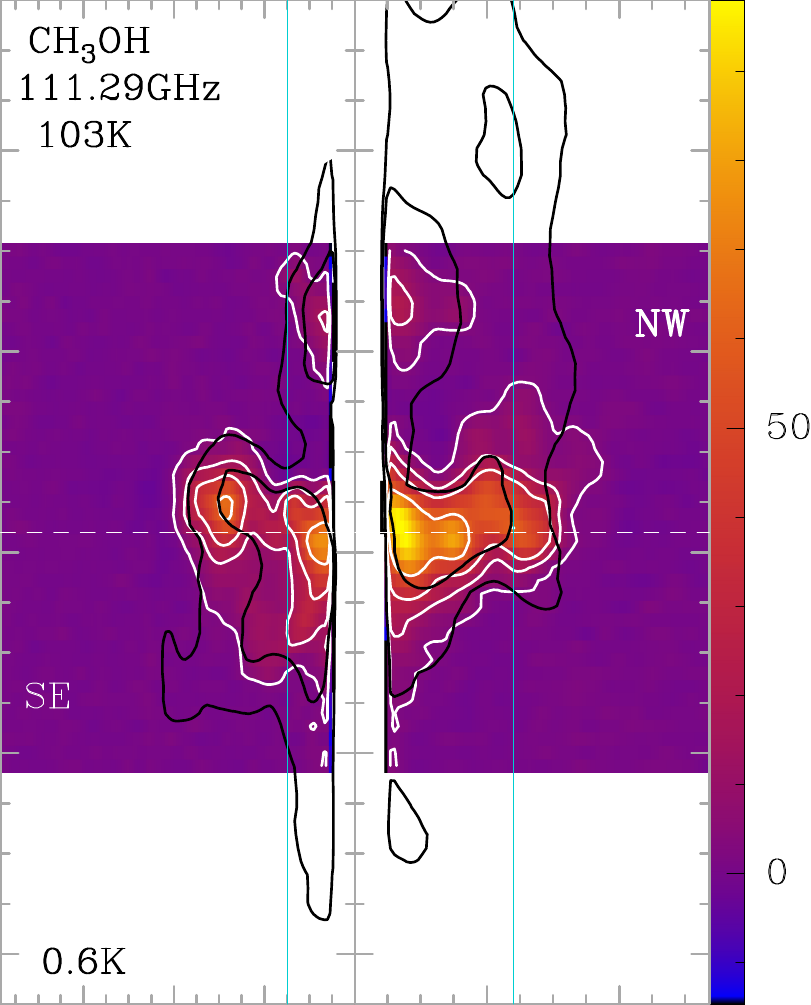}\hspace{0.23cm}
    \includegraphics[width=.219\textwidth]{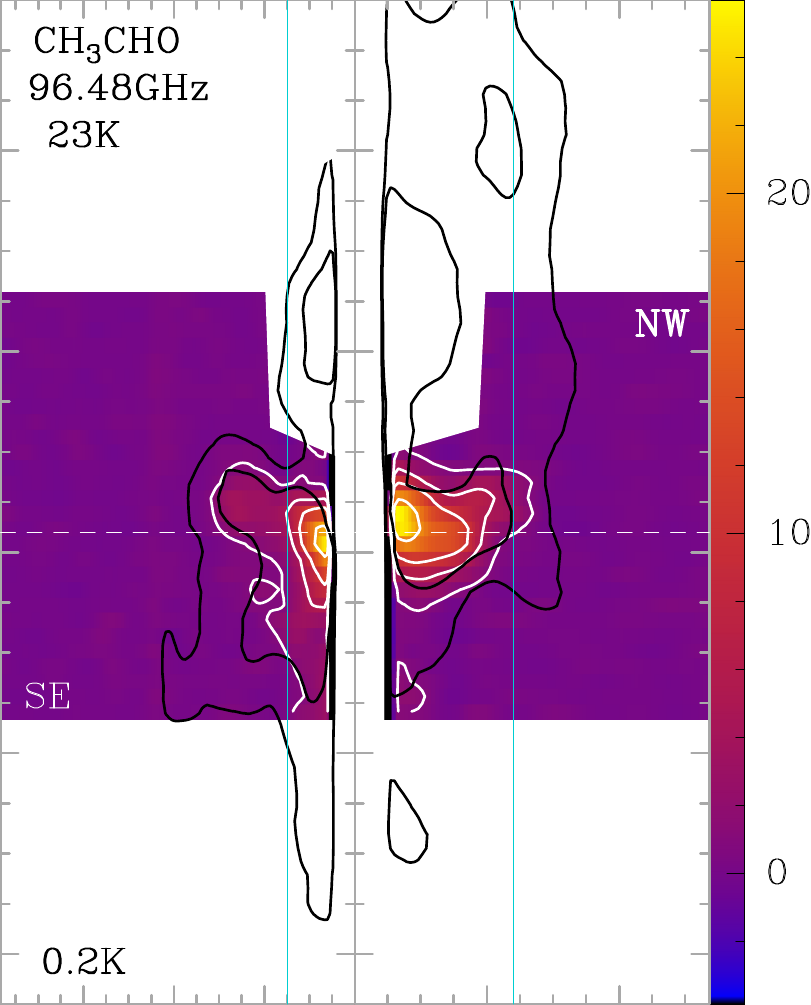}\\
    \includegraphics[width=.258\textwidth]{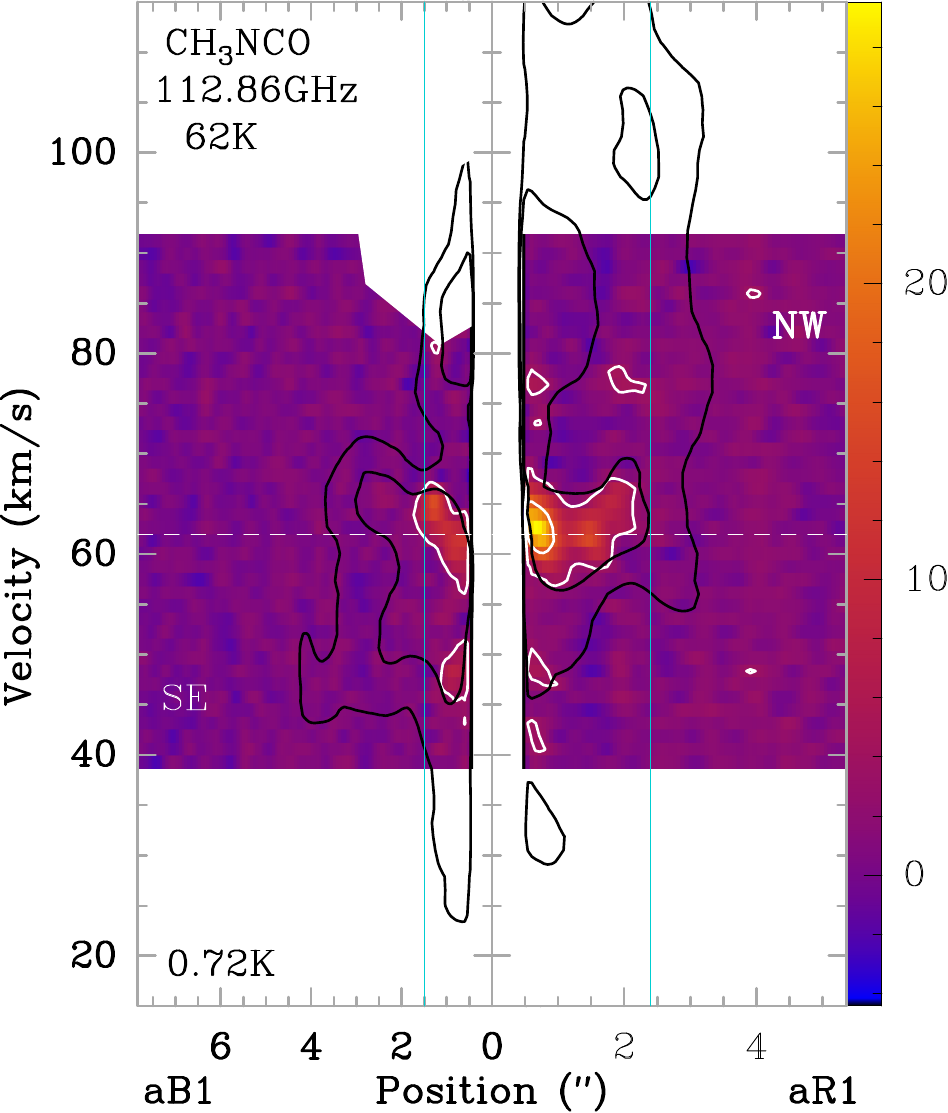}\hspace{0.2cm}
    \includegraphics[width=.221\textwidth]{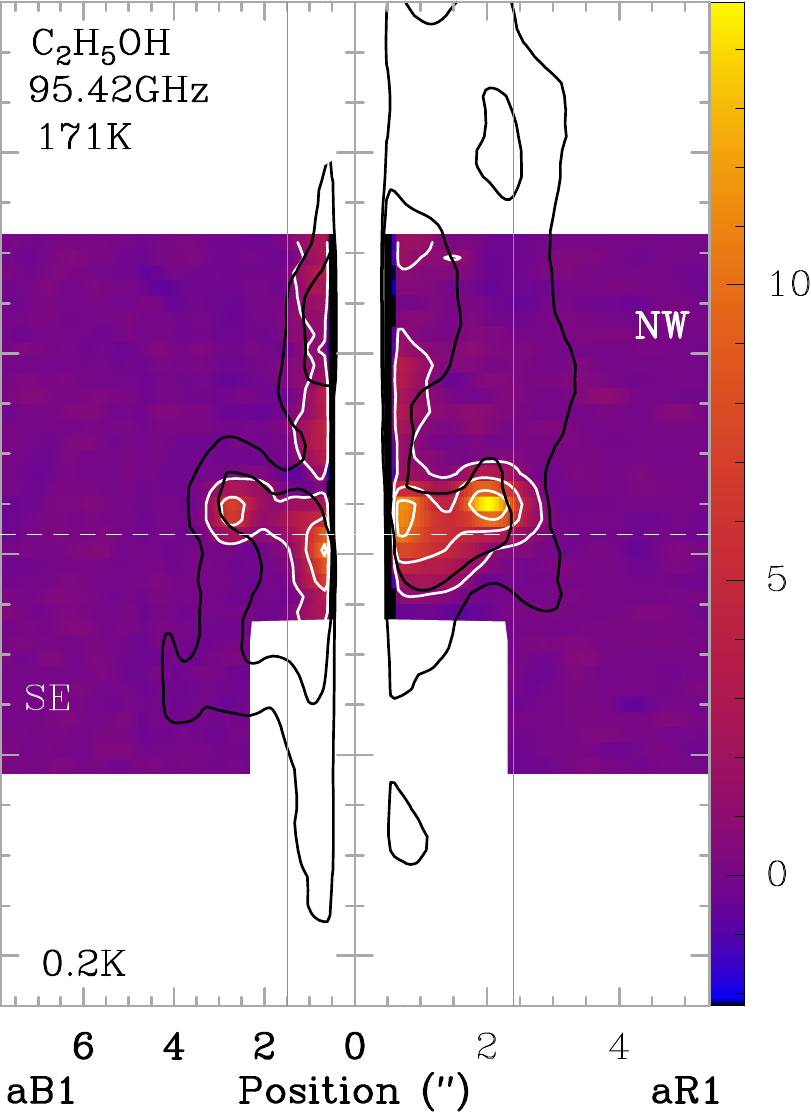}\hspace{0.2cm}
    \includegraphics[width=.221\textwidth]{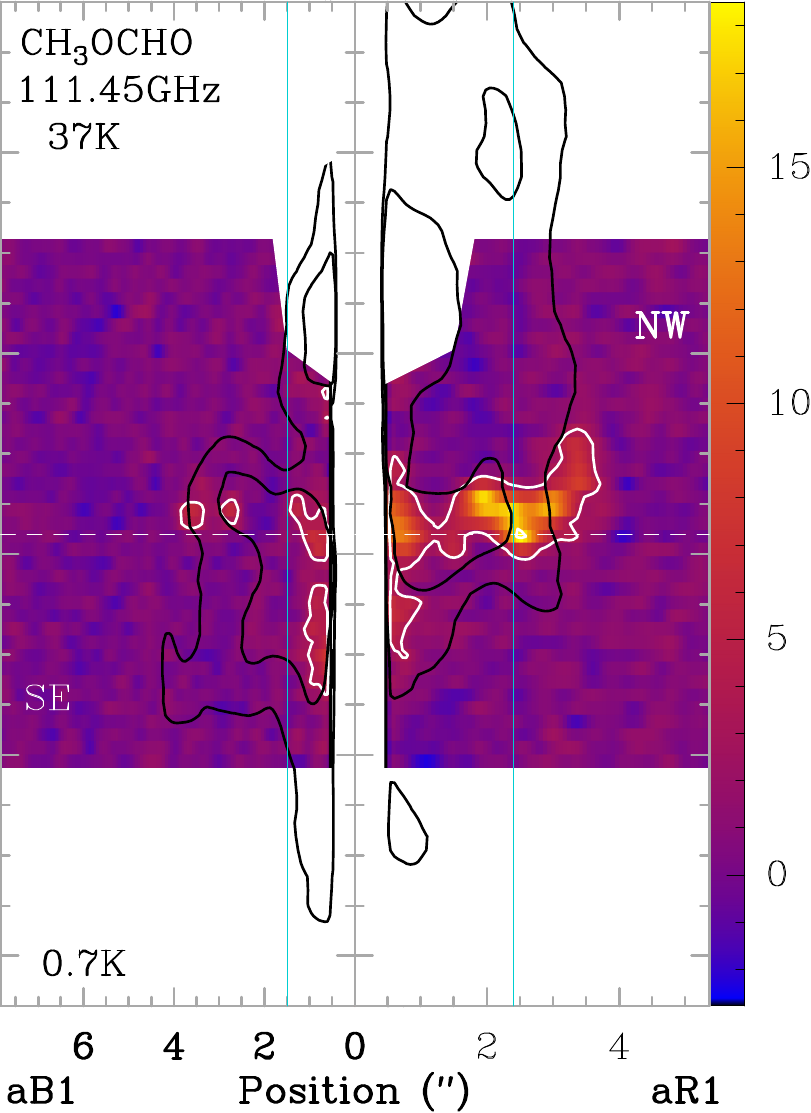}\hspace{0.2cm}
    \includegraphics[width=.221\textwidth]{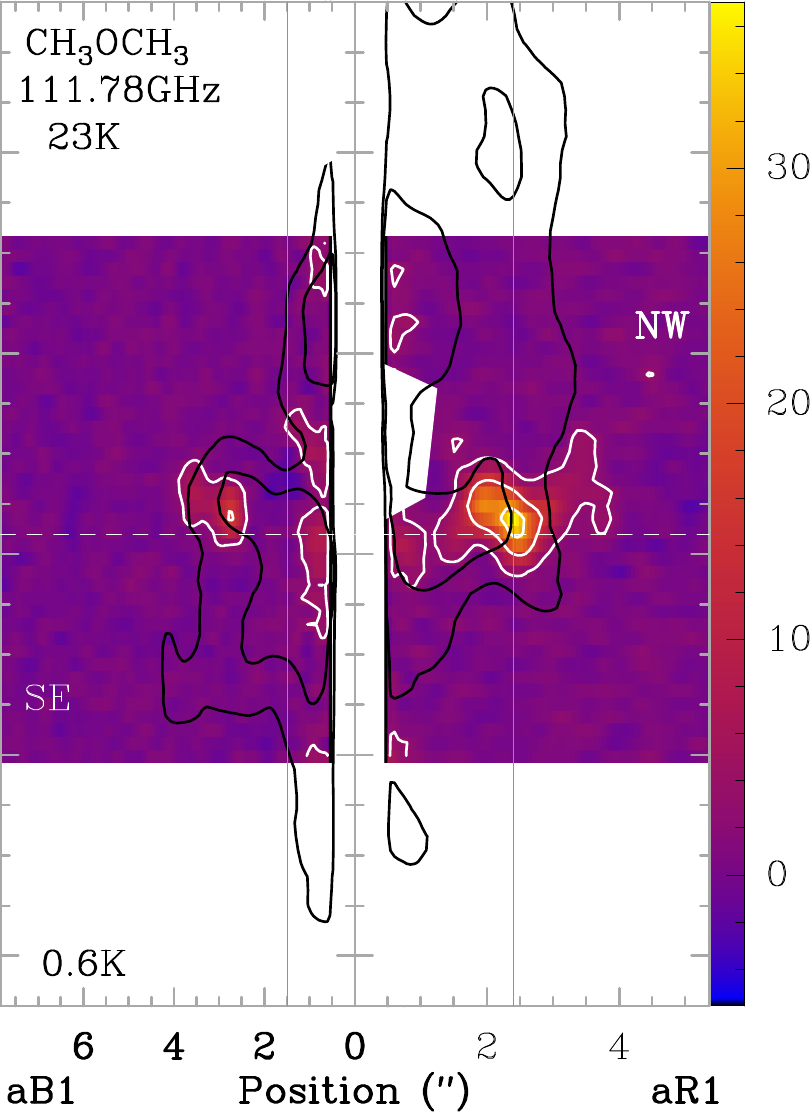}\\[-0.2cm]
    \caption{ Position-velocity diagrams along the solid black arrows labelled aB1 and aR1 in Fig.\,\ref{fig:so}. Here, the position labelled 0 corresponds to the centre of the hot core. The colour scale and white contours apply to the molecule written in the top-left corner together with the frequency and upper-level energy of the used transition. White contours are at 5$\sigma$, 25$\sigma$ and then increase by a factor 2, where $\sigma$ was measured in an emission-free region in the respective data cube and is written in the bottom-left corner. Although the region closest to the centre of the hot core and those identified with  contaminating emission from other species are masked in white, some weak emission from other molecules may still be included at distances $>$1\arcsec\,\,to the SW and NW. Black contours always show SO emission at 30$\sigma$ and 150$\sigma$. The maximum and minimum velocities shown in each map correspond to the outer integration limits used in Figs.\,\ref{fig:COMof}--\ref{fig:COMof3}, if existing. The white dashed line marks an average systemic velocity of 62\kms. Highlighted features in blue-shifted emission towards the southeast (SE, in the position-position maps) include B1 (elongated along velocity axis), B2 (elongated along both axes). Intensity peaks in red-shifted emission towards the northwest (NW) are labelled R1 and R2, and red-shifted emission close to the systemic velocity are labelled R3 (NW) and R4 (SE). Positions N1SE1 and N1NW3 are indicated with light-blue solid lines. }
    \label{fig:PV}
\end{figure*}

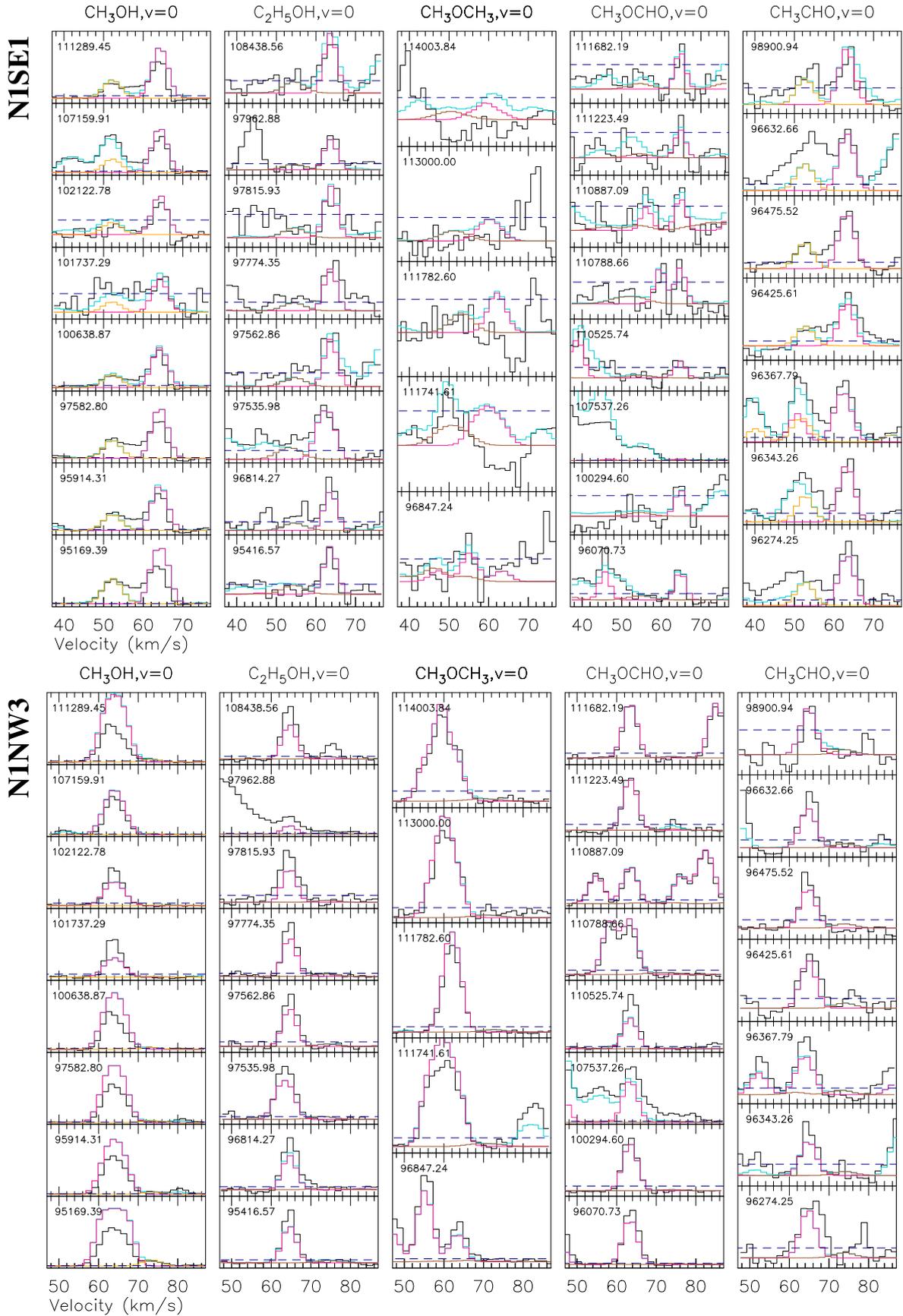
\begin{figure*}
    \centering
    \includegraphics[width=.8\textwidth]{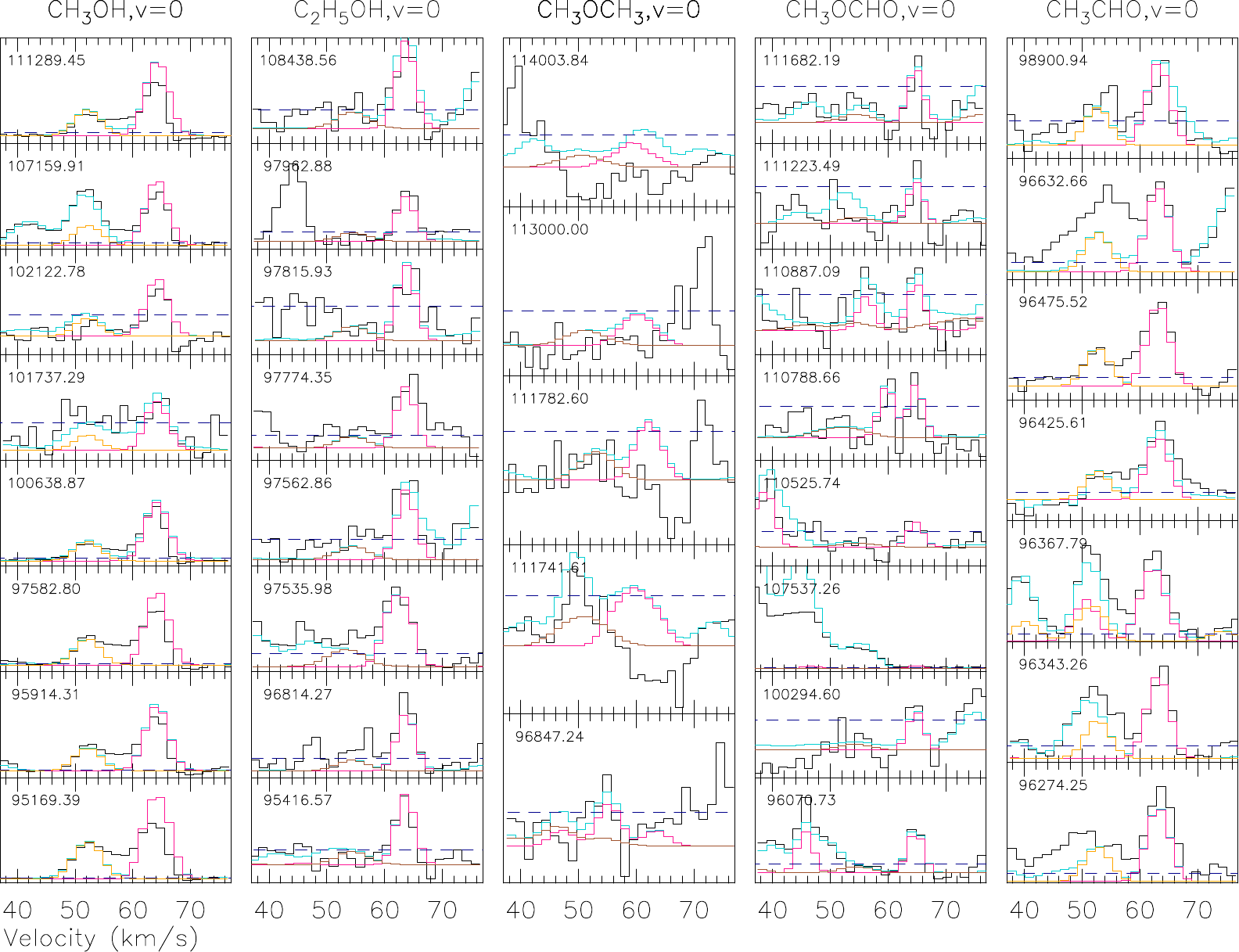}\\[0.2cm]
    \includegraphics[width=.8\textwidth]{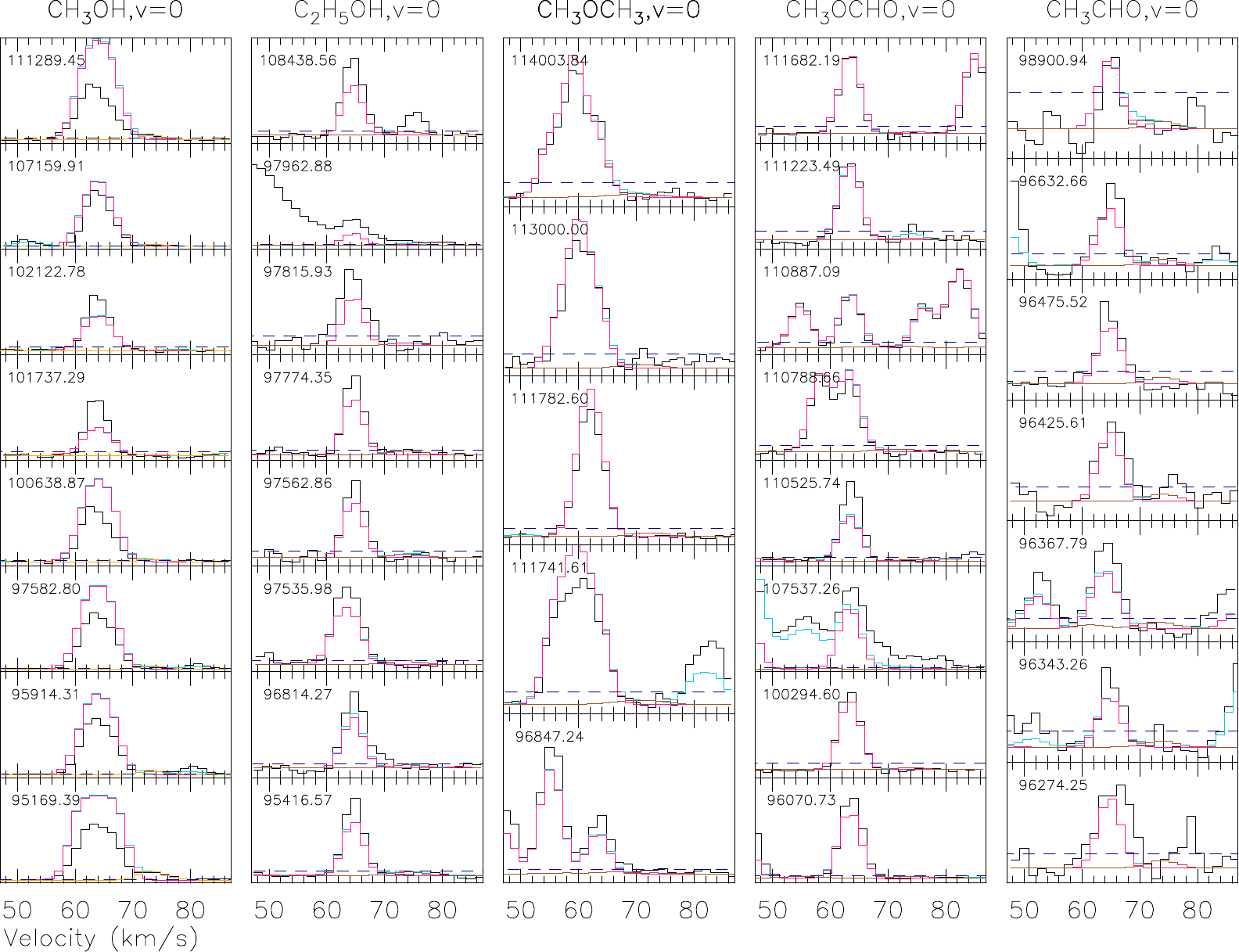}
    \begin{tikzpicture}[overlay]
       \node[rotate=90] at (-15.3,21.6) {\Large \textbf{N1SE1}};
       \node[rotate=90] at (-15.3,9.9) {\Large \textbf{N1NW3}};
    \end{tikzpicture}
    \caption{Observed spectra of O-bearing COMs (black) towards N1SE1 (\textit{top}) and N1NW3 (\textit{bottom}). The modelled spectra of the hot-core component are shown in pink (if detected) or purple (if not), those of the outflow component in orange (if detected) or brown (if not). The complete Weeds model is drawn in turquoise. The dashed line indicates the $3\sigma$ threshold for detection, where $\sigma$ is taken from Table\,2 in \citet{Belloche19}. The frequency in GHz at the reference velocity of 62\kms is shown in each panel.  }
    \label{fig:specs_o}
\end{figure*}

\begin{figure*}
    \centering
    \includegraphics[width=.9\textwidth]{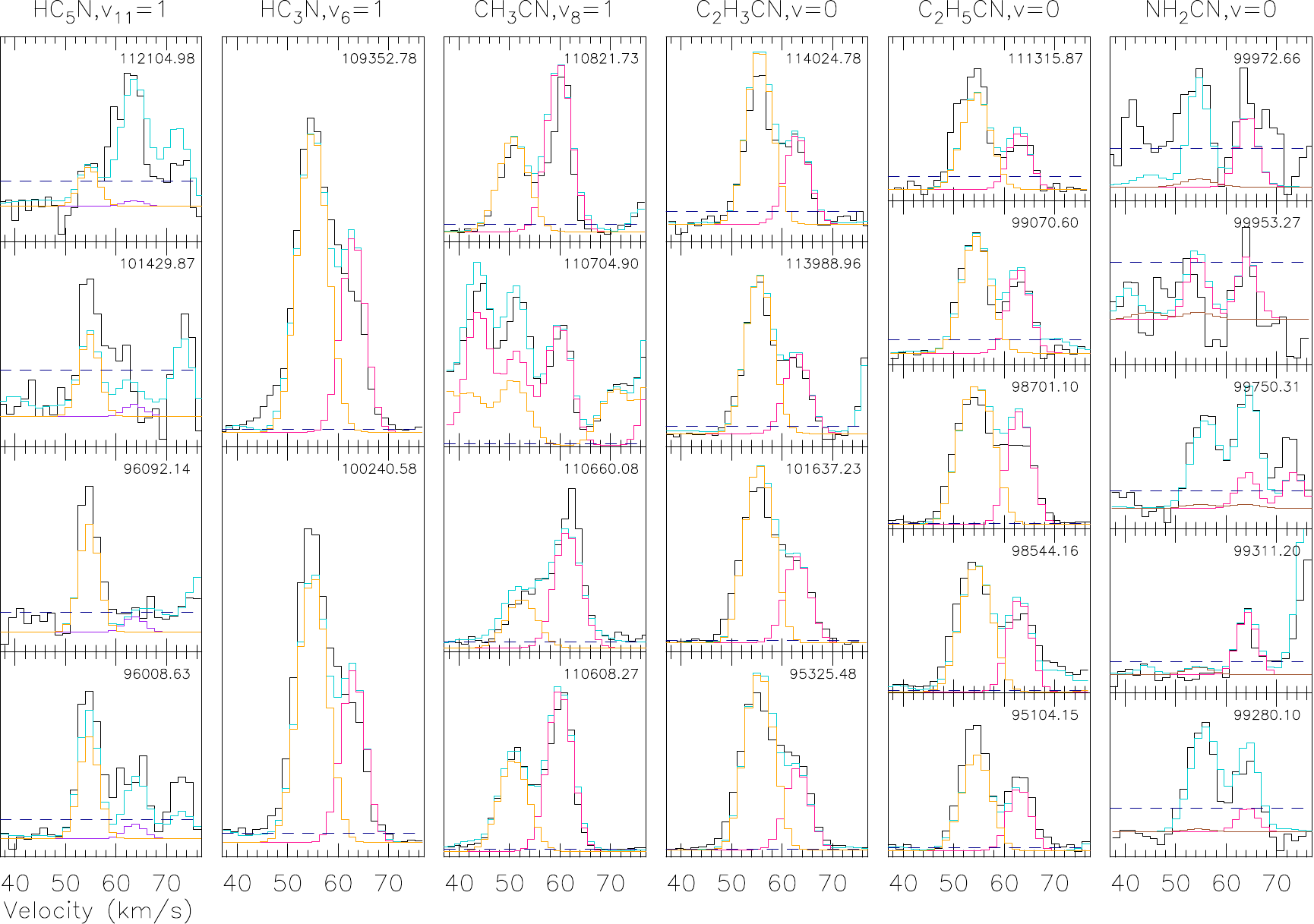}\\[0.2cm]
    \includegraphics[width=.9\textwidth]{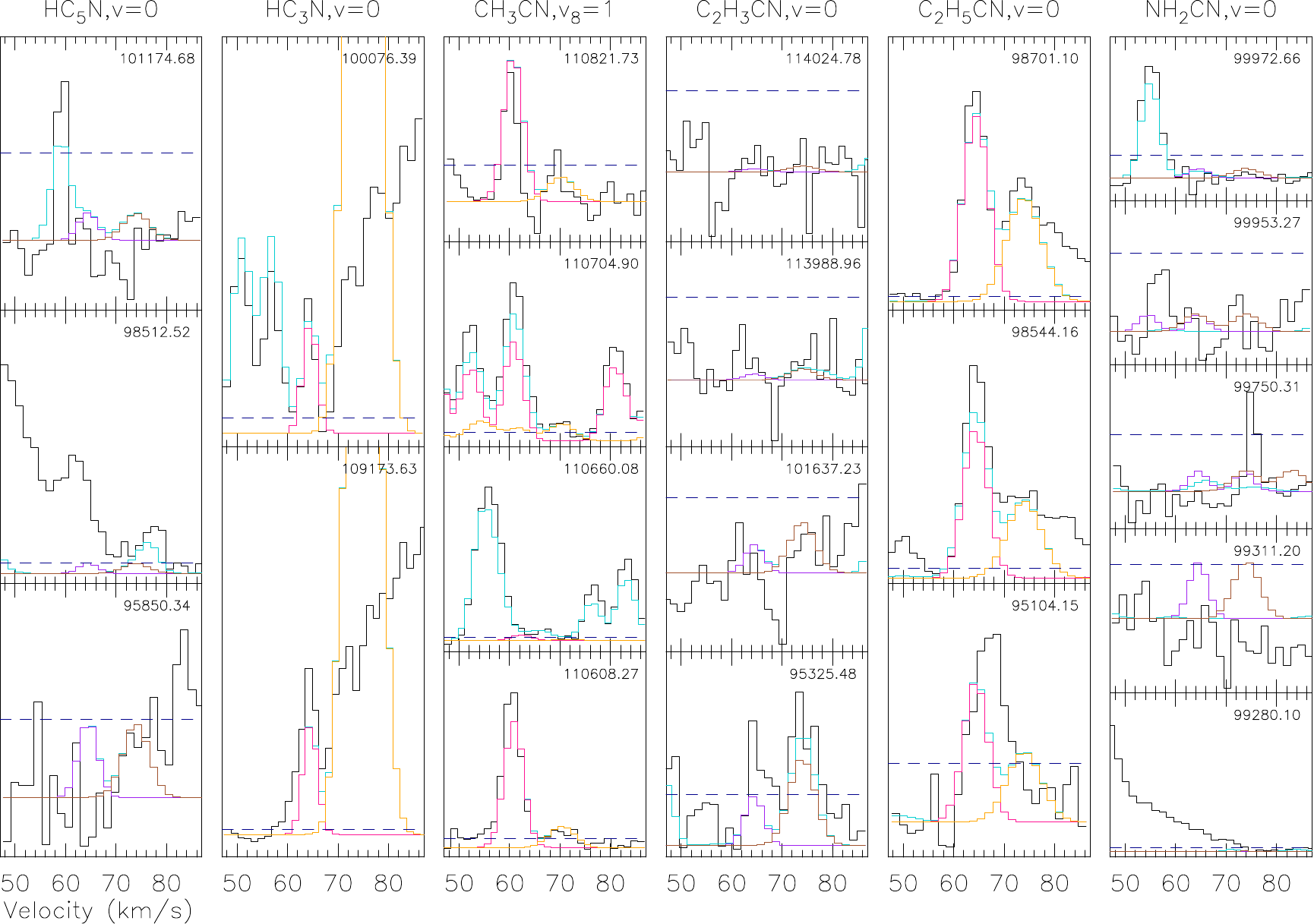}
    \begin{tikzpicture}[overlay]
       \node[rotate=90] at (-17.,22.3) {\Large \textbf{N1SE1}};
       \node[rotate=90] at (-17.,10.3) {\Large \textbf{N1NW3}};
    \end{tikzpicture}
    \caption{Same as Fig.\,\ref{fig:specs_o}, but for N-bearing molecules.  }
    \label{fig:specs_n}
\end{figure*}

\begin{figure*}
    \centering
    \includegraphics[width=.85\textwidth]{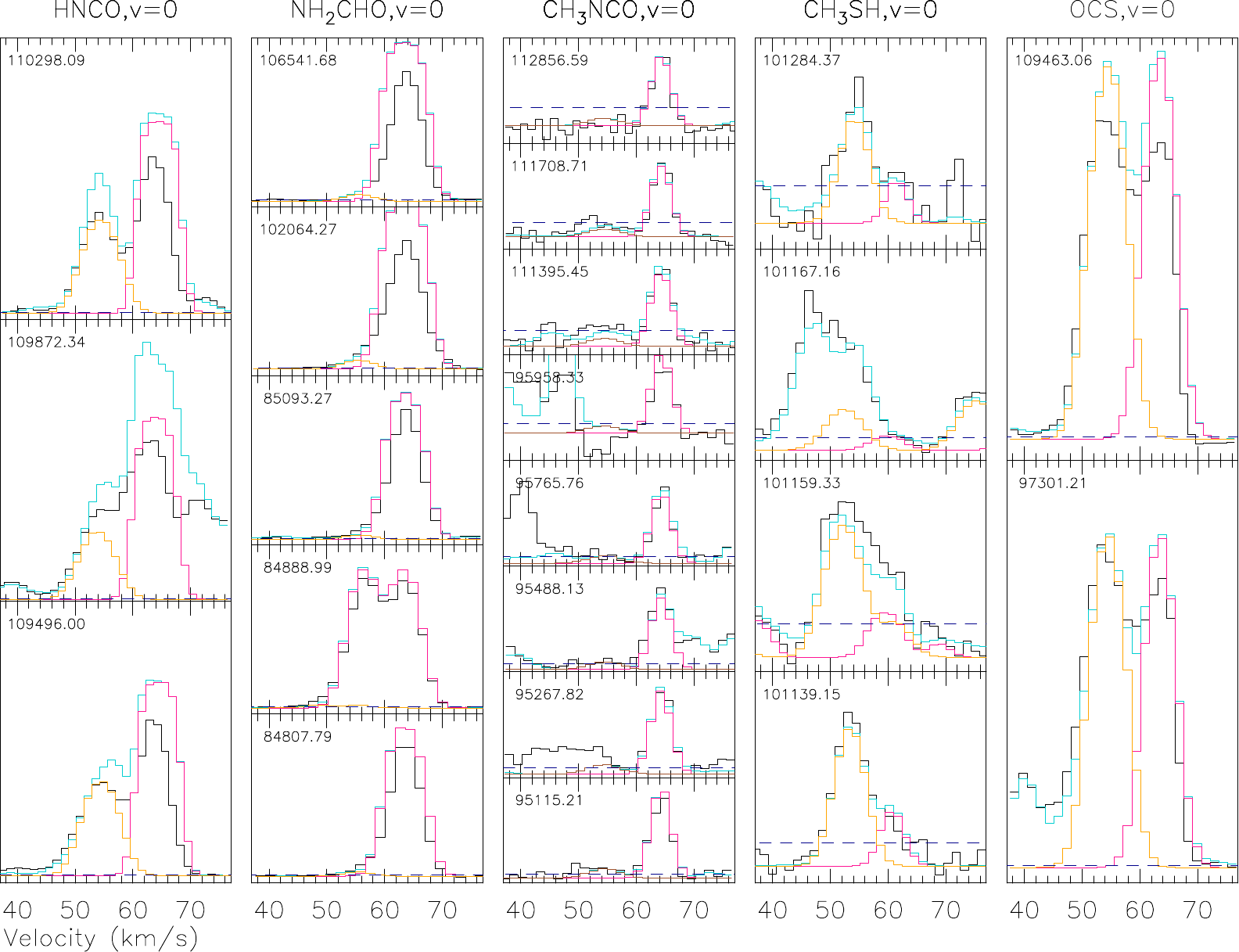}\\[0.2cm]
    \includegraphics[width=.85\textwidth]{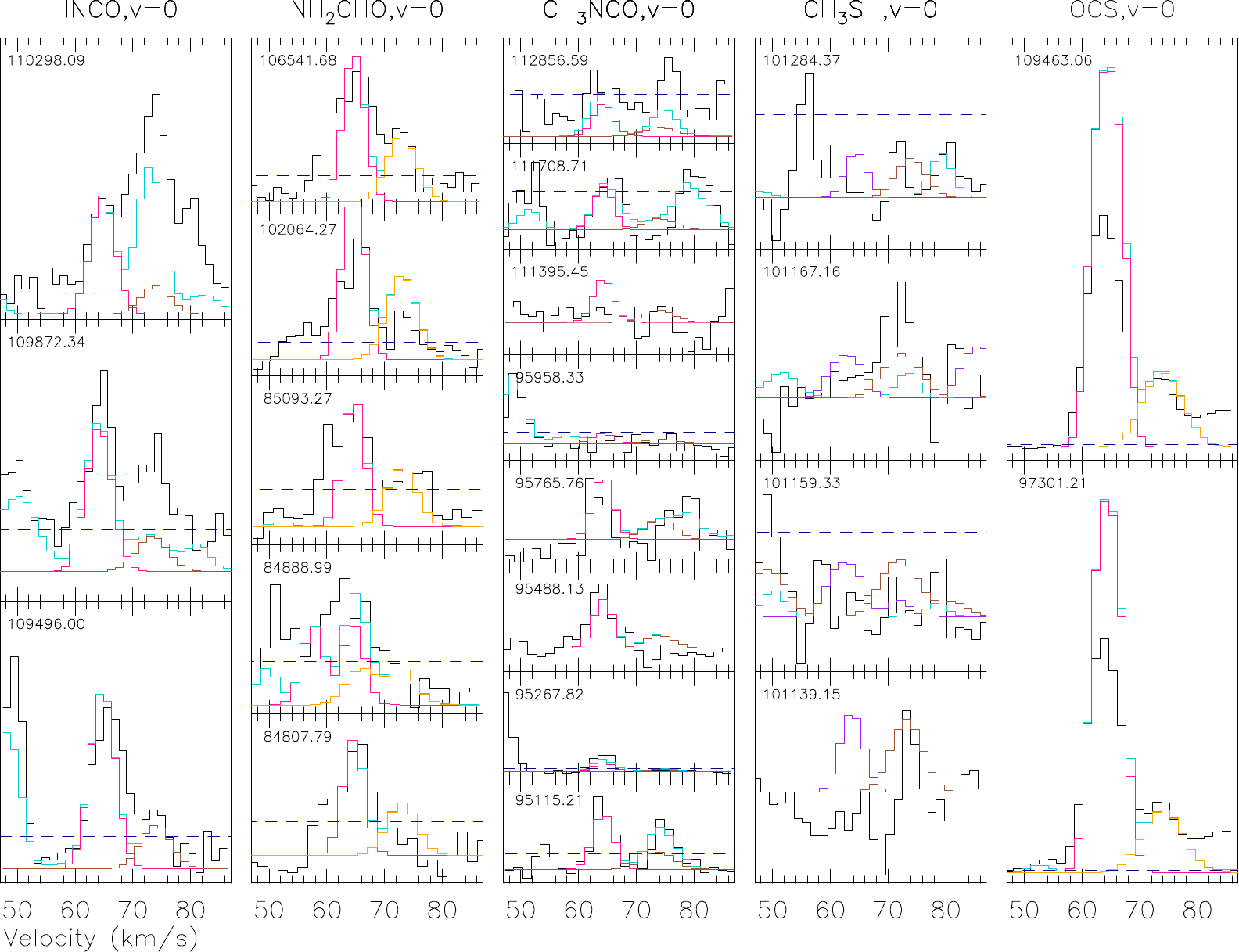}
    \begin{tikzpicture}[overlay]
       \node[rotate=90] at (-16.1,23) {\Large \textbf{N1SE1}};
       \node[rotate=90] at (-16.1,10.6) {\Large \textbf{N1NW3}};
    \end{tikzpicture}
    \caption{Same as Fig.\,\ref{fig:specs_o}, but for (N+O)- and S-bearing molecules.  }
    \label{fig:specs_nos}
\end{figure*}

\begin{figure*}[]
    \centering
    \includegraphics[width=.95\textwidth]{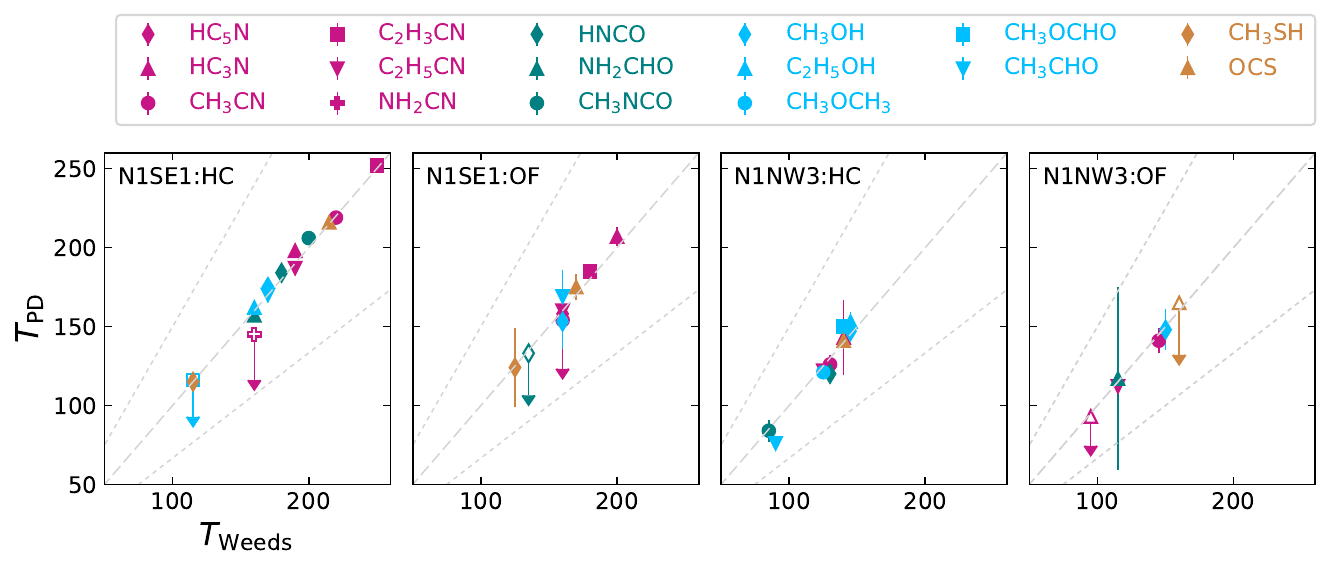}
    \caption{Comparison of rotational temperatures used for the Weeds models and derived in the population diagrams for the various molecules in the hot-core and outflow components at positions N1SE1 and N1NW3. Pink markers indicate N-bearing species, teal markers (N+O)-bearers, blue O-bearers, and orange S-bearers. Empty markers with arrows indicate upper limits. The grey dashed line shows where temperatures are equal. The two grey dotted lines indicate a factor 1.5 difference from unity.}
    \label{fig:Tweeds}
\end{figure*}

\begin{figure*}[htp]
    \centering
    \includegraphics[width=.9\textwidth]{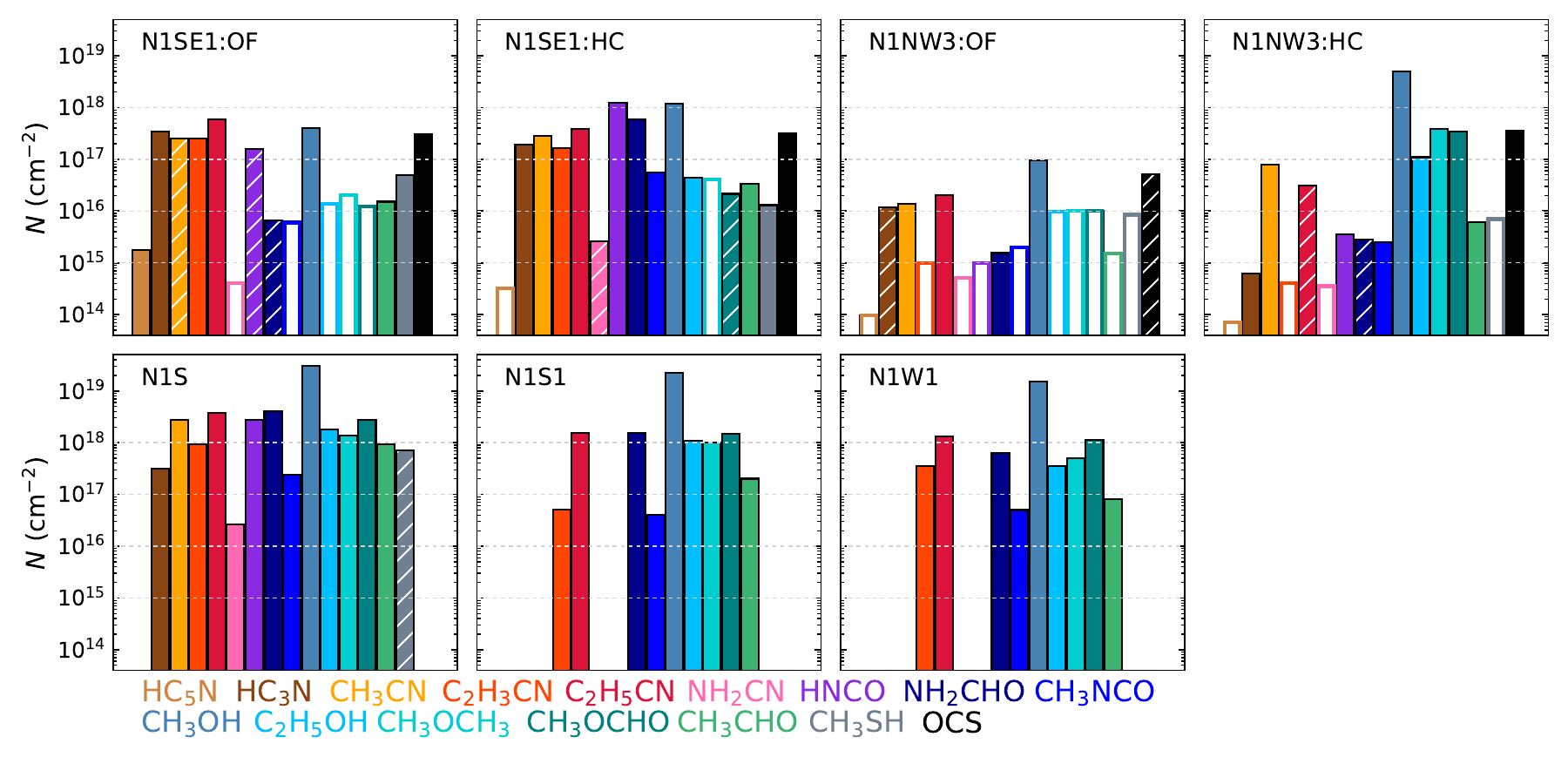}\\[-0.2cm]
    \caption{Column densities of various organic molecules for the outflow (OF) and hot-core (HC) components at positions N1SE1 and N1NW3, as well as at N1S, N1S1, and N1W1. Empty bars indicate upper limits. In all panels hatched bars indicate when the rotational temperature in a population diagram (PD) was fixed (NH\2CN at N1SE1:HC and \mmc at N1S), the molecule was detected but a PD could not be derived (\fmm at N1SE1:OF), or only an upper limit for the rotational temperature in the PD was derived. }
    \label{fig:Nbars}
\end{figure*}

\begin{figure*}[thp]
    \centering
    \includegraphics[width=.6\textwidth]{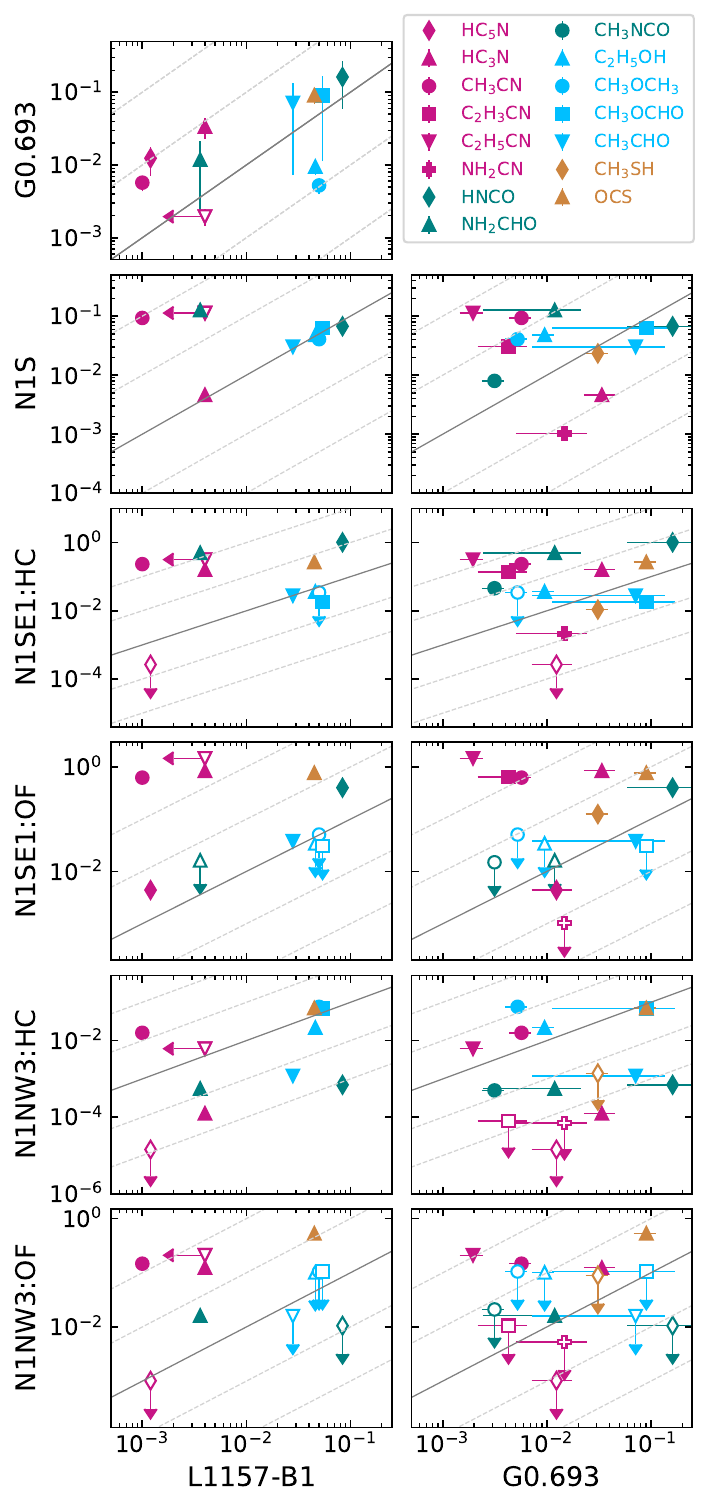}
    \caption{Same as Fig.\,\ref{fig:corner}, but for molecular abundances with respect to methanol and additionally showing component N1NW3:OF. }
    \label{fig:corner_met}
\end{figure*}

Position-velocity diagrams of various COMs and simpler molecules are presented in Fig.\,\ref{fig:PV}. In Figs.\,\ref{fig:specs_o}--\ref{fig:specs_nos}, we show observed and modelled spectra towards positions N1SE1 and N1NW3 of various transitions for those molecules for which abundances were derived in this work. 
In Fig.\,\ref{fig:Tweeds} temperatures used in the Weeds models are compared to the values obtained from the population-diagram analysis for both components at N1SE1 and N1NW3, respectively. Column densities of the molecules used in the Weeds models are shown in Fig.\,\ref{fig:Nbars} and the comparison of abundances with respect to methanol between Sgr\,B2\,(N1), G0.693, and L1157-B1 in Fig.\,\ref{fig:corner_met}. 



\section{Molecular abundances in G0.693 and L1157-B1 from the literature}\label{ass:sources}
\begin{table*}[h!]
    \caption{Molecular column densities in G0.693 and L1175--B1 reported by various authors.}
    \centering
    \begin{tabular}{lrrrrrr}
       \hline\hline\\[-0.2cm]
        & \multicolumn{4}{c}{G0.693} & \multicolumn{2}{c}{L1157--B1} \\\cmidrule(lr){2-5} \cmidrule(lr){6-7} 
        Molecule & Z18/Z23/RA21 & AA15 & RT08 & RT06 & HR\tablefootmark{*} (B1b) & LR\tablefootmark{**} \\[0.1cm]\hline\\[-0.3cm]
        HC\3N & $7.1\pm1.3(14)$ & $5.5\pm0.3(14)$ & -- & -- & $3.4(13)$\tablefootmark{c} & $4.0(12)$\tablefootmark{b} \\
        HC\5N & $2.6\pm0.1(14)$ & $3.5\pm1.2(13)$ & -- & -- & -- & $1.2(12)$\tablefootmark{b} \\
        CH\3CN & $11.5\pm0.3(13)$ & $28\pm3.0(13)$ & -- & -- & $1.0(13)$\tablefootmark{d} & $1.0\pm0.5(12)$\tablefootmark{e} \\
        C\2H\3CN & $9.0\pm0.1(13)$ & $5.1\pm1.6(13)$ & -- & -- & -- & -- \\
        C\2H\5CN & $4.1\pm0.4(13)$ & $8.1\pm1.6(13)$ & -- & -- & -- & $<4.0(12)$\tablefootmark{f} \\
        NH\2CN & $3.1\pm0.2(14)$ & $1.2\pm0.2(14)$ & -- & -- & -- & -- \\
        HNCO & $3.4\pm0.2(15)$ & $1.4\pm0.1(15)$ & -- & -- & -- & $3.3-8.4(13)$\tablefootmark{b} \\
        NH\2CHO & $6.3\pm1.4(14)/2.5\pm0.1(14)$ & $5.9\pm0.6(13)$ & -- & -- & $8.0(12)$\tablefootmark{a} &  $1.7-3.6(12)$\tablefootmark{b} \\
        CH\3NCO & $6.6\pm0.4(13)$ & -- & -- & -- & -- & -- \\
        CH\3OH & 1.6(16) & $1.2\pm0.1(17)$ & $2.1(16)$ & $3.2(16)$ &  $1.3\pm0.3(16)$\tablefootmark{a} & $1.0\pm0.2(15)$\tablefootmark{b} \\
        C\2H\5OH & 2.0(14) & $4.5\pm0.9(14)$ & -- & $1.3(15)$ & -- & $4.6\pm1.6(13)$\tablefootmark{b} \\
        CH\3OCH\3 & $1.1(14)$\tablefootmark{$\dagger$} & -- & -- & -- & -- & $5.0\pm1.4(13)$\tablefootmark{b} \\
        CH\3OCHO & -- & $\sim3.5(14)$ & $1.9(15)$ & $8.3(14)$ & -- & $5.4\pm1.4(13)$\tablefootmark{b} \\
        CH\3CHO & -- & $2-5(14)$ & -- & $1.5(15)$ & $7.0\pm3.0(13)$\tablefootmark{a} & $2.8\pm0.8(13)$\tablefootmark{b} \\
        CH\3SH & 6.5(14) & -- & -- & -- & -- & -- \\
        OCS & -- & $1.9\pm0.1(15)$ & -- & -- & $1.0(15)$\tablefootmark{a} & $4.5\pm1.0(13)$\tablefootmark{f} \\ 
        \hline\hline
    \end{tabular}
    \tablefoot{Column densities are written as $x(z) = x\times 10^z$\,\scm or $(x\pm y)(z) = (x\pm y)\times 10^z$\,\scm.\\ Z18 \citep[N-bearing molecules,][]{Zeng18}. Z23 \citep[\fmm,][]{Zeng23}. RA21 \citep[\met, \et, and \mmc,][]{Rodriguez-Almeida21a}. AA15: \cite{Armijos-Abendano15}. RT06: \cite{Requena-Torres06}. RT08: \citet{Requena-Torres08}. \tablefoottext{$\dagger$}{V.\,Rivilla (private communication).}\\ 
    \tablefoottext{*}{High angular resolution; Column densities were obtained from interferometric studies.} \tablefoottext{**}{Low angular resolurion;  Column densities were obtained from single-dish observations.}
    \tablefoottext{a}{SOLIS \citep[Seeds Of Life In Space,][]{Codella17,Codella20,Codella21}.}\tablefoottext{b}{ASAI \citep[Astrochemical Survey At IRAM,][]{Mendoza14,Mendoza18,Lefloch17}.}\tablefoottext{c}{\citet{Benedettini07}.}\tablefoottext{d}{\citet{Codella09}}\tablefoottext{e}{\citet{Arce08}}.\tablefoottext{f}{\citet{Sugimura11}}.} 
    \label{tab:G0693}
\end{table*}

\subsection{G0.693}

\cite{Requena-Torres06} performed observations with the IRAM\,30\,m telescope at an angular resolution of 10--25\arcsec\,towards multiple clouds located in the Galactic centre region. Their study was focussed on (mainly O-bearing) complex molecules. Amongst these sources, G0.693 was one of the chemically richest, and was observed again in a follow-up study by \cite{Requena-Torres08} with the Green Bank Telescope (GBT) at lower frequencies, where they confirmed the source as one of the largest repositories of O-bearing organic molecules in the Galactic centre region. Rotational temperatures were derived from rotational diagrams and were found to be $<$\,16\,K.
\cite{Armijos-Abendano15} performed a spectral line survey towards G0.693 at the same position as \cite{Requena-Torres08} with the Mopra\,22\,m radio telescope at an angular resolution of 30--40\arcsec covering a variety of simpler and more complex species. Based on a rotational-diagram analysis they derived rotational temperatures of 10--70\,K.
\cite{Zeng18} observed the same position with the IRAM\,30\,m telescope and the GBT at slightly higher angular resolution in order to perform a detailed study on N-bearing molecules. Excitation temperatures and column densities were obtained from an LTE analysis with \texttt{madcuba} or from rotational diagrams. The temperatures lie in a range of 10--30\,K. The column density of \fmm was recently updated by \citet{Zeng23} using new observational data obtained with the IRAM\,30\,m and the Yebes\,40\,m telescopes.
\citet{Rodriguez-Almeida21a} observed the source at the same position with the IRAM\,30m and the Yebes\,40\,m telescopes covering frequencies from 32 to 172\,GHz to study mainly S-bearing molecules. Based on modelling with \texttt{madcuba} they derived excitation temperatures of $\sim$15\,K. All studies above assume extended emission to derive column densities. 
The column densities reported for G0.693 by the various studies are summarised in Table\,\ref{tab:G0693}. \\
\\
In Fig.\,\ref{fig:corner} we compare abundances with respect to H\2 derived towards G0.693 and at positions in Sgr\,B2\,(N1), including N1S, N1SE1(:OF and HC), and N1NW3:HC. A detailed comparison of the chemical compositions is provided in the following. The key points are summarised in the main text in Sect.\,\ref{dss:sources}.
All molecules have higher abundances with respect to H\2 at N1S than in G0.693 by 1--2 orders of magnitude or even more, except for NH\2CN and HC\3N, whose abundances are similar to or higher by a factor of a few than in G0.693, respectively. 
The picture is not too different for N1SE1:HC. All cyanides and (N+O)-bearing molecules show higher abundances by about two orders of magnitude than in G0.693. Even abundances of HC\3N are enhanced by more than an order of magnitude in N1SE1:HC. The difference in abundance of O-bearing molecules and \mmc between N1SE1:HC and G0.693 is sightly smaller than for N1S. The abundance of OCS is higher by more than a factor 10 in N1SE1:HC and the upper limit for HC\5N in N1SE1:HC implies a much lower abundance than in G0.693.
In the outflow component at N1SE1, abundances of the cyanides and HC\3N behave similarly as in N1SE1:HC. The abundance of HC\5N is quite similar for G0.693 and N1SE1:OF, while NH\2CN is less abundant by a factor 10 in the latter. Sulphur-bearing molecules are more abundant by about a factor 10 and HNCO is still more abundant by a factor of a few in N1SE1:OF. For the other (N+O)-bearing molecules and most of the O-bearing species, we only derived upper limits, from which we cannot draw conclusions for the comparison to G0.693. 

The scatter between the families of molecules is larger in N1NW3:HC. Abundances of CH\3CN and \etc are higher by more than two orders of magnitude than in G0.693, while all other N-bearing species are similarly abundant or, more likely, less abundant. All detected O-bearing molecules, except for \ad, and OCS also have abundances that are higher by two orders of magnitude or more. The abundance of HNCO is similar in both positions, while \mic and \fmm abundances are higher by up to a factor 10 in N1NW3:HC.

Molecules that have been observed in G0.693 and that we do not detect in the outflow of Sgr\,B2\,(N1) include, for example, HC\7N \citep{Zeng18} and g--C\2H\5SH \citep{Rodriguez-Almeida21a}, with abundances with respect to H\2 of $\sim1\times10^{-10}$ and $\sim3\times10^{-10}$, respectively. We derive upper limits on the abundances in N1SE:OF of $\sim 6\times10^{-11}$ and $\sim3.8\times10^{-9}$, respectively, not including vibrational corrections.
In addition, we show a comparison of abundances with respect to methanol in the right column of Fig.\,\ref{fig:corner_met}. What stands out is, similar to the abundances with respect to H\2, the cyanides are significantly more abundant at N1S and in both components at N1SE1. 
In Fig.\,\ref{fig:corner_met} we also show component N1NW3:OF, however, with mostly upper limits on the abundances in this component, identifying similarities and differences to the chemical content of G0.693 is difficult.

\subsection{L1157-B1}
Abundances that were derived towards L1157-B1 from observations at low angular resolution originate from \citet{Arce08}, \citet{Lefloch17}, \citet{Mendoza14,Mendoza18}, and \citet{Sugimura11}. The first four studies carried out pointed observations towards L1157-B1 with the IRAM\,30\,m telescope, where the position is not always the same. The observations were performed at 2 and 1.3\,mm \citep{Arce08} or at 3\,mm. Rotational temperatures and column densities were derived based on a rotational-diagram analysis. Values of rotational temperature lie in a range of 10--30\,K for O- and (O+N)-bearing species and are higher for N-bearing molecules (50--60\,K or higher). \citet{Sugimura11} performed observations with the NRO\,45\,m telescope at 3\,mm and derived abundances based on the assumption of optically thin emission and a fixed rotational temperature that they took from \citet{Arce08}.

Interferometric observations were performed by \citet{Benedettini07} and \citet{Codella09} with the Plateau de Bure Interferometre (PdBI), run by IRAM, and by \citet{Codella17,Codella20,Codella21} with the array's successor NOEMA (NOrthern Extended Millimetre Array). The latter studies were carried out as part of the SOLIS (Seeds Of Life In Space) survey \citep{Ceccarelli17}. All observations yielded an angular resolution of 3--5\arcsec\,and three of them were done at 3\,mm, while for CH\3CN the $8_K-7_K$ at $\sim$147\,GHz was observed \citep[][]{Codella09}. 
Rotational temperatures were derived based on a rotational-diagram analysis, which yielded a value of $\sim$10\,K for \met and \ad \citep[][]{Codella20}, 73\,K for CH\3CN \citep{Codella09}, and $\geq$70\,K for OCS, which are in agreement with results derived from single-dish observations. \citet{Benedettini07} and \citet{Codella17} assumed a value of 80\,K and 10\,K, respectively, to derive column densities, which they took from previous studies. 
The column densities derived from single-dish (LR) and interferometric (HR) data are summarised in Table\,\ref{tab:G0693}. \\ 
\\
From the comparison of abundances with respect to H\2 and methanol in Figs.\,\ref{fig:corner} and \ref{fig:corner_met}, respectively, we saw that the gas-phase chemical inventory of L1157-B1 is similar to that in G0.693, the comparison to Sgr\,B2\,(N1) reveals similar differences in abundances with respect to H\2 in Fig.\,\ref{fig:corner}. 
Abundances of CH\3CN, \etc, and \fmm are higher by two orders of magnitude in N1S, while those of all O-bearing molecules, HNCO, and HC\3N are higher by factors of a few. 
At N1SE1:HC, abundances of O-bearing molecules are generally more similar to L1157-B1, while CH\3CN, \etc, \fmm, and here also HC\3N  have higher abundances by a factor 100 or more and HNCO and OCS abundances are higher by a factor 10 or more. The only N-bearing molecule that has a similar or lower abundance than in L1157-B1 is HC\5N. 
The picture is similar for N1SE1:OF, except that the the (N+O)-bearing species HNCO and \fmm are no longer significantly more abundant than in L1157-B1, and the O-bearing molecules are likely less abundant than in L1157-B1, except for \ad, which has similar abundances towards both positions. 
Abundances in N1NW3:HC behave again differently than in other hot-core components in Sgr\,B2\,(N1): all O-bearing molecules, except \ad, have higher abundances in N1NW3:HC than in L1157-B1 by an order of magnitude or more. These higher abundances are also seen for CH\3CN, \etc, and OCS. The abundance for HNCO is lower by factors of a few in N1NW3:HC, while the \fmm abundance is higher by a similar amount. \ad and HC\3N are similarly abundant in both sources.
The overall picture for abundances with respect to \met shown in Fig.\,\ref{fig:corner_met} is again similar to what we see for abundances with respect to H\2.

\section{Detailed comparison with hot-core models}\label{app:modelsRTG}

\begin{figure*}[htp]
    \centering
    \includegraphics[width=.9\textwidth]{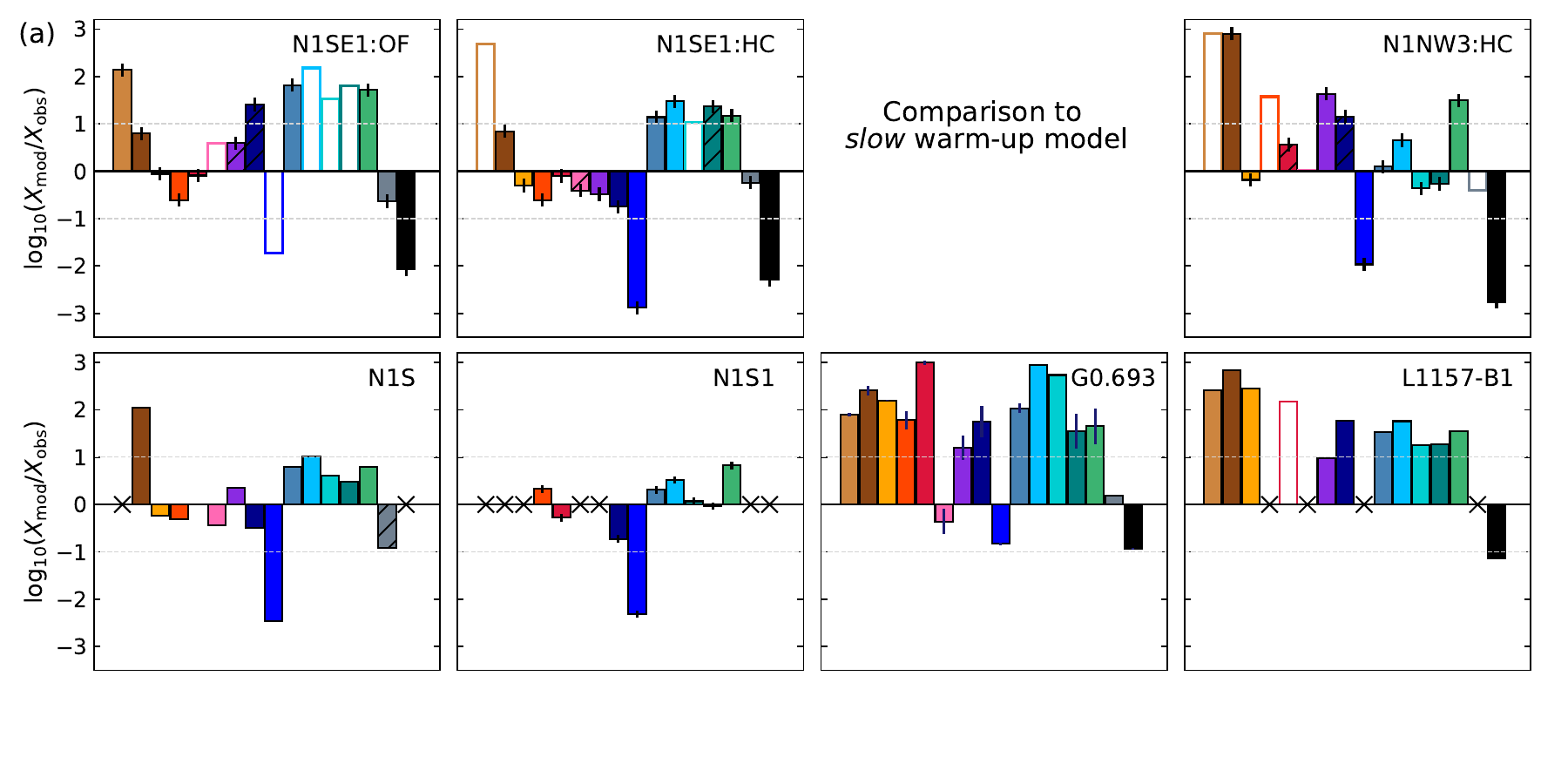}\\[-1cm]
    \includegraphics[width=.9\textwidth]{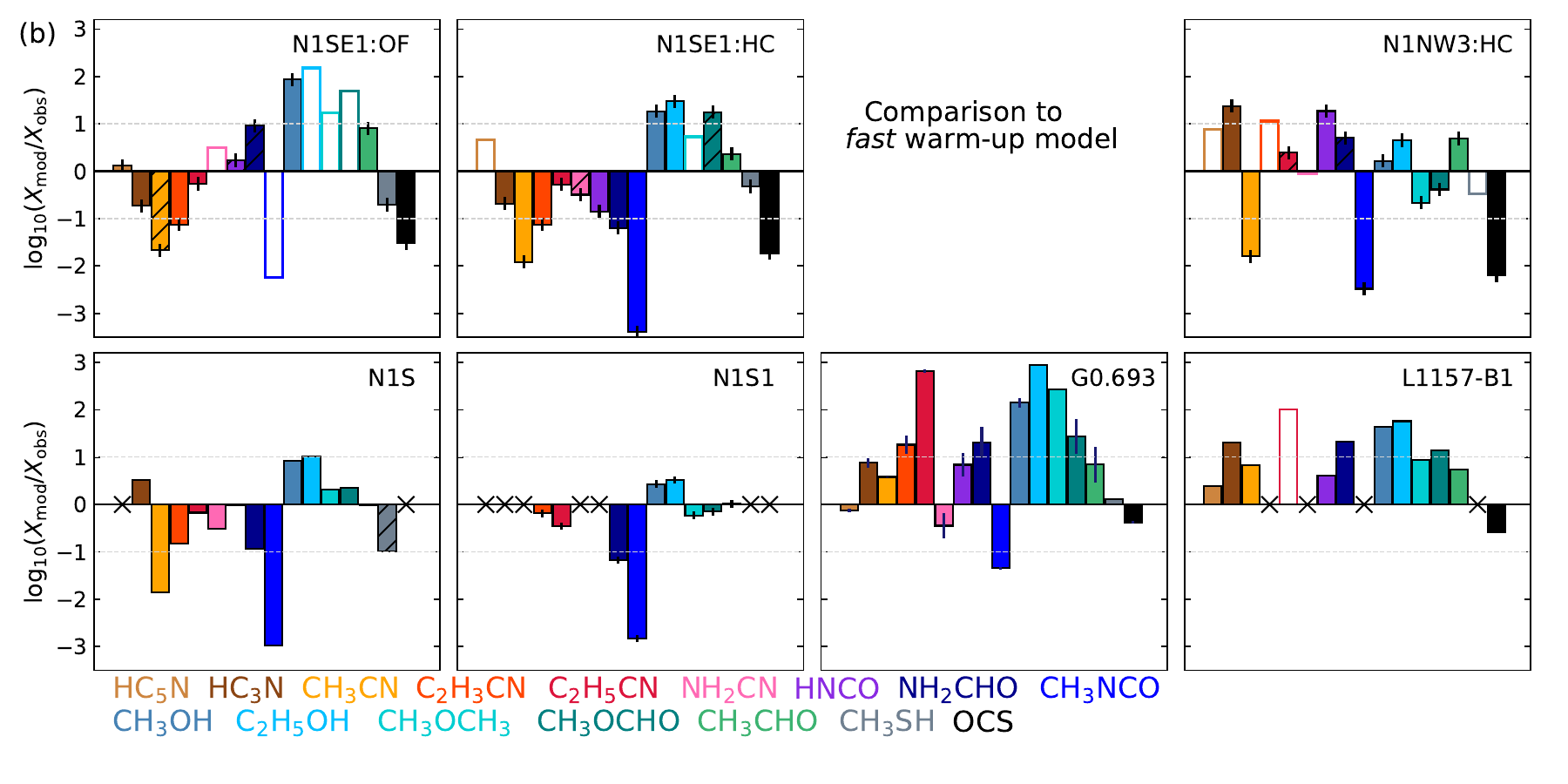}\\[-0.3cm]
    \caption{Comparison between modelled peak abundances with respect to H\2 and the observed abundances towards both components at N1SE1 and N1NW3:HC, N1S, N1S1, G0.693, and L1157-B1. For all observed values, H\2 column densities were derived from C$^{18}$O emission, except for N1S, for which dust emission was used. The modelled abundances $X_{\rm mod}$ are taken from Table\,17 in \citet{Garrod22}, where we use the values of the slow \textit{(a)} and fast \textit{(b)} warm-up models, and multiplied by a factor 2 to roughly convert to abundances with respect to H\2 (from total H). Empty bars indicate lower limits. Hatched bars indicate the same as in Fig.\,\ref{fig:Xbars-norm}. Black crosses indicate molecules for which no values are available. }
    \label{fig:cfmodel}
\end{figure*}

\begin{figure*}[htp]
    \centering
    \includegraphics[width=\textwidth]{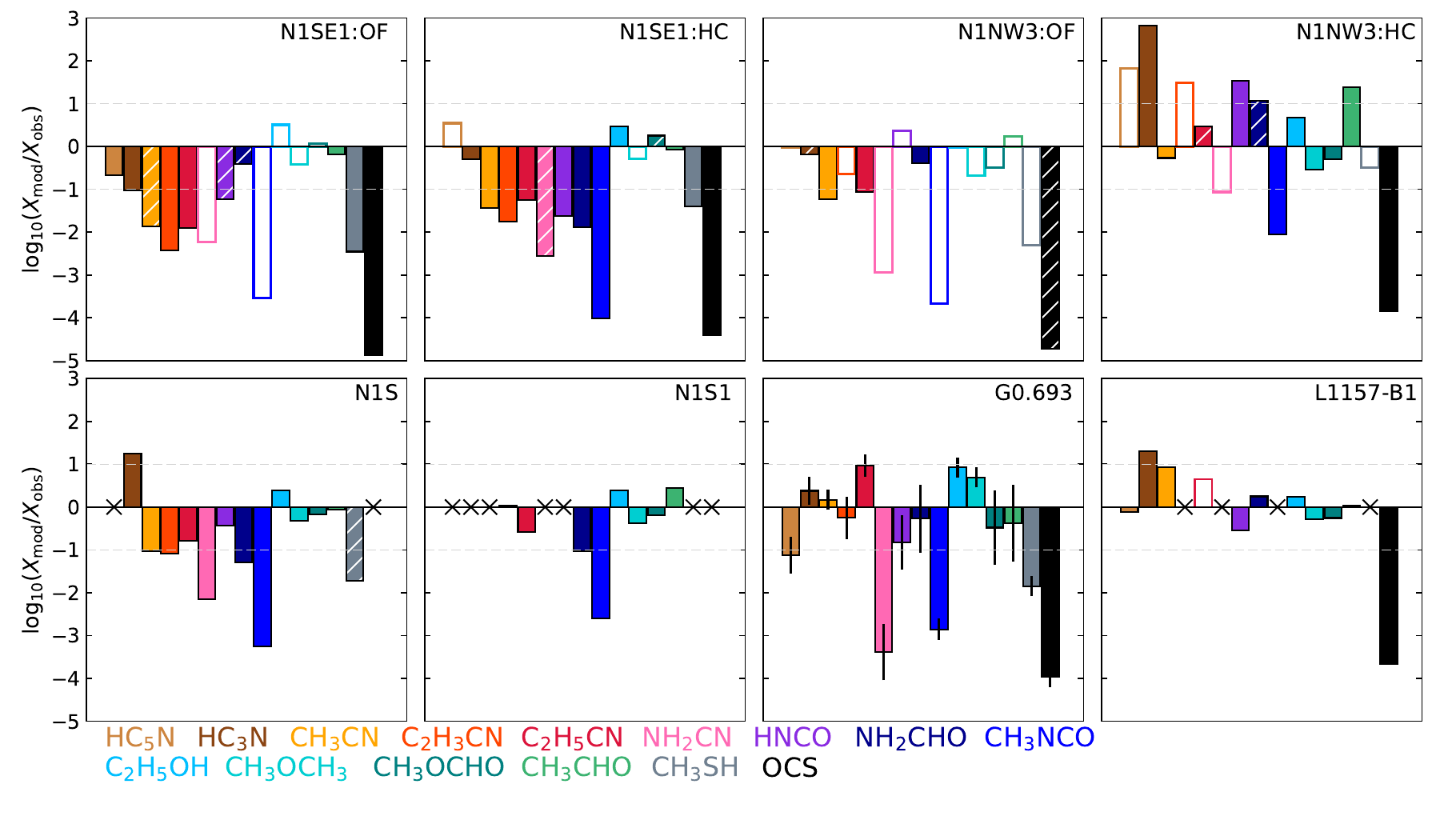}\\[-0.7cm]
    \caption{Comparison between modelled peak abundances with respect to methanol and the observed abundances towards both components at N1SE1 and N1NW3, N1S, N1S1, G0.693, and L1157-B1. The modelled abundances $X_{\rm mod}$ are taken from Table\,17 and 18 in \citet{Garrod22}, where we use the values of the slow warm-up model. Empty bars indicate lower limits. Hatched bars indicate the same as in Fig.\,\ref{fig:Xbars-norm}. Black crosses indicate molecules for which no values are available.}
    \label{fig:cfmodel_met}
\end{figure*}

In Sect.\,\ref{dss:modelsRTG} we introduced the astrochemical models conducted by \citet[][G22 hereafter]{Garrod22}, which have the goal to predict the chemistry in hot cores.
This section provides a detailed comparison between the modelled peak abundances and our observed abundances and a discussion on whether the underlying chemical network of each species used in the models can reproduce the observations or not.
In a similar way as in Fig.\,13 in Paper\,I and Fig.\,15 in G22, but using abundances with respect to H\2, we compare the modelled peak abundances with the observed ones in Fig.\,\ref{fig:cfmodel}, where we show the results for both components at positions N1SE1, for N1NW3:HC, for N1S and N1S1 (derived in Paper\,I), and those for G0.693 and L1157-B1 that were discussed in Sect.\,\ref{dss:sources}. 
The modelled abundances $X_{\rm mod}$ are taken from Table\,17 in G22, where we use the values of the slow warm-up (SWU hereafter) in Fig.\,\ref{fig:cfmodel}a, as we found these to best reproduce our observational results in Paper\,I, and the fast warm-up (FWU hereafter) in Fig.\,\ref{fig:cfmodel}b, as a shorter heating timescale may be closer to what we expect from a shock event. Black crosses indicate molecules for which no value is available. In Fig.\,\ref{fig:cfmodel_met} we additionally present the comparison to the model with SWU for abundances with respect to \met. However, we will focus the following discussion on the abundances with respect to H\2. 

\subsection{N1S and N1S1}\label{sss:n1s}

Abundances of most molecules at N1S and N1S1 have been derived and compared to the G22 model (assuming SWU) in Paper\,I. 
The observed abundance profiles of N- and (N+O)-bearing COMs showed either a plateau or an increasing behaviour at temperatures above 100\,K. In particular, the increase of abundances at high temperatures pointed to the COMs' formation partly or solely in the gas phase at these temperatures. 
The abundance of HC\3N at N1S was derived in this work and is overestimated in the model by factors of a few (FWU) or two orders of magnitude (SWU).
A better match with the model may be achieved if HC\3N was more efficiently converted to \vc and, eventually, \etc on dust grain surfaces. This was suggested as a way to increase the modelled abundances of these cyanides, because they were found to be underestimated by the model in Paper\,I. 
Methyl cyanide is predicted to be efficiently produced in the gas phase, mainly from the recombination of CH\3CNH$^+$ with an electron, where the protonated methyl cyanide is a product of the gas-phase reaction CH\3$^+$ + HCN. This seems plausible as the CH\3CN abundance, which was  previously derived at N1S \citep{Mueller21}, behaves similarly to \vc.
If we assume a FWU, for which CH\3CN abundances are largely underestimated, the production of the COM has to proceed more efficiently in the model or there are additional formation routes (likely in the gas phase) that are currently not considered. 
The value of NH\2CN column density at N1S was taken from an earlier ReMoCA result for N1S \citep{Kisiel22}. 
The molecule is predicted to be solely formed at earliest times during stage 1, that is when the ice mantles start to build. According to G22, a high observed abundance of this molecule suggests that photoprocessing of the ice mantles at these early times could have been significant. This formation at early times would explain the little difference between SWU and FWU.

The peak abundances for most O-bearing COMs were found at positions further away from Sgr\,B2\,(N1) than N1S in Paper\,I.
This can explain why the observed abundances at N1S are lower than the model predictions. In the FWU, there was less time to destroy \dme and \mf, explaining the better agreement with the model. A possible reason for the overproduction of \et was discussed in Paper\,I. 
Acetaldehyde is dominantly formed in the gas phase in the models. Therefore, the overproduction at N1S and N1S1 in the SWU models has to have another origin than for the other O-bearing species. Because the abundance is well reproduced for the FWU, maybe the heating timescale is the reason, which was also mentioned in Paper\,I.  
The chemical network for \mic is incomplete, which is likely the main reason for the heavy underestimation of its abundances not only at N1S and N1S1 but all positions shown in Fig.\,\ref{fig:cfmodel}. Formamide forms in both the solid and gas phases in the model, which was also indicated by the observed abundance profiles in Paper\,I. However, it also gets efficiently destroyed in the gas phase after its thermal desorption, which was not evident in the observed abundance profile. 
The abundance of HNCO was additionally derived in this work for N1S and seems to agree well with the modelled value. In the model, HNCO is a product of grain-surface chemistry and is efficiently photodissociated in the gas phase after desorption. This destruction pathway for HNCO cannot be inferred from our results, but, for example, it has been used as an explanation for observations towards Sgr\,B2\,(N) \citep[][]{Corbi2015}, where HNCO is depleted towards the H{\small II} regions.

Methyl mercaptan is purely a product of grain-surface chemistry and is formed mainly via H addition to CS during the collapse as well as the subsequent warm-up phase as can be seen from its panel in Fig.\,\ref{fig:G22}, which assumes a SWU. This kind of figure is an output of the G22 models and such plots were presented for a variety of molecules in Figs.\,13--14 in their article. These figures show qualitatively at what times (and temperature regime) of the pre- and protostellar evolution a molecule is dominantly formed (green) or destroyed (blue). The underestimation of \mmc abundances in the model may be the consequence of the fact that some input parameters related to sulphur chemistry are still uncertain in chemical models, for example the initial elemental abundance of S (G22) in general or, specifically for \mmc, the activation barriers for some of the reactions that eventually lead to the formation of the COM. 

\subsection{N1SE1}\label{dss:model_n1se1}

\begin{figure*}[hpt]
    \centering
    \includegraphics[width=.8\textwidth]{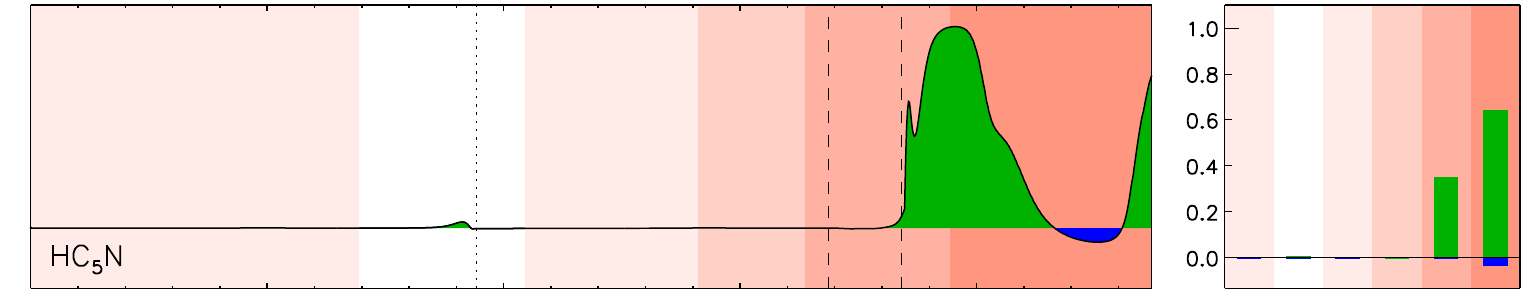}\\
    \includegraphics[width=.8\textwidth]{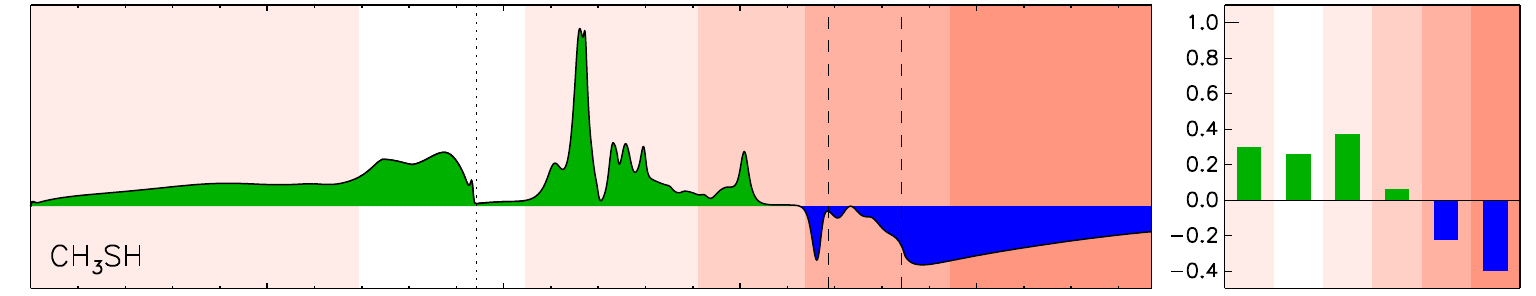}\\ 
    \includegraphics[width=.8\textwidth]{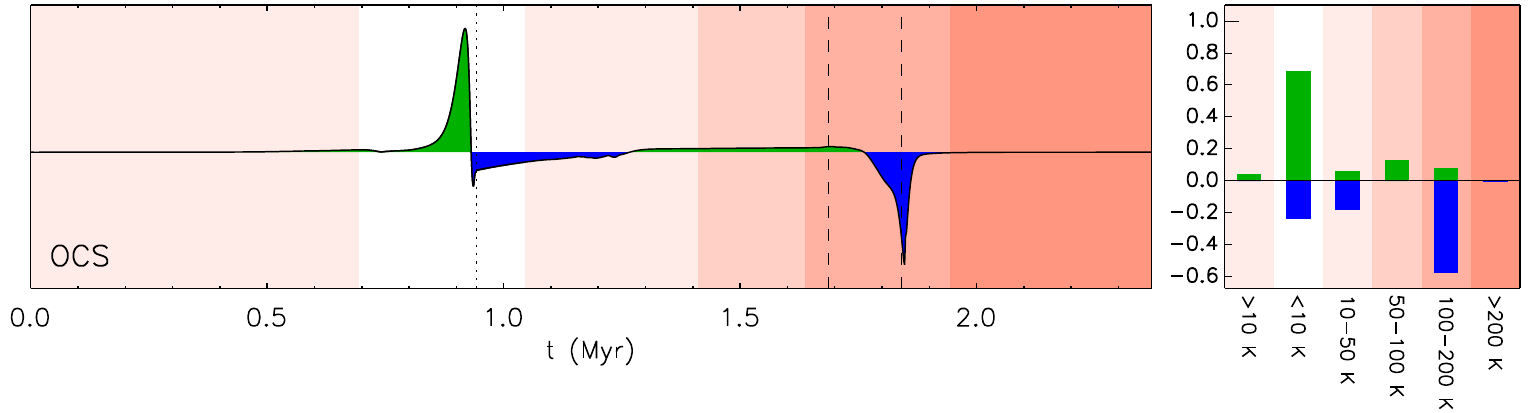}
    \caption{Same as Fig.\,13 in \citet{Garrod22}, but for HC\5N, \mmc, and OCS. It shows the net rate of change in the COM’s abundances, summed over all chemical phases, during the cold collapse phase and the subsequent slow warm-up. Green represents a net production of the molecule, blue a net destruction. The time axis is divided into several phases marked with different background colours that are associated with the temperature ranges labelled in the right panels. The dotted black line marks the start of the warm-up phase, the two dashed lines indicate the time of thermal desorption of water. }
    \label{fig:G22}
\end{figure*}

A more efficient conversion of HC\3N to \vc and \etc in the solid phase as proposed in Sect.\,\ref{sss:n1s} may help to better match the modelled and observed HC\3N abundances for N1SE1 in the case of the SWU (see Fig.\,\ref{fig:cfmodel}), however, it would lead to a larger discrepancy for the cyanides in this case and does not help at all the results for the FWU. 
The secure detection of HC\5N at N1SE1:OF may suggest that gas-phase chemistry plays an important role in this component. Not only this molecule but also HC\3N have both efficient formation pathways occurring at around 200\,K in the gas phase, where C\2H and HCN react to form HC\3N, which will again react with C\2H to form HC\5N. 
We show the rate of change of HC\5N abundance in the SWU in Fig.\,\ref{fig:G22}, where it is evident that there are two periods of efficient gas-phase production: at around 200\,K and at $\gtrsim$350\,K. At high temperatures during the second period, atomic N is sourced from NH\3 via H abstraction and reacts with hydrocarbons (e.g. C\5H\2), to form HC\5N. Our analysis did not reveal such high temperatures, however, a shock passage in the past could have caused a temporal temperature increase to 400\,K or more.
In the case of a FWU, there must be additional gas-phase formation routes for CH\3CN, \vc, HC\3N, and OCS to better match the observations and the models. A shock may be able to enhance the gas-phase chemistry or trigger additional reactions. 
This applies to both components at N1SE1.

As already seen for N1S and N1S1, \mic abundances are heavily underestimated in the model by multiple orders of magnitude, certainly at N1SE1:HC, likely mostly due to an incomplete chemical network. In Paper\,I we discussed additional formation routes that could potentially increase abundances in the model. 
Abundances of \fmm are underestimated in the model for N1S, N1S1, and N1SE1:HC, but it is overestimated for the outflow component in both the SWU and FWU. Despite predicted by the G22 models, there was no clear evidence for the destruction of \fmm at high temperatures in Paper\,I. Therefore we proposed that, for example, the COM's destruction should perhaps be less efficient in the model to better match the observed abundance. 
Abundances of HNCO behave similarly to those of \fmm at N1SE1, except that the over- or underproduction is not as severe. In the model, \fmm and HNCO are correlated in the sense that they are produced and destroyed at similar periods of time and temperature. A difference is however that HNCO is purely formed on dust grain surfaces. However, it can be a by-product of COM dissociation at high temperatures in the gas phase \citep{Garrod08,Tideswell08}. 

The overall larger overestimation of abundances of O-bearing COMs in the model suggests an enhanced destruction of these in N1SE1:HC and even more so in N1SE1:OF in the gas phase after their desorption from the dust grains. Again, this explanation does not apply to \ad as it is primarily a gas-phase product. Moreover, given that HNCO and NH\2CN are products of grain-surface chemistry and are efficiently destroyed in the gas phase in the model, we would expect them to behave more similarly to the O-bearing COMs. This suggests that either their destruction proceeds less efficiently or they have additional formation routes likely in the gas phase. 

As mentioned in Sect.\,\ref{sss:n1s}, the predictions for \mmc may have higher uncertainties due to unknowns in the sulphur chemistry in general and the chemistry of the COM in particular. 
The underestimation of the OCS abundances by the model is more severe. We show the rate of change for the SWU of the molecule's abundance during the protostellar evolution in Fig.\,\ref{fig:G22}. The molecule is mainly formed towards the end of the collapse phase, where the reaction S + CO produces OCS on the dust grain surfaces. In the current model, the activation barrier for this reaction is set to zero, while this assumption may not actually be correct \citep[][]{Loison12}. A non-zero activation barrier would however worsen the difference between the model and the observations. There  are a few reactions that produce OCS in the gas phase in the model. However, these are currently not able to reproduce the observed abundance. 

\subsection{N1NW3:HC}

Molecular abundances behave quite differently in the hot-core component at this position than at N1S or N1SE1.  Rotational temperatures at this position are between 80--150\,K (see Fig.\,\ref{fig:corner_T}), which is when we expect thermal desorption to proceed efficiently (see Paper\,I). This would explain the better agreement between the model and the observations for the O-bearing molecules in Fig.\,\ref{fig:cfmodel} and the result for NH\2CN, while gas-phase products such as \ad, \vc, and cyanopolyynes are overproduced by the model, especially for the SWU. This scenario does not explain the values for CH\3CN, which should behave similarly to, for example, \vc, and HNCO, which we would expect to behave similar to the other products of grain-surface chemistry, according to the model.

\subsection{G0.693 and L1157-B1}

Besides a few exceptions, abundances with respect to H\2 are all overproduced by the SWU and FWU models as can be seen in Fig.\,\ref{fig:cfmodel}, which may not be surprising, given that the model is designated to predict the chemistry of hot cores that are known to possess higher abundances of especially COMs than prestellar cores or low-mass protostars. One of the exceptions is NH\2CN, whose abundance is well predicted by the model in G0.693. This suggests that the molecule's formation at early times of stage 1 in the model may indeed be the dominant formation pathway. The other exceptions are OCS and \mic, for which we know that the chemical network is likely not complete.

\end{appendix}

%% file: tables/tab_n1se1hc.tex
\begin{table*}[htp!]
\caption{Weeds parameters used at N1SE1:HC and results obtained from the population diagrams.} 
\centering
\begin{tabular}{rrrcrrcrrrrr}
\hline\hline \\[-0.3cm] 
Molecule\tablefootmark{a} & Size\tablefootmark{b} & $T_{\rm rot,W}$\tablefootmark{c} & $T_{\rm rot,pd}$\tablefootmark{d} & $N_{\rm W}\tablefootmark{e}$ & $C_{\rm vib,W}$\tablefootmark{f} & $N_{\rm pd}\tablefootmark{g}$ & $C_{\rm vib,pd}$\tablefootmark{h} & $\Delta\varv$\tablefootmark{i} & $\varv_{\rm off}$\tablefootmark{j} \\ 
 & $(^{\prime\prime})$ & (K) & (K) & (cm$^{-2}$) & & (cm$^{-2}$) &  & (km\,s$^{-1}$) & (km\,s$^{-1}$) \\\hline \\[-0.3cm] 
HC$_5$N, $v=0$ & $2.0$ & $180$ &   -- & $\leq3.2(14)$ & $4.56$ &  -- &   -- & $4.6$ & $1.5$ \\ 
 \hline \\[-0.35cm] 
HC$_3$N, $v=0$ & $2.0$ & $190$ & $198\pm 12$ & $1.9(17)$ & $1.60$ & $(1.6\pm 0.6)(17)$ & $1.67$ & $5.5$ & $1.0$ \\ 
 $v_{7}=1$ & $2.0$ & $190$ & $198\pm 12$ & $1.9(17)$ & $1.60$ & $(1.6\pm 0.6)(17)$ & $1.67$ & $5.5$ & $1.0$ \\ 
 $v_{7}=2$ & $2.0$ & $190$ & $198\pm 12$ & $1.9(17)$ & $1.60$ & $(1.6\pm 0.6)(17)$ & $1.67$ & $5.5$ & $1.0$ \\ 
 $v_{6}=1$ & $2.0$ & $190$ & $198\pm 12$ & $1.9(17)$ & $1.60$ & $(1.6\pm 0.6)(17)$ & $1.67$ & $5.5$ & $1.0$ \\ 
 $v_{6}=2$ & $2.0$ & $190$ & $198\pm 12$ & $1.9(17)$ & $1.60$ & $(1.6\pm 0.6)(17)$ & $1.67$ & $5.5$ & $1.0$ \\ 
 $v_{4}=1$ & $2.0$ & $190$ & $198\pm 12$ & $1.9(17)$ & $1.60$ & $(1.6\pm 0.6)(17)$ & $1.67$ & $5.5$ & $1.0$ \\ 
 $v_{6}=1,v_{7}=1$ & $2.0$ & $190$ & $198\pm 12$ & $1.9(17)$ & $1.60$ & $(1.6\pm 0.6)(17)$ & $1.67$ & $6.0$ & $1.2$ \\ 
 $v_{6}=1,v_{7}=2$ & $2.0$ & $190$ & $198\pm 12$ & $1.9(17)$ & $1.60$ & $(1.6\pm 0.6)(17)$ & $1.67$ & $5.5$ & $1.0$ \\ 
 \hline \\[-0.35cm] 
CH$_3$CN, $v=0$ & $2.0$ & $220$ & $219\pm 11$ & $2.8(17)$ & $1.00$ & $(3.0\pm 0.6)(17)$ & $1.00$ & $5.5$ & $1.6$ \\ 
 $v_{8}=1$ & $2.0$ & $220$ & $219\pm 11$ & $2.8(17)$ & $1.00$ & $(3.0\pm 0.6)(17)$ & $1.00$ & $5.5$ & $1.6$ \\ 
 $v_{8}=2$ & $2.0$ & $220$ & $219\pm 11$ & $2.8(17)$ & $1.00$ & $(3.0\pm 0.6)(17)$ & $1.00$ & $5.5$ & $1.6$ \\ 
 \hline \\[-0.35cm] 
C$_2$H$_3$CN, $v=0$ & $2.0$ & $250$ & $252\pm 8$ & $1.6(17)$ & $1.09$ & $(1.5\pm 0.1)(17)$ & $1.09$ & $5.5$ & $1.0$ \\ 
 $v_{11}=1$ & $2.0$ & $250$ & $252\pm 8$ & $1.6(17)$ & $1.09$ & $(1.5\pm 0.1)(17)$ & $1.09$ & $5.5$ & $1.0$ \\ 
 $v_{15}=1$ & $2.0$ & $250$ & $252\pm 8$ & $1.6(17)$ & $1.09$ & $(1.5\pm 0.1)(17)$ & $1.09$ & $5.5$ & $1.0$ \\ 
 $v_{11}=2$ & $2.0$ & $250$ & $252\pm 8$ & $1.6(17)$ & $1.09$ & $(1.5\pm 0.1)(17)$ & $1.09$ & $5.5$ & $1.0$ \\ 
 \hline \\[-0.35cm] 
C$_2$H$_5$CN, $v=0$ & $2.0$ & $190$ & $187\pm 4$ & $3.8(17)$ & $1.73$ & $(4.4\pm 0.3)(17)$ & $1.70$ & $5.0$ & $1.0$ \\ 
 $v_{20}=1$ & $2.0$ & $190$ & $187\pm 4$ & $3.8(17)$ & $1.73$ & $(4.4\pm 0.3)(17)$ & $1.70$ & $5.0$ & $1.0$ \\ 
 $v_{13}=2$ & $2.0$ & $190$ & $187\pm 4$ & $3.8(17)$ & $1.73$ & $(4.4\pm 0.3)(17)$ & $1.70$ & $5.0$ & $1.0$ \\ 
 $v_{21}=2$ & $2.0$ & $190$ & $187\pm 4$ & $3.8(17)$ & $1.73$ & $(4.4\pm 0.3)(17)$ & $1.70$ & $5.0$ & $1.0$ \\ 
 $v_{12}=1$ & $2.0$ & $190$ & $187\pm 4$ & $3.8(17)$ & $1.73$ & $(4.4\pm 0.3)(17)$ & $1.70$ & $5.0$ & $1.0$ \\ 
 \hline \\[-0.35cm] 
NH$_2$CN, $v=0$ & $2.0$ & $160$ & $\leq145$ & $2.6(15)$ & $1.03$ & $(2.5\pm 0.7)(15)$ & $1.02$ & $4.6$ & $2.2$ \\ 
 \hline \\[-0.35cm] 
HNCO, $v=0$ & $2.0$ & $180$ & $184\pm 10$ & $1.2(18)$ & $1.02$ & $(1.0\pm 0.3)(18)$ & $1.02$ & $4.5$ & $2.5$ \\ 
 $v_{5}=1$ & $2.0$ & $180$ & $184\pm 10$ & $1.2(18)$ & $1.02$ & $(1.0\pm 0.3)(18)$ & $1.02$ & $4.5$ & $2.5$ \\ 
 $v_{6}=1$ & $2.0$ & $180$ & $184\pm 10$ & $1.2(18)$ & $1.02$ & $(1.0\pm 0.3)(18)$ & $1.02$ & $4.5$ & $2.5$ \\ 
 \hline \\[-0.35cm] 
NH$_2$CHO, $v=0$ & $2.0$ & $160$ & $157\pm 5$ & $6.0(17)$ & $1.09$ & $(5.7\pm 0.5)(17)$ & $1.08$ & $5.5$ & $1.5$ \\ 
 $v_{12}=1$ & $2.0$ & $160$ & $157\pm 5$ & $6.0(17)$ & $1.09$ & $(5.7\pm 0.5)(17)$ & $1.08$ & $5.5$ & $1.5$ \\ 
 \hline \\[-0.35cm] 
CH$_3$NCO, $v=0$ & $2.0$ & $200$ & $206\pm 20$ & $5.5(16)$ & $1.00$ & $(5.8\pm 1.3)(16)$ & $1.00$ & $4.0$ & $2.3$ \\ 
 $v=1$ & $2.0$ & $200$ & $206\pm 20$ & $5.5(16)$ & $1.00$ & $(5.8\pm 1.3)(16)$ & $1.00$ & $4.0$ & $2.3$ \\ 
 \hline \\[-0.35cm] 
CH$_3$OH, $v=0$ & $2.0$ & $170$ & $175\pm 5$ & $1.2(18)$ & $1.00$ & $(1.1\pm 0.1)(18)$ & $1.00$ & $4.5$ & $2.2$ \\ 
 $v=1$ & $2.0$ & $170$ & $175\pm 5$ & $1.2(18)$ & $1.00$ & $(1.1\pm 0.1)(18)$ & $1.00$ & $4.5$ & $2.2$ \\ 
 $v=2$ & $2.0$ & $170$ & $175\pm 5$ & $1.2(18)$ & $1.00$ & $(1.1\pm 0.1)(18)$ & $1.00$ & $4.5$ & $2.2$ \\ 
 \hline \\[-0.35cm] 
C$_2$H$_5$OH, $v=0$ & $2.0$ & $160$ & $162\pm 10$ & $4.4(16)$ & $1.38$ & $(4.9\pm 0.8)(16)$ & $1.39$ & $3.7$ & $1.7$ \\ 
 \hline \\[-0.35cm] 
CH$_3$OCH$_3$, $v=0$ & $2.0$ & $180$ &   -- & $\leq4.1(16)$ & $1.03$ &  -- &   -- & $4.6$ & $1.5$ \\ 
 \hline \\[-0.35cm] 
CH$_3$OCHO, $v=0$ & $2.0$ & $115$ & $\leq116$ & $2.2(16)$ & $1.08$ & $(2.2\pm 0.5)(16)$ & $1.08$ & $3.0$ & $2.8$ \\ 
 \hline \\[-0.35cm] 
CH$_3$CHO, $v=0$ & $2.0$ & $170$ & $170\pm 13$ & $3.6(16)$ & $1.08$ & $(1.9\pm 0.3)(16)$ & $1.02$ & $4.5$ & $1.2$ \\ 
 $v=1$ & $2.0$ & $170$ & $170\pm 13$ & $3.6(16)$ & $1.08$ & $(1.9\pm 0.3)(16)$ & $1.02$ & $4.5$ & $1.2$ \\ 
 \hline \\[-0.35cm] 
CH$_3$SH, $v=0$ & $2.0$ & $115$ & $115\pm 41$ & $1.4(16)$ & $1.08$ & $(1.1\pm 0.7)(16)$ & $1.00$ & $4.4$ & $-0.5$ \\ 
 \hline \\[-0.35cm] 
OCS, $v=0$ & $2.0$ & $215$ & $216\pm 11$ & $3.2(17)$ & $1.07$ & $(3.1\pm 0.4)(17)$ & $1.07$ & $6.0$ & $1.5$ \\ 
 $v_{2}=1$ & $2.0$ & $215$ & $216\pm 11$ & $3.2(17)$ & $1.07$ & $(3.1\pm 0.4)(17)$ & $1.07$ & $6.0$ & $1.5$ \\ 
 \hline\hline
\end{tabular}
\tablefoot{\tablefoottext{a}{COMs and vibrational states used to derive population diagrams. }\tablefoottext{b}{Assumed size of the emitting region.}\tablefoottext{c}{Rotational temperature used for the Weeds model.}\tablefoottext{d}{Rotational temperature derived from the population diagram.}\tablefoottext{e}{Column density used for the Weeds model.}\tablefoottext{f}{Vibrational state correction at temperature $T_{\rm rot,W}$ applied to the column density $N_{\rm W}$ when the partition function does not account for higher-excited vibrational states, where $C_{\rm vib,W}=C_{\rm vib,W}(T_{\rm rot,W})$.}\tablefoottext{g}{Column density derived from the population diagram.}\tablefoottext{h}{Vibrational state correction at temperature $T_{\rm rot,pd}$ applied to $N_{\rm pd}$, where $C_{\rm vib,pd}=C_{\rm vib,pd}(T_{\rm rot,pd})$.}\tablefoottext{i}{Linewidths ($FWHM$) of the transitions.}\tablefoottext{j}{Offset from the source systemic velocity, which was set to 62\kms.}\\ Values in parentheses show the decimal power, where $x(z) = x\times 10^z$ or $(x\pm y)(z) = (x\pm y)\times 10^z$. \\  Upper limits on $N_{\rm W}$ indicate that a population diagram could not be derived, either because too many transitions are contaminated or the molecule is not detected. Then, the temperature, $FWHM$, and offset velocity values used in Weeds correspond to medians derived from other molecules at this position. Upper limits on $T_{\rm rot,pd}$ indicate when upper limits on integrated intensity for non-detected transitions are used in the population diagrams.} 
\label{tab:n1se1hc}
\end{table*}

%% file: tables/tab_n1se1of.tex
\begin{table*}[htp!]
\caption{Same as Table\,\ref{tab:n1se1hc}, but for position N1SE1:OF.} 
\centering
\begin{tabular}{rrrcrrcrrrrr}
\hline\hline \\[-0.3cm] 
Molecule & Size & $T_{\rm rot,W}$ & $T_{\rm rot,pd}$ & $N_{\rm W}$ & $C_{\rm vib,W}$ & $N_{\rm pd}$ & $C_{\rm vib,pd}$ & $\Delta\varv$ & $\varv_{\rm off}$ \\ 
 & $(^{\prime\prime})$ & (K) & (K) & (cm$^{-2}$) & & (cm$^{-2}$) &  & (km\,s$^{-1}$) & (km\,s$^{-1}$) \\\hline \\[-0.3cm] 
HC$_5$N, $v=0$ & $2.0$ & $160$ & $158\pm 16$ & $1.8(15)$ & $3.53$ & $(2.2\pm 0.4)(15)$ & $3.44$ & $4.5$ & $-7.3$ \\ 
 $v_{11}=1$ & $2.0$ & $160$ & $158\pm 16$ & $1.8(15)$ & $3.53$ & $(2.2\pm 0.4)(15)$ & $3.44$ & $4.5$ & $-7.3$ \\ 
 $v_{11}=2$ & $2.0$ & $160$ & $158\pm 16$ & $1.8(15)$ & $3.53$ & $(2.2\pm 0.4)(15)$ & $3.44$ & $4.5$ & $-7.3$ \\ 
 \hline \\[-0.35cm] 
HC$_3$N, $v=0$ & $2.0$ & $200$ & $207\pm 6$ & $3.4(17)$ & $1.69$ & $(3.3\pm 0.5)(17)$ & $1.76$ & $6.5$ & $-7.0$ \\ 
 $v_{7}=1$ & $2.0$ & $200$ & $207\pm 6$ & $3.4(17)$ & $1.69$ & $(3.3\pm 0.5)(17)$ & $1.76$ & $6.5$ & $-7.0$ \\ 
 $v_{7}=2$ & $2.0$ & $200$ & $207\pm 6$ & $3.4(17)$ & $1.69$ & $(3.3\pm 0.5)(17)$ & $1.76$ & $6.5$ & $-7.0$ \\ 
 $v_{6}=1$ & $2.0$ & $200$ & $207\pm 6$ & $3.4(17)$ & $1.69$ & $(3.3\pm 0.5)(17)$ & $1.76$ & $6.5$ & $-7.0$ \\ 
 $v_{6}=2$ & $2.0$ & $200$ & $207\pm 6$ & $3.4(17)$ & $1.69$ & $(3.3\pm 0.5)(17)$ & $1.76$ & $6.5$ & $-7.0$ \\ 
 $v_{4}=1$ & $2.0$ & $200$ & $207\pm 6$ & $3.4(17)$ & $1.69$ & $(3.3\pm 0.5)(17)$ & $1.76$ & $6.5$ & $-7.0$ \\ 
 $v_{6}=1,v_{7}=1$ & $2.0$ & $200$ & $207\pm 6$ & $3.4(17)$ & $1.69$ & $(3.3\pm 0.5)(17)$ & $1.76$ & $6.5$ & $-7.0$ \\ 
 $v_{6}=1,v_{7}=2$ & $2.0$ & $200$ & $207\pm 6$ & $3.4(17)$ & $1.69$ & $(3.3\pm 0.5)(17)$ & $1.76$ & $6.5$ & $-7.0$ \\ 
 \hline \\[-0.35cm] 
CH$_3$CN, $v=0$ & $2.0$ & $160$ & $154\pm 0$ & $2.5(17)$ & $1.00$ & $(2.8\pm 0.5)(17)$ & $1.00$ & $6.5$ & $-7.5$ \\ 
 $v_{8}=1$ & $2.0$ & $160$ & $154\pm 0$ & $2.5(17)$ & $1.00$ & $(2.8\pm 0.5)(17)$ & $1.00$ & $6.5$ & $-7.5$ \\ 
 $v_{8}=2$ & $2.0$ & $160$ & $154\pm 0$ & $2.5(17)$ & $1.00$ & $(2.8\pm 0.5)(17)$ & $1.00$ & $6.5$ & $-7.5$ \\ 
 \hline \\[-0.35cm] 
C$_2$H$_3$CN, $v=0$ & $2.0$ & $180$ & $185\pm 4$ & $2.6(17)$ & $1.02$ & $(2.2\pm 0.2)(17)$ & $1.02$ & $5.8$ & $-6.5$ \\ 
 $v_{11}=1$ & $2.0$ & $180$ & $185\pm 4$ & $2.6(17)$ & $1.02$ & $(2.2\pm 0.2)(17)$ & $1.02$ & $5.8$ & $-6.5$ \\ 
 $v_{15}=1$ & $2.0$ & $180$ & $185\pm 4$ & $2.6(17)$ & $1.02$ & $(2.2\pm 0.2)(17)$ & $1.02$ & $5.8$ & $-6.5$ \\ 
 $v_{11}=2$ & $2.0$ & $180$ & $185\pm 4$ & $2.6(17)$ & $1.02$ & $(2.2\pm 0.2)(17)$ & $1.02$ & $5.8$ & $-6.5$ \\ 
 \hline \\[-0.35cm] 
C$_2$H$_5$CN, $v=0$ & $2.0$ & $160$ & $160\pm 2$ & $4.1(17)$ & $1.02$ & $(6.7\pm 0.3)(17)$ & $1.46$ & $6.5$ & $-7.5$ \\ 
 $v_{20}=1$ & $2.0$ & $160$ & $160\pm 2$ & $4.1(17)$ & $1.02$ & $(6.7\pm 0.3)(17)$ & $1.46$ & $6.5$ & $-7.5$ \\ 
 $v_{13}=2$ & $2.0$ & $160$ & $160\pm 2$ & $4.1(17)$ & $1.02$ & $(6.7\pm 0.3)(17)$ & $1.46$ & $6.5$ & $-7.5$ \\ 
 $v_{21}=2$ & $2.0$ & $160$ & $160\pm 2$ & $4.1(17)$ & $1.02$ & $(6.7\pm 0.3)(17)$ & $1.46$ & $6.5$ & $-7.5$ \\ 
 $v_{12}=1$ & $2.0$ & $160$ & $160\pm 2$ & $4.1(17)$ & $1.02$ & $(6.7\pm 0.3)(17)$ & $1.46$ & $6.5$ & $-7.5$ \\ 
 \hline \\[-0.35cm] 
NH$_2$CN, $v=0$ & $2.0$ & $160$ &   -- & $\leq4.1(14)$ & $1.03$ &  -- &   -- & $6.5$ & $-7.5$ \\ 
 \hline \\[-0.35cm] 
HNCO, $v=0$ & $2.0$ & $135$ & $\leq133$ & $1.6(17)$ & $1.00$ & $(1.6\pm 0.2)(17)$ & $1.00$ & $6.5$ & $-7.5$ \\ 
 \hline \\[-0.35cm] 
$^*$NH$_2$CHO, $v=0$ & $2.0$ & $160$ &   -- & $6.5(15)$ & $1.09$ &   -- &   -- & $6.0$ & $-6.5$ \\ 
 \hline \\[-0.35cm] 
CH$_3$NCO, $v=0$ & $2.0$ & $160$ &   -- & $\leq6.0(15)$ & $1.00$ &  -- &   -- & $6.5$ & $-7.5$ \\ 
 \hline \\[-0.35cm] 
CH$_3$OH, $v=0$ & $2.0$ & $160$ & $146\pm 11$ & $4.0(17)$ & $1.00$ & $(3.5\pm 0.8)(17)$ & $1.00$ & $5.2$ & $-9.8$ \\ 
 $v=1$ & $2.0$ & $160$ & $146\pm 11$ & $4.0(17)$ & $1.00$ & $(3.5\pm 0.8)(17)$ & $1.00$ & $5.2$ & $-9.8$ \\ 
 \hline \\[-0.35cm] 
C$_2$H$_5$OH, $v=0$ & $2.0$ & $160$ &   -- & $\leq1.4(16)$ & $1.38$ &  -- &   -- & $6.3$ & $-7.5$ \\ 
 \hline \\[-0.35cm] 
CH$_3$OCH$_3$, $v=0$ & $2.0$ & $160$ &   -- & $\leq2.0(16)$ & $1.02$ &  -- &   -- & $6.35$ & $-7.5$ \\ 
 \hline \\[-0.35cm] 
CH$_3$OCHO, $v=0$ & $2.0$ & $160$ &   -- & $\leq1.2(16)$ & $1.24$ &  -- &   -- & $6.5$ & $-7.5$ \\ 
 \hline \\[-0.35cm] 
CH$_3$CHO, $v=0$ & $2.0$ & $160$ & $169\pm 17$ & $1.5(16)$ & $1.01$ & $(8.7\pm 2.0)(15)$ & $1.02$ & $5.0$ & $-9.3$ \\ 
 $v=1$ & $2.0$ & $160$ & $169\pm 17$ & $1.5(16)$ & $1.01$ & $(8.7\pm 2.0)(15)$ & $1.02$ & $5.0$ & $-9.3$ \\ 
 \hline \\[-0.35cm] 
CH$_3$SH, $v=0$ & $2.0$ & $125$ & $124\pm 25$ & $5.0(16)$ & $1.00$ & $(4.8\pm 1.8)(16)$ & $1.00$ & $6.0$ & $-7.8$ \\ 
 \hline \\[-0.35cm] 
OCS, $v=0$ & $2.0$ & $170$ & $175\pm 8$ & $3.1(17)$ & $1.03$ & $(3.8\pm 0.5)(17)$ & $1.03$ & $6.5$ & $-7.5$ \\ 
 $v_{2}=1$ & $2.0$ & $170$ & $175\pm 8$ & $3.1(17)$ & $1.03$ & $(3.8\pm 0.5)(17)$ & $1.03$ & $6.5$ & $-7.5$ \\ 
 \hline\hline
\end{tabular}
\tablefoot{\tablefoottext{*}{The molecule is detected but a population diagram could not be derived due to contamination by the hot-core component.}} 
\label{tab:n1se1of}
\end{table*}

%% file: tables/tab_n1nw3hc.tex
\begin{table*}[htp!]
\caption{Same as Table\,\ref{tab:n1se1hc}, but for position N1NW3:HC.} 
\centering
\begin{tabular}{rrrcrrcrrrrr}
\hline\hline \\[-0.3cm] 
Molecule & Size & $T_{\rm rot,W}$ & $T_{\rm rot,pd}$ & $N_{\rm W}$ & $C_{\rm vib,W}$ & $N_{\rm pd}$ & $C_{\rm vib,pd}$ & $\Delta\varv$ & $\varv_{\rm off}$ \\ 
 & $(^{\prime\prime})$ & (K) & (K) & (cm$^{-2}$) & & (cm$^{-2}$) &  & (km\,s$^{-1}$) & (km\,s$^{-1}$) \\\hline \\[-0.3cm] 
HC$_5$N, $v=0$ & $2.0$ & $128$ &   -- & $\leq7.2(13)$ & $2.39$ &  -- &   -- & $4.3$ & $2.5$ \\ 
 \hline \\[-0.35cm] 
HC$_3$N, $v=0$ & $2.0$ & $140$ & $143\pm 24$ & $6.3(14)$ & $1.26$ & $(7.5\pm 2.5)(14)$ & $1.27$ & $3.3$ & $2.5$ \\ 
 $v_{7}=1$ & $2.0$ & $140$ & $143\pm 24$ & $6.3(14)$ & $1.26$ & $(7.5\pm 2.5)(14)$ & $1.27$ & $3.3$ & $2.5$ \\ 
 \hline \\[-0.35cm] 
CH$_3$CN, $v=0$ & $2.0$ & $130$ & $126\pm 6$ & $8.0(16)$ & $1.00$ & $(8.9\pm 2.2)(16)$ & $1.00$ & $4.3$ & $2.5$ \\ 
 $v_{8}=1$ & $2.0$ & $130$ & $126\pm 6$ & $8.0(16)$ & $1.00$ & $(8.9\pm 2.2)(16)$ & $1.00$ & $4.3$ & $2.5$ \\ 
 \hline \\[-0.35cm] 
C$_2$H$_3$CN, $v=0$ & $2.0$ & $128$ &   -- & $\leq4.0(14)$ & $1.00$ &  -- &   -- & $4.3$ & $2.5$ \\ 
 \hline \\[-0.35cm] 
C$_2$H$_5$CN, $v=0$ & $2.0$ & $125$ & $\leq122$ & $3.1(16)$ & $1.23$ & $(2.9\pm 0.2)(16)$ & $1.21$ & $5.7$ & $2.4$ \\ 
 \hline \\[-0.35cm] 
NH$_2$CN, $v=0$ & $2.0$ & $128$ &   -- & $\leq3.5(14)$ & $1.01$ &  -- &   -- & $4.3$ & $2.5$ \\ 
 \hline \\[-0.35cm] 
HNCO, $v=0$ & $2.0$ & $130$ & $120\pm 5$ & $3.5(15)$ & $1.00$ & $(3.2\pm 0.3)(15)$ & $1.00$ & $5.0$ & $2.8$ \\ 
 \hline \\[-0.35cm] 
NH$_2$CHO, $v=0$ & $2.0$ & $125$ & $\leq125$ & $2.7(15)$ & $1.00$ & $(3.4\pm 0.3)(15)$ & $1.04$ & $4.3$ & $2.7$ \\ 
 \hline \\[-0.35cm] 
CH$_3$NCO, $v=0$ & $2.0$ & $85$ & $84\pm 7$ & $2.5(15)$ & $1.00$ & $(2.5\pm 0.6)(15)$ & $1.00$ & $4.0$ & $2.3$ \\ 
 \hline \\[-0.35cm] 
CH$_3$OH, $v=0$ & $2.0$ & $145$ & $147\pm 5$ & $5.0(18)$ & $1.00$ & $(4.1\pm 0.6)(18)$ & $1.00$ & $5.7$ & $2.0$ \\ 
 $v=1$ & $2.0$ & $145$ & $147\pm 5$ & $5.0(18)$ & $1.00$ & $(4.1\pm 0.6)(18)$ & $1.00$ & $5.7$ & $2.0$ \\ 
 $v=2$ & $2.0$ & $145$ & $147\pm 5$ & $5.0(18)$ & $1.00$ & $(4.1\pm 0.6)(18)$ & $1.00$ & $5.7$ & $2.0$ \\ 
 \hline \\[-0.35cm] 
C$_2$H$_5$OH, $v=0$ & $2.0$ & $145$ & $153\pm 6$ & $1.1(17)$ & $1.29$ & $(1.2\pm 0.1)(17)$ & $1.33$ & $4.0$ & $2.6$ \\ 
 \hline \\[-0.35cm] 
CH$_3$OCH$_3$, $v=0$ & $2.0$ & $125$ & $121\pm 4$ & $3.8(17)$ & $1.00$ & $(3.8\pm 0.4)(17)$ & $1.00$ & $4.2$ & $1.4$ \\ 
 $v_{11}=1$ & $2.0$ & $125$ & $121\pm 4$ & $3.8(17)$ & $1.00$ & $(3.8\pm 0.4)(17)$ & $1.00$ & $4.2$ & $1.4$ \\ 
 \hline \\[-0.35cm] 
CH$_3$OCHO, $v=0$ & $2.0$ & $140$ & $150\pm 4$ & $3.5(17)$ & $1.16$ & $(4.5\pm 0.4)(17)$ & $1.19$ & $4.2$ & $1.4$ \\ 
 $v=1$ & $2.0$ & $140$ & $150\pm 4$ & $3.5(17)$ & $1.16$ & $(4.5\pm 0.4)(17)$ & $1.19$ & $4.2$ & $1.4$ \\ 
 \hline \\[-0.35cm] 
CH$_3$CHO, $v=0$ & $2.0$ & $90$ & $79\pm 4$ & $6.0(15)$ & $1.00$ & $(3.1\pm 0.3)(15)$ & $1.00$ & $4.5$ & $2.8$ \\ 
 $v=1$ & $2.0$ & $90$ & $79\pm 4$ & $6.0(15)$ & $1.00$ & $(3.1\pm 0.3)(15)$ & $1.00$ & $4.5$ & $2.8$ \\ 
 \hline \\[-0.35cm] 
CH$_3$SH, $v=0$ & $2.0$ & $128$ &   -- & $\leq7.0(15)$ & $1.00$ &  -- &   -- & $4.3$ & $2.5$ \\ 
 \hline \\[-0.35cm] 
OCS, $v=0$ & $2.0$ & $140$ & $141\pm 5$ & $3.5(17)$ & $1.01$ & $(3.2\pm 0.6)(17)$ & $1.01$ & $4.5$ & $2.4$ \\ 
 $v_{2}=1$ & $2.0$ & $140$ & $141\pm 5$ & $3.5(17)$ & $1.01$ & $(3.2\pm 0.6)(17)$ & $1.01$ & $4.5$ & $2.4$ \\ 
 \hline\hline
\end{tabular}
\label{tab:n1nw3hc}
\end{table*}

%% file: tables/tab_n1nw3of.tex
\begin{table*}[htp!]
\caption{Same as Table\,\ref{tab:n1se1hc}, but for position N1NW3:OF.} 
\centering
\begin{tabular}{rrrcrrcrrrrr}
\hline\hline \\[-0.3cm] 
Molecule & Size & $T_{\rm rot,W}$ & $T_{\rm rot,pd}$ & $N_{\rm W}$ & $C_{\rm vib,W}$ & $N_{\rm pd}$ & $C_{\rm vib,pd}$ & $\Delta\varv$ & $\varv_{\rm off}$ \\ 
 & $(^{\prime\prime})$ & (K) & (K) & (cm$^{-2}$) & & (cm$^{-2}$) &  & (km\,s$^{-1}$) & (km\,s$^{-1}$) \\\hline \\[-0.3cm] 
HC$_5$N, $v=0$ & $2.0$ & $129$ &   -- & $\leq9.7(13)$ & $2.42$ &  -- &   -- & $6.0$ & $12.0$ \\ 
 \hline \\[-0.35cm] 
HC$_3$N, $v=0$ & $2.0$ & $95$ & $93\pm 0$ & $1.2(16)$ & $1.07$ & $(1.3\pm 0.1)(16)$ & $1.07$ & $6.0$ & $13.0$ \\ 
 \hline \\[-0.35cm] 
CH$_3$CN, $v=0$ & $2.0$ & $145$ & $141\pm 8$ & $1.4(16)$ & $1.00$ & $(1.4\pm 0.3)(16)$ & $1.00$ & $6.0$ & $12.0$ \\ 
 $v_{8}=1$ & $2.0$ & $145$ & $141\pm 8$ & $1.4(16)$ & $1.00$ & $(1.4\pm 0.3)(16)$ & $1.00$ & $6.0$ & $12.0$ \\ 
 \hline \\[-0.35cm] 
C$_2$H$_3$CN, $v=0$ & $2.0$ & $129$ &   -- & $\leq1.0(15)$ & $1.00$ &  -- &   -- & $6.0$ & $12.0$ \\ 
 \hline \\[-0.35cm] 
C$_2$H$_5$CN, $v=0$ & $2.0$ & $115$ & $112\pm 10$ & $2.0(16)$ & $1.18$ & $(2.0\pm 0.2)(16)$ & $1.16$ & $8.0$ & $12.0$ \\ 
 \hline \\[-0.35cm] 
NH$_2$CN, $v=0$ & $2.0$ & $129$ &   -- & $\leq5.0(14)$ & $1.01$ &  -- &   -- & $6.0$ & $12.0$ \\ 
 \hline \\[-0.35cm] 
HNCO, $v=0$ & $2.0$ & $129$ &   -- & $\leq1.0(15)$ & $1.00$ &  -- &   -- & $6.0$ & $12.0$ \\ 
 \hline \\[-0.35cm] 
NH$_2$CHO, $v=0$ & $2.0$ & $115$ & $117\pm 60$ & $1.5(15)$ & $1.03$ & $(1.6\pm 1.2)(15)$ & $1.03$ & $6.0$ & $11.0$ \\ 
 \hline \\[-0.35cm] 
CH$_3$NCO, $v=0$ & $2.0$ & $129$ &   -- & $\leq2.0(15)$ & $1.00$ &  -- &   -- & $6.0$ & $12.0$ \\ 
 \hline \\[-0.35cm] 
CH$_3$OH, $v=0$ & $2.0$ & $150$ & $148\pm 13$ & $9.5(16)$ & $1.00$ & $(1.0\pm 0.2)(17)$ & $1.00$ & $5.5$ & $11.5$ \\ 
 \hline \\[-0.35cm] 
C$_2$H$_5$OH, $v=0$ & $2.0$ & $129$ &   -- & $\leq9.7(15)$ & $1.21$ &  -- &   -- & $6.0$ & $12.0$ \\ 
 \hline \\[-0.35cm] 
CH$_3$OCH$_3$, $v=0$ & $2.0$ & $129$ &   -- & $\leq1.0(16)$ & $1.00$ &  -- &   -- & $6.0$ & $12.0$ \\ 
 \hline \\[-0.35cm] 
CH$_3$OCHO, $v=0$ & $2.0$ & $129$ &   -- & $\leq1.0(16)$ & $1.12$ &  -- &   -- & $6.0$ & $12.0$ \\ 
 \hline \\[-0.35cm] 
CH$_3$CHO, $v=0$ & $2.0$ & $129$ &   -- & $\leq1.5(15)$ & $1.00$ &  -- &   -- & $6.0$ & $12.0$ \\ 
 \hline \\[-0.35cm] 
CH$_3$SH, $v=0$ & $2.0$ & $129$ &   -- & $\leq8.5(15)$ & $1.00$ &  -- &   -- & $6.0$ & $12.0$ \\ 
 \hline \\[-0.35cm] 
OCS, $v=0$ & $2.0$ & $160$ & $\leq165$ & $5.1(16)$ & $1.02$ & $(5.2\pm 0.6)(16)$ & $1.02$ & $8.0$ & $12.0$ \\ 
 \hline\hline
\end{tabular}
\label{tab:n1nw3of}
\end{table*}

%% file: table_lvine.tex
\begin{table*}[]
    \caption{Molecular transitions used to derive the LVINE maps shown in Figs.\,\ref{fig:lvine_coms} and \ref{fig:COMof}--\ref{fig:COMof3}.}
    \centering
    \begin{tabular}{lcrrrrrr}
       \hline\hline\\[-0.2cm]
        Molecule, vib. state & Transition & $\nu$\tablefootmark{a} & $E_{\rm ul}$\tablefootmark{b} & $A_{\rm ul}$\tablefootmark{c} & $g_{\rm ul}$\tablefootmark{d} & $\varv_r$\tablefootmark{e} & $\varv_b$\tablefootmark{f} \\
         & & (MHz) & (K) & ($10^{-5}$\,s$^{-1}$) & & (\kms) & (\kms) \\[0.1cm]\hline\\[-0.3cm]
        HC\3N, $v_7=1$ & $J=11-10, I=1e$ & 100322.411\,(0.002) & 349.7 & 7.7 & 23 & \multirow{2}{*}{100} & \multirow{2}{*}{30} \\
                       & $J=12-11, I=1f$ & 109598.818\,(0.020) & 355.0 & 0.1 & 25  \\
        HC\5N, $v=0$ & $J=36-35$ & 95850.335\,(0.000) & 85.1 & 9.5 & 219 & \multirow{2}{*}{90} & \multirow{2}{*}{40} \\
              & $J=42-41$ & 111823.024\,(0.000) & 115.4 & 15.1 & 255  \\
        C\2H\3CN, $v=0$ & $11(1,11)-10(1,10)$ & 101637.231\,(0.000) & 31.5 & 8.4 & 69 & \multirow{2}{*}{95} & \multirow{2}{*}{35} \\ 
                 & $10(2,8)-9(2,7)$ & 95325.476\,(0.005) & 33.8 & 6.7 & 63  \\
        C\2H\5CN, $v=0$ & $11(0,11)-10(0,10)$ & 96919.762\,(0.050) & 28.1 & 7.5 & 23 & \multirow{2}{*}{100} & \multirow{2}{*}{30} \\   
                 & $11(2,9)-10(2,8)$ & 99681.461\,(0.050) & 33.0 & 7.9 & 23  \\
        HNCO, $v=0$ & $5(1,5)-4(1,4)$ & 109495.996\,(0.006) & 59.0 & 1.7 & 11 & \multirow{2}{*}{90} & \multirow{2}{*}{35} \\
              & $5(1,4)-4(1,3)$ & 110298.089\,(0.005) & 59.2 & 1.7 & 11  \\
        NH\2CHO, $v=0$ & $5(1,5)-4(1,4)$ & 102064.267\,(0.000) & 17.7 & 7.1 & 11 & \multirow{2}{*}{85} & \multirow{2}{*}{40} \\ 
                & $5(1,4)-4(1,3)$ & 109753.503\,(0.000) & 18.8 & 8.8 & 11  \\ 
        CH\3OH, $v=0$ & $7(2,5)-8(1,8)\,A$ & 111289.453\,(0.013) & 102.7 & 0.3 & 60 & \multirow{2}{*}{90} & \multirow{2}{*}{40} \\ 
               & $8(0,8)-7(1,7)\,A$ & 95169.391\,(0.011) & 83.5 & 0.4 & 68  \\
        CH\3CHO, $v=0$ & $5(2,3)-4(2,2)\,E$ & 96475.524\,(0.003) & 23.0 & 2.5 & 22  & \multirow{2}{*}{85} & \multirow{2}{*}{45} \\
                & $5(1,4)-4(1,3)\,A$ & 98900.944\,(0.003) & 16.5 & 3.1 & 22  \\
        CH\3SH, $v=0$ & $4(0,4)-3(0,3)\,A+$ & 101139.150\,(0.001) & 12.1 & 0.9 & 9 & \multirow{3}{*}{85} & \multirow{3}{*}{45} \\  
                     & $4(0,4)-3(0,3)\,E$ & \multirow{2}{*}{101284.366\,(0.001)} & 13.6 & 0.9 & 9  \\
                     & $4(1,3)-3(1,2)\,E$ &  & 18.3 & 0.8 & 9  \\
        OCS, $v=0$ & $J=9-8$ & 109463.063\,(0.005) & 26.3 & 0.4 & 19 & \multirow{2}{*}{105} & \multirow{2}{*}{30} \\ 
            & $J=8-7$ & 97301.208\,(0.000) & 21.0 & 0.3 & 17  \\ 
        SO\2, $v=0$ & $12(4,8)-13(3,11)$ & 107843.470\,(0.002) & 111.0 & 0.3 & 25 & \multirow{2}{*}{105} & \multirow{2}{*}{25} \\
             & $7(3,5)-8(2,6)$ & 97702.333\,(0.002) & 47.8 & 0.2 & 15  \\
        SO, $v=0$ & $2(3)-1(2)$ & 109252.220\,(0.100) & 21.1 & 1.1 & 5 & 110 & 15 \\
        
        \hline\hline
    \end{tabular}
    \tablefoot{\tablefoottext{a}{Rest frequency with uncertainty in parentheses.} \tablefoottext{b}{Upper-level energy.} \tablefoottext{c}{Einstein A coefficient.} \tablefoottext{d}{Upper-level degeneracy.}\tablefoottext{e}{Outer integration limit for the red-shifted emission.}\tablefoottext{f}{Outer integration limit for the blue-shifted emission.}} 
    \label{tab:trans}
\end{table*}